\documentclass[12pt]{article}
\usepackage{graphics,cite,amssymb,epsfig,float}
\usepackage[usenames,dvips]{color}
\usepackage{amsmath}

\usepackage{titlesec,titletoc}

\makeatletter
\renewcommand{\theequation}{\thesection.\arabic{equation}}
\@addtoreset{equation}{section}
\makeatother

\def\lsim{\;\raise0.3ex\hbox{$<$\kern-0.75em\raise-1.1ex\hbox{$\sim$}}\;}
\def\gsim{\;\raise0.3ex\hbox{$>$\kern-0.75em\raise-1.1ex\hbox{$\sim$}}\;}
\def\ben{\begin{enumerate}}  \def\een{\end{enumerate}}
\def\bit{\begin{itemize}}    \def\eit{\end{itemize}}
\def\beq{\begin{equation}}   \def\eeq{\end{equation}}
\def\ba{\begin{array}}       \def\ea{\end{array}}
\def\bea{\begin{eqnarray}}   \def\eea{\end{eqnarray}}
\def\nn{\nonumber}
\def\noi{\noindent}
\def\nl{\newline}
\def\k{\kappa}
\def\l{\lambda}
\def\b{\beta}
\def\t{\theta}
\def\D0{D$0$\hskip -6pt/ }

\begin{document}

\begin{titlepage}
\begin{flushright}
LPT Orsay 09-76 \\ CFTP/09-032\\ LPTA/09-066\\
\end{flushright}


\begin{center}
\vspace{1cm}
{\Large\bf The Next-to-Minimal Supersymmetric Standard Model} \\
\vspace{2cm}

{\bf{Ulrich Ellwanger$^a$, Cyril Hugonie$^b$ and  Ana M. Teixeira$^{a,c}$}}
\footnote{Address after October 2009: LPC Clermont-Ferrand, UMR 6533,
CNRS/IN2P3, Universit\'e Blaise Pascal, 63177 Aubi\`ere, France}
\vspace{2cm}\\
\it  $^a$ LPT, UMR 8627, CNRS, Universit\'e de Paris--Sud,
91405 Orsay, France \\
\it $^b$ LPTA, UMR 5207, CNRS/IN2P3,\\
Universit\'e de Montpellier II, 34095 Montpellier, France\\
\it $^c$ CFTP, Departamento de F\'isica, Instituto Superior T\'ecnico,
1049-001 Lisboa, Portugal\\

\end{center}

\begin{abstract}
We review the theoretical and phenomenological aspects of the
Next-to-Minimal Supersymmetric Standard Model: the Higgs sector
including radiative corrections and the 2-loop $\b$-functions for
all parameters of the general NMSSM; the
tadpole and domain wall problems, baryogenesis; NMSSM phenomenology at
colliders, $B$ physics and dark matter; specific scenarios as the
constrained NMSSM, Gauge Mediated Supersymmetry Breaking,
$U(1)'$-extensions, CP and R-parity violation.
\end{abstract}

\end{titlepage}

\tableofcontents

\section{Introduction}

Supersymmetric extensions of the Standard Model (SM) are motivated by a
solution of the hierarchy problem
\cite{Witten:1981nf,Dimopoulos:1981zb,Witten:1981kv,Kaul:1981hi,
Sakai:1981gr},
an automatic unification of the running gauge couplings at a Grand
Unified (GUT) scale $M_\mathrm{GUT}$
\cite{Ellis:1990wk,Giunti:1991ta,Amaldi:1991cn,Langacker:1991an}, and
the possibility to explain the dark matter relic density in terms of a
stable neutral particle \cite{Pagels:1981ke,Goldberg:1983nd}. 

It is well known that a supersymmetric extension of the Higgs sector of
the SM \cite{Fayet:1974pd,Fayet:1976cr} requires the introduction of two
Higgs $SU(2)$-doublets $H_u$ and $H_d$, where vacuum expectation values
(vevs) of $H_u$ and $H_d$ generate masses for up-type quarks and
down-type quarks and charged leptons, respectively. The model with this
minimal field content in the Higgs sector is denoted as the Minimal
Supersymmetric Standard Model (MSSM) (for reviews see, e.\,g.,
\cite{Nilles:1983ge,Martin:1997ns,Chung:2003fi}). The Lagrangian of the
MSSM must contain a supersymmetric (SUSY) mass term $\mu$ for $H_u$
and $H_d$, which has to be of the order of the SUSY breaking scale
$M_\mathrm{SUSY}$ for phenomenological reasons (see below). This spoils
a potentially attractive property of supersymmetric extensions of the
SM: the electroweak scale generated by the Higgs vevs could depend only
on $M_\mathrm{SUSY}$, which would be the only scale asking for an
explanation to why it is far below $M_\mathrm{GUT}$ or the Planck scale
$M_\mathrm{Planck}$. The question how a supersymmetric mass parameter
$\mu$ can assume a value of the order of $M_\mathrm{SUSY}$ is denoted as
the ``$\mu$-problem''~\cite{Kim:1983dt} of the MSSM.

A simple and elegant way to solve this problem consists in generating an
effective (supersymmetric) mass term $\mu$ in a way similar to the
generation of quark and lepton masses in the SM: the mass term $\mu$ is
replaced by a Yukawa coupling of $H_u$ and $H_d$ to a scalar field, and
the  vev of the scalar field -- induced by the soft SUSY breaking terms
-- is of the desired order. Since the $\mu$~parameter carries no
$SU(3)\times SU(2)\times U(1)_Y$ quantum numbers, the field to be
introduced has to be a singlet $S$ (the complex scalar component of a
chiral superfield $\widehat{S}$), and the
resulting model is the Next-to-Minimal Supersymmetric Standard Model
(NMSSM), sometimes also denoted as the (M+1)SSM.

In fact, already the first attempts to construct supersymmetric
extensions of the SM employed such a singlet field 
\cite{Fayet:1974pd,Fayet:1976cr,Fayet:1977yc}. A singlet was also
present in most of the first globally supersymmetric GUT models
\cite{Sakai:1981gr,Ibanez:1982fr,Nanopoulos:1982wk,Ellis:1982fc,
Ellis:1982xz}. Then one realised that spontaneous supersymmetry breaking
in the framework of supergravity (SUGRA) leads in a simple way to the
desired soft SUSY breaking terms in the Lagrangian; see
\cite{Nilles:1983ge} for an early review. Within SUGRA, a $\mu$~term of
the order of $M_\mathrm{SUSY}$ can actually be generated if one assumes
the presence of a particular Higgs-dependent structure in the K\"ahler
potential \cite{Giudice:1988yz}. Still, the first locally
supersymmetric extensions of the SM
\cite{Barbieri:1982eh,Nilles:1982dy,Frere:1983ag} as well as most GUT
models within SUGRA 
\cite{Chamseddine:1982jx,Nath:1982zq,Ferrara:1982ke,Nath:1983aw,
Arnowitt:1983ah,Veselov:1985gd,Dragon:1985dq,Ellwanger:1985hb}
used a singlet field in the Higgs sector leading to variants of the
NMSSM at the weak or SUSY breaking scale $\lsim 1$~TeV. (See also SUGRA
models motivated by string theory
\cite{Cohen:1985kf,Derendinger:1985cv,Binetruy:1985xm,Ellis:1985yc,
Ibanez:1986si,Costa:1986jb,Antoniadis:1987dx,Campbell:1987gz,
Antoniadis:1987tv,Ellis:1988tx}.)

Expanding around the vacuum with non-vanishing vevs of the neutral
CP-even components of $H_u$, $H_d$ and $S$, one finds that the scalar
components of $\widehat{S}$ mix with the neutral scalar components of
$\widehat{H}_u$ and $\widehat{H}_d$ leading, in the absence of complex
parameters (corresponding to
the absence of explicit CP violation), to three CP-even and two CP-odd
neutral scalars (see \cite{Ellis:1988er,Drees:1988fc,Maniatis:2009re}
for some reviews). Likewise, the
fermionic superpartner of $\widehat{S}$ mixes with the neutral fermionic
superpartners of $\widehat{H}_u$, $\widehat{H}_d$ (and the neutral
electroweak gauginos)
leading to five neutralinos. As a consequence, both the Higgs and
the neutralino sectors of the NMSSM can get considerably modified
compared to the MSSM.

In the Higgs sector, important alterations with respect to the MSSM are
a possibly larger mass of the Higgs scalar with SM-like couplings to
gauge bosons, and additional possibly light states with reduced
couplings to gauge bosons. Notably a light CP-odd scalar with vanishing
couplings to two gauge bosons like all CP-odd scalars (but with possibly
even enhanced couplings to quarks and leptons) can appear in the Higgs
spectrum, allowing for new Higgs-to-Higgs decays. Under these
circumstances, the detection of Higgs bosons at colliders can become
considerably more complicated. At present it is not even guaranteed that
a single Higgs scalar can be observed at the LHC within the NMSSM, see
Section~\ref{sec:5}. In addition, a light CP-odd scalar can affect
``low energy'' observables in $B$~physics,
$\Upsilon$~physics and the anomalous magnetic moment of the muon.

The modifications within the neutralino sector are particularly relevant
if the additional singlet-like neutralino is the lightest one and,
simultaneously, the lightest supersymmetric particle (LSP). This would
have an important impact on all decay chains of supersymmetric particles
(sparticles), and hence on their signatures at colliders. For instance,
the next-to-lightest supersymmetric particle (NLSP) can have a long life
time leading to displaced vertices. Also, the LSP relic density has to
be reconsidered in this case.

Hence, apart from the theoretical motivations for the NMSSM, its
phenomenological consequences must be worked out in order not to miss
(or misinterpret) both Higgs and sparticles signals -- or the absence
thereof -- at past, present and future colliders.

In the present paper we review theoretical and phenomenological aspects
of the NMSSM: Higgs masses and couplings including 
radiative corrections, issues related to the NMSSM Higgs potential
as the nature of the electroweak phase transition, Higgs and sparticle
spectra within specific assumptions on the origin of supersymmetry
breaking as minimal SUGRA and gauge mediated supersymmetry breaking
(GMSB), NMSSM specific Higgs and sparticle signatures at colliders,
possible impact on low energy observables, and
the dark matter relic density and its detection. Possible variants of
the simplest NMSSM as explicit CP~violation in the Higgs sector,
R-parity violation (in connection with neutrino masses) and extra
$U(1)'$~gauge symmetries are sketched briefly. Relevant formulae like
Feynman rules, renormalisation group (RG) equations and details of the
radiative corrections to the Higgs masses are given in the Appendices.

Clearly, we cannot present all details of all results that have been
obtained within the NMSSM up to now. However, in all Sections we attempt
to reference to the complete available literature, where the various
subjects are discussed.

Let us conclude the Introduction by recalling the arguments for a
$\mu$~parameter of the order of $M_\mathrm{SUSY}$, whose necessity
constitutes the main motivation for the NMSSM:
both complex Higgs scalars $H_u$ and $H_d$ of the MSSM have to be
components of chiral superfields which contain, in addition, fermionic
$SU(2)$-doublets $\psi_u$ and $\psi_d$. The Lagrangian of the MSSM can
contain supersymmetric mass terms for these fields, i.e.
identical positive masses squared $\mu^2$ for $|H_u|^2$ and $|H_d|^2$,
and a Dirac mass $\mu$ for $\psi_u$ and~$\psi_d$. In the presence of a
SUSY mass term $\sim \mu$ in the Lagrangian, a soft SUSY breaking mass
term $B\mu\,H_u H_d$ can also appear, where the soft SUSY breaking
parameter $B$ has the dimension of a mass.

For various reasons the mass parameter $\mu$ \emph{cannot} vanish.
First, a Dirac mass $\mu$ for $\psi_u$ and~$\psi_d$ is required for
phenomenological reasons: both fermionic $SU(2)$-doublets $\psi_u$ and
$\psi_d$ contain electrically charged components. Together with with the
fermionic superpartners of the $W^\pm$~bosons, they constitute the
so-called chargino sector (two charged Dirac fermions) of SUSY
extensions of the SM. Due to the fruitless searches for a chargino
at LEP, the lighter chargino has to have a mass above $\sim 103$~GeV
\cite{LEPSUSYWG}. Analysing the chargino mass matrix, one finds that
this lower limit implies that the Dirac mass $\mu$ for $\psi_u$
and~$\psi_d$ -- for arbitrary values of the other parameters -- has to
satisfy the constraint $|\mu| \gsim 100$~GeV.

Second, an analysis of the Higgs potential shows that a non-vanishing
term $B\mu\,H_u H_d$ is a necessary condition for that \emph{both}
neutral components of $H_u$ and $H_d$ are non-vanishing at the minimum.
This, in turn, is required in order to generate masses for up-type
quarks, down-type quarks and leptons by the Higgs mecanism. Moreover,
the numerical value of the product $B\mu$ should be roughly of the order
of the electroweak scale ($M_Z^2$).

Third, $\mu=0$ would generate a Peccei-Quinn symmetry in the Higgs
sector, and hence an unacceptable massless axion \cite{Kim:1983dt}.

However, $|\mu|$ must not be too large: the Higgs potential must be
unstable at its origin $H_u = H_d = 0$ in order to generate the
electroweak symmetry breaking. Whereas the soft SUSY breaking mass
terms for $H_u$ and $H_d$ of the order of the SUSY breaking scale
$M_\mathrm{SUSY}$ can generate such a desired instability, the
$\mu$-induced masses squared for $H_u$ and $H_d$ are always positive,
and must \emph{not} dominate the negative soft SUSY breaking mass terms.
Consequently the $\mu$~parameter must obey $|\mu| \lsim
M_\mathrm{SUSY}$. Hence, both ``natural'' values $\mu = 0$ and very
large $\mu$ ($\sim M_\mathrm{GUT}$ or $\sim M_\mathrm{Planck}$) are
ruled
out, and the need for an explanation of $\mu \approx M_\mathrm{SUSY}$ is
the $\mu$-problem.

Within the NMSSM, where $\mu$ is generated by the vev $\left< S\right>$
of a singlet~$S$, $\left< S\right>$ has to be of the order of
$M_\mathrm{SUSY}$; this is easy to obtain with the help of soft SUSY
breaking negative masses squared (or trilinear couplings) of the order
of $M_\mathrm{SUSY}$ for $S$. Then, $M_\mathrm{SUSY}$ is the only scale
in the theory. In this sense, the NMSSM is the simplest supersymmetric
extension of the SM in which the weak scale is generated by the
supersymmetry breaking scale $M_\mathrm{SUSY}$ only.

\section{Lagrangian of the general NMSSM}
\label{sec:2}

In this Section, we present the scalar potential (at tree level) and the
mass matrices in the general NMSSM. Their diagonalization and
conventions for the mixing matrices are deferred to the Appendix A,
where we also list the Feynman rules for the physical eigenstates.

\subsection{Tree level potential and mass matrices}
\label{sec:2.1}

As in any softly broken supersymmetric theory, the Lagrangian of the
NMSSM is specified by the supersymmetric gauge interactions, the
superpotential and the soft supersymmetry breaking gaugino masses,
scalar masses and trilinear couplings.

To begin with, we consider the general NMSSM defined as the MSSM with an
additional gauge singlet chiral superfield $\widehat{S}$, including the
most general renormalisable couplings in the superpotential and the
corresponding soft SUSY breaking terms in ${\cal L}_\mathrm{soft}$.
(Here we limit ourselves, however, to the R-parity and CP conserving
case. Chiral superfields are denoted by hatted capital
letters; unhatted capital letters indicate their complex scalar
components.)
In the general NMSSM, the terms in the superpotential $W_\mathrm{Higgs}$
depending exclusively on the Higgs superfields $\widehat{H}_u$,
$\widehat{H}_d$ and $\widehat{S}$ are (here we follow the SLHA2
conventions in \cite{Allanach:2008qq} where, however, $\widehat{H}_u$
is denoted as $\widehat{H}_2$, and $\widehat{H}_d$ as $\widehat{H}_1$):

\beq\label{2.1e}
W_\mathrm{Higgs} = (\mu + \lambda \widehat{S})\,\widehat{H}_u \cdot 
\widehat{H}_d + \xi_F \widehat{S} + \frac{1}{2} \mu' \widehat{S}^2 +
\frac{\kappa}{3} \widehat{S}^3
\eeq
where the terms $\sim \l$, $\k$ are dimensionless Yukawa couplings, the
terms $\sim \mu$, $\mu'$ are supersymmetric mass terms, and $\xi_F$
of dimension $mass^2$ parametrizes a (supersymmetric) tadpole term.
To (\ref{2.1e}) we have to add the Yukawa couplings of the quark and
lepton superfields:

\beq\label{2.2e}
W_\mathrm{Yukawa} = h_u\, \widehat{Q} \cdot \widehat{H}_u\;
\widehat{U}^c_R + h_d\, \widehat{H}_d \cdot \widehat{Q}\;
\widehat{D}^c_R + h_e\, \widehat{H}_d \cdot \widehat{L}\;
\widehat{E}_R^c
\eeq
where the Yukawa couplings $h_u$, $h_d$, $h_e$ and the superfields
$\widehat{Q}$, $\widehat{U}^c_R$, $\widehat{D}^c_R$, $\widehat{L}$ and
$\widehat{E}_R^c$ should be understood as matrices and vectors in
family space, respectively.

In (\ref{2.1e}) and (\ref{2.2e}), the $SU(2)$~doublets are
\beq\label{2.3e}
\widehat{Q} = \left(\ba{c} \widehat{U}_L \\ \widehat{D}_L
\ea\right) , \
\widehat{L} = \left(\ba{c} \widehat{\nu}_{L} \\ \widehat{E}_L
\ea\right) , \
\widehat{H}_u = \left(\ba{c} \widehat{H}_u^+ \\ \widehat{H}_u^0
\ea\right) , \
\widehat{H}_d = \left(\ba{c} \widehat{H}_d^0 \\ \widehat{H}_d^-
\ea\right) ,
\eeq
and the products of two $SU(2)$ doublets are defined as, e.\,g.,
\beq\label{2.4e}
\widehat{H}_u \cdot \widehat{H}_d = \widehat{H}_u^+ \widehat{H}_d^- 
- \widehat{H}_u^0 \widehat{H}_d^0\ .
\eeq

The corresponding soft SUSY breaking masses and couplings are, again in
the SLHA2 conventions \cite{Allanach:2008qq},
\bea\label{2.5e}
-{\cal L}_\mathrm{soft} &=&
m_{H_u}^2 | H_u |^2 + m_{H_d}^2 | H_d |^2 
+ m_{S}^2 | S |^2+m_Q^2|Q^2| + m_U^2|U_R^2| \nn \\
&&+m_D^2|D_R^2| +m_L^2|L^2| +m_E^2|E_R^2|
\nn \\
&&+ (h_u A_u\; Q \cdot H_u\; U_R^c - h_d A_d\; Q \cdot H_d\; D_R^c 
- h_{e} A_{e}\; L \cdot H_d\; E_R^c\nn \\ &&
+\lambda A_\lambda\, H_u \cdot H_d\; S + \frac{1}{3} \kappa A_\kappa\,
S^3 + m_3^2\, H_u \cdot H_d + \frac{1}{2}m_{S}'^2\, S^2 + \xi_S\, S 
+ \mathrm{h.c.}) \; .
\eea
(Sometimes the definitions $m_3^2 = B\mu,\ m_{S}'^2 = B'\mu'$
are used.)

Clearly, the dimensionful supersymmetric parameters $\mu$, $\mu'$ and
$\xi_F$ in the superpotential~(\ref{2.1e}) (and the associated soft SUSY
breaking
parameters $m_3^2$, $m_{S}'^2$ and $\xi_S$ in (\ref{2.5e})) have
to be of the order of the weak or SUSY breaking scale, in contradiction
to one of the theoretical motivations for the NMSSM mentioned in the
Introduction. Although some of these terms are non-vanishing in various
scenarios, one considers mostly the simpler NMSSM with a scale
invariant superpotential where $\mu = \mu' = \xi_F = 0$,
\beq\label{2.6e}
W_\mathrm{sc.inv.} = \lambda \widehat{S}\,\widehat{H}_u \cdot 
\widehat{H}_d + \frac{\kappa}{3} \widehat{S}^3
\eeq
and vanishing parameters $m_3^2$, $m_{S}'^2$ and $\xi_S$ in
(\ref{2.5e}). Then, a vev $s$ of $\widehat{S}$ of the order of
the weak or SUSY breaking scale generates an effective $\mu$-term with
\beq\label{2.7e}
\mu_\mathrm{eff} = \l s\; ,
\eeq
which solves the $\mu$-problem of the MSSM.

As any supersymmetric theory with a scale invariant (cubic)
superpotential, the complete Lagrangian -- including the soft SUSY
breaking terms -- specified by (\ref{2.6e}) possesses an accidental
$\mathbb{Z}_3$-symmetry corresponding to a multiplication of all components of
all chiral superfields by a phase $e^{2\pi i/3}$. In the following
we denote the version with the scale invariant
superpotential (\ref{2.6e}) as the ``$\mathbb{Z}_3$-invariant NMSSM''. Any of the
dimensionful terms in  the general superpotential (\ref{2.1e}) breaks
the $\mathbb{Z}_3$-symmetry explicitly. Subsequently, the version corresponding
to the general superpotential (\ref{2.1e}) will be denoted as the
``general NMSSM''. In the literature, ``NMSSM'' stands mostly for the
$\mathbb{Z}_3$-invariant NMSSM. In the following we will retain this convention.

The Higgs sector of the $\mathbb{Z}_3$-invariant NMSSM is specified by the seven
parameters \beq\label{2.8e}
\l,\ \k,\ m_{H_u}^2,\ m_{H_d}^2,\ m_{S}^2,\
A_\lambda\ \mathrm{and}\ A_\kappa\; .
\eeq
Expressions for the Higgs mass matrices in the $\mathbb{Z}_3$-invariant NMSSM can
be found, e.\,g., in \cite{Ellis:1988er,Drees:1988fc,Franke:1995tc,
Miller:2003ay,Ellwanger:2004xm}; in the following we discuss, for
completeness, the general NMSSM from which the $\mathbb{Z}_3$-invariant NMSSM can
always be obtained by setting $m_3^2 = m_{S}'^2 = \xi_S = \mu = \mu' =
\xi_F = 0$.

From the SUSY gauge interactions, the $F$- and the soft SUSY breaking
terms one obtains the Higgs potential:
\bea
V_\mathrm{Higgs} & = & \left|\lambda \left(H_u^+ H_d^- - H_u^0
H_d^0\right) + \kappa S^2 + \mu' S +\xi_F\right|^2 \nn \\
&&+\left(m_{H_u}^2 + \left|\mu + \lambda S\right|^2\right) 
\left(\left|H_u^0\right|^2 + \left|H_u^+\right|^2\right) 
+\left(m_{H_d}^2 + \left|\mu + \lambda S\right|^2\right) 
\left(\left|H_d^0\right|^2 + \left|H_d^-\right|^2\right) \nn \\
&&+\frac{g_1^2+g_2^2}{8}\left(\left|H_u^0\right|^2 +
\left|H_u^+\right|^2 - \left|H_d^0\right|^2 -
\left|H_d^-\right|^2\right)^2
+\frac{g_2^2}{2}\left|H_u^+ H_d^{0*} + H_u^0 H_d^{-*}\right|^2\nn \\
&&+m_{S}^2 |S|^2
+\big( \lambda A_\lambda \left(H_u^+ H_d^- - H_u^0 H_d^0\right) S + 
\frac{1}{3} \kappa A_\kappa\, S^3 + m_3^2 \left(H_u^+ H_d^- - H_u^0
H_d^0\right) \nn \\
&& +\frac{1}{2} m_{S}'^2\, S^2 + \xi_S\, S + \mathrm{h.c.}\big)
\label{2.9e}
\eea
where $g_1$ and $g_2$ denote the $U(1)_Y$ and $SU(2)$ gauge couplings,
respectively.

The neutral physical Higgs fields (with index~R for the CP-even, index~I
for the CP-odd states) are obtained by expanding the full
scalar potential (\ref{2.9e}) around the real neutral vevs $v_u$, $v_d$
and $s$ as
\beq\label{2.10e}
H_u^0 = v_u + \frac{H_{uR} + iH_{uI}}{\sqrt{2}} , \quad
H_d^0 = v_d + \frac{H_{dR} + iH_{dI}}{\sqrt{2}} , \quad
S = s + \frac{S_R + iS_I}{\sqrt{2}}\; ;
\eeq
where the vevs have to be obtained from the minima of
\bea
V_\mathrm{Higgs} & = & \left(-\lambda v_u v_d + \kappa s^2 + \mu' s 
+\xi_F\right)^2 +\frac{g_1^2+g_2^2}{8}\left(v_u^2 - v_d^2\right)^2 
\nn \\
&&+\left(m_{H_u}^2 + \left(\mu + \lambda s\right)^2\right) v_u^2 
+\left(m_{H_d}^2 + \left(\mu + \lambda s\right)^2\right) v_d^2
\nn\\
&&+m_{S}^2\, s^2 -2 \lambda A_\lambda\, v_u v_d s +  \frac{2}{3}
\kappa A_\kappa\, s^3 - 2m_3^2\, v_u v_d + m_{S}'^2\, s^2 +
2\xi_S\, s \;,
\label{2.11e}
\eea

The signs of some parameters in a Lagrangian have no physical meaning,
since they can be changed by field redefinitions $\phi \to -\phi$.
Analysing all possible field redefinitions in the Lagrangian above, one
finds that one can choose positive Yukawa couplings $\l$, $h_t$, $h_b$,
$h_\tau$ (the latter corresponding to the (3,3) components in family
space of $h_u$, $h_d$ and $h_e$ in (\ref{2.2e})) and positive vevs $v_u$
and $v_d$, whereas $\kappa$, $s$ and all dimensionful
parameters can have both signs. In addition, in the general NMSSM one of
the dimensionful $S$-dependent parameters can be removed by a constant
shift of the real component of $S$. In particular, one can put the
MSSM-like $\mu$-term in the superpotential~(\ref{2.1e}) to zero by a
redefinition $s \to s-\mu/\lambda$ (assuming $\lambda \neq 0$) and
corresponding redefinitions of the other dimensionful parameters; in the
following we will assume, for simplicity, that this convention is used
(which does \emph{not} imply, in general, that $m_3^2=0$\,!).

We define, as usual,
\beq\label{2.12e}
\tan\beta = \frac{v_u}{v_d}
\eeq
(positive by assumption), and we have
\beq\label{2.13e}
M_Z^2 = g^2 v^2\quad \mathrm{where}\
g^2 \equiv \frac{g_1^2+g_2^2}{2}\; ,\quad 
v^2 = v_u^2 + v_d^2 \simeq (174\ \text{GeV})^2\; .
\eeq
Furthermore it is convenient to define, together with $\mu_\mathrm{eff}$
as in (\ref{2.7e}),
\beq\label{2.14e}
B_\mathrm{eff} = A_\lambda+ \kappa s,
\quad \widehat{m}_3^2 = m_3^2 + \lambda(\mu' s + \xi_F)
\eeq
where we have used the convention $\mu=0$. $B_\mathrm{eff}$ plays the
r\^ole of the MSSM-like $B$-parameter, and $\widehat{m}_3^2$ vanishes
in the $\mathbb{Z}_3$-invariant NMSSM.

It is possible to use the three minimisation equations of the
potential (\ref{2.11e}) with respect to $v_u$, $v_d$ and $s$ in order to
replace the three parameters $m_{H_u}^2$, $m_{H_d}^2$ and $m_{S}^2$ by
$v_u$, $v_d$ and $s$. The minimisation equations are given by
\bea
&v_u \Big(m_{H_u}^2+\mu_\mathrm{eff}^2+\l^2\,v_d^2
+\frac{g_1^2+g_2^2}{4}(v_u^2-v_d^2)\Big)
-v_d \left(\mu_\mathrm{eff} B_\mathrm{eff} +\widehat{m}_3^2\right)= 0\;
,\nn \\
&v_d \Big(m_{H_d}^2+\mu_\mathrm{eff}^2+\l^2\,v_u^2
+\frac{g_1^2+g_2^2}{4}(v_d^2-v_u^2)\Big) 
-v_u\left(\mu_\mathrm{eff} B_\mathrm{eff} +\widehat{m}_3^2\right) = 0\;
,\nn \\
&s\Big(m_{S}^2 +m_{S}'^2 +\mu'^2 +2\k\xi_F 
+\k A_\k s +2\k^2 s^2 +3\k s\mu'
+\l^2(v_u^2+v_d^2)-2\l\k v_u v_d \Big)\nn \\
&+\xi_S+\xi_F \mu' -\l v_u v_d(A_\l+\mu') = 0\; .
\label{2.15e}
\eea
From the first two of these equations one can derive
\beq\label{2.16e}
\frac{v_u v_d}{v^2} \equiv \frac{1}{2}\sin 2\b = \frac{\mu_\mathrm{eff}
B_\mathrm{eff} +\widehat{m}_3^2}
{m_{H_u}^2+m_{H_d}^2+2\mu_\mathrm{eff}^2+\l^2\,v^2} \; ;
\eeq
hence, in order to have both $v_u$ and $v_d$ different from zero
($\tan\b \neq 0,\ \infty$), we need $\mu_\mathrm{eff} B_\mathrm{eff}
+\widehat{m}_3^2 \neq 0$ in the general NMSSM, and $\mu_\mathrm{eff}
B_\mathrm{eff}\neq 0$ in the $\mathbb{Z}_3$-invariant NMSSM.

Given $M_Z$, one can choose as six independent parameters in the Higgs
sector of the $\mathbb{Z}_3$-invariant NMSSM
\beq\label{2.17e}
\lambda,\ \kappa,\ A_\lambda,\ A_\kappa,\ \tan\beta,\ 
\mu_\mathrm{eff},
\eeq
to which one has to add in the general NMSSM (in the convention $\mu
= 0$) the five parameters
\beq\label{2.18e}
m_3^2,\ \mu',\ m_{S}'^2,\ \xi_F\ \mathrm{and}\ \xi_S\; .
\eeq

Let us consider the conditions arising from a phenomenologically
acceptable minimum of the potential: both $v_u$ and $v_d$ must not
vanish, and -- in the absence of a $\mu$-term -- $s$ must be large
enough to generate a sufficiently large effective $\mu$-term
$\mu_\mathrm{eff}=\l s \gsim 100$~GeV (see the Introduction). In the
$\mathbb{Z}_3$-invariant NMSSM, the dominant terms for large $s$ in the potential
(\ref{2.11e}) are
\beq\label{2.19e}
V_\mathrm{Higgs}(s) \sim m_{S}^2\, s^2 + \frac{2}{3}\k A_\k\, s^3
+ \k^2 s^4\; .
\eeq
One easily finds that $A_k^2 \gsim 8\, m_{S}^2$ is a condition
for $s \neq 0$ \cite{Derendinger:1983bz,Ellwanger:1996gw}, and
\beq\label{2.20e}
A_k^2 \gsim 9\, m_{S}^2
\eeq
a condition for an absolute minimum with 
\beq\label{2.21e}
s \simeq \frac{1}{4\k} \left( -A_\k - \sqrt{A_\k^2 - 8 m_{S}^2}
\right)\; .
\eeq

We note that in the case of a scale invariant superpotential
(\ref{2.6e}) and without \emph{any} soft terms for the singlet -- as it
can happen in GMSB models \cite{Dine:1993yw} -- a vev $s \neq 0$ is
still triggered by the $s$-dependent terms in the potential
(\ref{2.11e}) neglected in (\ref{2.19e}). However, the resulting value
of $s$ is too small in order to give $\mu_\mathrm{eff} \gsim
100$~GeV~\cite{Dine:1993yw}.

Depending on the parameters, the Higgs potential of the $\mathbb{Z}_3$-invariant
NMSSM can possess several local minima (see, e.\,g.,
\cite{Maniatis:2006jd}). Notably, one should verify whether vacua where
one of the vevs $v_u$, $v_d$ or $s$ vanishes (typically preferred
during the cosmological evolution) are not deeper. A general analysis
(taking the radiative corrections to the effective potential into
account) is quite involved, but typically one obtains upper bounds on
$\k$ at least of the type $\k^2 < \l^2$~\cite{Ellwanger:1996gw}.

\medskip

The tree level Higgs mass matrices are obtained by expanding the full
scalar potential~(\ref{2.9e}) around the real neutral vevs $v_u$, $v_d$
and $s$ as in (\ref{2.10e}).
Then, the elements of the $3 \times 3$ CP-even mass matrix ${\cal
M}_S^2$ read in the basis $(H_{dR}, H_{uR}, S_R)$ after the elimination
of $m_{H_d}^2$, $m_{H_u}^2$ and $m_{S}^2$ (still in the general NMSSM,
but assuming $\mu = 0$)
\bea
{\cal M}_{S,11}^2 & = & g^2 v_d^2 + (\mu_\mathrm{eff}\, B_\mathrm{eff} +
\widehat{m}_3^2)\,\tan\beta\;, \nn\\
{\cal M}_{S,22}^2 & = & g^2 v_u^2 + (\mu_\mathrm{eff}\, B_\mathrm{eff} +
\widehat{m}_3^2)/\tan\beta\;, \nn\\
{\cal M}_{S,33}^2 & = & \l (A_\l + \mu') \frac{v_u v_d}{s}
+ \k s (A_\k + 4\k s+ 3 \mu') - (\xi_S + \xi_F \mu')/s\;, \nn\\
{\cal M}_{S,12}^2 & = & (2\l^2 - g^2) v_u v_d - 
\mu_\mathrm{eff}\, B_\mathrm{eff} - \widehat{m}_3^2 \;, \nn\\ 
{\cal M}_{S,13}^2 & = & \l (2 \mu_\mathrm{eff}\, v_d -
(B_\mathrm{eff} + \k s + \mu')v_u)\;, \nn\\
{\cal M}_{S,23}^2 & = & \l (2 \mu_\mathrm{eff}\, v_u -
(B_\mathrm{eff} + \k s + \mu')v_d)\;.
\label{2.22e}
\eea

Rotating the upper left $2 \times 2$ submatrix by an angle $\beta$, one
finds that one of its diagonal elements reads
\beq\label{2.23e}
M_Z^2\left(\cos^2 2\beta + \frac{\lambda^2}{g^2} \sin^2 2\beta\right)
\eeq
which constitutes an upper bound on the lightest eigenvalue of ${\cal
M}_S^2$. The additional positive contribution $\sim \l^2\sin^2 2\b$ (as
compared to the MSSM) in the NMSSM is highly welcome in view
of the present lower LEP bound of $\sim 114$~GeV on the mass of a Higgs
scalar with SM-like couplings to gauge bosons \cite{Schael:2006cr}.
However, this additional contribution is relevant only for not too large
$\tan\b$; in fact, the expression inside the parenthesis in
(\ref{2.23e}) is larger than one only for $\l^2 > g^2$, in which case it
is maximal for small $\tan\b$. Moreover, the actual lightest eigenvalue
of ${\cal M}_S^2$ is smaller than the value given in (\ref{2.23e}) in
general, see the discussion in Section~\ref{sec:3.2}.

\medskip

In the general NMSSM, the elements of the $3 \times 3$ CP-odd mass
matrix ${\cal M'}_P^2$ read in the basis
$(H_{dI}, H_{uI}, S_I)$
\bea
{\cal M'}_{P,11}^2 & = & (\mu_\mathrm{eff}\, B_\mathrm{eff} +
\widehat{m}_3^2)\,\tan\beta\;, \nn\\
{\cal M'}_{P,22}^2 & = & (\mu_\mathrm{eff}\, B_\mathrm{eff} +
\widehat{m}_3^2)/\tan\beta\;, \nn\\
{\cal M'}_{P,33}^2 & = & \l (B_\mathrm{eff}+3\k s +\mu')\frac{v_u
v_d}{s} -3\k A_\k s  -2 m_{S}'^2 -\k \mu' s 
-\xi_F\left(4\k + \frac{\mu'}{s}\right) -\frac{\xi_S}{s}\; ,
 \nn\\
{\cal M'}_{P,12}^2 & = & \mu_\mathrm{eff}\, B_\mathrm{eff} +
\widehat{m}_3^2\; , \nn\\
{\cal M'}_{P,13}^2 & = & \l v_u (A_\l - 2\k s - \mu')\;, \nn\\
{\cal M'}_{P,23}^2 & = & \l v_d (A_\l - 2\k s - \mu')\;.
\label{2.24e}
\eea
${\cal M'}_{P}^2$ contains always a massless Goldstone mode ${G}$.
Next we rotate this mass matrix into the basis (${A}, {G},
S_I$), where ${A} = \cos\b\, H_{uI}+ \sin\b\, H_{dI}$:
\beq\label{2.25e}
\left(\ba{c}H_{dI} \\  H_{uI} \\ S_I \ea\right) = 
 \left(\ba{ccc} \sin\b & -\cos\b & 0 \\ 
 \cos\b & \sin\b & 0 \\
 0 & 0 & 1 \ea\right)
\left(\ba{c} {A} \\ {G} \\  S_I \ea\right)\; .
\eeq
Dropping the Goldstone mode, the remaining $2 \times 2$ mass matrix
${\cal M}_{P}^2$ in the basis ($A, S_I$) has the elements
\bea
{\cal M}_{P,11}^2 & = & \frac{2 (\mu_\mathrm{eff}\, B_\mathrm{eff} +
\widehat{m}_3^2)}{\sin 2\b}\; , \nn\\
{\cal M}_{P,22}^2 & = & \l (B_\mathrm{eff}+3\k s +\mu')\frac{v_u
v_d}{s} -3\k A_\k s  -2 m_{S}'^2 -\k \mu' s 
-\xi_F\left(4\k + \frac{\mu'}{s}\right) -\frac{\xi_S}{s}\; , \nn\\
{\cal M}_{P,12}^2 & = &\l (A_\l - 2\k s - \mu')\, v\;.
\label{2.26e}
\eea

In the $\mathbb{Z}_3$-invariant NMSSM, ${\cal M}_{P}^2$ simplifies to
\bea
{\cal M}_{P,11}^2 & = & \frac{2\, \mu_\mathrm{eff}\,
B_\mathrm{eff}}{\sin 2\b}\; , \nn\\
{\cal M}_{P,22}^2 & = & \l (B_\mathrm{eff}+3\k s)\frac{v_u v_d}{s} -3\k
A_\k\, s\; , \nn\\
{\cal M}_{P,12}^2 & = &\l (A_\l - 2\k s)\, v
\label{2.27e}
\eea
with $B_\mathrm{eff}$ as given in (\ref{2.14e}). The matrix element
${\cal M}_{P,11}^2$ corresponds to the mass squared $M_A^2$ of the (only
physical) CP-odd scalar $A$ of the MSSM. 

\medskip

Finally the charged Higgs mass matrix in the basis $(H_u^+, H_d^{-*} =
H^+_d)$ is given by
\beq\label{2.28e}
{\cal M'}_\pm^2 = \left(\mu_\mathrm{eff}\, B_\mathrm{eff} +
\widehat{m}_3^2 + v_u v_d (\frac{g_2^2}{2} - \l^2)\right)
\left(\ba{cc} \cot\b & 1 \\ 1 & \tan\b \ea\right)\; .
\eeq
It contains one massless Goldstone mode, and one eigenstate with mass
\beq\label{2.29e}
{\cal M}_\pm^2 = \frac{2 (\mu_\mathrm{eff}\, B_\mathrm{eff} +
\widehat{m}_3^2)}{\sin 2\b} + v^2 (\frac{g_2^2}{2} - \l^2)\; .
\eeq

Due to the term $\sim \l^2$, the charged Higgs mass in the NMSSM can be
somewhat smaller than in the MSSM (for a given value of $M_A^2 \equiv
{\cal M}_{P,11}^2$). In contrast to the MSSM it is not even guaranteed
within the NMSSM that $U(1)_\mathrm{e m}$ remains unbroken:
considering again the $\mathbb{Z}_3$-invariant NMSSM where $\widehat{m}_3^2 =
0$, the expression for the charged Higgs mass squared becomes negative
for $s = \mu_\mathrm{eff} = 0$, $\l^2 > g_2^2/2$, indicating a possible
minimum in field space where the charged Higgs has a vev. Although
radiative corrections have to be added and the depth of this minumum
has to be compared to the physical one with $s \neq 0$, $\l$ is bounded
from above by the absence of a charged Higgs vev.

The diagonalization of all scalar mass matrices is carried out in
Appendix A (together with the Feynman rules); next we consider the
fermionic sector.

\medskip

First, we have to consider the soft SUSY breaking gaugino mass terms,
which do not differ from the MSSM. Denoting the $U(1)_Y$ gaugino by
$\l_1$, the $SU(2)$ gauginos by $\l_2^i$ ($i=1,2,3$) and the $SU(3)$
gauginos by $\l_3^a$ ($a=1,\dots,8$), the soft SUSY breaking gaugino
mass terms in the Lagrangian read
\beq\label{2.30e}
{\cal L}  =  \frac{1}{2} M_1 \l_1 \l_1 + \frac{1}{2} M_2 \l_2^i \l_2^i
 + \frac{1}{2} M_3 \l_3^a \l_3^a\; .
\eeq

In the neutralino sector, $\l_1$ and $\l_2^3$ mix with the neutral
higgsinos $\psi_d^0, \psi_u^0, \psi_S$ and generate a symmetric $5
\times 5$ mass matrix ${\cal M}_0$. In the basis $\psi^0 = (-i\l_1 ,
-i\l_2^3, \psi_d^0, \psi_u^0, \psi_S)$, the resulting mass terms in the
Lagrangian read
\beq\label{2.31e}
{\cal L} = - \frac{1}{2} (\psi^0)^T {\cal M}_0 (\psi^0) + \mathrm{h.c.}
\eeq
where
\beq\label{2.32e}
{\cal M}_0 =
\left( \ba{ccccc}
M_1 & 0 & -\frac{g_1 v_d}{\sqrt{2}} & \frac{g_1 v_u}{\sqrt{2}} & 0 \\
& M_2 & \frac{g_2 v_d}{\sqrt{2}} & -\frac{g_2 v_u}{\sqrt{2}} & 0 \\
& & 0 & -\mu_\mathrm{eff} & -\l v_u \\
& & & 0 & -\l v_d \\
& & & & 2 \k s + \mu'
\ea \right)
\eeq
and the term $\mu'$ in the (5,5) element appears in the general NMSSM
only.

As in the MSSM, the charged $SU(2)$ gauginos are
$\l^- = \frac{1}{\sqrt{2}}\left(\l_2^1 + i \l_2^2\right)$, and
$\l^+ = \frac{1}{\sqrt{2}}\left(\l_2^1 - i \l_2^2\right)$, which mix
with the charged higgsinos $\psi_u^+$ and $\psi_d^-$. Defining 
\beq\label{2.33e}
\psi^+ = \left(\ba{c} -i\l^+ \\ \psi_u^+ \ea\right)\ , \qquad
\psi^- =  \left(\ba{c} -i\l^- \\ \psi_d^- \ea\right)\ ,
\eeq
the corresponding mass terms in the Lagrangian can be written as
\beq\label{2.34e}
{\cal L} = -\frac{1}{2} (\psi^+ , \psi^-)
\left(\ba{cc} 0 & X^T \\ X & 0 \ea\right)
\left(\ba{c} \psi^+ \\ \psi^- \ea\right) + \mathrm{h.c.} 
\eeq
with
\beq\label{2.35e}
X = \left(\ba{cc} M_2 & g_2 v_u \\ g_2 v_d & \mu_\mathrm{eff} \ea\right)
\ .
\eeq
Again, the diagonalization of the neutralino and chargino mass matrices
will be described in Appendix A.

\medskip

Finally we give the top and bottom squark and $\tau$ slepton
mass-squared matrices. $\tilde{t}_L$, $\tilde{t}_R$,
$\tilde{b}_L$ and $\tilde{b}_R$ denote the scalar
components of the third generation quark superfields
$\widehat{U}_{L_3}$,
$\widehat{U}_{R_3}$, $\widehat{D}_{L_3}$ and $\widehat{D}_{R_3}$;
$\widetilde{\nu}_{\tau_L}$ and $\widetilde{\tau}_L$
the scalar components of the third generation lepton superfields
$\widehat{\nu}_{L_3}$ and $\widehat{E}_{L_3}$, and $\widetilde{\tau}_R$
the scalar component of the third generation lepton superfield
$\widehat{E}_{R_3}$.  (The superfields were
given in (\ref{2.2e}) and (\ref{2.3e}).) $m_T^2 \equiv m_{U_3}^2$,
$m_B^2 \equiv m_{D_3}^2$, $m_{Q_3}^2$, $m_{E_3}^2$ and
$m_{L_3}^2$ are the soft SUSY breaking masses squared for the third
generation (see ${\cal L}_\mathrm{soft}$ in (\ref{2.5e})), assumed to be
diagonal in family space.
\medskip

The top squark mass matrix reads in the basis ($\tilde{t}_R$, 
$\tilde{t}_L$):
\beq\label{2.36e} 
\left(\ba{cc} m_T^2 + h_t^2 v_u^2-(v_u^2-v_d^2)\frac{g_1^2}{3}
& h_t (A_t v_u - \mu_\mathrm{eff} v_d)  \\
 h_t (A_t v_u - \mu_\mathrm{eff} v_d) & 
 m_{Q_3}^2 + h_t^2 v_u^2 +
 (v_u^2-v_d^2)\left( \frac{g_1^2}{12}-\frac{g_2^2}{4}\right)
 \ea \right)\; .
\eeq
 
The bottom squark mass matrix reads in the basis ($\tilde{b}_R$, 
$\tilde{b}_L$):
\beq\label{2.37e}
\left(\ba{cc} m_B^2 + h_b^2 v_d^2 +(v_u^2-v_d^2)\frac{g_1^2}{6}
& h_b  (A_b v_d-\mu_\mathrm{eff} v_u) \\
h_b (A_b v_d-\mu_\mathrm{eff} v_u) & m_{Q_3}^2 + h_b^2 v_d^2 +
 (v_u^2-v_d^2)\left( \frac{g_1^2}{12}+\frac{g_2^2}{4}\right)
\ea \right)\; .
\eeq
 
The tau slepton mass matrix reads in the basis ($\widetilde{\tau}_R$, 
$\widetilde{\tau}_L$):
\beq\label{2.38e}
\left(\ba{cc} m_{E_3}^2 + h_\tau^2 v_d^2 +(v_u^2-v_d^2)\frac{g_1^2}{2}
& h_\tau  (A_\tau v_d-\mu_\mathrm{eff} v_u) \\
h_\tau (A_\tau v_d-\mu_\mathrm{eff} v_u) & m_{L_3}^2 + h_\tau^2 v_d^2 -
 (v_u^2-v_d^2)\left( \frac{g_1^2 - g_2^2}{4}\right)
\ea \right)\; .
\eeq
 
The tau sneutrino ($\widetilde{\nu}_{\tau_L}$) mass squared is:
\beq\label{2.39e}
m_{L_3}^2 -(v_u^2-v_d^2)\left(\frac{g_1^2+g_2^2}{4} 
\right)\; .
\eeq

\medskip

Herewith we conclude the presentation of the tree level Lagrangian;
clearly one has to add radiative corrections to the Lagrangian and all
the resulting mass matrices. For the couplings/Feynman rules we
refer to Appendix A.

\subsection{Limiting cases: the effective MSSM and approximate global
symmetries}
\label{sec:2.2}

\subsubsection{The effective MSSM}
\label{sec:2.2.1}

As it becomes clear from the superpotential (\ref{2.1e}) of the general
NMSSM, all couplings between the components of the singlet superfield
$\widehat{S}$ and the MSSM Higgs superfields $\widehat{H}_u$ and 
$\widehat{H}_d$ vanish for $\l \to 0$. In order to generate a reasonably
large value for $\mu_\mathrm{eff}$ (in the absence of an
MSSM-like $\mu$-term in the superpotential), one should keep $\l s$
$\gsim 100$~GeV in this limit. From (\ref{2.21e}) one finds that
the vev $s$ scales as $1/\k$, hence a reasonable decoupling limit is
\beq\label{2.40e}
\l \sim \k \to 0,\quad s \sim 1/\k \to \infty,
\eeq
while keeping all dimensionful parameters fixed.

The corresponding parameters of the effective MSSM can be easily deduced
from (\ref{2.7e}), (\ref{2.14e}) as well as the 1-2 components of the
CP-even and CP-odd Higgs mass matrices (\ref{2.22e}) and (\ref{2.24e})
(and the charged Higgs and neutralino mass matrices):
\bea
\mu_\mathrm{eff} &=& \l s\; ,\nn \\
m_{3\,\mathrm{eff}}^2 &=& \widehat{m}_3^2 + \mu_\mathrm{eff} (A_\l + \k
 s)\; .
\label{2.41e}
\eea
In the $\mathbb{Z}_3$-invariant NMSSM $\widehat{m}_3^2$ vanishes, and one has
\bea
m_{3\, \mathrm{eff}}^2 &=& \mu_\mathrm{eff} B_\mathrm{eff} \; ,\nn \\
B_\mathrm{eff} &=& A_\l + \k s\; .
\label{2.42e}
\eea

In the limit (\ref{2.40e}) all couplings between the MSSM sector
(including quarks, leptons etc.) and the CP-even, CP-odd and fermionic
singlet states vanish; it seems \emph{\`a priori} impossible to
distinguish the NMSSM from the MSSM, since the singlet-like states would
never be produced. However, in this limit the singlino-like neutralino
$\chi_S^0$ with its mass given by $({\cal M}_0)_{55} = 2\k s$ (in the
$\mathbb{Z}_3$-invariant NMSSM) can easily be the LSP. Then, assuming R-parity
conservation and a small non-vanishing value of $\l$, all sparticle
decay cascades will first proceed as in the MSSM into the MSSM-like NLSP
(which could be a charged slepton!) which, at the end, will decay into
$\chi_S^0$ + SM-particles. The final decay of the MSSM-like NLSP
involves necessarily a coupling $\sim \l$, implying a possibly very long
lifetime of the NLSP leading to displaced vertices
\cite{Ellwanger:1997jj}. Then, the difference between the NMSSM and the
MSSM can be spectacular even in the decoupling limit (\ref{2.40e}).

\subsubsection{The Peccei-Quinn symmetry limit}
\label{sec:2.2.2}


If the term $\l\, \widehat{S}\,\widehat{H}_u \cdot \widehat{H}_d$ would
be the only $\widehat{S}$ dependent term in the superpotential of the
$\mathbb{Z}_3$-invariant NMSSM, the Lagrangian would be invariant under a
Peccei-Quinn-like symmetry
\beq\label{2.43e}
H_u \to H_u\, e^{i\varphi_{PQ}},\quad H_d \to H_d\, e^{i\varphi_{PQ}},
\quad S  \to S\, e^{-2i\varphi_{PQ}}
\eeq
which allows to solve the strong CP-problem
\cite{Peccei:1977ur,Peccei:1977hh}.
Since this global symmetry is spontaneously broken by the vevs $v_u$,
$v_d$ and $s$, a massless Nambu-Goldstone boson (the Peccei-Quinn axion)
would appear in the CP-odd scalar sector, as can be verified by
computing the determinant of ${\cal M}_P^2$ in the basis $(A, S_I)$ in
(\ref{2.27e}) for $\k=0$.

The decomposition of the Peccei-Quinn axion $A_\text{PQ}$ in terms of
the weak eigenstates $H_{uI}$, $H_{dI}$ and $S_I$ is given by
\bea
A_\text{PQ} &=& \frac{1}{N}\left(v\,\sin 2\b\, A -2\,s\, S_I\right)\;
,\quad\mathrm{where}\nn \\
N&=&\sqrt{v^2\sin^2 2\b+4s^2}\;,\quad
A =\cos\b\, H_{uI} + \sin\b\, H_{dI}\; .
\label{2.44e}
\eea
Hence, in most of the parameter space where $s \gg v\,\sin 2\b$,
$A_\text{PQ}$ is dominantly (but never purely) singlet-like.

Apart from the strong cosmological constraints on a Peccei-Quinn axion,
it is not straightforward to stabilise the potential for the vev $s$ of
the NMSSM in the Peccei-Quinn limit; obviously the approximate
expressions in (\ref{2.19e}) and (\ref{2.21e}) are no longer appropriate
for $\k = 0$. In \cite{Chun:1994ct}, couplings to additional singlets
have been introduced in order to stabilise the potential for $s$. In
\cite{Ciafaloni:1997gy,Hall:2004qd, Feldstein:2004xi}, the limit $\l \ll
1$ has been considered, the potential including dominant radiative
corrections in \cite{Miller:2003hm}, and the Peccei-Quinn limit for
large $\l$ in \cite{Barbieri:2007tu}. In any case, the stability of the
scalar potential imposes strong constraints on the parameters of the
NMSSM in the strict Peccei-Quinn limit $\k = 0$ \cite{Ciafaloni:1997gy,
Miller:2003ay, Miller:2003hm, Hall:2004qd,Feldstein:2004xi,
Barbieri:2007tu}.

If one prefers to avoid the cosmological constraints on a very light
Peccei-Quinn axion, one can consider the situation where the
Peccei-Quinn symmetry is explicitly broken by small additional terms in
the superpotential of the NMSSM such as a small non-vanishing value for
$\k$ in the $\mathbb{Z}_3$-invariant NMSSM \cite{Miller:2003ay} as it
can be obtained in constructions of the NMSSM from the heterotic string
\cite{Lebedev:2009ag}. Then the axion
acquires a mass at tree level (without QCD contributions); if it is
still light, it can lead to distinctive signatures at colliders due to
the possible decay of the SM-like CP-even Higgs scalar into two
pseudo-Goldstone bosons \cite{Dobrescu:2000jt}. (For a scenario with an
approximate Peccei-Quinn symmetry and large $\l$, see
\cite{Barbieri:2007tu}.)

In any case, the couplings of $A_\text{PQ}$ are important for its
phenomenological signatures: the couplings of $A_\text{PQ}$ to gauge
bosons, quarks and leptons are induced by its doublet components
indicated in (\ref{2.44e}), which are small for $s \gg v\,\sin2\b$. As
a result, its couplings $g_{Add}$ to down-type quarks and leptons are
not enhanced for large $\tan\b$, but given by
\beq\label{2.45e}
g_{Add} \sim \frac{v}{s}\, g_{Hdd}^\text{SM}
\eeq
where $g_{Hdd}^\text{SM}$ is the coupling of the SM-like CP-even Higgs
scalar to down-type quarks and leptons. The couplings of $A_\text{PQ}$
to all CP-even scalars $H_i$ vanish in the strict Peccei-Quinn limit,
but become non-zero as soon as the Peccei-Quinn symmetry is explicitly
broken. Then, while small, these couplings can still induce a branching
fraction for $H_i \to 2\,A_\text{PQ}$ which is larger than the branching
fraction for $H_i\to b\bar{b}$ \cite{Dobrescu:2000jt}. 

\subsubsection{The R-symmetry limit}


For $A_\l,\ A_\k \to 0$, the Higgs sector of the $\mathbb{Z}_3$-invariant NMSSM
(specified by the superpotential (\ref{2.6e}) and the soft terms in
(\ref{2.8e})) is invariant under an R-symmetry under which the scalar
fields transform as
\beq\label{2.46e}
H_u \to H_u\, e^{i\varphi_{R}},\quad H_d \to H_d\, e^{i\varphi_{R}},
\quad S  \to S\, e^{i\varphi_{R}}\; .
\eeq
Again, this global symmetry is spontaneously broken by the vevs $v_u$,
$v_d$ and $s$, and a massless Nambu-Goldstone boson (now an R-axion)
would appear in the CP-odd scalar sector, as can be verified by
computing the determinant of ${\cal M}_P^2$ in (\ref{2.25e}) for
$B_\mathrm{eff}=\k s$, $A_\l=A_\k=0$. The decomposition of the R-axion
$A_{R}$ in terms of the weak eigenstates $H_{uI}$, $H_{dI}$ and $S_I$ is
given by
\beq
A_{R} = \frac{1}{N}\left(v\,\sin 2\b\, A +\,s\,
S_I\right)\; ,\quad
\mathrm{where}\quad N=\sqrt{v^2\sin^2 2\b+s^2}\
\label{2.47e}
\eeq
and $A$ as in (\ref{2.44e}). Again, in most of the
parameter space where $s \gg v\,\sin 2\b$, $A_{R}$ is dominantly (but
never purely) singlet-like. However, in the full Lagrangian this
R-symmetry is explicitly broken by the gaugino mass terms, which
induce non-vanishing values for $A_\l$ (and subsequently for $A_\k$)
through radiative corrections, as can be seen from the $\b$-functions
for $A_\l$ and $A_\k$ in Appendix~B.3. Even if the statement
$A_\l=A_\k=0$ is scale dependent (and hence unnatural), one can
still consider the situation where both trilinear couplings are
relatively small 
\cite{Dobrescu:2000yn,Dermisek:2006wr,Morrissey:2008gm} implying a light
``pseudo'' R-Goldstone boson given approximately by $A_R$ as in
(\ref{2.47e}).

The couplings of $A_R$ to down-type quarks and leptons (in the
R-symmetry limit and for large $\tan\b$) differ by a factor 2 from
(\ref{2.45e}) and are given by
\beq\label{2.48e}
g_{Add} \sim \frac{2v}{s}\, g_{Hdd}^\text{SM}\; ,
\eeq
while the couplings to CP-even scalars $H_i$ depend on the values of
$A_\l$ and $A_\k$. Again, this scenario can allow for possible (even
dominant)  $H_i \to 2\,A_{R}$ decays with important consequences for
Higgs searches at colliders~\cite{Dobrescu:2000yn,Dermisek:2006wr,
Morrissey:2008gm}.

\medskip

Hence, light pseudoscalars can easily appear in the NMSSM in the form of
(pseudo-)\linebreak Nambu-Goldstone bosons; however, light pseudoscalars
in the NMSSM can also result from accidential relations among the
parameters in which case they would not suffer from suppressed couplings
to down-type quarks and leptons. Phenomenological consequences of light
pseudoscalars will be discussed in Sections~\ref{sec:5} and \ref{sec:6}.

\section{Radiative corrections}
\label{sec:3}

\subsection{Renormalisation group equations}
\label{sec:3.1}

Supersymmetry allows to formulate models at a very high scale, such as
the GUT or Planck scale, while avoiding quadratically divergent quantum
corrections involving an ultraviolet cutoff $\Lambda \sim
M_\mathrm{GUT}$ or $\Lambda \sim M_\mathrm{Planck}$. However, assuming
$M_\mathrm{SUSY} \sim M_\mathrm{weak}$, the quantum corrections still
generate large logarithms
$\sim\ln\left(\Lambda/M_\mathrm{SUSY}\right)$. 

Fortunately, these can be summed up by the introduction of
scale dependent parameters in the Lagrangian, where the scale dependence
is described by the renormalisation group equations (RGEs) or
$\beta$-functions. If the parameters are assumed to be given at a large
scale $\Lambda \sim M_\mathrm{GUT}$ or $\Lambda \sim M_\mathrm{Planck}$,
the large logarithms are accounted for by the integration of the RGEs
down to a low scale $M_\mathrm{SUSY} \sim M_\mathrm{weak}$.

In the case of the three gauge couplings of the SM, this procedure
allows for a successful unification at $M_\mathrm{GUT} \sim 2\times
10^{16}$~GeV in the MSSM \cite{Ellis:1990wk,Giunti:1991ta,Amaldi:1991cn,
Langacker:1991an}. It should be underlined that gauge coupling
unification remains valid in the NMSSM:
the additional gauge singlet field has no effect on the one-loop
$\beta$-functions of the gauge couplings, and the additional Yukawa
couplings $\l$ and $\k$ only appear in the two-loop terms of the gauge
$\beta$-functions (see (\ref{b.1e}) in Appendix B). The resulting
effect on the numerical values of the gauge couplings at the GUT scale
is negligibly small (of the order of the unknown threshold effects at
the GUT scale).

However, this is true only if the running Yukawa couplings $\l$, $\k$,
$h_t$ and $h_b$ remain sufficiently small so that perturbation
theory remains valid; typically one requires that Yukawa couplings
remain below 1 or $\sqrt{4\pi}$. Once the running Yukawa couplings have assumed
values of ${\cal{O}}(1)$ at a given scale, na\"ive extrapolations of the
perturbative RGEs result in singularities (so-called Landau
singularities) for the running couplings very close to that scale; as a
result, upper bounds of 1 or $\sqrt{4\pi}$ have nearly the same consequences in
practice. Since the running Yukawa couplings increase towards large
scales (unless the one-loop contributions from gauge couplings are
dominant, see the $\b$-functions in Appendix~B), avoiding Landau
singularities implies upper bounds on the couplings at the
scale $M_\mathrm{SUSY} \sim M_\mathrm{weak}$.

The $\b$-functions for the Yukawa couplings have been derived for the
NMSSM to one-loop order in \cite{Derendinger:1983bz}, and to two-loop
order in~\cite{King:1995vk,Masip:1998jc}, see Appendix~B.1.
(Studies of analytic solutions, quasi-fixed points and RG invariants
have been performed in \cite{Mambrini:2001wt,Nevzorov:2001vj,
Nevzorov:2001jh, Mambrini:2001sj, Demir:2004aq}; the impact of the NMSSM
specific Yukawa couplings on $h_b$-$h_\tau$ unification was studied in
\cite{Allanach:1994zd}.) Upper bounds on the Yukawa couplings at the
scale $M_\mathrm{weak}$ are given in
\cite{Drees:1988fc,Durand:1988rg,Durand:1988wn,Binetruy:1991mk,
Espinosa:1991gr,Moroi:1992zk,Elliott:1993uc,Elliott:1993bs,
Yeghian:1999kr} to one-loop order, and in \cite{King:1995vk} to two-loop
order, with the following results: The top quark Yukawa coupling $h_t$
at the scale $m_t$ depends not only on the top quark mass $m_t$, but
increases with decreasing $\tan\b$ as given in (\ref{a.11e}). Hence, for
small $\tan\b$, $h_t$ can become too large ($\gsim \sqrt{4\pi}$) at the GUT
scale. Depending on the precise value of $m_t$ and on the threshold
corrections between the scales $m_t$ and $M_\mathrm{SUSY}$, one obtains
a lower bound on $\tan\b$ close to unity (typically between 1.5 and 2)
both in the MSSM and the NMSSM. A similar reasoning based on $h_b$ leads
to an upper bound on $\tan\b$, $\tan\b \lsim 80$.

The NMSSM specific Yukawa coupling $\l$ plays an important r\^ole for
the upper bound on the mass of the lightest CP-even Higgs scalar, see
(\ref{2.23e}). The validity of perturbation theory up to the GUT scale
implies $\l < 0.7 - 0.8$ at the weak scale, depending on the precise
value of $m_t$, on the threshold corrections, on $\k$ and notably on
$\tan\b$: a large value of $h_t$ (i.\,e. small values of $\tan\b$)
amplifies the increase of $\l$ towards large scales, thus implying a
decreasing upper limit on $\l$ at low scales. Larger values of $\k$ have
the same effect on $\l$. 

\begin{figure}[h!]
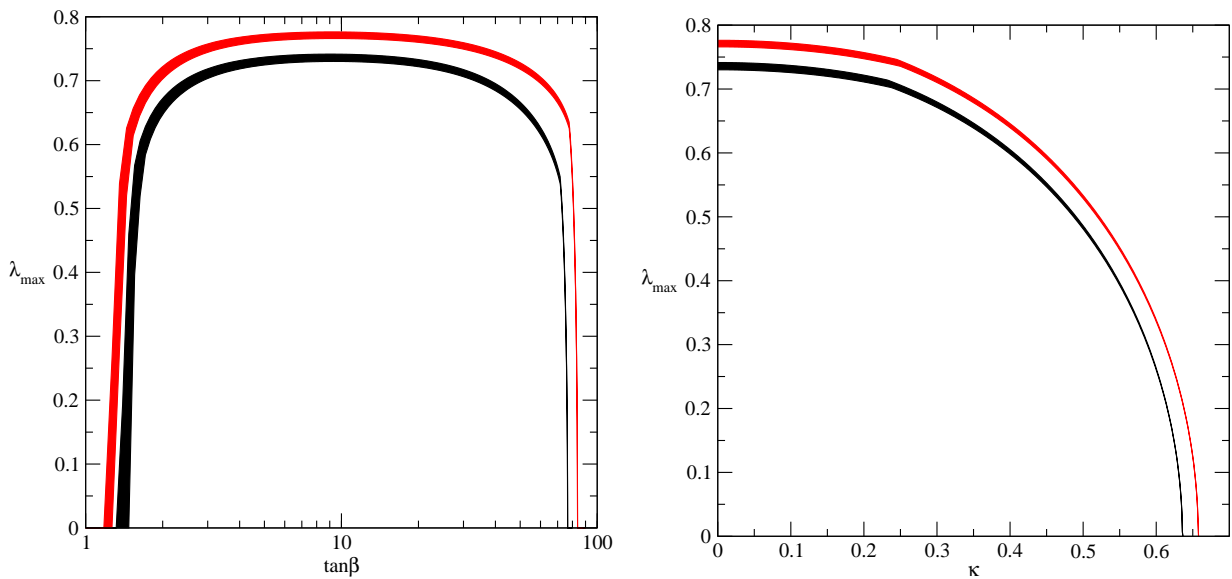

\begin{center}
\begin{tabular}{cc}
\hspace*{-10mm}
\psfig{file=figures/fig_1.eps, clip=,scale=0.45} \ \  &
\hspace*{-5mm}
\psfig{file=figures/fig_2.eps, clip=,scale=0.45}
\end{tabular}
\caption{Left panel:
upper bound on $\l$ ($\lambda_\text{max}$) as a function of
$\tan{\beta}$ for fixed $\k=0.01$. Right panel: $\lambda_\text{max}$ as
a function of $\k$ for fixed $\tan\b=10$. Black (lower) bands: light
spectrum, red (upper) bands: heavy spectrum. Inside the bands the top
quark mass is $171.2 \pm 2.1$~GeV.}
\label{fig:lmax}
\end{center}
\end{figure}

These features are visible in Fig.~\ref{fig:lmax}, where we plot
$\l_\text{max}$ as function of $\tan\b$ and $\k$ (using NMSSMTools
\cite{Ellwanger:2004xm,Ellwanger:2005dv}, see Appendix~D). The black
(dark) and red (grey) bands correspond to different choices of the Higgs
and sparticle spectrum: the black bands correspond to a light spectrum
with sfermion masses of 200~GeV, $\mu_\text{eff}=100$~GeV, a bino mass
$M_1$ of 50~GeV (with $M_2$, $M_3$ from (\ref{3.1e}) below) and Higgs
masses below $\sim 200$~GeV. The red
bands correspond to a heavy spectrum with sfermion masses,
$\mu_\text{eff}$ and heavy Higgs bosons of 1~TeV, and
$M_1=200$~GeV. (The squark/slepton trilinear couplings are chosen to
vanish in both cases for simplicity.) Inside the black and red bands,
the top quark mass is $171.2 \pm 2.1$~GeV~\cite{Amsler:2008zzb}, a
larger top mass leading to a slightly lower upper bound $\l_\text{max}$.
However, the variation of $\l_\text{max}$ with the threshold corrections
of the running gauge and Yukawa couplings induced by the unknown
sparticle and (heavy) Higgs spectrum is numerically more important.

On the left-hand side in Fig.~\ref{fig:lmax}, $\k$ is fixed to 0.01; one
sees both the decrease of the upper bound on $\l$ for $\tan\b \sim 1$
and very large $\tan\b$ induced by very large values of $h_t$, $h_b$,
respectively. On the right-hand side, $\tan\b=10$ and one sees the
decrease of the upper bound on $\l$ with increasing $\k$.

\medskip

Motivated by simple models for supersymmetry breaking,
the soft SUSY breaking parameters are often assumed to satisfy
particularly simple relations at the GUT or the Planck scale such as
universal gaugino masses $M_{1/2}$, universal trilinear couplings $A_0$
and universal scalar masses $m_0$ (where, sometimes, the singlet mass
and trilinear coupling are allowed to play a special r\^ole, see 
Section~\ref{sec:ss.1.3}). Then, the RG evolution of these parameters
from the high to the weak scale plays an important r\^ole. 

The one-loop $\b$-functions for all soft SUSY breaking parameters of the
MSSM were computed in \cite{Inoue:1982pi,Inoue:1983pp}, for the NMSSM in
\cite{Derendinger:1983bz,King:1995vk}, and for a general supersymmetric
model in \cite{Gato:1984ya,Falck:1985aa}. The general two-loop
$\b$-functions for the soft terms can be found in
\cite{Martin:1993zk,Yamada:1994id}, from which those of the (general)
NMSSM can be deduced. They are given in Appendix~B.

Universal gaugino masses $M_1=M_2=M_3=M_{1/2}$ at the GUT  scale imply
specific relations among these parameters (defined in (\ref{2.30e})) at
a low scale. At 1~TeV one finds (for squark masses $\lsim$~1~TeV)
\beq\label{3.1e}
M_3 : M_2 : M_1 \sim 5.5 : 1.9 : 1
\eeq
with somewhat larger ratios (depending on threshold effects) below
1~TeV. These values hold for the MSSM as well as for the NMSSM.

In the case of universal scalar masses $m_0$ at the GUT scale (with
$m_0^2 \geq 0$), the RG evolution down to the weak scale is essential
for the Higgs masses: at the weak scale, at least one Higgs mass squared
has to be negative in order to trigger electroweak symmetry breaking.
Fortunately $m_{H_u}^2 < 0$ at the weak scale is induced nearly
automatically by the large top Yukawa coupling \cite{Ibanez:1982fr},
given the large top quark mass.

In models with GMSB, the
soft SUSY breaking terms are induced radiatively at a messenger scale
$M_\mathrm{mess}$, which is typically far above the weak scale. In
principle, all possible terms in the Lagrangian of the general NMSSM
can be generated (see \cite{Ellwanger:2008py} and
Section~\ref{sec:ss.2}). Then, all these parameters have to be evolved by
the RGEs from the messenger scale down to the weak scale. For
completeness, we give the 2-loop $\b$-functions for all parameters of
the general NMSSM in Appendix~B.

\subsection{Radiative corrections to the Higgs masses, and the upper
bound on the lightest CP-even Higgs mass}\label{sec:3.2}

The Higgs mass matrices in Section~\ref{sec:2}, and notably the upper
bound (\ref{2.23e}) on the mass of the lightest CP-even Higgs scalar,
have been derived from the tree level potential (\ref{2.9e}), where
all quartic terms are determined by supersymmetry through the
superpotential and (electroweak) supersymmetric gauge interactions.
This hypothesis would be justified only if the scale of supersymmetry
breaking $M_\mathrm{SUSY}$ would be smaller than the Higgs vevs (or
$M_Z$), which is obviously not the case.

However, for $M_\mathrm{SUSY} > M_Z$ the deviation of the Higgs
potential from (\ref{2.9e}) is calculable -- as a
function of the mass splittings among the superpartners -- in the form
of quantum corrections involving scales (momenta) $Q^2$ with $M_Z^2
\lsim Q^2 \lsim M_\mathrm{SUSY}^2$. The dominant contributions to
$V_\text{Higgs}$ originate
from top quark/squark loops, since these particles have the largest
couplings to Higgs fields (to $H_u$), and lead to an increase of the
upper bound~(\ref{2.23e}) on the mass of the lightest CP-even Higgs
scalar. 
This phenomenon had first been discussed in the MSSM to one-loop and
subsequently to two-loop order in \cite{Okada:1990vk,
Ellis:1990nz,Okada:1990gg,Haber:1990aw,Ellis:1991zd,Yamada:1991ih,
Espinosa:1991fc} where, at tree level, (\ref{2.23e}) with $\l = 0$ would
imply a mass of the lightest CP-even Higgs scalar below $M_Z$. 

An approximate formula for the mass $M_\text{SM}$ of the SM-like Higgs
scalar in the NMSSM in the limit $\k s \gg |A_\k|$,
$|A_\l|$ (corresponding to a heavy singlet-like scalar), 
including the dominant top/stop radiative corrections, is given by
\bea
M^2_\text{SM} &\simeq& M_Z^2 \cos^2 2\b 
+\l^2 v^2 \sin^2 2\b -\frac{\l^2}{\k^2}v^2(\l-\k\sin2\b)^2\nn \\
&+& \frac{3m_t^4}{4\pi^2 v^2}
\left(\ln\left(\frac{m_T^2}{m_t^2}\right) +\frac{A_t^2}{m_T^2}
\left(1-\frac{A_t^2}{12m_T^2}\right)\right)
\label{3.2e}
\eea
where $v$ is defined in (\ref{2.12e}), 
the soft SUSY breaking stop masses squared in (\ref{2.36e}) are assumed
to satisfy $m_T^2 \sim m^2_{Q_3} \gg m_t^2$, $A_t$ is
the stop trilinear coupling assumed to satisfy $|A_t| \gg m_t,\
\mu_\text{eff}$; the terms $\sim \l^2$ are specific to
the NMSSM, and the last term in the first line originates from the
mixing with the singlet-like scalar. In the MSSM, where $\l=0$, the LEP
bound on $ M_\text{SM}$ implies that $\tan\b$ has to be large such that
$\cos 2\b \sim 1$, $m_T$ above $\sim 300$~GeV for maximal mixing ($A_t^2
\sim 6\, m_T^2$, maximising the second line in (\ref{3.2e})), or
$\gsim 1$~TeV otherwise. 

In order to maximise $M_\text{SM}$ in the NMSSM, $\l$ should be as large
as possible, and $\tan\b$ should be small in order to avoid a
suppression from $\sin^2 2\b$. (As discussed before, $\l$ is bounded
from above by $\l \lsim 0.7-0.8$ if one requires the absence of a Landau
singularity below the GUT scale.) However, the negative contribution
from the mixing with the singlet-like scalar should vanish;
without neglecting $A_\l$, the relevant mixing term is proportional to
$(\l - \sin 2\b(\k + A_\l/(2 s)))^2$ \cite{Ellwanger:2006rm}. If this
expression is not small, a larger value of $\l$ can even
generate a decrease of the mass of the Higgs scalar with SM-like
couplings to the $Z$ boson in the NMSSM.

The resulting upper bound on the lightest CP-even Higgs mass in the
NMSSM has been studied in the leading log approximation
in~\cite{Espinosa:1991fc, Ellwanger:1991bq,
Binetruy:1991mk,Espinosa:1991gr,Moroi:1992zk, Ellwanger:1992jp,
Elliott:1993ex,Diaz:1994bv,Asatrian:1995be}. Full one-loop calculations
of the corresponding upper bound involving top/bottom quark/squark loops
have been carried out in~\cite{Ellwanger:1993hn,Elliott:1993uc,
Elliott:1993bs,Pandita:1993tg,Pandita:1993hx,Ellwanger:1995ru,
King:1995vk,Franke:1995xn,Ham:1996mi,Daikoku:2000eq,Nevzorov:2000uv}.
(Analyses at large values of $\tan\b$ have been performed in
\cite{Ananthanarayan:1995xq, Ananthanarayan:1995zr,
Ananthanarayan:1996zv}, and upper bounds for more general
supersymmetric Higgs sectors have been considered in
\cite{Kane:1992kq,Espinosa:1992hp,Espinosa:1998re}.)

At present, additional known radiative corrections to the Higgs mass
matrices in the NMSSM include MSSM-like electroweak together with the
NMSSM-specific Higgs one-loop contributions
\cite{Ellwanger:2005fh,Degrassi:2009yq} and dominant two-loop terms
\cite{Diaz:1995yv, Masip:1998jc,Yeghian:1999kr, Ellwanger:1999ji,
Degrassi:2009yq}. In order to discuss these in detail, it is convenient
to separate the quantum corrections involving scales $Q^2$ with $Q^2
\gsim M_\mathrm{SUSY}^2$ from those with scales $Q^2 \lsim
M_\mathrm{SUSY}^2$.

The result of the quantum corrections with $Q^2 \gsim M_\mathrm{SUSY}^2$
is still a supersymmetric effective Lagrangian (including soft SUSY
breaking terms), where all running parameters (couplings and masses) are
defined, within a given subtraction scheme, at the scale $Q^2 \sim
M_\mathrm{SUSY}^2$. (If desired, the parameters at the scale $Q^2$ can
be obtained in terms of parameters
at a higher scale with the help of the RGEs.) Subsequently, the quantum
corrections with $Q^2 \lsim M_\mathrm{SUSY}^2$ (i.\,e. with an
ultraviolet cutoff $M_\mathrm{SUSY}^2$) have to be evaluated, generating
a non-supersymmetric effective action
including an effective Higgs potential, effective couplings of fermions
and wave function normalisation constants. From the effective potential
and couplings one can derive the so-called running masses, which still
differ somewhat from the physical pole masses (the poles of the
propagators).

The terms in the effective action can be computed in a systematic
expansion in powers of coupling constants and large logarithms where, to
start with, the couplings are defined at the scale $Q^2 \sim
M_\mathrm{SUSY}^2$. Hence, some additional effort is necessary in order
to relate the gauge couplings to those measured at the scale $M_Z^2$,
the Higgs vevs to $M_Z$, and the Yukawa couplings to the quark
and lepton pole masses: the gauge and Yukawa couplings at the scale $Q^2
\sim M_\mathrm{SUSY}^2$ have to be obtained from the measured couplings
at lower scales by the integration of RGEs where, however, all possible
threshold effects have to be taken into account.

The electroweak and NMSSM specific Higgs one-loop contributions, and the
two-loop contributions $\propto\, h_t^2\, \alpha_s$, have recently been
computed in~\cite{Degrassi:2009yq} without an expansion in large
logarithms. However, subsequently we confine ourselves to
electroweak and NMSSM-specific Higgs one-loop contributions involving
large logarithms, and to two-loop contributions proportional to two
powers of large logarithms.
In Appendix~C we summarise the corresponding formulae, which allow
to determine the lightest CP-even Higgs mass in the NMSSM with the
following accuracy (considering the tree level mass squared following
from the first line in (\ref{3.2e}) to be of order $g^2$ or
$\l^2$, where $g$ stands for $g_1$ or $g_2$):

\noindent
(i) One-loop contributions from top/bottom quarks/squarks are fully
included, which generate contributions of order $h_t^4$, $h_b^4$
and $g^2\,h_t^2$ etc., possibly multiplied by a large logarithm
$\ln\left(M_\mathrm{SUSY}^2/m_t^2\right)$. (Here and below we identify
$M_\mathrm{SUSY}^2$ with an average value of the soft SUSY breaking
squark mass terms, $M_\mathrm{SUSY}^2 = m_{Q_3}\, m_T$. It
is possible to allow for different definitions for $M_\mathrm{SUSY}^2$
\cite{Ellwanger:2004xm} in cases where other soft terms are larger than
$m_{Q_3}\,, m_T$, which we do not consider here for simplicity.) 

\noindent
(ii) Additional one-loop corrections
considered are of the orders $g^4$, $g^2\l^2$, $g^2\k^2$, $\l^4$ and
$\k^4$, if multiplied by a large logarithm. Here, loops from squarks
and sleptons, charginos and neutralinos, Higgs and gauge bosons
provide relevant contributions.

\noindent
(iii) At two-loop
order, only dominant logarithms $\sim
\ln^2\left(M_\mathrm{SUSY}^2/m_t^2\right)$ multiplied by $h_t^6$ or
$\alpha_s\,h_t^4$ (and $h_t$ replaced by $h_b$, possibly relevant at
large $\tan\b$) are considered, including some subdominant
effects due to the evaluation of $h_t$ and $\alpha_s$ at the scale
$M_\mathrm{SUSY}$.

\begin{figure}[t]
\begin{center}
\epsfig{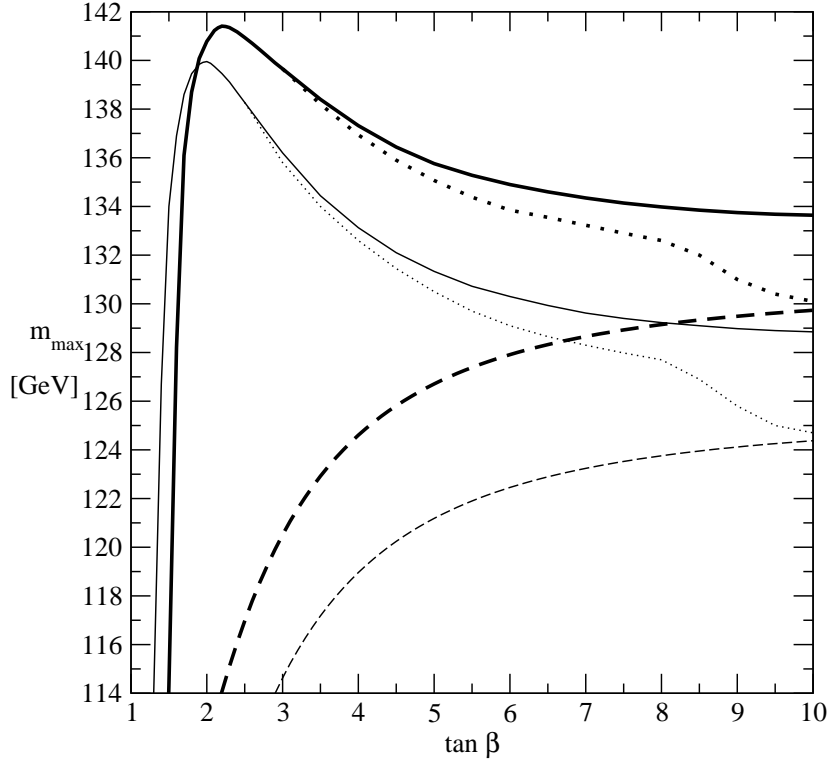}
\end{center}
\caption{Upper bound on the lightest Higgs mass in the NMSSM as a
function of $\tan\b$ for $m_{t}=178$~GeV ($M_A$ arbitrary: thick full
line, $M_A = 1$~TeV: thick dotted line) and $m_{t} = 171.4$~GeV (thin
full line: $M_A$ arbitrary, thick dotted line: $M_A = 1$~TeV) and in the
MSSM (with $M_A = 1$~TeV) for $m_{t}=178$~GeV (thick dashed line) and
$m_{t} = 171.4$~GeV (thin dashed line). Squark and gluino masses are
1~TeV and $A_{t}=A_b=2.5$~TeV. (From \cite{Ellwanger:2006rm}.)}
\label{fig:mhvstb}
\end{figure}

Taking these loop corrections into account, and requiring perturbative
running Yukawa couplings $h_t$, $\l$ and
$\k$ below the GUT scale, the upper bound on the lightest CP-even Higgs
mass has been studied in \cite{Ellwanger:2006rm} as a function of
$\tan\b$ and for different values of $m_t$ in the NMSSM, and compared to
the MSSM with the result shown in Fig.~\ref{fig:mhvstb}. (In
Fig.~\ref{fig:mhvstb}, the upper bound is denoted as $m_\text{max}$.)
The squark mass terms (and hence $M_\mathrm{SUSY}$) have been chosen as
1~TeV; the upper bound would still increase slowly (logarithmically)
with $M_\mathrm{SUSY}$. In order to maximise the one-loop top/bottom
(s)quark contributions to the lightest CP-even Higgs mass for these
squark masses, the trilinear soft couplings are chosen as
$A_{t}=A_b=2.5$~TeV. The threshold effects depend
somewhat on the gaugino masses, which are $M_1=150$~GeV, $M_2=300$~GeV
and  $M_3 = 1$~TeV (roughly in agreement with (\ref{3.1e})); the
remaining parameters $\l$, $\k$, $A_\l$, $A_\k$ and $\mu_\mathrm{eff}$
of the $\mathbb{Z}_3$-invariant NMSSM have been chosen such that the
upper bound is maximised, without encountering Landau singularities
(which requires $\k$ as small as possible) nor violating other
constraints such as an unstable potential, which forbids $\k \to 0$.

The lower dashed lines in Fig.~\ref{fig:mhvstb} refer to the MSSM, where
the mass of the CP-odd scalar $M_A$ -- which can be chosen as the other
independent parameter in the Higgs sector apart from $\tan\b$ -- is set
to $M_A = 1$~TeV. The other parameters (and approximations) are the same
as described above. In the MSSM, the increase of the upper bound with
$\tan\b$ originates from the tree level term (the first term $\sim
\cos^2 2\b$ in (\ref{3.2e})), according to which $m_\text{max}$ is
maximised for large $\tan\b$. Due to the one-loop top (s)quark
contributions, the upper bound $m_\text{max}$ increases with $m_t$.
Numerically, a variation $\Delta m_t$ of $m_t$ implies nearly the same
variation $\Delta m_\text{max}$ for large $\tan\b$.

In the NMSSM, the second term $\sim \sin^2 2\b$ in the tree level
expression (\ref{3.2e}) dominates the first one for sufficiently large
$\l$, and accordingly $m_\text{max}$ is maximal for low values of
$\tan\b$. On the other hand, the absence of a Landau singularity for
$\l$ below the GUT scale implies a decrease of the maximally allowed
value of $\l$ at $M_\mathrm{SUSY}$ with increasing $h_t$, i.\,e. with
increasing $m_t$ and decreasing $\tan\b$, see Fig.~\ref{fig:lmax}.
(At large $\tan\b$, arbitrary variations of the NMSSM parameters $\l$,
$\k$, $A_\l$, $A_\k$ and $\mu_\mathrm{eff}$ can imply a mass $M_A$ of
the MSSM-like CP-odd scalar far above 1~TeV. For comparison with the
MSSM, $m_\text{max}$ in the NMSSM with $M_A \leq 1$~TeV is depicted as
dotted lines in Fig.~\ref{fig:mhvstb}.)

Numerically, for $m_t = 171.4$~GeV, $m_\text{max}$ in the NMSSM
(thin full line in Fig.~\ref{fig:mhvstb}) is 140~GeV, and assumed for
$\tan\b=2$, $\l = 0.70$ and $\k = 0.05$, whereas for $m_t = 178$~GeV
(thick full line in Fig.~\ref{fig:mhvstb}), $m_\text{max}$
increases just to 141.5~GeV for $\tan\b=2.2$, $\l = 0.68$ and $\k =
0.07$.

It must be emphasised, however, that the non-observation of a Higgs
boson in standard search channels with a mass below $\sim 140-145$~GeV
(allowing for larger $M_\mathrm{SUSY}$ and remaining theoretical
uncertainties) would \emph{not} rule out the NMSSM: here the lightest
CP-even Higgs scalar could have small couplings to all quarks, leptons
and gauge bosons if it is dominantly singlet-like, in which case the
mass of the Next-to-lightest Higgs scalar with SM-like couplings to
gauge bosons is bounded from above, and can be $\sim 20$~GeV larger
\cite{Ellwanger:1999ji}, see also Section~\ref{sec:5.1}. Also, the
lightest CP-even Higgs scalar could have unconventional decay channels 
which can make its discovery at colliders quite
difficult. In such a case one could possibly detect one of the heavier
CP-even Higgs scalars of the NMSSM, whose masses and couplings can vary
over a wide range. (Radiative corrections to their masses are also given
in the Appendix~C.) 

Finally, $m_\text{max}$ in the NMSSM obviously increases if one
allows for larger values of $\l$ \cite{Barbieri:2006bg, 
Cavicchia:2007dp}. Since this would imply a Landau singularity below
$M_\text{GUT}$ for the particle content of the NMSSM, one can
consider modifications of the theory at higher scales. For
instance, one can allow for new strong gauge interactions at
intermediate scales \cite{Harnik:2003rs,Chang:2004db,Delgado:2005fq},
large extra dimensions implying a lower effective ultraviolet cutoff
\cite{Birkedal:2004zx}, or replace the singlet $S$ with its coupling
$\l$ by an unparticle operator \cite{Deshpande:2007jy} (in which case
the Higgs content would not be NMSSM-like). Modifications of the
$\b$-functions due to additional $SU(5)$ multiplets with masses in the
TeV range have been considered in \cite{Barbieri:2007tu}: additional
$SU(5)$ multiplets lead to larger gauge couplings at larger scales
which, in turn, affect the $\b$-function for $\l$ such that larger
values of $\l$ at the weak scale are allowed. In~\cite{Batra:2004vc} an
extra $SU(2)$ gauge symmetry has been introduced, acting on the third
generation and the Higgs doublets only and allowing for $\l \sim 1$ at
low energy, still consistent with perturbativity and GUT-scale gauge
coupling unification. This, together with small values for $\tan\b$
(possibly below~1) leads to an upper bound on the lightest Higgs mass of
$m_\text{max} \lesssim 250$ GeV. The phenomenology of large-$\l$
scenarios with $\l$ up to 1.5 has been studied in \cite{Cao:2008un},
again with the result that $\tan\b$ must be small for $\l \to 1.5$.

\subsection{Radiative corrections to coupling constants}
\label{sec:3.3}

Radiative corrections to coupling constants can be ultraviolet
divergent or (ultraviolet) finite. In SUSY extensions of the
Standard Model, all ultraviolet divergent radiative corrections can be
absorbed into redefinitions of the SUSY preserving or soft
SUSY breaking masses and couplings, which are effectively
described by the RG equations for these parameters (valid between
$M_\mathrm{SUSY}$ and, e.\,g., $M_\text{GUT}$).

However, ultraviolet finite quantum corrections involving scales between
$M_\text{weak}$ and $M_\mathrm{SUSY}$ can induce new couplings (or
modify SUSY relations between existent couplings) which \emph{cannot}
be described in terms of an effective supersymmetric theory plus soft
terms. Amongst others, such ultraviolet finite quantum corrections to
the Higgs mass terms are responsible for the increase of the upper limit
on the lightest CP-even Higgs mass with the logarithm of the top squark
masses, as discussed in the previous subsection and in Appendix~C.

Subsequently we briefly discuss quantum corrections to $h_b$ and
comment on the coupling $Z
b\bar{b}$. Radiative corrections to the Higgs self-couplings are given
at the end of Appendix~A. We leave aside the threshold effects for the
running gauge and Yukawa couplings at the various sparticle and Higgs
mass scales; for a recent NMSSM specific analysis
see~\cite{Wang:2009vv}.

\bigskip
\noi {\it Large tan\,$\beta$ corrections to $h_b$}
\medskip

The superpotentials of the MSSM and the NMSSM contain couplings of $H_u$
to up-type quarks, and of $H_d$ to down-type quarks and leptons. It is
well known \cite{Hall:1993gn,Hempfling:1993kv,Carena:1994bv} that
squark-gluino- and squark-chargino loops can induce a coupling of $H_u$
to down-type quarks and leptons giving rise to an effective Lagrangian
\cite{Carena:1999bh}
\beq\label{3.3e}
{\cal L}_\mathrm{eff} = h_b H_d^0 b\,\bar{b} + \Delta h_b H_u^0
b\,\bar{b} +\ ... \; .
\eeq

As a result, the physical $b$-quark mass $m_b$ is given by
\beq\label{3.4e}
m_b = h_b\, v_d + \Delta h_b\, v_u \equiv h_b\, v_d (1+\Delta{m_b})
\eeq
with
\beq\label{3.5e}
\Delta{m_b} = \frac{\Delta h_b}{h_b} \tan\b\;.
\eeq
Notably for large $\tan\b$, large values of $\mu$ (or $\mu_\mathrm{eff}$
in the NMSSM), large gluino masses and/or large values of $|A_t|$,
the quantity $\Delta m_b$ can be sizable; explicit expressions for
$\Delta m_b$ in the MSSM can be found
in~\cite{Hall:1993gn,Hempfling:1993kv,Carena:1994bv,Carena:1999bh}.
These hold also in the NMSSM, provided $\mu$ is replaced by
$\mu_\mathrm{eff}$. As a first consequence, the formula for the Yukawa
coupling $h_b$ as function of $m_b$ in (A.11) has to be corrected and
should read
\beq\label{3.6e}
h_b^{\mathrm{corr}} = \frac{m_b}{v \cos\b (1+\Delta{m_b})}\; .
\eeq
The corresponding value for $h_b$ has to be used as boundary condition
at the weak scale for the integration of the RGEs up to the GUT scale.
Moreover, the couplings of the Higgs bosons to $b$-quarks in (A.12) have
to be corrected as follows:
\bea
S_i b_L b_R^c & : & \frac{h_b^{\mathrm{corr}}}{\sqrt{2}} \left(S_{i1}
+ \Delta{m_b} \tan\b\, S_{i2}\right) \nn \\
A_i b_L b_R^c & : & i\frac{h_b^{\mathrm{corr}}}{\sqrt{2}} P_{i1} \nn \\
H^- t_L b_R^c & : & -h_b^{\mathrm{corr}} \sin\b \; ,
\label{3.7e}
\eea
where the couplings of the CP-odd and charged Higgs bosons follow from
the $SU(2)$-invariant completion of the effective Lagrangian
(\ref{3.3e}).

In the NMSSM, squark-gluino- and squark-chargino-loops also induce
direct couplings of the singlet $S$ to quarks and leptons
\cite{Hodgkinson:2006yh}. They give rise to additional terms $\sim
\Delta^S h_b\; S b\,\bar{b}$ in the effective Lagrangian (\ref{3.3e}),
which have been worked out in the limit of large $\tan\b$ in
\cite{Hodgkinson:2006yh}, together with the additional corrections to
the Higgs $b$-quark couplings (\ref{3.7e}). The terms $\Delta^S h_b\; S
b\,\bar{b}$ are proportional to the singlet-components of the Higgs mass
eigenstates and to $v_{u,d}/s$; hence they are small for $s \gg
v_{u,d}$.

\newpage
\noi {\it The coupling $Z b \bar{b}$}
\medskip

In view of the discrepancy between the Standard Model and the
forward-backward asymmetry
$A_{FB}^b$ in $b$-quark production at SLC and LEP \cite{:2005ema}, radiative
corrections to the coupling $Z b \bar{b}$ are of particular interest.
However, given that the measurements of $R_b = \Gamma(Z \to
b\bar{b})/\Gamma(Z \to \mathrm{hadrons})$ agree quite well with the
Standard Model, explanations of the discrepancies only in the
asymmetries are not easy. A recent analysis of radiative corrections to
the $Z b \bar{b}$ vertex in supersymmetric extensions of the SM
including the NMSSM is given in \cite{Cao:2008rc}, including possible
NMSSM specific contributions to other electroweak precision observables.
The net result of this study is, however, that the NMSSM cannot improve
the agreement with the measurements, since NMSSM specific contributions
have the opposite sign of the one required.

\section{The tadpole and domain wall problems, baryo\-ge\-nesis,
hybrid inflation}
\label{sec:4}

\subsection{The tadpole problem}

In order to avoid a fine tuning problem in the NMSSM, none of the
dimensionful terms in the superpotential $W_\mathrm{Higgs}$
(\ref{2.1e}) or in ${\cal L}_\mathrm{soft}$ in
(\ref{2.5e}) should be much larger than the weak or SUSY breaking scale.

However, if a singlet superfield $\widehat{S}$ couples in the most
general way to heavy fields (as it is possible in GUTs or GMSB),
radiative corrections can induce very large terms in the effective
action which are linear in $\widehat{S}$ in the superpotential (denoted
by $\xi_F$ in (\ref{2.1e})) or linear in $S$  in ${\cal
L}_\mathrm{soft}$ (denoted by $\xi_S$ in (\ref{2.5e})). These terms are
called tadpole terms, and if they are too large, the corresponding model
has a ``tadpole problem''. This problem has been discussed in the
context of globally supersymmetric theories with soft supersymmetry
breaking terms -- as induced by a hidden sector in supergravity -- in
\cite{Polchinski:1982an,Ferrara:1982ke,Nilles:1982mp,Lahanas:1982bk,
Hall:1983iz,AlvarezGaume:1983gj}, using general power counting rules
and/or by explicitly evaluating Feynman diagrams. 

Once non-renormalisable supergravity interactions suppressed by powers
of $M_\mathrm{Planck}$ are taken into account, power counting rules
signal again a potential tadpole problem \cite{Ellwanger:1983mg}, as
confirmed by explicit calculations in \cite{Bagger:1993ji,Jain:1994tk,
Bagger:1995ay}.

In the case of soft supersymmetry breaking terms $\sim M_\mathrm{SUSY}$
induced by a hidden sector in supergravity, the orders of magnitude of
the radiatively induced tadpole terms are
\bea
\xi_F &\sim& \Lambda\; M_\mathrm{SUSY}\; ,\nn \\
\xi_S &\sim& \Lambda\; M_\mathrm{SUSY}^2\; ,
\label{4.1e}
\eea
where $\Lambda \sim M_\mathrm{GUT}$ in the presence of general couplings
of $\widehat{S}$ to GUT fields, or $\Lambda \sim M_\mathrm{Planck}$ in
the presence of the most general non-renormalisable supergravity
interactions of $\widehat{S}$. In both cases, the induced orders of
magnitude of $\xi_F$ and $\xi_S$ are too large. 

In the case of GMSB (see Section~\ref{sec:ss.2}), the phenomenologically
required soft supersymmetry breaking mass terms of sparticles are
induced radiatively; the source of supersymmetry breaking are so-called
messenger fields $\widehat{\varphi}$ with SM gauge quantum numbers and
supersymmetric mass terms $\sim M_\mathrm{mess}$, but whose real and
imaginary scalar components have different masses, split by a scale
$\widehat{m}$. This kind of supersymmetry breaking is also denoted as
$F$-type splitting, since it can be represented in terms of a
non-vanishing $F$~component of a spurion superfield which couples to the
messenger fields $\widehat{\varphi}$. Then, $M_\mathrm{SUSY}$ in
(\ref{4.1e}) is given by $M_\mathrm{SUSY} \sim
\widehat{m}^2/M_\mathrm{mess}$, and $\Lambda$ by $M_\mathrm{mess}$.
Hence, if $M_\mathrm{mess}$ as well as the $F$-type splitting
$\widehat{m}$ are not much larger than the weak scale, the singlet
tadpole problem is absent
\cite{Nemeschansky:1983gw,Ellwanger:1984bi,Dragon:1985dq,
Ellwanger:1985hb,Ellwanger:1994uu}; if these scales are larger than the
weak scale, the tadpole diagrams can be suppressed by small Yukawa
couplings, see~\cite{Ellwanger:2008py} and Section~\ref{sec:ss.2}.

In order to circumvent the tadpole problem in the NMSSM in the case of
soft supersymmetry breaking terms induced by a hidden sector in
supergravity, we note that the dangerous terms (\ref{4.1e}) with 
$\Lambda \sim M_\mathrm{GUT},\ M_\mathrm{Planck}$ can be generated only
if $\widehat{S}$ is a singlet with respect to \emph{all} continuous and
discrete symmetries of the full theory. As soon as the heavy/hidden
sector is invariant under a discrete symmetry under which $\widehat{S}$
transforms, the terms (\ref{4.1e}) are absent if the discrete symmetry
is unbroken, or $\Lambda$ is of the order of the scale of the
breakdown of the discrete symmetry (possibly multiplied by high powers
of coupling constants and loop factors $1/16\pi^2$). Hence, if the NMSSM
is embedded in a GUT or supergravity theory with an exact or approximate
discrete symmetry such that $\Lambda \lsim M_\mathrm{SUSY}$, the tadpole
problem is avoided.

\subsection{The domain wall problem}

Discrete symmetries can generate another problem, 
however~\cite{Vilenkin:1984ib}: once they are spontaneously broken after
a symmetric phase in the hot early universe, domain walls are generated
which can dominate the energy density of the universe, creating
unacceptably large anisotropies of the cosmic microwave background
radiation and spoiling successful
nucleosynthesis~\cite{Vilenkin:1984ib}. 

In particular, the NMSSM with a scale invariant superpotential is
invariant under a $\mathbb{Z}_3$-symmetry (see the discussion following
(\ref{2.7e})). Then, after electroweak symmetry breaking, the universe
would consist of different ``bubbles'' of the same vacuum energy, but in
each of which the phases of $\left<H_u\right>$, $\left<H_d\right>$ and
$\left<S\right>$ would differ by a $\mathbb{Z}_3$-transformation; these bubbles
are separated by domain
walls~\cite{Ellis:1986mq,Gelmini:1988sf,Rai:1992xw,Abel:1995uc,
Abel:1995wk}.

However, since the $\mathbb{Z}_3$-symmetry is just an accidental symmetry of any
scale invariant superpotential, it is not expected that it will be
preserved by Planck scale suppressed (non-renormalisable) gravitational
interactions. Whereas such a violation of the $\mathbb{Z}_3$-symmetry is
typically sufficient to avoid the domain wall problem (since the vacuum
energy within different bubbles would be slightly different, leading to
a collapse of the ones with higher energy)
\cite{Ellis:1986mq,Gelmini:1988sf,Rai:1992xw,Abel:1995uc}, it may again
lead to dangerously large tadpole diagrams \cite{Ellwanger:1983mg,
Bagger:1993ji,Jain:1994tk,Bagger:1995ay,Abel:1995wk}. It has been
believed \cite{Abel:1995wk}, that the conflict between the domain wall
and tadpole problems cannot be solved.

Subsequently solutions of this problem have been proposed in 
\cite{Abel:1996cr,Kolda:1998rm,Panagiotakopoulos:1998yw}: it is possible
to impose constraints on $\mathbb{Z}_3$-symmetry breaking non-renormalisable
interactions or (renormalisable) hidden sectors in the form of various
additional symmetries, such that $\mathbb{Z}_3$-symmetry breaking
\emph{renormalisable} terms -- as the tadpole terms above -- are
generated radiatively, but with very small coefficients. These
$\mathbb{Z}_3$-symmetry breaking terms can still solve the domain wall problem of
the otherwise $\mathbb{Z}_3$-invariant NMSSM, without having a visible impact on
its phenomenology.

In \cite{Panagiotakopoulos:1999ah} a $\mathbb{Z}_5$-R-symmetry of the
non-renormalisable interactions has been considered, as a consequence of
which a tadpole term $\sim \xi_S$ is generated only at the six-loop
level. If the order of magnitude of $\xi_S$ is $\xi_S \ll
M_\mathrm{SUSY}^3$, it constitutes one of the possible terms that can
solve the domain wall problem of the otherwise $\mathbb{Z}_3$-invariant NMSSM.
However, one can also assume that $\xi_S \sim M_\mathrm{SUSY}^3$
\cite{Panagiotakopoulos:1999ah}; in this case the trilinear term
$\sim\frac{\k}{3}\widehat{S}^3$ in the superpotential of the NMSSM is
not even phenomenologically required and can be omitted. The resulting
model has been denoted as the New Minimal MSSM (or nMSSM
\cite{Panagiotakopoulos:2000wp}, if a term $\xi_F \sim
M_\mathrm{SUSY}^2$ is added to the superpotential as well), which we
will discuss in more detail in Section~\ref{sec:ss.3}.

\subsection{Electroweak baryogenesis}
\label{sec:4.3}

A strongly first order electroweak phase transition is a necessary
(albeit not sufficient) condition for baryogenesis relying on anomalous
baryon number violating Standard Model processes \cite{Kuzmin:1985mm}:
during a first order phase transition bubbles are formed, within which
the Higgs fields assume non-vanishing values corresponding to the
absolute minimum of the zero temperature effective potential. These
bubbles are separated by expanding ``walls'' from the phase with
vanishing Higgs vevs corresponding to the minimum of the effective potential at
high temperature. If the dynamics of the processes across the
expanding walls sufficiently violates CP and baryon number, the
baryon asymmetry of the present universe could be explained
\cite{Kuzmin:1985mm}.

A first order electroweak phase transition is difficult to achieve in
the SM and in simple supersymmetric extensions as the MSSM, given the
present lower bounds on Higgs and stop masses (which play an important
r\^ole for the one-loop corrections to the effective potential); see,
e.\,g., \cite{Carena:2002ss}. On the
other hand, the more complicated tree level Higgs potential in various
versions of the NMSSM can easily give rise to a sufficiently strong
electroweak phase transition \cite{Pietroni:1992in, Davies:1996qn,
Huber:1998ck, Huber:1999sa, Huber:2000mg, BasteroGil:2000bw,
Menon:2004wv,Ham:2004nv, Funakubo:2005pu,Huber:2006wf}. 
However, the trajectory in the space of fields $S$, $H_u^0$ and $H_d^0$
must be studied carefully in order to avoid a phase transition in two
steps where first $S$ alone assumes a vev (implying an
insufficient violation of CP) and subsequently $H_u^0$ and $H_d^0$
assume their vevs as in the MSSM with fixed $\mu_\mathrm{eff}$. To this
end, the vevs of $S$, $H_u^0$ and $H_d^0$ should be of the same order
along the trajectory in field space \cite{Huber:1998ck,Funakubo:2005pu},
without violating present bounds on the Higgs sector. This seems easier
to achieve in the presence of additional terms in the superpotential
and/or ${\cal L}_\mathrm{soft}$ beyond those of the $\mathbb{Z}_3$-invariant
NMSSM \cite{Davies:1996qn,Huber:1999sa,Huber:2000mg}. Simultaneously,
such additional terms explicitly break the $\mathbb{Z}_3$-symmetry avoiding the
domain wall problem. The simplest additional terms are actually the
tadpole terms $\sim \xi_S$ and/or $\sim \xi_F$, which allow to omit the
terms $\sim \k$ of the $\mathbb{Z}_3$-invariant NMSSM; the resulting model is the
nMSSM mentioned above and discussed in Section~\ref{sec:ss.3}.
Electroweak baryogenesis in the nMSSM has been studied 
in~\cite{Menon:2004wv,Ham:2004nv, Huber:2006wf}.

In addition to a first order phase transition responsible for the
formation of bubble walls, the processes inside the expanding bubble
walls have to violate CP in order to generate a baryon asymmetry.
CP~violation can originate from the chargino sector (through a
non-vanishing phase of $M_2\times \mu_\text{eff}$ in the NMSSM) or the
Higgs sector (the prospects for CP~violation in the Higgs sector of the
NMSSM are reviewed in Section~\ref{sec:var.1}). In both cases, strong
constraints from the upper bound on the (neutron) electric dipole
moments
have to be respected, which is possible in the
nMSSM~\cite{Menon:2004wv,Huber:2006wf}.

The possibility of spontaneous CP~violation inside the bubble walls
without CP~violation at zero temperature has been advocated in
\cite{Comelli:1994ew,Comelli:1994rt}; then, bounds on the neutron
electric dipole moment are trivially satisfied. A similar mechanism
inside the domain walls separating phases related by the $\mathbb{Z}_3$-symmetry
of the $\mathbb{Z}_3$-invariant NMSSM has been considered in~\cite{Abel:1995uc}.
Electroweak baryogenesis in a $U(1)'$ model with a secluded
$U(1)'$-breaking sector (see Section~\ref{sec:var.3}), where one can
have a sufficiently strong first order electroweak phase transition, has
been studied in~\cite{Kang:2004pp,Kang:2009rd}. 
Once the criteria for successful baryogenesis are fulfilled in the
NMSSM, the background density of gravitational waves produced during the
electroweak phase transition can be within reach of  the gravitational
wave experiment LISA~\cite{Apreda:2001tj, Apreda:2001us}.

\subsection{Hybrid inflation}

A variant of the NMSSM without domain wall problem, solving the strong
CP problem via an invisible axion, and allowing for hybrid inflation has
been proposed and discussed in\cite{BasteroGil:1997vn,BasteroGil:1998te,
BasteroGil:2004ae, EytonWilliams:2004bm,EytonWilliams:2005bg,
Antusch:2005kf}. Two singlets $\phi$ and $N$ are introduced, and the
relevant part of the superpotential is given by
\beq
W = \l\phi H_u H_d + \k \phi N^2\; .
\eeq

The soft SUSY breaking mass squared of $\phi$ is assumed to be very
small and negative, and both Yukawa couplings $\l$ and $\k$ are assumed
to be $\l \sim \k \sim 10^{-10}$ which can be motivated by brane
constructions and/or type~I string theory. The model has a Peccei-Quinn
symmetry (see Section~2.2.2) with Peccei-Quinn charges -2 for $\phi$ and
+1 for $N$. During the inflationary epoch the vevs of $N$, $H_u$ and
$H_d$ vanish, and the inflaton $\phi$ rolls slowly along an almost flat
direction of the scalar potential. Only near the end of inflation $N$,
$H_u$ and $H_d$ develop vevs. Domain walls, created by the spontaneous
breaking of the Peccei-Quinn symmetry, are diluted by the inflation.
The vev of $\phi$ after inflation is automatically of the correct order
such that $\mu_\mathrm{eff} = \l\phi$ is of the order of
the SUSY breaking scale, and the constraints on the parameters
associated with inflation (number of $e$-folds, curvature perturbations
and the spectral index) can be satisfied.

\section{NMSSM phenomenology at colliders}
\label{sec:5}

Due to the additional singlet superfield $\widehat{S}$, the
phenomenology of the NMSSM can differ strongly from the MSSM in the
Higgs and neutralino sectors. Subsequently we discuss these subjects
separately.

\subsection{The Higgs sector}
\label{sec:5.1}

The mass matrices of the three CP-even and the two (physical) CP-odd
neutral Higgs sectors are given in Section~\ref{sec:2.1} at tree level,
and the dominant radiative corrections in Appendix~C. In general, after
the diagonalization of the mass matrices as in Appendix~A, the $SU(2)$
doublets mix with the singlet states in both sectors. As a consequence,
the reduced couplings 
\beq\label{5.1e}
\xi_i = \sin\b\, S_{i2}+\cos\b\, S_{i1} 
\eeq
of the 3 CP-even mass eigenstates $H_i$ to the electroweak gauge bosons
(normalised with respect to the couplings of the SM Higgs scalar) can be
very small; however, they always satisfy the sum rule
\beq\label{5.2e}
\sum_{i=1}^3 \xi_i^2 = 1\; .
\eeq
Another useful sum rule involving the eigenvalues $M_{H_i}^2$ of the
CP-even mass matrix ${\cal M}_S^2$ has been given in
\cite{Panagiotakopoulos:2000wp}:
\bea
\sum_{i=1}^{3} \xi_i^2 M_{H_i}^2 &=& 
\cos^2\b \left({\cal M}_S^2\right)_{11}
+2\cos\b \sin\b \left({\cal M}_S^2\right)_{12}
+\sin^2\b \left({\cal M}_S^2\right)_{22}\nn \\
&=& M_Z^2\left(\cos^2 2\beta + 
\frac{\lambda^2}{g^2} \sin^2 2\beta\right) +\ \text{rad. corrs.}\, ,
\label{5.3e}
\eea
where the last line is the upper bound $m_\text{max}^2$ on the lightest
CP-even Higgs mass in the NMSSM (corresponding to (\ref{2.23e}) at tree
level) including the radiative corrections which lift it up to $\sim
140$~GeV, see Fig.~\ref{fig:mhvstb}. Using $M_{H_3} \geq M_{H_2}$ (by
definition) and eliminating $\xi_2^2+\xi_3^2$ with the help of
(\ref{5.2e}), one can derive from (\ref{5.3e})~\cite{Kamoshita:1994iv}
\beq\label{5.4e}
M_{H_2}^2 \leq \frac{1}{1-\xi_1^2}\left(m_\text{max}^2 - \xi_1^2
M_{H_1}^2\right)\, ,
\eeq
which will be useful below.

The extended CP-even and CP-odd Higgs sectors allow for the
possibility of additional Higgs-to-Higgs decays compared to the MSSM.
Subsequently we will discuss separately the regions in the parameter
space of the NMSSM where Higgs-to-Higgs are allowed/disallowed.

\subsubsection{LEP and $e^+\,e^-$ colliders}
\label{sec:5.1.1}

To start with, the constraints from LEP1 and LEP2 must be translated
carefully into constraints on the parameter space of the NMSSM. Early
studies  \cite{Ellwanger:1995ru, King:1995ys, Ham:1996dc, Gunion:1996fb,
Ham:1996sf, Ellwanger:1999ji} confined themselves to the standard Higgs
search channels
\beq\label{5.5e}
e^+\,e^- \to H\,Z,\qquad H\to b\bar{b},\ \tau^+\tau^-\; ,
\eeq
neglecting possible unconventional Higgs decay modes.

\begin{figure}[ht!]
\begin{center}
\includegraphics[scale=0.5,angle=-0]{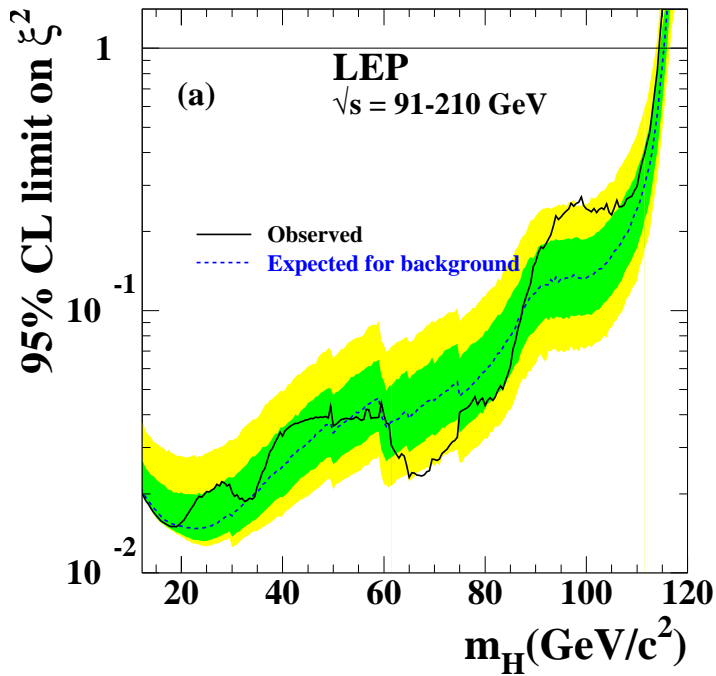}
\end{center}
\caption{Upper bounds on $\xi^2$ from LEP \cite{Schael:2006cr}, where SM
branching ratios $H \to b\bar{b}$ and $H \to \tau^+\,\tau^-$ are
assumed. Full line: observed limit; dashed line: expected limit; dark
(green) shaded band: within 68\% probability; light (yellow) band:
within 95\% probability.}
\label{fig:5.1}
\end{figure}

Since the lightest CP-even scalar $H_1$ can have a dominant singlet
component, its reduced coupling $\xi_1$ to $Z$ bosons can be smaller
than in the MSSM with, possibly, $\xi_1 \to 0$. The LEP Working group
for Higgs Boson Searches \cite{Schael:2006cr} has published upper bounds
on $\xi^2$ as function of $M_{H}$ (valid for any $H_i$)
combining results from the ALEPH, DELPHI, L3 and OPAL collaborations,
as shown in Fig.~\ref{fig:5.1}. Clearly, even very light CP-even
scalars are not ruled out if $\xi_1$ is sufficiently small, and such
scenarios can indeed be realised within the NMSSM
\cite{Ellwanger:1993xa,Kamoshita:1994iv, Ellwanger:1995ru,
Franke:1995xn, King:1995ys, Ham:1996dc, Gunion:1996fb, Ham:1996sf,
Krasnikov:1997nh, Demir:1999mm, Ellwanger:1999ji,BasteroGil:2000bw}. 

In the limit $\xi_1 \to 0$ corresponding to a light singlet-like CP-even
Higgs scalar, (\ref{5.4e}) shows that the upper bound on the
Higgs mass discussed in Section~\ref{sec:3.2} applies now to the
Next-to-lightest CP-even Higgs scalar $H_2$, which helps to establish
no-lose theorems at colliders (stating that at least one Higgs boson is
detectable). In the case $\xi_1 \neq 0$, but $\xi_1$ small, the lightest
CP-even scalar $H_1$ would still be very difficult to detect at the
Tevatron and at the LHC, and Higgs searches will only be sensitive to
$H_2$ (and possibly $H_3$) in the CP-even sector. For $\xi_1 \neq 0$,
but $\xi_1^2 \leq \xi_2^2$, the upper bound on the mass $M_{H_2}$ of the
observable $H_2$ is somewhat
alleviated~\cite{Kamoshita:1994iv,Ellwanger:1999ji}. Using again
(\ref{5.4e}) and the upper bound on $\xi_1^2$ as a function of $M_{H_1}$
from Fig.~\ref{fig:5.1} (assuming SM decay branching ratios for $H_1$),
one finds that the maximal possible value of $M_{H_2}$ is assumed for
$\xi_1^2 \sim \xi_2^2 \sim 1/2$, in which case Fig.~\ref{fig:5.1} gives
$M_{H_1} \gsim 110$~GeV implying $M_{H_2} \lsim 162$~GeV.

The relevance of Higgs-to-Higgs decays in the NMSSM for Higgs searches
was first mentioned in \cite{Gunion:1996fb}, and underlined in
\cite{Dobrescu:2000jt,Dobrescu:2000yn} in the framework of scenarios
with an approximate Peccei-Quinn or R-symmetry (see 
Section~\ref{sec:2.2})
where a CP-odd scalar $A_1$ can be very light allowing for $H_i \to
A_1\,A_1$ decays. Then the branching ratios for $H_i \to b\bar{b}$ and
$H_i \to \tau^+\,\tau^-$ would be suppressed, and the upper bounds on
$\xi^2$ in Fig.~\ref{fig:5.1} should be re-interpreted
\cite{Ellwanger:2004xm, Dermisek:2005ar, Ellwanger:2005uu,
Dermisek:2005gg}. In particular, a CP-even scalar $H$ with a SM coupling
to the $Z$ boson ($\xi=1$) but with $M_H < 114$~GeV can be compatible
with the constraints from LEP \cite{Dermisek:2005ar, Ellwanger:2005uu,
Dermisek:2005gg, Chang:2005ht, Graham:2006tr, Dermisek:2006wr, 
Dermisek:2007yt,Dermisek:2008id,Morrissey:2008gm,Dermisek:2008uu}.

On the other hand, once Higgs-to-Higgs decays as $H_i \to A_1\,A_1$
(possibly also $H_i \to H_1\,H_1$) are kinematically allowed, many
additional Higgs search channels studied at LEP can be relevant. The
constraints on the corresponding cross sections, after combining results
from the ALEPH, DELPHI, L3 and OPAL collaborations, have also been
worked out by the LEP Working group for Higgs Boson Searches
\cite{Schael:2006cr}, and those particularly relevant for the NMSSM have
been summarised recently in \cite{Chang:2008cw}:
\begin{enumerate}
\item Searches for $e^+\,e^- \to H\,Z$ independent of the $H$ decay
mode, looking for a peak of the $M_X$ recoil mass distribution in
$e^+\,e^- \to X\,Z$, by OPAL \cite{Abbiendi:2002qp}; these give $M_H >
82$~GeV if $H$
couples to $Z$ with SM strength ($\xi\sim 1$).
\item Searches for $H\to \Phi\,\Phi$ and $\Phi\to b\,\bar{b}$ ($\Phi=$
CP-even or CP-odd Higgs) by OPAL \cite{Abbiendi:2004ww} and DELPHI
\cite{Abdallah:2004wy}; once combined, these give $M_H > 110$~GeV for
$\xi\sim 1$ \cite{Schael:2006cr}.
\item Searches for $H\to \Phi\,\Phi$ and $\Phi\to g\,g$, $\Phi\to
c\,\bar{c}$, $\Phi\to \tau^+\,\tau^-$ by OPAL \cite{Abbiendi:2002in}
exclude 45~GeV $<M_H<86$~GeV for $\xi\sim 1$. A new
analysis of ALEPH data~\cite{Schael:2010aw} implies stronger limits on
$H\to \Phi\,\Phi\to 4\,\tau$ as $M_H \gsim 109$~GeV for $M_\Phi =
10$~GeV and $\xi^2\times BR(H\to\Phi\,\Phi)\times BR(\Phi\to \tau\tau)^2
\sim 1$. \end{enumerate}

Also possibly relevant could be \cite{Schael:2006cr} $e^+\,e^- \to H\,Z$
with $H\to jj$ (two jets), $H\to \gamma\,\gamma$ and $H$ decaying
invisibly (as, e.\,g., into neutralinos), and $e^+\,e^- \to H\,A$
together with $H\, A\to 3A \to 6b$, all of which are considered in the
public code NMHDECAY \cite{Ellwanger:2004xm,Ellwanger:2005dv}, which is
part of NMSSMTools (see Appendix~D).

In Fig.~\ref{fig:5.1} one can note a slight excess of events for $M_H
\sim 95-100$~GeV of $2.3\,\sigma$ statistical significance
\cite{Barate:2003sz}. In the NMSSM this excess could be explained by 
\begin{itemize}
\item a SM-like CP-even scalar of corresponding mass and $\xi \sim 1$,
but with reduced $BR(H \to b\,\bar{b}) \sim 0.1$ \cite{Dermisek:2005gg,
Dermisek:2006wr,  Dermisek:2007yt} due to a dominant decay $H \to A\,A$
with a branching ratio $\sim 0.8$. This scenario could alleviate the
little fine tuning problem related to the non-observation of
a CP-even scalar at LEP (see Section~\ref{sec:ss.4}), but requires $M_A
< 10.5$~GeV due to the constraints (2) on $H\to \Phi\,\Phi$ and $\Phi\to
b\,\bar{b}$;
\item a CP-even scalar $H_1$ of corresponding mass
with $\xi \lsim 0.4$ \cite{Dermisek:2007ah} as it can appear, e.\,g., in
the constrained NMSSM (see Section~\ref{sec:ss.1})
\cite{Djouadi:2008yj,  Djouadi:2008uj} or for large $\l\approx 0.7-0.8$
\cite{Cavicchia:2008fn}. Then, another CP-even scalar $H_2$ with $\xi
\gsim 0.9$ and a mass not far above 115~GeV is present in the spectrum.
\end{itemize}
In the first scenario, difficulties for Higgs searches at the Tevatron
and at the LHC are foreseeable (see below), whereas probably only $H_2$
would be detectable at hadron colliders within the second scenario.

If $H_1$ is singlet-like ($\xi_1^2 \leq \xi_2^2$) \emph{and} decays
unconventionally, the upper bound on $M_{H_2}$ deduced above ($M_{H_2}
\lsim 162$~GeV from $M_{H_1} \gsim 110$~GeV) is no longer valid. Using
the upper bound on $M_{H_1}$ from constraints on $H_1 \to A_1\, A_1 \to
4b$, (\ref{5.4e}) gives $M_{H_2} \lsim 167$~GeV, whereas the weaker
upper bound on $M_{H_1}$ independent of its decay mode (using only the
$Z$ recoil mass) allows for $M_{H_2} \lesssim192$~GeV.
\medskip

The prospects for NMSSM Higgs searches at $e^+\,e^-$ colliders of higher
energy, as an ILC, have been investigated in \cite{Kamoshita:1994iv,
Ham:1996dc, Espinosa:1998xj, Ellwanger:2003jt, Miller:2004uh,
Weiglein:2004hn,Ham:2004nv}; for a review see~\cite{Gunion:2003fd}. Due
to the sum rule (\ref{5.2e}), one can derive a theoretical lower limit
on the production cross section $\sigma_i$ for at least one CP-even
Higgs boson, obtained as $\sigma_i > 0.04$~pb for $\sqrt{s}= 300$~GeV in
\cite{Kamoshita:1994iv}. (Production cross sections at $\sqrt{s}=$ 500,
1000 and 2000~GeV were computed in \cite{Ham:1996dc}.)

As at LEP, one can search for Higgs bosons at an ILC independently from
their decay modes using the recoil spectrum (the invariant mass $M_X$)
in $e^+\,e^- \to Z\, X$, $Z \to e^+\,e^-$ and $Z \to
\mu^+\,\mu^-$. This allows to establish a no-lose theorem
for arbitrary singlet extensions of the MSSM (assuming perturbativity of
all Yukawa couplings up to $10^{16}$~GeV, implying upper bounds on the
Higgs masses) for $\sqrt{s}= 500$~GeV and an integrated luminosity
${\cal L} = 100$~fb$^{-1}$~\cite{Espinosa:1998xj}. Also the $ZZH$
couplings can be determined by this method~\cite{Gunion:2003fd}. The
clean environment of $e^+\,e^-$ colliders would allow for the study of
NMSSM specific Higgs-to-Higgs decays~\cite{Ellwanger:2003jt}, and for
the discovery of light Higgs states that would be difficult to observe
at the LHC~\cite{Miller:2004uh, Weiglein:2004hn}.

\subsubsection{Tevatron and LHC}

Next we turn to hadron colliders, starting with regions in the NMSSM
parameter space where Higgs-to-Higgs decays are kinematically forbidden
or occur only with small branching ratios. Then, the search for Higgs
bosons in the NMSSM can proceed as in the MSSM (see
\cite{Djouadi:2005gj} and references therein); of course, Higgs
boson production cross sections and decay branching fractions in the
NMSSM have to be rescaled (for tree level couplings) or re-evaluated
(for loop corrected or loop induced couplings), employing the tree level
couplings given in Appendix~A.2. Important search channels are
\begin{enumerate}
\item $g\,g \to H$ (gluon-gluon fusion) with $H \to W\,W^{(*)} \to
l^+l^-\nu\bar{\nu}$, $H \to Z\,Z^{(*)} \to 4$~leptons, $H\to
\gamma\,\gamma$ (the latter also for $A$ instead of $H$); 
\item Vector boson fusion corresponding to $q\,\bar{q} \to
q\,\bar{q}\,V\,V$ with $V\,V \to H$ at the Tevatron, and $q\,{q} \to
q\,{q}\,V\,V$ with $V\,V \to H$ at the LHC (where $V=W^\pm,\,Z$), 
and $H\to\gamma\,\gamma$, 
$H \to \tau^+\,\tau^-$, $H \to W\,W^{(*)}$, $H$
decaying invisibly or $H \to b\,\bar{b}$ (the latter decay suffers from
a too large background at the LHC);
\item $q\,\bar{q}\to W^{*} \to W\,H$ and $q\,\bar{q}\to Z^{*} \to 
Z\,H$ (Higgs-strahlung);
\item Associated production with heavy quark pairs
$q\,\bar{q}/g\,g \to Q\bar{Q}\,H$, where $Q=t,\,b$.
\end{enumerate}

In some regions of the parameter space of the NMSSM, more Higgs bosons
than in the MSSM could be observed (e.\,g. in the two photon channel
\cite{Moretti:2006sv}), which would allow to distinguish the two models.
On the other hand, since all $\xi_i$ can be simultaneously less than 1
(respecting the sum rule (\ref{5.2e})), it was not clear at first
instance whether any of the $H_i$ could always be detected after
combining the results for Higgs searches at LEP2 with the prospects for
the LHC~\cite{Gunion:1996fb,Krasnikov:1997nh}, even if Higgs-to-Higgs
decays are assumed to be kinematically forbidden. Later on, the analysis
of Higgs searches at the LHC was refined and additional Higgs production
channels (notably vector boson fusion) have been considered. On the
theoretical side, radiative corrections to the Higgs masses have been
improved by dominant two-loop corrections; these lower the CP-even Higgs
masses, increasing the region in the NMSSM parameter space excluded by
LEP2. As a result, a no-lose theorem for the detection of at least one
CP-even Higgs scalar in the NMSSM at the LHC with an integrated
luminosity of 600~fb$^{-1}$ (two detectors with 300~fb$^{-1}$ each)
could be established \cite{Ellwanger:2001iw,
Ellwanger:2003jt,Ellwanger:2005uu}, however, under the assumption that
Higgs-to-Higgs decays are kinematically forbidden.

\medskip

Once Higgs-to-Higgs decays are possible (or even dominant), Higgs
searches at the Tevatron and the LHC can become considerably more
complicated; see \cite{Ellwanger:2005uu,
Chang:2005ht,Aglietti:2006ne,Barbieri:2007tu,
Djouadi:2008uw,Chang:2008cw,Rottlander:2009zz} for discussions
of possible scenarios and proposals for search channels. In principle,
CP-even scalars $H_2$ can decay into a pair of CP-even $H_1$ in
LEP-allowed regions of the NMSSM parameter space \cite{Ellwanger:2005uu,
Djouadi:2008uw, Djouadi:2008uj} (see \cite{Cavicchia:2007dp} for very
large $\l \approx 2$ where $M_{H_1} \approx 200-300$~GeV, $M_{H_2}
\approx 350 - 700$~GeV); but most of the recent studies concentrated on
decays into light CP-odd scalars $A$ motivated by an approximate
Peccei-Quinn or R-symmetry, and/or the $H \to AA$ explanation of the
light excess of events at LEP.

For $M_A > 2\,m_b$ (in which case LEP constraints require $M_H \gsim
110$~GeV for SM-like couplings of $H$), the cascade $H \to A\,A$ would
end mostly in a $4\,b$ final state. This final state has a very large
background at hadron colliders (which seemed to make it invisible at
least at the Tevatron, unless the $H$ production rate is enhanced
relative to the SM \cite{Stelzer:2006sp}). Hence, a first study proposed
to
consider the subdominant $H \to A\,A \to b\bar{b}\, \tau^+\tau^-$ final
state, with $H$ produced via vector boson fusion at the LHC
\cite{Ellwanger:2003jt,Ellwanger:2004gz}. However, here the visibility
of the signal depends strongly on the poorly known background from
$t\,\bar{t}$ production.
Subsequently the Higgs-strahlung process has been added, where the
lepton(s) from $W$, $Z$ decays can help to trigger on the relevant
events \cite{Moretti:2006hq,Cheung:2007sva,Carena:2007jk}. Then, both
$4b$ and $2b\,2\tau$ final states can be relevant for both the Tevatron
and the LHC~\cite{Carena:2007jk} (the $2b\,2\tau$ final state still
being more promising at the LHC).

For $M_A < 2\,m_b$, the cascade $H \to A\,A$ would end mostly in a
$4\tau$ final state; clearly, the visible $\tau$-decay products would
not generate narrow peaks in the invariant masses of $M_A$ or $M_H$. At
the Tevatron, the prospects to detect the Higgs scalar in this case look
pretty dim \cite{Graham:2006tr} unless the $H$ production cross section
is enhanced; the $4\gamma$ decay mode would require enhanced branching
ratios for $A\to \gamma\gamma$ \cite{Chang:2006bw}. At the LHC
particular efforts would also be required. Proposals for signals and
cuts
appropriate for the $A\,A \to 4\,\tau \to 2\,\mu + 2\,$jets final state
have been made in \cite{Belyaev:2008gj}; with 100~fb$^{-1}$ of
integrated luminosity, the expected rates after cuts are $\sim 8\times
10^3$ from  $H$~production via vector boson fusion, and $\sim 10^3$ from
$H$~production via Higgs-strahlung ($W^{\pm\ *} \to H+W^\pm$) where one
can trigger on a lepton from $W^\pm$ decays.
In \cite{Forshaw:2007ra} it has been
proposed to consider diffractive Higgs production ($pp \to pp+H$) in
order to be sensitive to $H \to 4\tau$, which requires to
install additional forward detectors. Using a track-based analysis in
which all events with more than 6 tracks in the central region are
discarded, a viable signal seems possible after accumulating
300~fb$^{-1}$ of integrated luminosity.
In \cite{Lisanti:2009uy,Dermisek:2009fd}, the
subdominant $H \to A\,A \to 2\tau\,2\mu$ final state was discussed: in
spite of the small branching fraction (with $2\mu$ from direct $A$
decays) it was argued that, for $M_H \sim 102$~GeV, the Tevatron
can see a signal over background for an integrated luminosity ${\cal
L}\sim 10$~fb$^{-1}$, and the LHC already for ${\cal L} \sim
1$~fb$^{-1}$, with $H$ being produced via gluon-gluon fusion.

Light $A$ production in association with charginos \cite{Arhrib:2006sx}
(which requires $\l \gsim 1$ to be observable), and from neutralino
decays \cite{Cheung:2008rh} has also been considered. The LHC discovery
potential for $\l=2$ -- where $M_{H_1}$ can be as large as 200 to
300~GeV -- has been analysed in \cite{Cavicchia:2007dp}. Further details
of current ATLAS and CMS studies of benchmark scenarios including the
$H \to A\,A \to 4\tau$ final state can be found
in~\cite{Djouadi:2008uw}.

The charged Higgs decays $H^\pm \to W^\pm + H$ or $H^\pm \to W^\pm + A$
play only a marginal r\^ole in the MSSM. In the NMSSM, larger regions in
the parameter space exist (at low $\tan\b$) where $M_{H^\pm} \lsim m_t$,
but where these processes are kinematically allowed and the
corresponding branching fractions are important \cite{Drees:1999sb}. In
the case of a light $A$ with a non-negligible $SU(2)$ doublet component,
the second process (and also $pp \to H^\pm+A \to W^\pm + AA$) can be
observable \cite{Akeroyd:2007yj}.

\medskip

Higgs searches at the Tevatron have started to test regions of the NMSSM
parameter space. Recall that, in the NMSSM, a CP-even Higgs $H_2$ with
nearly SM-like couplings can have a mass up to 160~GeV (if a light
singlet-like $H_1$ exists, escaping LEP constraints), or even up to
$\sim 190$~GeV, if the singlet-like $H_1$ decays unconventionally (see
above). The present results of searches for a SM Higgs
at the Tevatron (CDF and \D0) from various production and decay channels
have recently been combined \cite{:2009pt}, and exclude a Higgs scalar
with SM couplings in the mass range $160 < m_H < 170$~GeV. While such a
Higgs scalar would be impossible in the MSSM, the corresponding mass
range touches the parameter space of the NMSSM.

In \cite{Abazov:2008zz} the \D0 collaboration has looked for CP-even or
CP-odd scalars $\Phi$ produced in association with $b$ quarks and
decaying into $\tau^+\,\tau^-$, assuming a branching ratio of a SM-like
Higgs scalar. In the mass range 90~GeV$ <M_\Phi<$ 150~GeV, very large
values of $\tan\b \gsim 40-80$ (which would imply an enhanced
production cross section) could be excluded. This result is relevant for
both the MSSM and the NMSSM, although one usually expects $\tan\b
\lsim 60$.

The \D0 collaboration has also searched for $H \to A_1\,A_1$ decays in
the range $M_{A_1} < 2\,m_\tau$ where the branching ratio $A_1 \to
\mu^+\,\mu^-$ is not too small. The $4\,\mu$ signature has been
searched, and an upper limit of about 10~fb on $\sigma(p\bar{p} \to
H\,X) \times BR(H \to A\,A) \times \left(BR(A \to \mu^+\mu^-)\right)^2$
has been set \cite{Scodellaro:2009sm}. Assuming $M_H = 120$~GeV and
$BR(H \to A\,A) =100\%$, the negative result implies an upper bound on
the $BR(A \to \mu^+\mu^-) \lsim 10\%$. The final state
$\mu^+\mu^-\,\tau^+\tau^-$ (relevant for $M_{A_1} > 2\,m_\tau$) has also
been investigated, but the limits are still a factor $\sim 4$ above the
theoretical expectations.
Further results from the Tevatron can be expected in the near
future.

\subsection{The neutralino sector}

The neutralino mass matrix in the general NMSSM is given in
(\ref{2.32e}) as function of the vevs $v_u$, $v_d$ and $s$. In the
$\mathbb{Z}_3$-invariant NMSSM, the last term $2\mu'$ in the $(5,5)$ element
is absent. In \cite{Pandita:1994ms, Pandita:1994vw,Ellwanger:1997jj,
Choi:2004zx} approximate formulae for the mass eigenvalues and mixing
matrices have been derived. From $M_{\chi_1^0}^2 \leq M_1^2 + M_Z^2
\sin^2\t_W$, and assuming the relations (\ref{3.1e}) for the gaugino
masses, an upper bound on the lightest neutralino mass as function of
the gluino mass $M_3$ can be deduced~\cite{Pandita:1994zp}. In the 
nMSSM (where $\k=0$, see Section \ref{sec:ss.3}), the bound
$M_{\chi_1^0} \leq \l v$ (with $v$ as in (\ref{2.13e}))
holds~\cite{Hesselbach:2007te}.

The full set of neutralino couplings in the $\mathbb{Z}_3$-invariant NMSSM,
necessary to study neutralino production processes at $e^+\,e^-$
colliders, in sfermion (squark/slep\-ton) decays, chargino/neutralino
cascade decays and Higgs decays, is given in \cite{Franke:1995tc}. 

Clearly, in the decoupling limit $\l \sim \k \to 0$, $s \sim 1/\k \to
\infty$ of the $\mathbb{Z}_3$-invariant NMSSM (see Section~\ref{sec:2.2.1}), it
becomes practically impossible to distinguish the neutralino sector of
the NMSSM from the one of the MSSM: in this limit, the singlino-like
neutralino $\chi^0_S$ ceases to mix with the MSSM-like neutralinos, and
all production cross sections as well as partial widths of decays into
$\chi^0_S$ tend to zero. An important exception to this rule is the
scenario where $\chi^0_S$ is the LSP (for $\k s$ small enough), and
R-symmetry is conserved: then all sparticle decay chains will first
proceed as in the MSSM into the NLSP (which can now be charged!), and
only at the end of the decay chain the NLSP finally decays into
$\chi^0_S$. This additional decay (leading typically to extra leptons in
the final state) should allow to distinguish the NMSSM from the MSSM in
the singlino-LSP regime. Due to a possibly tiny value of the
NLSP$-\chi^0_S$ coupling, the NLSP lifetime can be quite long leading to
displaced vertices at colliders, see below. Subsequently we briefly
discuss neutralino production processes at $e^+\,e^-$ colliders, in
sfermion and cascade decays.

\subsubsection{Neutralino pair production at $e^+\,e^-$ colliders}

Neutralino pair production at $e^+\,e^-$ colliders can proceed via a $Z$
boson in the $s$~channel and via the exchange of a selectron in the 
$t/u$~channels. Formulae for the corresponding cross sections are given
in \cite{Franke:1995tf} where, however, the ``photino/zino''-basis in
the gaugino sector is employed, differing from the one used here. In
\cite{Choi:2004zx}, the formulae for the cross sections are given in the
``bino-wino''-basis employed here.

Present constraints on the neutralino sector originate from its
contribution to the invisible $Z$ width and from bounds on $e^+\,e^- \to
\chi^0_i\,\chi^0_j$ ($i \neq j$) from LEP. Note that bounds on the
mass of the lightest neutralino $\chi^0_1$ relying on MSSM-like
relations between neutralino and chargino masses cannot be applied to
the NMSSM. For earlier discussions of LEP constraints on the neutralino
sector of the NMSSM see \cite{Drees:1990fs,Franke:1994hj}.

Assuming 3 massless neutrinos, additional contributions to the invisible
$Z$ width $\Gamma_{Z}^\mathrm{inv}$ should not exceed $\sim 2$~MeV (see
the Section on $\tilde{\nu}$ mass limits in \cite{Amsler:2008zzb}). This
limit has to be compared to the partial $Z$ width into neutralinos (if
$m_{\chi^0_i} < M_Z/2$), given by
\beq\label{5.6e}
\Gamma_{Z\to \chi^0_i\chi^0_i} = \frac{M_Z^3\, G_F}{12\sqrt{2}\pi}
(N_{i3}^2-N_{i4}^2)^2 
\left(1-\frac{4 m_{\chi^0_i}^2}{M_Z^2}\right)^{3/2}\; ,
\eeq
where the mixing matrix elements $N_{ij}$ are defined in (\ref{a.7e}).
Clearly, the mixing matrix elements $N_{i3}$ and $N_{i4}$ (proportional
to the higgsino components of the neutralinos) are tiny for singlino
and/or gaugino like neutralinos, which are thus not much constrained.

Signals from $e^+\,e^- \to \chi^0_i\,\chi^0_j$ with 
subsequent $\chi^0_i$ decays ($i > 1$) have been searched for by DELPHI
\cite{Abdallah:2003xe} and OPAL \cite{Abbiendi:2003sc} at LEP, with
$\sqrt{s}$ up to 209~GeV. The non-observation of such signals imposes
upper bounds on the production cross sections $\sigma(e^+e^- \to
\chi^0_1\chi^0_i) \lsim 10^{-2}$~pb and $\sigma(e^+e^- \to
\chi^0_i\chi^0_j) \lsim 10^{-1}$~pb ($i,j > 1$).

Phenomenological analyses of pair production of neutralinos in the NMSSM
at $e^+\,e^-$ colliders at higher energies have been
performed in \cite{Franke:1995tf,Choi:2001ww,Franke:2001nx,Choi:2004zx, 
MoortgatPick:2005vs, Basu:2007ys}. Since the information on the
neutralino sector from the LHC will be quite limited, an ILC-like
$e^+\,e^-$ collider can be crucial to distinguish the NMSSM
neutralino sector from the one of the MSSM \cite{MoortgatPick:2005vs},
although it cannot be guaranteed that the difference is visible if one
is close to the decoupling limit mentioned above. This
question has also been addressed in the radiative production of the
lightest neutralino pair, $e^+\,e^- \to \chi^0_1\,\chi^0_1\,\gamma$,
at an ILC with $\sqrt{s} = 500$~GeV in \cite{Basu:2007ys}.

\subsubsection{Decays into and of neutralinos}

As in the MSSM, in the NMSSM sfermions $\tilde{f}_i$ (squarks and
sleptons) can decay into neutralinos or charginos via $\tilde{f}_i \to
f_i\,\chi^0_j$ \cite{Choi:2004zx} or $\tilde{f}_i \to f_j\,\chi^\pm$,
where the latter decay does not exhibit any NMSSM specific features.
However, due to the modified neutralino sector, the processes
$\tilde{f}_i \to f_i\,\chi^0_j$ can be quite different in the NMSSM:
even a mostly singlino-like neutralino can be produced in decays of
squarks and sleptons of the third generation, if a non-negligible
higgsino component induced by a not too small value of $\l$
\cite{Kraml:2005nx} is present. In addition, gaugino components of
the neutralinos also allow for their production in sfermion decays of
all generations \cite{Choi:2004zx}.

Higgs bosons could also decay invisibly into neutralinos; these
processes can still be interesting for Higgs searches via the recoil
spectrum of $Z$ decay products in $Z^* \to Z\,H$, see above. An
exception are GMSB models (see Section~\ref{sec:ss.2}), where the
neutralinos can subsequently decay into a gravitino and a photon; the
present constraints on and the future prospects of this Higgs decay mode
have been studied in \cite{Mason:2009qh}.

Once produced, neutralinos can decay in many different ways (except for
the LSP):
\begin{itemize}
\item into a fermion + sfermion (if kinematically allowed);
\item via two-body decays (if kinematically allowed) $\chi^0_i \to
\chi^0_j\,Z$, $\chi^0_i \to \chi^\pm\,W^\mp$, $\chi^0_i \to
\chi^0_j\,\Phi$, where $\Phi$ is a CP-even or CP-odd Higgs,
and via the radiative decay $\chi^0_i \to\chi^0_j\,\gamma$;
\item via three body decays $\chi^0_i \to
\chi^0_j\,l^+\,l^-$, $\chi^0_i \to
\chi^0_j\,q\,\bar{q}$ and $\chi^0_i \to
\chi^0_j\,\nu\,\bar{\nu}$ (via $Z$, Higgs and sfermion exchange).
\end{itemize}
Formulae for the relevant partial widths can be found in 
\cite{Franke:1995tf,Ellwanger:1997jj,Choi:2004zx}. Note that the decay
$\chi^0_i \to \chi^0_j\,A_1$ can be relevant for the
search for a light pseudoscalar Higgs \cite{Cheung:2008rh}. If the 
$m_{\chi^0_i} - m_{\chi^0_j}$ mass difference is small (e.\,g. in the
case $\chi^0_i \equiv \chi^0_2 \sim$~bino, $\chi^0_j \equiv \chi^0_1
\sim$~singlino), soft leptons from $\chi^0_2 \to \chi^0_1\,l^+\,l^-$ can
be an important signal for the NMSSM, but their detection at the LHC may
require low cuts on lepton transverse momenta \cite{Kraml:2008zr}. 

\subsubsection{Displaced vertices}\label{sec:5.2.3}

As stated above, for small $\l$ the couplings between the NLSP and a
singlino-like LSP in the NMSSM can be so small that the large NLSP 
lifetime leads to macroscopically displaced vertices. This phenomenon
has been studied first in the case of a bino NLSP in the scale invariant
NMSSM in \cite{Ellwanger:1997jj}, and applied to searches at LEP in
\cite{Ellwanger:1998vi}. Such additional cascades in sparticle decays
(as compared to the MSSM), with possibly displaced vertices, can also
occur in more general extensions of the MSSM by singlets
\cite{Hesselbach:2000qw, Martin:2000eq, Barger:2006kt} as, for instance,
in models designed to explain the CDF multi-muon
events\cite{Domingo:2008gh} where the cascades originate from
Higgs decays.

Considering the case of a bino-like NLSP ${\chi}^0_2$ (with a mass $\sim
M_1$) and a singlino-like LSP ${\chi}^0_1$, approximations for the
relevant neutralino mixing parameters can be derived for small $\l$, and
an approximate expression for the (typically dominant) bino to singlino
decay width via right-handed slepton ($\widetilde{E}_R$) exchange can be
obtained~\cite{Ellwanger:1997jj}:
\beq\label{5.7e} 
\Gamma({\chi}^0_2
\stackrel{\widetilde{E}_R}\longrightarrow {\chi}^0_1 l^+ l^-) \simeq
2\times 10^{-6} \lambda^2 M_1
\left(\frac{M_Z}{m_{\widetilde{E}_R}}\right)^4 I(\eta,\omega)
\eeq
where $\eta = {m_{\chi_1^0}}/{m_{\chi_2^0}}$, $\omega =
m_{\chi_2^0}/m_{\widetilde{E}_R}$ and the phase space integral
$I(\eta,\omega)$ is of ${\cal O}(10^{-1})$ for masses compatible with
universal soft terms at the GUT scale \cite{Ellwanger:1997jj}. The
important point is that the bino decay width is proportional to $\l^2$,
implying a long lifetime for $\l \to 0$.

In the fully constrained NMSSM with its nearly unique sparticle spectrum
(once a dark matter relic density complying with WMAP constraints has
been imposed, see Section~\ref{sec:ss.1}), the LSP is always
singlino-like \cite{Djouadi:2008yj,Djouadi:2008uj}, and the NLSP is the
lighter stau. The stau mass must only be a few GeV above the LSP mass in
order to allow for a sufficient reduction of the relic density via
coannihilation, see
section~\ref{sec:dm}. Then, again using approximations for the relevant
neutralino mixing parameters as in \cite{Ellwanger:1997jj}, the
expression for the stau decay width is approximately given
as~\cite{Djouadi:2008uj}
\begin{equation}\label{5.8e}
\Gamma(\tilde \tau_1 \to \chi_1^0 \tau)\, \approx \,
\lambda^2\,\frac{\sqrt{\Delta m^2 - m_\tau^2}}{4 \pi m_{\tilde \tau_1}} 
\,(\alpha \Delta m - \beta m_\tau)\; ,
\end{equation}
where the coefficients $\alpha$ and $\beta$ still depend on
$m_{\tilde \tau_1}$ and $m_{\chi_1^0}$ (of the order  $0.01\gsim \alpha
\sim \beta \gsim 0.0001$, decreasing with $m_{\tilde \tau_1}$), and 
$\Delta m \equiv m_{\tilde \tau_1} - m_{\chi_1^0}$. The stau decay
width can be small due to a small value of $\l$ \emph{and} a possible
phase space suppression for small $\Delta m - m_\tau$. For $\Delta m
\sim$~a few GeV and $\lambda \sim 10^{-4}$, stau lengths of flight
of ${\cal{O}}$(mm) are possible \cite{Djouadi:2008uj}.

In the near future it will be important to study the possible impact of
NMSSM specific displaced vertices for sparticle searches at hadron
colliders.

\section{$\pmb{b}$ physics and the anomalous magnetic moment of the
muon}
\label{sec:6}

In this Section we briefly discuss NMSSM specific effects in
$B$~physics, $\Upsilon$~physics and for the anomalous magnetic moment of
the muon. In most cases an extensive literature exists on contributions
from MSSM-like supersymmetric extensions of the Standard Model, which
are beyond the scope of the present review. In general, the
contributions in the NMSSM can differ from those of the MSSM due to
the different Higgs and neutralino sectors. Correspondingly, most of the
results within the MSSM -- like constraints on the parameters or a
possible improved agreement between theory and experiment -- are valid
in the NMSSM as well.

A notable exception is the region in the NMSSM parameter space
corresponding to a light CP-odd Higgs scalar $A_1$. Here, important new
contributions ``beyond the MSSM'' can arise, which lead to constraints
on this part of parameter space and/or to possible new phenomena (in the
case of $\Upsilon$/$\eta_b$~physics). 

\subsection*{6.1\quad $\pmb{B}$~physics}
\addtocounter{subsection}{1}
\addcontentsline{toc}{subsection}{\numberline{6.1}$B$~physics}

The following $B$~physics processes have been studied in the NMSSM:
$BR(\bar{B} \to X_s \gamma)$ in \cite{Hiller:2004ii,Domingo:2007dx},
$BR(\bar{B}_s\to\mu^+\mu^-)$ and $\Delta M_{s (d)}$ in
\cite{Hiller:2004ii,Domingo:2007dx,Hodgkinson:2008qk}, 
$\bar{B}^+\to\tau^+\nu_{\tau}$ in \cite{Domingo:2007dx}, $BR(\bar{B} \to
X_s l^+ l^-)$ in \cite{Hiller:2004ii,Heng:2008rc}, and $BR(\bar{B} \to
\gamma l^+ l^-)$ in \cite{Heng:2008rc}. 

In most cases, the NMSSM specific contributions to the $B$~physics
observables originate essentially from a possibly light CP-odd Higgs
scalar $A_1$, and the fact that a flavour violating vertex
$b$-$q$-$A_1$ (with $q=s,d$) is generated, amongst others, by
squark-chargino loops (see, e.\,g., \cite{Buras:2002vd,Hiller:2004ii}).
Here and below, minimal flavour violation is assumed, i.\,e. the
only flavour violation originates from the Yukawa sector and is
parametrized by the CKM matrix. For large $\tan\b$ and $|A_t|$, the
loop-generated flavour violating vertex $b$-$q$-$A_1$ is
roughly proportional to $\tan^2\b\, |A_t|$; as a consequence, the NMSSM
specific contributions induced by a light CP-odd Higgs scalar $A_1$
increase strongly with $\tan\b$ (similar to most of the other MSSM-like
contributions). Subsequently we briefly discuss the various $B$~physics
processes.

\bigskip
\noindent{\it $BR(\bar{B} \to X_s \gamma)$ }
\medskip

The branching ratio $BR(\bar{B} \to X_s \gamma)$ is one of the most
intensively studied quantity in $B$~physics. In the past, constraints
from $b \to s\gamma$ have been particularly severe, since the
experimental world average for $BR(\bar{B} \to X_s \gamma)$ was somewhat
below the Next-to-leading order (NLO) SM prediction, while the
beyond the SM (BSM) contribution involving a charged Higgs boson in the
relevant diagram is positive. 

This situation has changed during the last years: the present world
average estimated by the Heavy Flavour Averaging Group
\cite{Barberio:2008fa} reads (for a lower cut on the photon energy
$E_\gamma > E_0\; =\; 1.6$~GeV)
\beq\label{6.1e}
\left.BR(\bar{B} \to X_s \gamma)\right|_\text{exp} = (3.52 \pm
0.23 \pm 0.09) \times 10^{-4}\;.
\eeq

The SM Next-to-Next-to-Leading Order ${\cal O}(\alpha_s^2)$ corrections
to the total $BR(\bar{B} \to X_s \gamma)$ branching fraction have
recently been combined \cite{Misiak:2006zs,Misiak:2006ab}, giving
\beq\label{6.2e}
\left.BR(\bar{B} \to X_s \gamma)\right|_\text{SM} = (3.15 \pm
0.23)\times 10^{-4}\;.
\eeq

In \cite{Becher:2006pu} the cut $E_\gamma > 1.6$~GeV
on the photon energy has been treated differently, leading to an even
lower SM prediction:
\beq\label{6.3e}
\left.BR(\bar{B} \to X_s \gamma)\right|_\text{SM} = (2.98 \pm
0.26)\times 10^{-4}.
\eeq
This result can be interpreted as a (weak) hint for positive BSM
contributions to $b \to s\gamma$; in any case constraints on the
parameter space of supersymmetric models have become less stringent.

In supersymmetric extensions of the SM, positive contributions to
$BR(\bar{B} \to X_s \gamma)$ arise from loops with charged Higgs bosons.
Additional contributions involving stop quarks and charginos are roughly
proportional to $\tan\b$ and to the trilinear coupling $A_t$, and can
have either sign. For an analysis of the corresponding constraints on
the parameter space of the MSSM see, e.g., \cite{Carena:2006ai}; in the
meantime, the full two-loop SUSY QCD corrections of the MSSM
contributions have been computed in \cite{Degrassi:2006eh}. 

NMSSM specific contributions to $BR(\bar{B} \to X_s \gamma)$ from the
extended Higgs and neutralino sectors arise only at the two-loop level.
These were considered in \cite{Hiller:2004ii} and, in more detail, in
\cite{Domingo:2007dx} with the result that the effects are negligibly
small, at least for a fixed charged Higgs mass (which can be somewhat
lower in the NMSSM due to the negative term ~$\sim \l^2$ in
(\ref{2.29e})). Even contributions from a light CP-odd Higgs $A_1$ are
not very important here, since $A_1$ appears only in loops and not in a
possibly resonant $s$-channel.

Also in the case of $\bar{B}^+\to\tau^+\nu_{\tau}$ the NMSSM specific
effects are merely indirect: this process is dominated by $W^+$ exchange
at tree level in the SM; in the MSSM and NMSSM, additional contributions
come from charged Higgs exchange, which can be somewhat lighter in the
NMSSM.

\bigskip
\noindent{\it $BR(\bar{B}_s\to\mu^+\mu^-)$}
\medskip

The most up-to-date experimental result on this process is an upper
bound (at $95\%$ confidence level) from CDF\ \cite{cdfmumunew}:
\begin{equation}\label{6.4e}
\left.BR(\bar{B}_s \to \mu^+ \mu^-)\right|_\text{exp} < 5.8\times
10^{-8}\;.
\end{equation}
This is still one order of magnitude beyond the SM estimate
\cite{Buchalla:1995vs}
\begin{equation}\label{6.5e}
\left.BR(\bar{B}_s \to \mu^+ \mu^-)\right|_\text{SM} = (3.8 \pm
0.1)\times 10^{-9}\;,
\end{equation}
so that there is some room for potentially large new physics
contributions.

\begin{figure}[ht!]
\begin{center}
\includegraphics[scale=0.4,angle=-0]{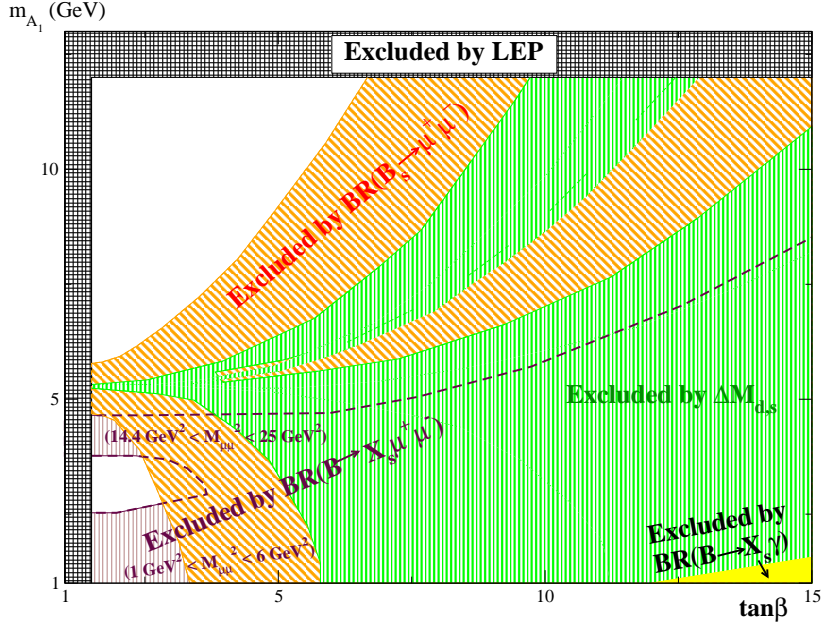}
\end{center}
\caption{Excluded regions in the $\tan\beta$-$M_{A_1}$ plane for
$A_t=-2500$~GeV: the gridded region is excluded by LEP, the green
(vertically shaded) region is excluded by $\Delta M_{d,s}$, the brown
(diagonally shaded) region by $BR(\bar{B}_s\to\mu^+\mu^-)$, the region
inside the dashed lines by $BR(\bar{B}\to X_s\mu^+\mu^-)$ with
$1~\mbox{GeV}<M_{\mu^+\,\mu^-}<\sqrt{6}~\mbox{GeV}$ or
$\sqrt{14.4}~\mbox{GeV}$ $<M_{\mu^+\,\mu^-}<m_b$, and the lower yellow
corner at $\tan\b \gsim 12$ and small $M_{A_1}$ by $BR(\bar{B} \to X_s
\gamma)$ (from \cite{Domingo:2007dx}).}
\label{fig:bsgfig8}
\end{figure}

The SM contributions originate from box and gauge-penguin diagrams,
which are small with respect to the experimental bound. The
corresponding MSSM contributions can be found in \cite{Buras:2002vd} and
have been generalized to the NMSSM in
\cite{Hiller:2004ii,Domingo:2007dx,Hodgkinson:2008qk}. Here, the
flavour changing $b$-$s$-$A_1$ vertex $\sim\tan^2\b$ can lead to
significant contributions arising from penguin diagrams, notably if the
CP-odd scalar $A_1$ in the $s$-channel (decaying into $\mu^+\,\mu^-$) is
close to its mass shell, i.e. if its mass is close to $M_{\bar{B}_s}$.
(Since the $A_1$-$\mu^+$-$\mu^-$ vertex is
proportional to $\tan\b$, the $A_1$ contribution to $BR(\bar{B}_s \to
\mu^+ \mu^-)$ is proportional to $\tan^6\b$.) In fact, CP-odd Higgs
masses equal to $M_{\bar{B}_s}$ are always excluded by the experimental
bound on this process, but the width of the forbidden region depends on
$\tan\b$ and $|A_t|$. An example for a forbidden region in the
($M_{A_1}$, $\tan\b$)-plane is shown in Fig. \ref{fig:bsgfig8} below
(taken from \cite{Domingo:2007dx}); for a study at very large $\tan\b
\sim 50$, see \cite{Hodgkinson:2008qk}.

\bigskip
\noindent{\it $\Delta M_{s (d)}$}
\medskip

Information on the mass differences $\Delta M_{s,d}\equiv
m_{\bar{B}_{s,d}}-m_{B_{s,d}}$ originates from measurements of $B$~meson
oscillations. The present result for $\Delta M_s$ obtained by the CDF
collaboration~\cite{cdf} is
\begin{equation}
\Delta M_s^\text{exp}= 17.77 \pm 0.12\ \text{ps}^{-1}\; .
\end{equation}
The SM prediction depends on the hadronic factor $f_{B_s}$ and CKM
matrix elements; for a comparison with BSM contributions, the CKM matrix
elements should be determined by tree level measurements such that their
values are not ``polluted'' by BSM effects \cite{ball}. Using
$|V_{ts}^* V_{tb}^{\phantom{*}}| = (41.3 \pm 0.7)\times 10^{-3}$ and
$f_{B_s} \sqrt{\hat{B}_{B_s}} = 0.281 \pm 0.021$~GeV given by the HPQCD
collaboration \cite{dalgic}, one obtains \cite{Domingo:2007dx}
\begin{equation}
\Delta M_s^\text{SM} = 20.5 \pm 3.1\ \text{ps}^{-1}\;.
\end{equation}
Hence, despite the large uncertainty, a negative contribution from new
physics seems favoured.

The measurement of $\Delta M_d$ is also quite precise 
\cite{hfagnew}:
\begin{equation}
\Delta M_d^\text{exp} = 0.507 \pm 0.004\ \text{ps}^{-1}\; .
\end{equation}
However, again the SM prediction suffers from large uncertainties:
\begin{equation}
\Delta M_d^\text{SM} =0.59 \pm 0.19\ \text{ps}^{-1}\;,
\end{equation}
obtained with a tree level determination of $|V_{td}^*
V_{tb}^{\phantom{*}}| = (8.6 \pm1.4)\times 10^{-3}$ \cite{ball} and a
hadronic factor $f_{B_d} \sqrt{\hat{B}_{B_d}}$ calculated from $f_{B_s}
\sqrt{\hat{B}_{B_s}}/ \left(f_{B_d} \sqrt{\hat{B}_{B_d}}\right) = 1.216
\pm 0.041$~\cite{okamoto}.

Supersymmetric contributions arise from box diagrams at the one-loop
level, but also from double-penguin diagrams involving two 
flavour changing vertices like $b$-$s(d)$-$A_1$ 
\cite{Hiller:2004ii,Domingo:2007dx,Hodgkinson:2008qk}, which are
particularly important if $A_1$ is exchanged in the $s$-channel. Now the
$A_1$-contribution is proportional to $\tan^4\b$, and leads again
to an exclusion of $A_1$~masses near $M_B \sim 5$~GeV, as shown in 
Fig.~\ref{fig:bsgfig8}. In the MSSM, a relation between
$\Delta M_{s (d)}$ and $BR(\bar{B}_s\to\mu^+\mu^-)$ at large $\tan\b$
can be deduced \cite{Buras:2002wq}; as pointed out 
in~\cite{Hiller:2004ii}, this relation is spoiled by the
$A_1$-exchange diagrams in the $s$-channel in the NMSSM.

\bigskip
\noindent{\it$BR(B\to X_sl^+l^-)$ }
\medskip

This branching ratio (for $l=e$ or $\mu$) has been measured by Babar
\cite{Aubert:2004it} and Belle \cite{Iwasaki:2005sy}. In order to avoid
contributions from $c$-quark resonances, the regions 
$1~\mbox{GeV}^2<M^2_{l^+l^-}<6~\mbox{GeV}^2$ (low) and
$14.4~\mbox{GeV}^2<M^2_{l^+l^-}<25~\mbox{GeV}^2$ (high) are considered
separately:
\begin{align}
BR(\bar{B}\to X_s l^+ l^-)_{\mathrm{low}}=
\begin{cases}
(1.493\pm0.504^{+0.411}_{-0.321})\times 10^{-6} \ \ &\mbox{Belle}\\
(1.8\pm0.7\pm0.5)\times 10^{-6}\ &\mbox{BaBar}\\
(1.60\pm 0.50)\times 10^{-6}&\mbox{average} 
\end{cases}
\end{align}
\begin{align}
BR(\bar{B}\to X_s l^+ l^-)_{\mathrm{high}}=
\begin{cases}
(0.418\pm0.117^{+0.061}_{-0.068})\times 10^{-6} \ \ &\mbox{Belle}\\
(0.5\pm0.25^{+0.08}_{-0.07})\times 10^{-6}\ &\mbox{BaBar}\\
(0.44\pm 0.12)\times 10^{-6}&\mbox{average} 
\end{cases}
\end{align}

The SM analysis has become quite refined (for a review, see
\cite{Hurth:2008jc}):
\begin{align}
\begin{cases}
BR(\bar{B}\to X_s \mu^+
\mu^-)_{\mathrm{low}}^\text{SM}&=(1.59\pm0.11)\times 10^{-6}\\
BR(\bar{B}\to X_s \mu^+
\mu^-)_{\mathrm{high}}^\text{SM}&=2.40\times10^{-7}(1^{+0.29}_{-0.26})
\end{cases}\\
\begin{cases}
BR(\bar{B}\to X_se^+e^-)_{\mathrm{low}}^\text{SM}&=(1.64\pm0.11)\times
10^{-6}\\
BR(\bar{B}\to X_s e^+
e^-)_{\mathrm{high}}^\text{SM}&=2.09\times10^{-7}(1^{+0.32}_{-0.30})
\end{cases}
\end{align}
These values are well within $2\,\sigma$ of the experimental
measurements.

In the NMSSM, the process $B\to X_sl^+l^-$ is also sensitive to a light
CP-odd scalar \cite{Hiller:2004ii,Heng:2008rc}: as in the case of the
$BR(\bar{B}_s\to\mu^+\mu^-)$, the flavour changing $b$-$s$-$A_1$ vertex
proportional to $\tan^2\b$ (with $A_1$ decaying into the $l^+ l^-$ pair)
can lead to significant contributions, and again exclude regions of
light $A_1$ masses as shown in Fig.~\ref{fig:bsgfig8}. The contributions
to the total
branching ratio ($BR(B\to X_s\mu^+\mu^-)^\text{exp} = (4.3 \pm
1.2)\times 10^{-6}$ according to \cite{Yao:2006px}) were studied in
\cite{Heng:2008rc}, but only the dependency on $\tan\b$ was shown
explicitly. Other processes such as $BR(\bar{B}\to X_s\tau^+\tau^-)$
and $BR(\bar{B}_s\to l^+l^-\gamma)$ are interesting for
Higgs phenomenology as well \cite{Heng:2008rc}, but here experimental
data is not yet available.

\bigskip
\noindent{\it Combined constraints from B~physics}
\medskip

The combined constraints from $BR(\bar{B}_s\to\mu^+\mu^-)$, $\Delta
M_{d,s}$ and $BR(\bar{B} \to X_s \gamma)$
on the mass of a light CP-odd scalar in the NMSSM have been
studied in \cite{Domingo:2007dx}, focusing on the dependency
on parameters like $\tan\b$, $M_{A_1}$ and $A_t$. (In addition, LEP
constraints on the Higgs sector were taken into account.) In
Fig.~\ref{fig:bsgfig8} we show the excluded domains for the mass of a
light CP-odd Higgs scalar as a function of $\tan\b$ for $A_t =
-2500$~GeV where, in addition, constraints from $BR(\bar{B}\to
X_s\mu^+\mu^-)$ are indicated.

As mentioned before, constraints from $BR(\bar{B}_s\to\mu^+\mu^-)$ and
$\Delta M_{d,s}$ exclude domains for a mass of a light CP-odd Higgs
scalar around $M_B \sim 5$~GeV for all $\tan\b$. Apart from the visible
strong increase of the
excluded region with $\tan\b$, the excluded region decreases for smaller
values of $|A_t|$ due to the smaller flavour violating $A_1$-quark
couplings induced by stop-chargino loops. 

Constraints from $BR(\bar{B}\to X_s\mu^+\mu^-)$ from
$1~\mbox{GeV}<M_{\mu^+\,\mu^-}<\sqrt{6}~\mbox{GeV}$ and
$\sqrt{14.4}~\mbox{GeV}$ $<M_{\mu^+\,\mu^-}<m_b$ exclude always masses
of a light CP-odd Higgs scalar inside these domains (where $A_1$ would
be on-shell). For larger $\tan\b$, larger regions of $M_{A_1}$ are
excluded by this observable, although constraints from $\bar{B}_s\to
\mu^+\mu^-$ are typically more significant.

\subsection*{6.2\quad $\pmb{\Upsilon}$ and $\pmb{\eta_b}$ physics}

\addtocounter{subsection}{1}
\addcontentsline{toc}{subsection}{\numberline{6.2}${\Upsilon}$ 
and ${\eta_b}$ physics}

\medskip
\noindent{\it Radiative $\Upsilon$ decays}
\medskip

Light CP-odd scalars could also be produced in radiative $\Upsilon$
decays $\Upsilon(nS)\to\gamma\, A_1$ with $A_1\to l^+l^-$, $l=\mu,\tau$
\cite{Hiller:2004ii,McElrath:2005bp,Dermisek:2006py,Fullana:2007uq,
Mangano:2007gi, Hodgkinson:2008ei,McKeen:2008gd,Domingo:2008rr,
Domingo:2009tb}. The
branching ratio depends essentially, apart from $M_{A_1}$, on the
coupling of $A_1$ to $b$-quarks. It is useful to introduce a reduced
coupling $X_d$, which denotes the coupling of $A_1$ to $b$-quarks
relative to the corresponding coupling of the SM Higgs boson. Using the
decomposition of the mass eigenstate $A_1$ according to (A.2) in
Appendix~A (where $P'_{ij}$ are the elements of an orthogonal $2\times
2$ matrix) and the $\b$-dependence of $h_b$ as in (A.11), one obtains
\beq\label{6.14e}
X_d = \tan\b P'_{11}
\eeq
where $P'_{11}$ can be of ${\cal O}(1)$. (Sometimes $P'_{11}$ is denoted
as $\cos\t_A$.) Hence, for $\tan\b \gg 1$, $X_d$ can also be much
larger than 1.

The $BR \left(\Upsilon(1S) \to \gamma A_1\right)$ is given by the
Wilczek formula~\cite{Wilczek:1977zn}
\beq\label{6.15e}
\frac{BR \left(\Upsilon(1S) \to \gamma A_1\right)}
{BR \left(\Upsilon(1S) \to \mu^+\mu^-\right)} =
\frac{G_F m_b^2 X_d^2}
{\sqrt{2}\pi\alpha}\biggl(1-\frac{m_{A_1}^2}{m_{\Upsilon(1S)}^2}\biggr)
\times  F 
\eeq
where $\alpha$ denotes the fine structure constant, and $F$ is a
correction factor which includes three kinds of corrections to the
leading-order Wilczek formula (the relevant formulae are summarised in
\cite{guide}): bound state, QCD and relativistic corrections.

The CLEO collaboration recently published their negative results on the
search for a CP-odd scalar (with a width below $10$~MeV) \cite{:2008hs}.
Using the Wilczek formula (\ref{6.15e}), these results allow to put
upper limits on $X_d$ \cite{Dermisek:2006py,Domingo:2008rr}. 

Two caveats must be mentioned, however: first, the various ({e.g.}
relativistic) corrections parametrized by $F$ in (\ref{6.15e}) become
large and unreliable for $m_{A_1}\gsim 8.8$~GeV, where $F$ vanishes if
the presently known corrections are extrapolated na\"ively. Second, the
experimental search assumed a width of $A_1$ below $\sim 10$~MeV which
could be violated for $m_{A_1}$ close to the $\Upsilon$ mass and/or for
very large $X_d$.

Nevertheless, CLEO results impose strong constraints on CP-odd scalars
with masses below $\sim 8.8$~GeV, which are shown as function of
$m_{A_1}$ and $X_d$ (together with other constraints) in
Fig.~\ref{fig:xd_ma1} in Section~\ref{sec:6.4}.

\bigskip
\noindent{\it Mixing between the light CP-odd Higgs and $\eta_b$
states}
\medskip

If the mass of the CP-odd Higgs is above $\sim 9$~GeV (as favoured by
CLEO constraints), but below the $B \bar{B}$ threshold of $\sim
10.5$~GeV, a mixing between CP-odd hadronic resonances $\eta_b(nS)$ and
the $A_1$ can become relevant in case of a large coupling $X_d$
\cite{Drees:1989du,SanchisLozano:2004gh, Fullana:2007uq,
Domingo:2009tb}. Such a mixing could have a direct impact on the masses
of the $\eta_b$ system. 

The BABAR collaboration has recently determined the $\eta_b(1S)$ mass
$m_{\eta_b(1S)}$ with an error of only a few MeV from radiative decays
$\Upsilon \to \gamma\,\eta_b$ of excited $\Upsilon$ states and the
observation of peaks in the photon energy spectrum. The result from
$\Upsilon(3S)$ decays is $m_{\eta_b(1S)} = 9388.9^{+3.1}_{-2.3}\
(\mathrm{stat}) \pm 2.7\ (\mathrm{syst})$~MeV \cite{:2008vj}, and the
result from $\Upsilon(2S)$ decays is $m_{\eta_b(1S)} =
9392.9^{+4.6}_{-4.8}\ (\mathrm{stat}) \pm 1.9\ (\mathrm{syst})$~MeV
\cite{:2009pz}. The average gives \cite{:2009pz}
$m_{\eta_b(1S)} = 9390.9 \pm 3.1\ \mathrm{MeV}$,
implying a hyperfine splitting $E_{hfs}(1S) = m_{\Upsilon(1S)} - 
m_{\eta_b(1S)}$ of 
\beq
E_{hfs}^\text{exp}(1S) = 69.9 \pm 3.1\ \mathrm{MeV}\; .
\eeq

This result can be compared to predictions from QCD. Recent results
based on perturbative QCD are in good agreement with each other and give
$E_{hfs}^\text{QCD}(1S) = 44 \pm 11\ \mathrm{MeV}$
\cite{Recksiegel:2003fm} and $E_{hfs}^\text{QCD}(1S) = 39 \pm 14\
\mathrm{MeV}$ \cite{Kniehl:2003ap} (see \cite{Penin:2009wf} and refs.
therein; in phenomenological models as quark models $E_{hfs}(1S)$ varies
over a wider range \cite{Brambilla:2004wf}). The discrepancy between
$E_{hfs}^\text{exp}(1S)$ and $E_{hfs}^\text{QCD}(1S)$ could easily be
cured in the presence of a mixing of the observed $\eta_b$ with a CP-odd
scalar $A_1$ with a mass somewhat above $\sim 9.4$~GeV
\cite{Domingo:2008rr}.

On the one hand, a CP-odd Higgs scalar $A_1$ with a mass very close to
$9.389$~GeV (before mixing) \emph{and} a strong $\eta_b - A_1$ mixing is
excluded, since then the mass of the physical eigenstate (after mixing,
i.e. after the diagonalization of the $2\times 2$ mass matrix) could 
\emph{not} be given by $9.389$~GeV. This implies an upper bound on
$X_d$ for $M_{A_1}$ near $9.389$~GeV \cite{Domingo:2008rr}, which is
also shown in Fig.~\ref{fig:xd_ma1}.

\begin{figure}[ht!]
\begin{center}
\includegraphics*[width=0.55\linewidth]{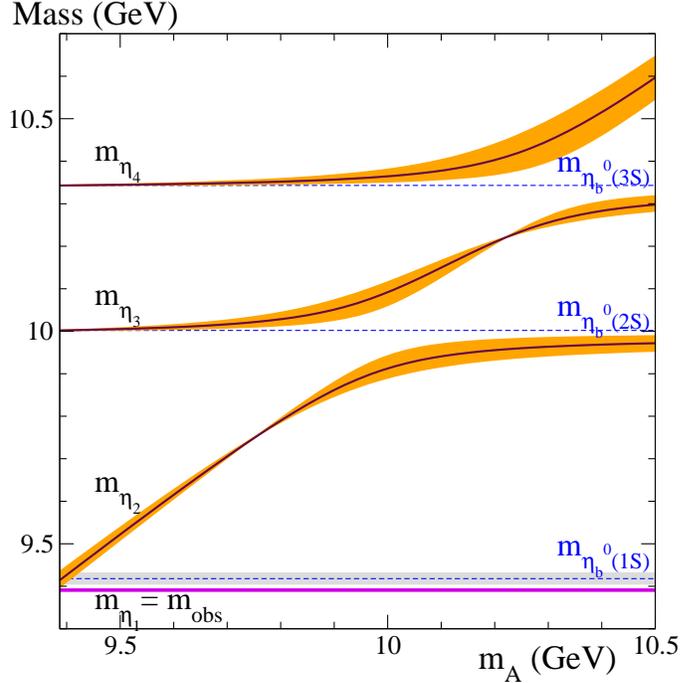}
\end{center}
\caption{Masses of all eigenstates $\eta_i$ of the $\eta_b(nS)- A_1$
system as function of $M_{A_1}$, once $m_{\eta_1}$
is forced to coincide with the BABAR result for $m_{\eta_b(1S)}$
(denoted by $m_\text{obs}$; from~\cite{Domingo:2009tb}).}
\label{fig:metamA3}
\end{figure}

On the other hand, a CP-odd Higgs scalar $A_1$ with a mass below
10.5~GeV can mix with all $\eta_b(nS)$ states ($n=1,2,3$), possibly
generating both a distorted spectrum as well as unusually large
branching ratios into $\tau^+\,\tau^-$ in the $\eta_b(nS)- A_1$ system
\cite{Domingo:2009tb}. Assuming that the large observed hyperfine
splitting $E_{hfs}^{exp}(1S)$ compared to $E_{hfs}^{QCD}(1S)$ is induced
by a mixing with $A_1$, $X_d$ can be determined as a function of
$M_{A_1}$ (leading to an increasing $X_d$ with $M_{A_1}$), and
predictions for the masses of the $\eta_b(nS)- A_1$ system can be made
(within errors). Denoting the 4~eigenstates of the $\eta_b(nS)- A_1$
system by $\eta_i$, $i=1\dots 4$, these masses are shown in
Fig.~\ref{fig:metamA3} as function of $M_{A_1}$ \cite{Domingo:2009tb}.
($M_{A_1}$ is denoted by $m_{A}$ in Fig.~\ref{fig:metamA3}.)
For clarity we have indicated the masses $m_{\eta_b^0(nS)}$ before
mixing as horizontal dashed lines. For $M_{A_1} \lsim 9.8$~GeV, the
state $\eta_2$ has a large $A_1$ component, but for $M_{A_1} \gsim
9.8$~GeV its mass drops below $M_{A_1}$ due to the strong mixing with
$\eta_b(2S)$.

\subsection{The anomalous magnetic moment of the muon}
\label{sec:6.3}

The measurement of the anomalous magnetic moment of the muon
$a_{\mu}=(g_{\mu}-2)/2$ by the E821 experiment at BNL has reached a
level of precision which is sensitive to supersymmetric contributions.
The latest experimental value obtained is \cite{Bennett:2006fi}:
\begin{equation}\label{6.17e}
a_{\mu}^\text{exp}=11\,659\,208.0(5.4)(3.3)\times 10^{-10}\;.
\end{equation}

However, the evaluation of $a_{\mu}$ in the SM suffers from a relatively
large uncertainty of the hadronic contribution (see, e.g.,
\cite{Davier:2007ua,Miller:2007kk,Jegerlehner:2007xe} and refs.
therein). If it is determined using data from $e^+e^-\to\mbox{hadrons}$,
one obtains a $\sim 3\,\sigma$ deviation between the SM and the
experimental value \cite{Davier:2007ua,Miller:2007kk,Jegerlehner:2007xe}
as, for instance, in \cite{Domingo:2008bb}:
\begin{equation}\label{6.18e}
a_{\mu}^\text{exp}-a_{\mu}^\text{SM}(e^+e^-) =(27.7\pm 9.3)\times
10^{-10}\;,
\end{equation}
where the leading uncertainty originates both from the experimental
measurement and the estimate of the hadronic contribution from $e^+e^-$
data. Hence, a positive contribution to $a_{\mu}$ seems desirable.

However, if the hadronic contribution to $a_{\mu}$ is estimated from
hadronic $\tau$ decays
\cite{Davier:2007ua,Jegerlehner:2007xe,Miller:2007kk,Zhang:2008pka}, the
discrepancy is reduced. The disadvantage of this
approach is that it relies on assumptions on the pion form factor,
isospin violating effects and vector meson mixings. On the other hand,
recent (preliminary) measurements using the radiative return of
$e^+e^-\to\pi^+\pi^-$ seem in better agreement with the estimate from
hadronic $\tau$ decays \cite{Davier:2009ag,Davier:2009zi}, according to
which the discrepancy is reduced to $\sim 1.9\,\sigma$.

The various contributions to $a_{\mu}$ in MSSM-like supersymmetric
models are reviewed and summarised in \cite{Stockinger:2006zn}.
In the MSSM, the (dominant) one-loop contributions to $a_{\mu}$
originate from chargino/sneutrino or neutralino/smuon loops, which
increase linearly with $\tan\b$ and are positive for a positive
$\mu$-term of the MSSM. This phenomenon persists in the NMSSM with $\mu$
replaced by $\mu_{\mathrm{eff}}$; the effects of the extended neutralino
sector in the NMSSM are small \cite{Konig:1991tr}. 

Studies of $a_{\mu}$ in the NMSSM including a possibly light CP-odd
Higgs scalar were performed in 
\cite{Gunion:2005rw,Domingo:2008bb,Gunion:2008dg}; details of the
necessary generalizations of all MSSM-like one- and two-loop
contributions (see \cite{Stockinger:2006zn} and refs. therein) are given
in \cite{Domingo:2008bb}. The relevance of light CP-odd Higgs scalars
for $a_{\mu}$ was actually first discussed in the context of general
Two-Higgs-doublet extensions of the SM
\cite{Chang:2000ii,Cheung:2001hz,Krawczyk:2002df}: it is remarkable that
positive contributions to $a_\mu$ from two-loop diagrams involving a
closed fermion loop can dominate the negative contributions from
one-loop Higgs-diagrams. The sum of both contributions from a light
$A_1$ to $a_\mu$ has a (positive) maximum for $M_{A_1}\sim 6$~GeV. Below
$M_{A_1}\sim 3$~GeV, the negative one-loop contribution dominates, which
increases the discrepancy with respect to the measurement. Since the
contributions of $A_1$ to $a_\mu$ are proportional to $X_d^2$ (its
reduced coupling to charged leptons), one obtains upper limits on $X_d$
for $M_{A_1}\lsim 3$~GeV from (\ref{6.18e}), even after allowing for
$2\,\sigma$ deviations, which are shown in Fig.~\ref{fig:xd_ma1} below.

\subsection{Combined constraints on a light CP-odd Higgs}
\label{sec:6.4}

\begin{figure}[ht!]
\begin{center}
\includegraphics[
width=9cm,height=9cm,
clip=,]{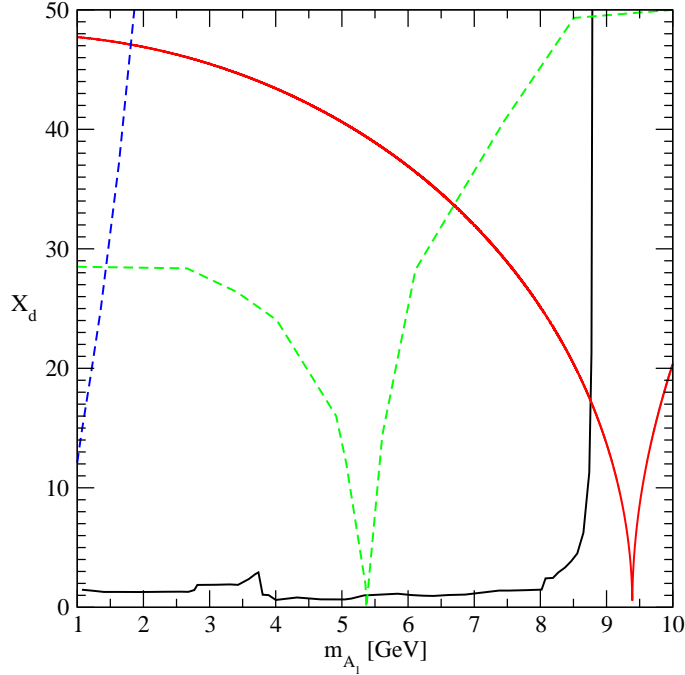} 
\caption{Upper
bounds on $X_d$ versus the $A_1$ mass for all parameters scanned over
(see \cite{Domingo:2008rr} for details). Indicated are constraints from
$B_s \to \mu^+\mu^-$ and $\Delta M_q$, $q=d,s$ as a green dashed line,
constraints from $a_\mu$ as a blue dashed line, the latest bounds
from CLEO on $BR\left(\Upsilon \to \gamma \tau^+ \tau^-\right)$ as a
black line and constraints due to the measured $\eta_b(1S)$ mass by
BABAR as a red line.}
\vspace*{-5mm}
\label{fig:xd_ma1}
\end{center}
\end{figure}

It is interesting to compare the constraints on a light $A_1$ in the
NMSSM from CLEO, $B$~physics, $a_\mu$ and from the measured
$\eta_b(1S)$ mass. Clearly, constraints from $B$~physics and $a_\mu$
(notably at large $\tan\b$) also depend on MSSM-like parameters as
$M_{H^\pm}$, chargino and slepton masses, which we will not review
here and for which we refer to the corresponding literature. To some
extent, the MSSM-like parameters of the NMSSM also affect the
constraints from $B$~physics and $a_\mu$ on a light $A_1$ in the
NMSSM. It is possible, however, to scan over these parameters and
obtain regions in the ($X_d$, $A_1$)-plane, which are \emph{always}
excluded. The results are shown in Fig.~\ref{fig:xd_ma1} taken
from \cite{Domingo:2008rr}, where more details can be found.

We learn from Fig.~\ref{fig:xd_ma1} that, for $M_{A_1} \lsim 9$~GeV,
CLEO constraints are generally stronger than those from $B_s \to
\mu^+\mu^-$, $\Delta M_q$ ($q=d,s$) and $a_\mu$, with the exception
of a small window for $M_{A_1}$ near $M_B \sim 5$~GeV. In the range
9~GeV~$\lsim M_{A_1} \lsim 10$~GeV, constraints due to the measured
$\eta_b(1S)$ mass by BABAR play a r\^ole, but we recall that $m_{A_1}$
slightly above 9.4~GeV can even have a desirable effect in the form of a
negative shift of the $\eta_b(1S)$ mass, which would be interpreted as
an unusually large $\Upsilon(1S)-\eta_b(1S)$ hyperfine splitting. In
addition, constraints from LEP2 on a light CP-odd scalar below
the $B-\bar{B}$ threshold are quite weak (cf. Section~\ref{sec:5}),
which makes this mass range particularly interesting.

Future searches for a light $A_1$ in this mass range in radiative
$\Upsilon$ decays will be challenging due to large backgrounds and the
softness of the emitted photon. This problem can be overcome
if one studies a breakdown of lepton universality in the inclusive
leptonic decays of the $\Upsilon$ system as advocated in 
\cite{SanchisLozano:2002pm,SanchisLozano:2003ha,Domingo:2008rr}: since
$A_1$ would decay dominantly into $\tau^+\;\tau^-$, the branching
fraction into $2\tau$ of an $\Upsilon(nS)$ state, decaying partially via
$A_1$, would be enhanced.

For a very light $A_1$ ($M_{A_1} \lsim 1$~GeV), the decay $A_1 \to
\mu^+\,\mu^-$ could be dominant. The CLEO bounds mentioned above are
valid for $A_1$ masses down to $\sim 250$~MeV, but the recent BABAR
constraints from the absence of dimuon signals in $\Upsilon(3S),\ 
\Upsilon(2S)\to \gamma\,A$ can be more relevant. These read 
\cite{Aubert:2009cp}
\beq\label{6.19e}
BR(\Upsilon(3S) \to \gamma\,A)\times BR(A \to \mu^+\,\mu^-) \lsim
5.2\times 10^{-6}\quad \mathrm{for} \quad  0.212 \leq M_A \leq 9.3\
\mathrm{GeV}
\eeq
and give \cite{Bai:2009ka}
\beq\label{6.20e}
X_d \lsim 0.4\quad \mathrm{if}\quad BR(A \to \mu^+\,\mu^-) \sim 1\; .
\eeq
Very light CP-odd scalars can also show up in $\Omega$,
$\Sigma$, $K$, $\eta$ and $\pi$ decays, which have been studied in
the NMSSM in \cite{Hiller:2004ii,He:2006fr,Mangano:2007gi,
Xiangdong:2007vv,He:2008zw,Chang:2008np,Bai:2009ka}. For
$M_{A_1} \lsim 360$~MeV, the decay $K^+ \to \pi^+ A_1$ is possible; for
$M_{K^+}-M_{\pi^+} \gsim M_{A_1} \gsim 2\,M_\mu$, the result of
the HyperCP collaboration \cite{Park:2001cv}
\beq\label{6.21e}
BR(K^+ \to \pi^+ \mu^+\mu^-) = 9.8 \pm 1.0 \pm 0.5 \times 10^{-8}
\eeq
implies $ X_d\left(1-\tan^{-2}\b\right) \lsim 0.06 $ \cite{Bai:2009ka}
within this specific range for $M_{A_1}$.

\section{Specific scenarios}

\subsection{The constrained NMSSM}
\label{sec:ss.1}

Even if one confines oneself to the $\mathbb{Z}_3$-invariant NMSSM with a
scale invariant superpotential, the number of independent parameters in
the Higgs sector at tree level (six according to (\ref{2.17e})) is
considerably larger than the two parameters of the Higgs sector of the
MSSM (typically chosen as $\tan\b$ and $M_A$). However, one can make the
assumption that the soft SUSY breaking terms are universal at a large
scale like the GUT scale (not very different from the Planck scale), as
it would be the case in mSUGRA with spontaneous supersymmetry breaking
in a hidden sector and the mediation of supersymmetry breaking to the
observable sector via flavour blind gravitational interactions
\cite{Nilles:1983ge}. The corresponding version of the
$\mathbb{Z}_3$-invariant NMSSM is denoted as the constrained NMSSM or
cNMSSM.

Here, as in the constrained MSSM (cMSSM), all soft scalar masses
squared, trilinear soft couplings and gaugino masses
are assumed to be given by common values $m_0^2$,  $A_0$ and 
$M_{1/2}$, respectively, at the GUT scale. In fact, the number of
independent parameters in the cNMSSM is the same as in the cMSSM: the
additional parameters $\mu$ and $B$ of the cMSSM \cite{Chung:2003fi} are
replaced by the Yukawa couplings $\l$ and $\k$ in the
superpotential~(\ref{2.6e}). Apart from the gauge and quark/lepton
Yukawa couplings, the Lagrangian of the cNMSSM depends on just five
parameters $m_0^2$, $A_0$, $M_{1/2}$, $\l$ and $\kappa$, and the correct
value of $M_Z$ reduces the dimension of the parameter space from five to
four.

It is possible that the gravitational couplings (in the K\"ahler
potential in supergravity) of the singlet $S$ in the NMSSM differ from
those of all other fields; then the singlet-dependent soft SUSY breaking
terms could differ from universality (at the GUT or Planck scale).
Therefore it is interesting to analyse models with relaxed universality
conditions of the soft terms, which are slightly more general than the
fully constrained NMSSM.

Below we will discuss first the cNMSSM, where the singlet-dependent soft
SUSY breaking terms $m_S$ and $A_\k$ are assumed to coincide with $m_0$
and $A_0$, respectively, at the GUT scale. Subsequently we will turn to
the cNMSSM with relaxed universality constraints, where the values of
$m_S$ and/or $A_\k$ at the GUT scale are allowed to differ from $m_0$
and $A_0$.

First analyses of the phenomenologically acceptable regions of the
cNMSSM parameter space \cite{Derendinger:1983bz, Ellis:1988er,
Drees:1988fc} were handicapped by the still unknown top quark mass,
which plays an important r\^ole due to the impact of the top Yukawa
coupling on the RG equations \cite{Derendinger:1983bz} and the radiative
corrections to the Higgs masses \cite{Ellwanger:1993hn}. Somewhat later
-- but still before all present LEP2 constraints were available -- the
phenomenologically acceptable regions in this parameter space were
investigated in \cite{Abel:1992ts,Ellwanger:1993xa,Elliott:1994ht,
Ellwanger:1995ru, King:1995vk, Ellwanger:1996gw} (see \cite{Brax:1994ae}
for an analysis of non-universal soft terms from orbifold string
theory, and \cite{Kraniotis:1995gk} for constraints on soft terms in
dilaton-dominated SUSY breaking scenarios).

LEP2 constraints on the parameters of the Higgs sector of the cNMSSM
have been analysed in~\cite{Ellwanger:1999ji,Ellwanger:2006rm}.
Concerning these,
it is important to recall that larger values of $\l$ do
\emph{not} necessarily imply an increase of the mass of the SM-like
CP-even Higgs scalar: as indicated in and discussed below (\ref{3.2e}),
the mixing of the SM-like Higgs with a heavy singlet-like Higgs (also
proportional to $\l$) leads to a decrease of the SM-like Higgs mass.
In the constrained parameter space of the cNMSSM, the off-diagonal
matrix elements cannot be fine tuned to~0. As a consequence, LEP
constraints on the Higgs sector lead to an \emph{upper} bound on $\l$
within the cNMSSM:
\beq\label{ss.1e}
\l \lsim 0.3\;.
\eeq
Then the lower bound on $|\mu_\mathrm{eff}| \equiv |\l s|$ of 
$|\mu_\mathrm{eff}| \gsim 100$~GeV from the non-observation of a
chargino lighter than $\sim$~100~GeV at LEP implies
\beq\label{ss.2e}
|s| \gsim 300\ \mathrm{GeV}\; .
\eeq

For $|s| \gg v_u,\,v_d$, the dominant $s$-dependent terms in the
potential (\ref{2.11e}) are given by (\ref{2.19e}) leading to the
condition (\ref{2.20e}) for an absolute minimum with $s \neq 0$
\cite{Derendinger:1983bz,Ellwanger:1996gw}:
\beq\label{ss.3e}
A_\k^2 \gsim 9\, m_{S}^2\; .
\eeq

For small $\l$ and hence small $\k$ from vacuum
stability~\cite{Ellwanger:1996gw}, the parameters $A_\kappa$ and $m_S$
are hardly renormalised between the GUT and the electroweak scales in
the cNMSSM, and the above condition translates to (assuming $m_0^2
\geq 0$)
\begin{equation}\label{ss.4e}
m_0 \,\lesssim\, \frac{1}{3} |A_0|\;.
\end{equation}

Next, we consider the CP-odd Higgs boson mass matrix (\ref{2.27e}) whose
diagonal matrix element ${\cal M}_{P,22}^2 \sim -3\k A_\k s$ (for large
$s$ as relevant here) must be positive. For positive $s$ and $\kappa$
this implies negative trilinear couplings
\begin{equation}\label{ss.5e}
A_\kappa\, \sim\, A_0 < 0\;.
\end{equation}

In the region of the parameter space of the cNMSSM where $\l,\,\k \ll
1$, its phenomenology would be very close to the one of the cMSSM with
corresponding soft SUSY breaking terms, since
the additional Higgs and neutralino states of the NMSSM decouple and
would never be produced -- unless the additional singlino-like
neutralino is the LSP (see below). For larger $\l\sim 0.3$ and low
$\tan\b \sim 2$, the
mass of the lightest CP-even Higgs scalar with substantial couplings to
gauge bosons and fermions can be about 10~GeV larger than in the cMSSM.
However, the lightest CP-even Higgs scalar can also be singlet-like
\cite{Ellwanger:1993xa,Elliott:1994ht, Ellwanger:1995ru,
King:1995vk,Ellwanger:1996gw,Ellwanger:1999ji}, and escape LEP
constraints due to its reduced coupling to the $Z$~boson in spite of its
small mass. The search for such a Higgs scalar would prove to be quite
difficult at the LHC.

\subsubsection{Constraints from the absence of charge and colour
breaking minima}

Additional constraints on the parameter space follow from the absence of
charge and/or colour breaking (CCB) minima of the potential, where
squark and/or slepton vevs do not vanish. A dangerous direction in field
space is along the $D$-flat direction $|E_{R_1}|=|L_1|=|H_d|$ (with
sleptons $E_{R_1}$ and $L_1$ of the first generation with the smallest
Yukawa coupling $h_e$). A deeper minimum is avoided if
\beq\label{ss.6e}
\frac{1}{3}A_e^2 < m_{E_1}^2 + m_{L_1}^2 + m_{H_d}^2
\eeq
at the corresponding scale $A_e/h_e$ 
\cite{Derendinger:1983bz,Stephan:1997ds,
Ellwanger:1999bv}. (In the MSSM, an additional term $\mu^2$ has to be
added to the right-hand side of (\ref{ss.6e}) \cite{Frere:1983ag}; in
the NMSSM, $s$ and hence $\mu_\mathrm{eff}$ can vanish in the dangerous
minimum of the potential.) In principle, another dangerous direction in
field space corresponds to $|T_R|=|Q_3|=|H_u|$; however, the
corresponding constraints are never more relevant than the ones
following from (\ref{ss.6e}) in the cNMSSM.

In terms of the universal soft SUSY breaking parameters $A_0$, $M_{1/2}$
and $m_0$, (\ref{ss.6e}) at the scale $A_e/h_e$ becomes
\cite{Ellwanger:1999bv} (using the corresponding RG equations)
\beq\label{ss.7e}
\left(A_0-0.5\, M_{1/2}\right)^2 \lsim 9\, m_0^2+2.67 
M_{1/2}^2
\eeq
implying an upper bound on $|A_0|$.

Equally delicate could be the so-called unbounded-from-below (UFB)
directions in field space, which are both $D$-flat and $F$-flat. 
In~\cite{Stephan:1997ds,Ellwanger:1999bv}, it has
been clarified that such dangerous directions in the field space of the
MSSM are still present in the NMSSM, although the singlet vev $s$ gives
an additional positive contribution to the potential. Analytic
approximations to the potential along such dangerous directions have
been studied in \cite{Abel:1998ie}, with the conclusion that the
inequality 
\beq\label{ss.8e}
m_0 \gsim (0.3 - 1.0)\, M_{1/2}
\eeq 
(where 1.0 corresponds to low $\tan\b$, 0.3 to large $\tan\beta$) is an
approximate condition for the absence of deeper minima in these
directions. However, since the decay rate of the standard vacuum is
usually much larger than the age of the universe \cite{Abel:1998ie},
(\ref{ss.8e}) can be violated if we assume that the early cosmology
(temperature-induced positive masses squared for the squarks and
sleptons) places us into the local standard minimum of the scalar
potential. 

\subsubsection{The constrained NMSSM with dark matter constraints}

Next one can require that the LSP of the cNMSSM provides the correct
dark matter relic density, see Section~\ref{sec:dm}. Early studies of correspondingly allowed
regions of the cNMSSM parameter space have been performed in
\cite{Abel:1992ts,Stephan:1997rv}; in~\cite{Stephan:1997ds}, the bounds
(\ref{ss.7e}) and (\ref{ss.8e}) from the absence of CCB and UFB minima
were taken into account. In the meantime, constraints on Higgs and
sparticle masses as well on the dark matter relic density
\cite{Spergel:2006hy,Tegmark:2006az} have become tighter (and
RGEs/radiative corrections are known to a higher accuracy), with the
result that the regions in the parameter space of the cNMSSM considered
in these early studies are no longer phenomenologically viable.

As discussed above, the UFB constraint (\ref{ss.8e}) can in principle be
violated, and in the cNMSSM this is necessary in view of the up-to-date
experimental constraints~\cite{Djouadi:2008yj,Djouadi:2008uj}. On the
other hand, the constraint on $m_0$ (\ref{ss.4e}) (from $s \neq 0$) has
to be respected in the cNMSSM, and it turns out that only regions where
$m_0 \ll M_{1/2}$ are phenomenologically viable.

In the cMSSM \cite{Chung:2003fi}, small values of $m_0 \lsim
\frac{1}{5}M_{1/2}$ result in a (charged) stau LSP $\widetilde{\tau}_1$,
which is excluded. In the cNMSSM, the additional
singlino-like neutralino $\chi^0_1$ can still be lighter than the
lightest stau and be the true LSP -- this allows for very small (or
vanishing) values of $m_0$. However, the $\chi^0_1$ annihilation rate in
the early universe must be large enough in order to avoid an excess of
singlino-like dark matter. It turns out that, within the restricted
parameter space of the cNMSSM, no $s$-channel resonances (Higgs or $Z$)
with a mass twice the $\chi^0_1$ mass are present, which could enhance
the $\chi^0_1-\chi^0_1$ annihilation rate (see Section~\ref{sec:dm}).
The only possibility to reduce the $\chi^0_1$ abundance is then via
coannihilation with
$\widetilde{\tau}_1$ (now the NLSP), whose mass has to be just somewhat
above the $\chi^0_1$ mass (see Section~\ref{sec:dm}). Then the
$\chi^0_1$ abundance is reduced by the process $\chi^0_1 + X \to
\widetilde{\tau}_1 + X'$, and the
$\widetilde{\tau}_1 - \widetilde{\tau}_1$ annihilation cross section is
generally large enough. It should be mentionned that the singlino-like
neutralino $\chi^0_1$ would be practically invisible in direct or
indirect dark matter detection experiments in this scenario.

The condition that the mass of $\widetilde{\tau}_1$ is close to (just
somewhat above) the $\chi^0_1$ mass leads to $m_0 \lsim$ $\frac{1}{10}
M_{1/2}$ and $A_0 \sim -\frac{1}{4}M_{1/2}$
\cite{Djouadi:2008yj,Djouadi:2008uj}; with these restrictions on the
parameters, the LEP constraints on the Higgs sector are even more severe
and require
\beq\label{ss.9e}
\l \lsim 0.02\; .
\eeq
(As discussed in \cite{Djouadi:2008yj,Djouadi:2008uj} and in
Section~\ref{sec:dm}, the successful coannihilation of $\chi^0_1$ with
$\widetilde{\tau}_1$ implies a lower bound on $\l$ of $\sim 10^{-5}$.) 

With $A_0$ being determined in terms of $M_{1/2}$ and the strong upper
bounds on $m_0$ and $\l$, the spectrum of the cNMSSM is nearly
completely determined by $M_{1/2}$. $\tan\b$, which is no longer a free
parameter, comes out relatively large with $\tan\b > 25$.  

The SM-like Higgs mass is 115-120~GeV for $M_{1/2} \gsim 400$~GeV,
increasing with $M_{1/2}$. Within this range of $M_{1/2}$, all sparticle
masses satisfy lower bounds from direct searches and from
precision observables as those from $B$~physics.
If one requires that the SUSY contribution to the anomalous
magnetic moment of the muon explains the $\sim 3\,\sigma$ deviation
between the SM and the experimental value (see Section~\ref{sec:6.3}),
the value of $M_{1/2}$ should be below $\sim 1$~TeV
\cite{Djouadi:2008yj,Domingo:2008bb,Djouadi:2008uj}.

\begin{figure}[t!]
\begin{tabular}{cc}\hspace*{-6mm}\vspace*{-3mm}
\psfig{file=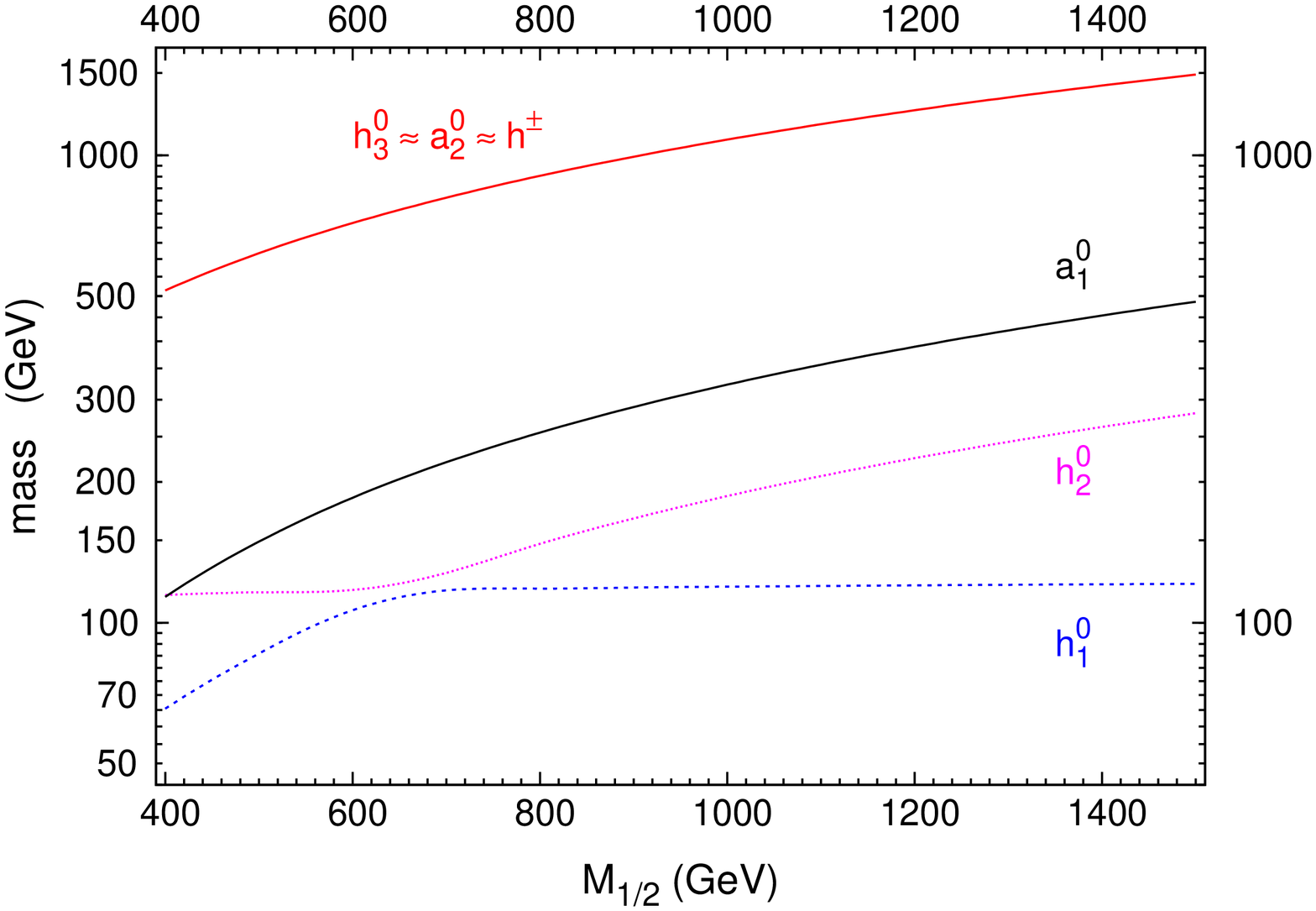, clip=, angle=270, 
width=82mm} \hspace*{-3mm}&\hspace*{-5mm}
\psfig{file=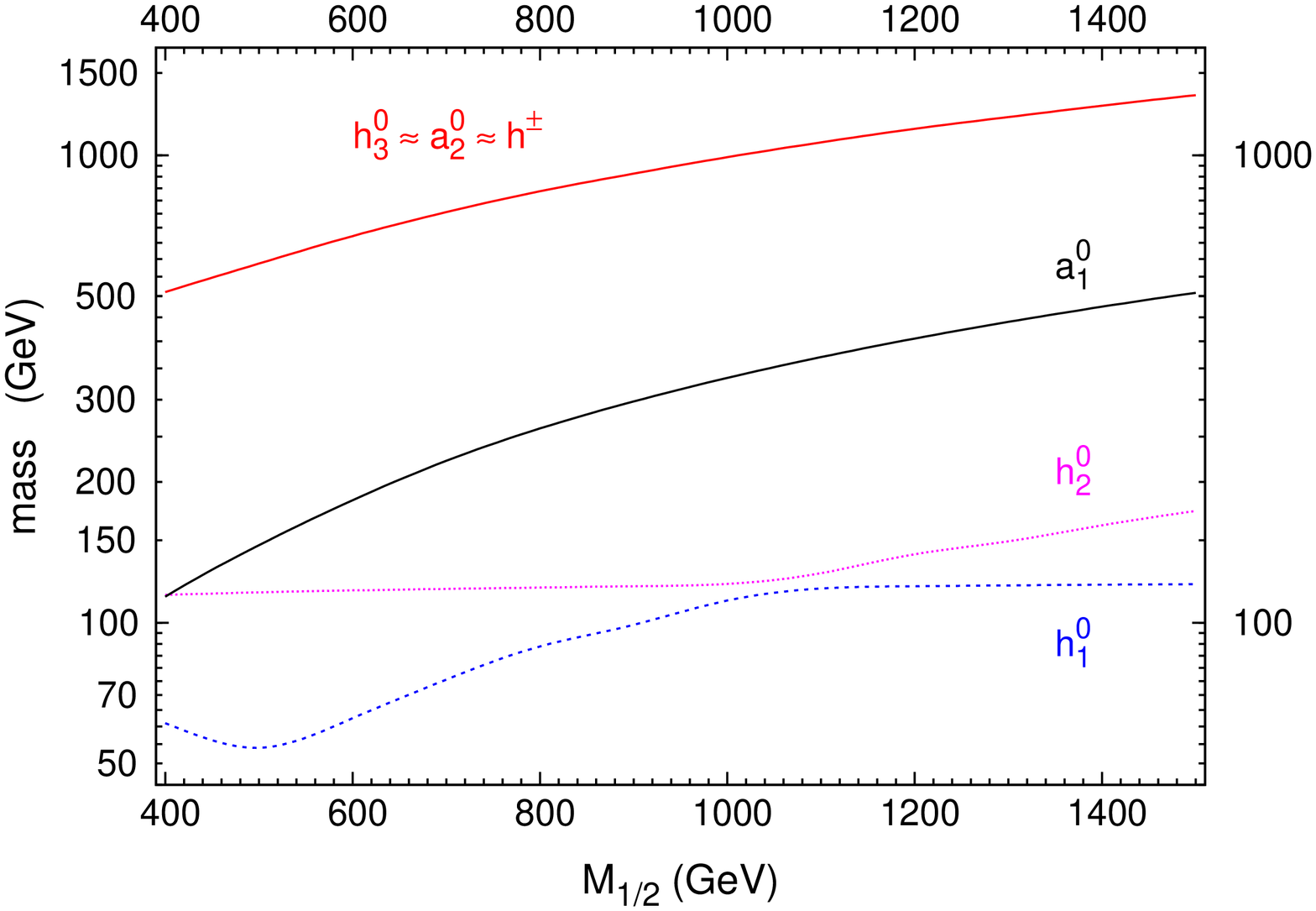, clip=, angle=270, 
width=82mm}
\end{tabular}
\caption{The Higgs masses in the cNMSSM as a function of $M_{1/2}$ (in
GeV). $m_0= 0$ in the left panel, while $m_0\sim M_{1/2}/10$ in the
right panel. From below, the displayed lines correspond to the states
$h^0_1$ (blue/dotted), $h^0_2$ (pink/dashed), $a^0_1$ (full/black) and
$a^0_2$ (full/red) which is degenerate with  the $h^0_3$ and $h^\pm$
states (from~\cite{Djouadi:2008uj}).}
\label{fig:m12:phenoH}
\end{figure}

In Fig.~\ref{fig:m12:phenoH} we display the masses of the neutral
CP-even, CP-odd, and charged Higgs bosons as a function of the parameter
$M_{1/2}$. On the left-hand side, we take $m_0=0$, while on the
right-hand side we take the maximally viable value (from the dark matter
relic density) $m_0 \sim \frac{1}{10}M_{1/2}$.

The CP-even Higgs boson with the dominant singlet component is the only
Higgs state whose mass depends -- apart from $M_{1/2}$ -- on $m_0$. For
small $M_{1/2}$ it is lighter than the SM-like Higgs boson, escaping LEP
constraints due to the very small coupling to the $Z$~boson. For
increasing values of $M_{1/2}$, its mass increases until it becomes
nearly degenerate with the SM-like CP-even Higgs state. In this region
of parameter space, the singlet-like and SM-like Higgs states strongly
mix; for a mass of the singlet-like Higgs state of $\sim 100$~GeV, the
excess of events at this mass observed at LEP (cf.
Section~\ref{sec:5.1.1}) could be explained~\cite{Djouadi:2008uj}.

\begin{figure}[b!]
\vspace*{-5mm}
\begin{tabular}{cc}\hspace*{-6mm}
\psfig{file=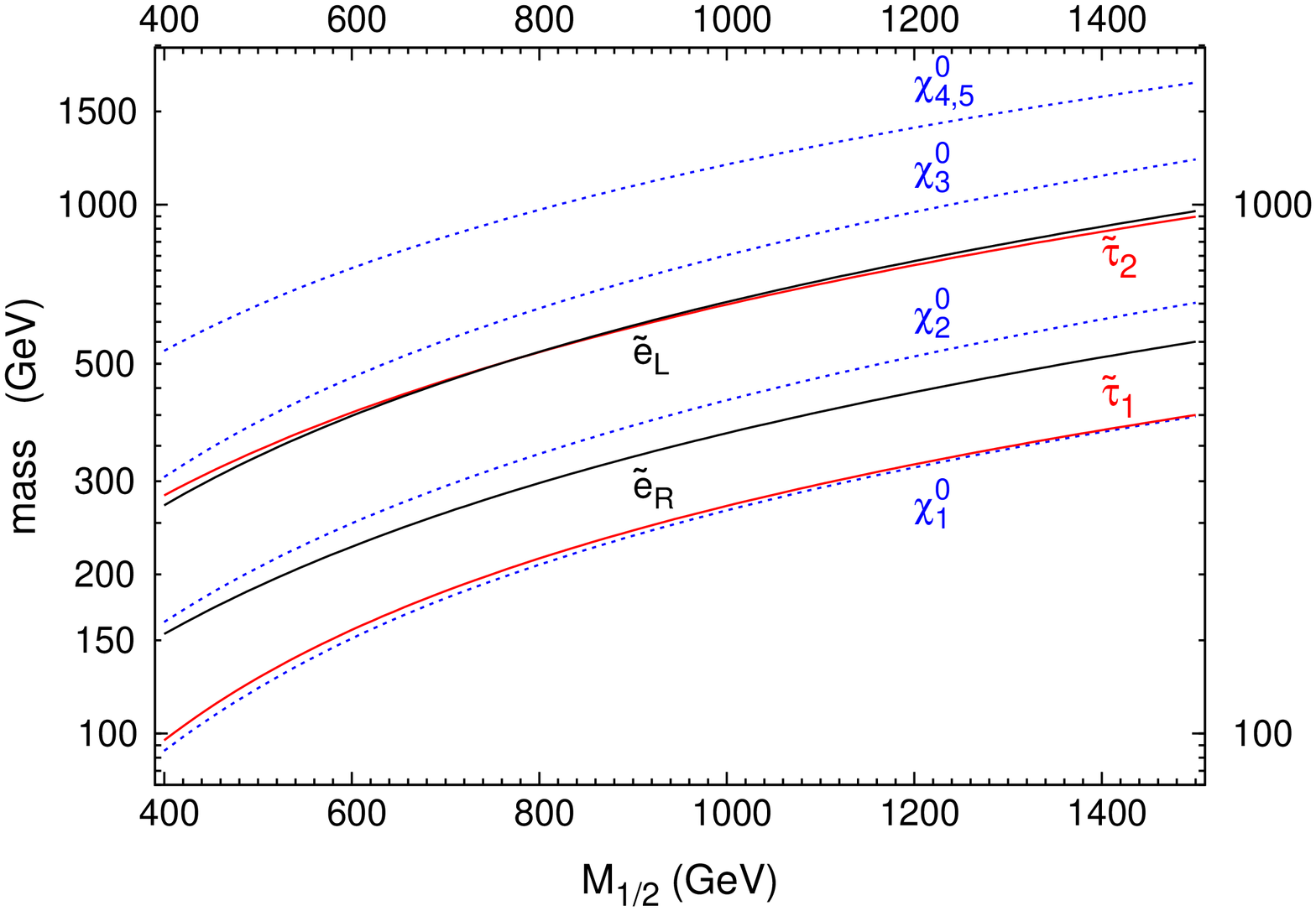, clip=, angle=270, 
width=83mm} \hspace*{-5mm}&
\hspace*{-5mm}
\psfig{file=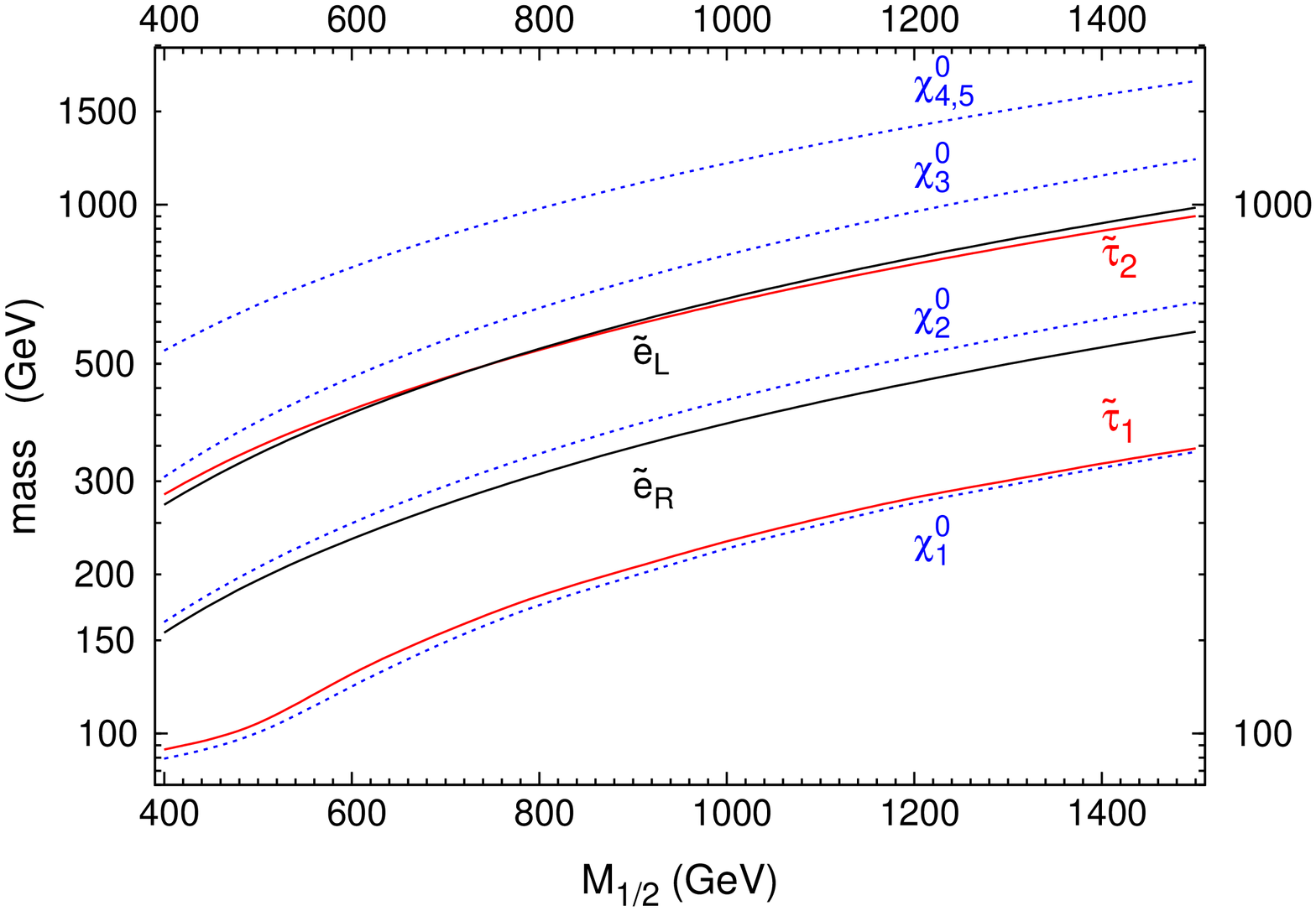, clip=, angle=270, 
width=83mm}
\end{tabular}
\caption{Neutralino (blue/dotted lines), selectron (black/full lines) 
and stau (red/full lines) masses in the cNMSSM as a function of
$M_{1/2}$ (in GeV); on the left-hand side $m_0=0$, while on the
right-hand side $m_0\sim M_{1/2}/10$. In both panels the states are
ordered in mass as $m_{\chi^0_1} \lesssim m_{\tilde \tau_1} < m_{\tilde
e_R} < m_{\chi^0_2} < m_{\tilde \tau_2} \lesssim m_{\tilde e_L} <
m_{\chi^0_3} < m_{\chi^0_{4,5}}$. The charginos $\chi_1^\pm$ and
$\chi_2^\pm$ are degenerate in mass with, respectively, $\chi_3^0$ and
$\chi_{4,5}^0$ (from~\cite{Djouadi:2008uj}).}
\label{fig:m12:phenoX0st}
\end{figure}

The neutralino and slepton mass spectra are shown in
Fig.~\ref{fig:m12:phenoX0st}. As for the Higgs bosons, $m_0=0$ in the
left-hand panel, and $m_0 \sim \frac{1}{10} M_{1/2}$ in the right-hand
panel.  The two nearly degenerate sets of lower lines in both panels 
correspond to the masses
of  the $\chi_1^0$ singlino-like LSP (blue/dotted) and the lighter stau
$\widetilde{\tau}_1$ NLSP (red/full). The mass difference between these
two states is smaller than $\sim 8$~GeV, as required in order to
obtain a cosmological relic density for the singlino $\chi_1^0$ 
compatible with WMAP. 
The pattern for the masses of the charginos and the heavier neutralinos
(blue/dotted lines) follows the one of the MSSM, once the proper relabeling
of the states is made. Since the effective higgsino mass
parameter $\mu_{\rm eff}$ is generally quite large, $\mu_{\rm eff}\gsim 
M_2$, the heavier  neutralino states $\chi_{4}^0$ and $\chi_{5}^0$ are
higgsino-like with masses $\sim \mu_{\rm eff}$. 
The states $\chi_2^0$ and $\chi_3^0$  are, respectively,
bino and wino-like with masses $m_{\chi_3^0} \approx 2 m_{\chi_2^0}
\approx M_2$ (with $M_2\approx 0.75\, M_{1/2}$). 
The charginos $\chi_1^\pm$ and $\chi_2^\pm$ are nearly degenerate in
mass with, respectively, the wino-like $\chi_3^0$ and  the higgsino-like
$\chi_{4,5}^0$ states.

In any case, the ``smoking gun'' for the cNMSSM would be the
$\widetilde{\tau}_1$ NLSP: all sparticle branching ratios into the
singlino-like LSP are tiny for $\l$ satisfying (\ref{ss.9e}),
hence all sparticles would decay at first into the $\widetilde{\tau}_1$
NLSP. Only then the $\widetilde{\tau}_1$ will decay into the $\chi^0_1$
LSP and a $\tau$ lepton, which will thus appear in \emph{every}
sparticle decay chain. For very small $\l$ (still larger than $10^{-5}$,
however) or a very small $\widetilde{\tau}_1-\chi^0_1$ mass difference,
the $\widetilde{\tau}_1$ life time can be so large that its decay
vertices are visibly displaced \cite{Djouadi:2008yj, Djouadi:2008uj}
(see Section~\ref{sec:5.2.3}).
This signal would definitively be spectacular and serve to distinguish
the cNMSSM from other supersymmetric extensions of the Standard Model.

\subsubsection{The constrained NMSSM with relaxed universality
conditions}
\label{sec:ss.1.3}

The constrained NMSSM with relaxed universality conditions in the
singlet sector, allowing for both $m_S^2 \neq m_0^2$ and $A_\k \neq A_0$
at the GUT scale, has been first analysed -- without requiring a dark
matter relic density in agreement with WMAP constraints --
in~\cite{Ellwanger:2006rn}. Then the bounds (\ref{ss.4e}) (from the
bound (\ref{ss.3e}) on  $m_S$, $A_\k$) and (\ref{ss.5e}) no longer apply;
notably $m_0$ is no longer bounded from above.
Also the upper limit (\ref{ss.1e}) on $\l$ is relaxed to $\l \lsim
0.55$. The freedom in the choice of $A_\k$ allows for a possibly quite
light CP-odd Higgs scalar $A_1$. This opens new phenomenologically
viable regions in parameter space, where the lightest CP-even Higgs mass
$M_{H_1}$ is well below 114~GeV, but LEP constraints are satisfied since
$H_1$ decays dominantly into a pair of CP-odd Higgs scalars (see
Section~\ref{sec:5.1}). On the other hand, the value of $A_\k$ at the
GUT scale must be chosen inside a narrow window in order to obtain $0 <
M_{A_1} < \frac{1}{2}\,M_{H_1}$ \cite{Ellwanger:2006rn,Djouadi:2008uw}.

Since the detection of Higgs-to-Higgs decays will be particularly
challenging at the LHC, some particular points in the parameter space of
the semi-constrained NMSSM have been proposed as benchmark points
in~\cite{Djouadi:2008uw}. These include scenarios corresponding to a
light CP-odd Higgs scalar $A_1$; allowing, in addition, the
Higgs masses $m_{H_u}$ and $m_{H_d}$ to differ from $m_0$ at the GUT
scale, scenarios with light CP-even singlet-like Higgs scalars can be
obtained as well. Both cases require particular search strategies at the
LHC, which are briefly surveyed in~\cite{Djouadi:2008uw} and discussed
in Section~\ref{sec:5.1}.

A complete analysis of the parameter space of the semi-constrained
NMSSM, including dark matter constraints combined with those from LEP
and other colliders on Higgs and sparticle searches, $B$-physics and the
anomalous magnetic moment of the muon, has been performed
in~\cite{Hugonie:2007vd,Belanger:2008nt}. In contrast to
the fully constrained NMSSM, where the small values for $m_0$ imply a
$\widetilde\tau_1$ lighter than the bino-like neutralino, the LSP can
now either be bino-, higgsino- or singlino-like. This, and the possible
presence of light Higgs states in $s$-channels, lead to many additional
regions in parameter space where the WMAP constraints on the dark
matter relic density can be satisfied. The corresponding processes, as
well as the prospects for the direct detection of the neutralino
LSP~\cite{Belanger:2008nt}, will be discussed in more detail in
Section~\ref{sec:dm}.

The potential of the LHC to unambiguously identify the semi-constrained
NMSSM by observing at least 3 neutral Higgs states is confined to the
regime of large values of $\tan\b \sim 50$ and $\l \sim 0.1$, where one
can benefit both from the enhanced couplings of the heavy Higgs states
to $b$-quarks and from substantial Higgs doublet/singlet
mixings~\cite{Belanger:2008nt}.

The semi-constrained NMSSM with $m_S \neq m_0$, but imposing $A_\k =
A_0$ at the GUT scale, has been investigated in~\cite{Balazs:2008ph},
with the result that the phenomenologically viable regions in parameter
space are still considerably larger than in the fully constrained NMSSM.

\subsection{The NMSSM and Gauge Mediated Supersymmetry Breaking}
\label{sec:ss.2}

The $\mu$-problem of the MSSM is particularly acute in the framework of
GMSB (see \cite{Giudice:1998bp} for a
review). The essential ingredients of GMSB models are a hidden or
sequestered sector, where supersymmetry is spontaneously broken in such
a way that the component $F_X$ of at least one superfield $\widehat{X}$
-- often considered as a spurion without kinetic term -- does not
vanish. In addition, a messenger sector $\widehat{\varphi}_i$ with
vector-like Standard Model gauge quantum numbers exists, which couples
to $\widehat{X}$ in the superpotential:
\beq\label{ss.10e}
W = \widehat{X} \widehat{\varphi}_i \widehat{\varphi}_i +\dots
\eeq
The messengers $\widehat{\varphi}_i$ have a super\-symmetric mass
$M_\text{mess}$
(the messenger scale), which can be parametrized by a vev $\left< X
\right> = M_\text{mess}/2$ of the scalar component of $\widehat{X}$. The
non-vanishing vev of $F_X$ induces a SUSY breaking mass splitting
$\widehat{m}$ between the scalar/pseudoscalar components of
$\widehat{\varphi}_i$. Typically, dynamical supersymmetry breaking is
assumed as the origin of $F_X \neq 0$ \cite{Giudice:1998bp}, but
K\"ahler potentials in supergravity of the No-Scale type, together with
Giudice-Masiero-like terms for the messengers, can also lead to models
with GMSB \cite{Ellwanger:1994uu}. Since the messengers 
$\widehat{\varphi}_i$ carry Standard Model gauge quantum numbers --
often in the form of a complete $SU(5)$ multiplet -- they induce SUSY
breaking gaugino masses of the order
\beq\label{ss.11e}
M_i \sim \frac{\alpha_i}{4\pi} \frac{F_X}{M_\text{mess}}
\sim \frac{\alpha_i}{4\pi} \frac{\widehat{m}^2}{M_\text{mess}}
\eeq
at one loop, and SUSY breaking scalar masses squared of the order
\beq\label{ss.12e} 
m_i^2 \sim \left(\frac{\alpha_i}{4\pi}
\frac{\widehat{m}^2}{M_\text{mess}} \right)^2
\eeq 
at two loops. However, neither a $\mu$-term nor a $B\mu$-term are
generated, which would lead to serious phenomenological problems as
summarised in the Introduction. The simplest solution to this problem is
the introduction of a singlet superfield $\widehat{S}$ with the scale
invariant superpotential (\ref{2.6e}). However, since $\widehat{S}$ is a
gauge singlet, the radiative corrections mentioned before will
\emph{not} generally
generate a soft mass $m_S^2$ and/or a trilinear coupling
$A_\k$ of the order $M_\mathrm{SUSY} \sim M_i$ \cite{Dine:1993yw,
deGouvea:1997cx}, whereas at least one of these parameters (with $m_S^2
< 0$) is required in order to trigger a sufficiently large vev of $s$,
cf. (\ref{2.19e}) -- (\ref{2.21e}).

Numerous proposals have been put forward in order to successfully solve
the $\mu$-problem in GMSB models within the NMSSM or slight
modifications thereof \cite{Dine:1993yw, Dine:1994vc, Ellwanger:1994uu,
Dine:1995ag, Dvali:1996cu, Ciafaloni:1996zh, Ciafaloni:1997gy,
Agashe:1997kn, Giudice:1997ni, deGouvea:1997cx, Giudice:1998bp,
Langacker:1999hs, Dubovsky:1999xc, Han:1999jc, Chacko:2001km,
Suematsu:2003si,
Delgado:2007rz, Giudice:2007ca, Dine:2007dz, Liu:2008pa,
Ellwanger:2008py, Morrissey:2008gm, Komargodski:2008ax, Mason:2009iq}:

(i) Several gauge singlets (or non-renormalisable self-interactions
of $\widehat{S}$) could trigger a vev $s$ which is large enough
\cite{Dine:1994vc, Dine:1995ag, Dvali:1996cu, deGouvea:1997cx,
Langacker:1999hs, Han:1999jc, Giudice:2007ca, Dine:2007dz,
Komargodski:2008ax, Mason:2009iq}. Since gauge symmetries do not
constrain the possible couplings of singlets, the choice of the
terms kept in the superpotential should preferably be justified by a
discrete (R-)symmetry.

(ii) If the messenger scale $M_\text{mess}$ is large enough, the soft
mass $m_S^2$ can become negative at low scales -- as desired -- due to a
positive $\beta$-function and the RG evolution; however, additional
positive terms in the $\beta$-function of $m_S^2$ beyond the ones of the
NMSSM (see (B.6)) are generally necessary to this end. Such terms appear
once $\widehat{S}$ couples to additional light matter $\widehat{Q}_i$
with vector-like Standard Model gauge quantum numbers in the form
$\widehat{S} \widehat{Q}_i \widehat{Q}_i$, provided the soft SUSY
breaking scalar masses of $Q_i$ are large enough~\cite{Dine:1994vc,
Agashe:1997kn, deGouvea:1997cx, Morrissey:2008gm}. Alternatively, the
positive terms in the $\beta$-function of $m_S^2$ within the NMSSM (B.6)
can already be large enough, if one allows for more general soft SUSY
breaking Higgs (and squark) masses. This requires more than one spurion
$\widehat{X}$ and $SU(5)$-breaking spurion-messenger couplings
\cite{Liu:2008pa, Mason:2009iq}.

(iii) The singlet $\widehat{S}$ can couple directly to the messengers
$\widehat{\varphi}_i$. The simplest coupling of the form $\eta\,
\widehat{S} \widehat{\varphi}_i \widehat{\varphi}_i$ generates tadpole
terms $\xi_F$ and $\xi_S$ \cite{Dine:1994vc, Ellwanger:1994uu,
Dvali:1996cu, Ciafaloni:1997gy, Ellwanger:2008py} (defined in
(\ref{2.1e}) and (\ref{2.5e})) of the order
\bea
\xi_F &\sim& \frac{\eta}{16\pi^2} F_X \sim \eta\; M_\mathrm{SUSY}
M_\text{mess}\; ,\nn \\
\xi_S &\sim& \frac{\eta}{16\pi^2} \frac{F_X^2}{M_\text{mess}} \sim \eta\;
M_\mathrm{SUSY}^2 M_\text{mess}
\label{ss.13e}
\eea
(see (\ref{4.1e}) with $\Lambda \sim M_\text{mess}$). On the one hand,
such terms would trigger a vev~$s$ even for negligibly small parameters
$m_S^2$ and $A_\k$; on the other hand, the values of $\xi_F$ and $\xi_S$
can easily be too large for a messenger scale $M_\text{mess} \gg
M_\mathrm{SUSY}$ -- this is nothing but the singlet tadpole problem
already discussed in Section~\ref{sec:4}. However, a small Yukawa
coupling $\eta \sim 10^{-5}$ can render such a scenario
phenomenologically acceptable \cite{Ellwanger:2008py}.

(iv) The singlet $\widehat{S}$ can couple directly to a generalized
messenger sector invariant under a discrete symmetry such that tadpole
diagrams are forbidden \cite{Giudice:1997ni}. Then parameters $A_\k$ and
$m_S^2 < 0$ of the desired order are radiatively generated to one- and
two-loop order, respectively, leading to a consistent phenomenology
\cite{Delgado:2007rz, Giudice:2007ca, Ellwanger:2008py}.

\medskip

The phenomenological consequences of the diverse scenarios can be
quite different. First, the squark and slepton spectrum depends on the
quantum numbers of the messengers and is generally easy to distinguish
from, e.g., mSUGRA. Second, additional light matter fields
$\widehat{Q}_i$ as in (ii) above would typically be observable at the
LHC. Finally, an approximate $U(1)_\text{R}$~symmetry in the NMSSM Higgs sector
\cite{Dobrescu:2000jt, Dobrescu:2000yn, Schuster:2005py,
Dermisek:2006wr, Dermisek:2007yt} in the context of GMSB 
\cite{Ellwanger:2008py, Morrissey:2008gm} can imply light CP-odd Higgs
scalars, into which the SM-like Higgs scalar can decay; as discussed in
Section~\ref{sec:5}, this scenario would have an important impact on the
search for Higgs bosons at the LHC.

\medskip

Finally we should mention that the $\mu$-problem is also present in
models with supersymmetry breaking from extra dimensions, Anomaly
Mediated Supersymmetry Breaking and models where the MSSM sector couples
to a superconformal field theory; again, the NMSSM and its variants as
described above have been proposed as a way out
\cite{Chacko:1999am,Kaplan:1999ac,ArkaniHamed:2001mi, 
Kobayashi:2001kz,Kitano:2004zd, Carpenter:2005tz,Choi:2005uz}.

\subsection{The nMSSM}
\label{sec:ss.3}

As discussed in Section~4, a solution of the domain wall problem of the
scale invariant NMSSM invariant under a discrete $\mathbb{Z}_3$~symmetry
consists in assuming the presence of tadpole terms $\sim \xi_S$ and/or
$\sim \xi_F$ in the soft SUSY breaking Higgs potential and/or the
superpotential. In order to solve the domain wall problem, these
parameters can be very small such that their impact on the phenomenology
can be neglected. On the other hand it has been proposed in
\cite{Panagiotakopoulos:1999ah} that notably $\xi_S$ could be of the
order $\xi_S \sim M_\mathrm{SUSY}^3$ and, in
\cite{Panagiotakopoulos:2000wp}, $\xi_F \sim M_\mathrm{SUSY}^2$. Then,
the tadpole terms in the potential trigger a vev $s \sim
M_\mathrm{SUSY}$ even for $\k \to 0$, i.e. the terms
$\frac{\k}{3}\widehat{S}^3$ in the superpotential and $\frac{1}{3}\k
A_\k S^3$ in ${\cal L}_\mathrm{soft}$ can be omitted. The resulting
version of the NMSSM, where the singlet only appears in the term $\l 
\widehat{S}\widehat{H}_u\cdot \widehat{H_d}$ in the superpotential and
in the tadpole terms $\sim \xi_S$ and/or $\sim \xi_F$, has been denoted
as the Minimal Non-minimal or new Minimal Supersymmetric Standard Model.

The phenomenological consequences of this class of models consist in a
possibly quite light charged Higgs boson (lighter than the SM-like Higgs
boson) into which top quarks can decay~\cite{Panagiotakopoulos:2000wp,
Panagiotakopoulos:2001zy}, and a neutralino LSP which has always a large
singlino component \cite{Panagiotakopoulos:2000wp, Dedes:2000jp}. Even
so, as shown in \cite{Menon:2004wv,Barger:2005hb,
Balazs:2007pf,Cao:2009ad}, the dark matter relic density can well be of
the required magnitude: the neutralino LSP can annihilate either through
the $Z$~resonance (due to its small, but non-vanishing higgsino
component) or through a CP-odd Higgs resonance in the $s$-channel (see
Section~\ref{sec:dm}). 
Bounds from $Z \to \chi_1^0 \chi_1^0$ have been checked; interestingly,
the SM-like Higgs scalar could dominantly decay invisibly into $\chi_1^0
\chi_1^0$ \cite{Menon:2004wv}. 
All additional phenomenological constraints on
supersymmetric models can be satisfied in this quite ``economic'' class
of models \cite{Balazs:2007pf,Cao:2009ad}, and the prospects of
neutralino, chargino and Higgs discoveries at the LHC and the ILC in the
nMSSM have been worked out in \cite{Balazs:2007pf}. Finally we recall
that the nMSSM also allows for successful electroweak baryogenesis
\cite{Menon:2004wv,Ham:2004nv, Huber:2006wf} (see Section~\ref{sec:4}).

\subsection{The fine tuning problem in the NMSSM}
\label{sec:ss.4}

One of the most important motivations of supersymmetric extensions of
the SM is the solution of the hierarchy problem: if one assumes that the
Lagrangian of the SM is valid up to a very high scale as $M_\text{GUT}$,
due to quadratically divergent radiative corrections the Higgs mass
squared has to be tuned with a relative precision of
$(M_\text{weak}/M_\text{GUT})^2 \sim 10^{-28}$ in order to obtain a weak
scale (given by the Higgs vev) 14 orders of magnitude below the
GUT~scale.

In supersymmetric extensions of the SM, the weak scale is naturally
given by the SUSY breaking scale $M_\mathrm{SUSY}$ (if the $\mu$ problem
is solved). The details of this relation depend, however, on the
supersymmetric model considered: on its field content, and on the values
of the unknown parameters. It is well known that the lower LEP bounds on
a Higgs scalar with a SM-like coupling to the $Z$~boson (and a large
branching ratio into $b\bar{b}$), as well as the present lower bounds on
sparticle masses, induce a ``little fine tuning problem'' in the MSSM:
its parameters have to be tuned with a relative precision of $\sim
10^{-2}$ in order to satisfy these constraints. It has been suggested
that the NMSSM (or variants thereof) can alleviate this problem, which
we will discuss below (see \cite{Dermisek:2009si} for a recent review).

There are different ways to quantify the amount of fine tuning within a
given model: an analytic approach consists in identifying the
independent parameters $p_a$ of the fundamental Lagrangian, in terms of
which all particle masses including $M_Z$ are determined. To each
parameter one associates a fine tuning measure $\Delta_a$ defined by
\beq\label{ss.14e}
\Delta_a = \left| \frac{d\ln{M_Z}}{d\ln{p_a}}\right|
\eeq
(or $M_Z$ replaced by $M_Z^2$), and the ``amount of fine tuning'' is
given by the maximum of all
$\Delta_a$. For the SM with a fundamental Lagrangian defined at the GUT
scale one would obtain $\Delta_\text{max} \sim 10^{14}$, but
all values much larger than 1 are considered as unnatural. In practice,
after radiative corrections,
the dependence of $M_Z$ on all $p_a$ is so complicated that the
various $\Delta_a$ can only be approximatively estimated which is,
however, sufficient to identify a potential fine tuning problem.

Alternatively one can use codes which compute the particle
spectrum numerically in terms of input parameters: then one can perform
scans over a large number of input parameters, and study which
proportion of the parameter space is still consistent with present
experimental constraints. However, the result of such an analysis is
related to the required amount of fine tuning only if the input
parameters correspond to the independent parameters $p_a$ of the
fundamental Lagrangian. Then, although the result depends somewhat on
the measure and on the initial range used to scan over the parameter
space, its physical interpretation is clearer than the one of the formal
fine tuning parameters $\Delta_a$ introduced above.

Let us now consider the origins of the ``little fine tuning
problem'', starting with the generation of the weak scale in terms of
the soft SUSY breaking Higgs masses $m_{H_u}$, $m_{H_d}$
and $\mu_\mathrm{eff}$ (in the NMSSM): neglecting the radiative
corrections (which play no important r\^ole here), the minimisation
equations~(\ref{2.15e}) of the Higgs potential imply
\beq\label{ss.15e}
M_Z^2 \simeq -2\mu_\mathrm{eff}^2 +\frac{2(m_{H_d}^2 -\tan^2\b\,
m_{H_u}^2)}{\tan^2\b -1}\; .
\eeq
In the absence of fine tuning, all terms on the right-hand side of
(\ref{ss.15e}) should be of comparable magnitude, and no large
cancellations should occur; hence both $\mu_\mathrm{eff}^2$ and
$-m_{H_u}^2$ should
not be much larger than ${\cal O}(M_Z^2)$. The bound
$|\mu_\mathrm{eff}| \gsim 100$~GeV from the lower bound on chargino
masses already starts to generate a slight tension, but the main problem
stems typically from the value of $m_{H_u}^2$ once the soft SUSY
breaking terms are generated at a large scale as $M_\mathrm{GUT}$: 

Fortunately, given the large value of the top Yukawa coupling, the
dominant positive terms proportional to the stop masses squared in the
one-loop RGE for $m_{H_u}^2$ (cf. (\ref{b.6e})) generate easily
$m_{H_u}^2 < 0$ at the weak scale as desired~\cite{Ibanez:1982fr}.
However, one
typically obtains $m_{H_u}^2 \sim -m^2_T$, which is often much larger
(in absolute value) than $M_Z^2$: from the RGEs for $m^2_T$ one finds
that $m_T$ is never much smaller than the gluino mass $M_3$ (at
$M_\text{SUSY}$) which, in turn, is bounded from below by $\sim 300$~GeV
from searches at the Tevatron (see \cite{Amsler:2008zzb} and refs.
therein).

Clearly these arguments are the same for the MSSM and the NMSSM, and
could be alleviated in the case of the generation of the soft SUSY
breaking terms at a much lower scale than $M_\mathrm{GUT}$, as
possible in GMSB. (See \cite{Ciafaloni:1996zh, Ciafaloni:1997gy,
Agashe:1997kn, deGouvea:1997cx, Kobayashi:2006fh, Mason:2009iq} for
discussions of the fine tuning problem in GMSB and proposals for
solutions within the NMSSM or variants thereof.) 

At least within the MSSM, another strong argument for large values of
$m_T$ originates from the LEP bound on the SM-like Higgs scalar mass,
whose theoretical expression was given approximately in (\ref{3.2e})
according to which it increases proportional to
$\ln\left(\frac{m_T^2}{m_t^2}\right)$. In the MSSM, large values for
$m_T$ are unavoidable in order to satisfy the LEP bound. Albeit large
stop masses are consistent with the non-observation of stops,
they would generate a too large value for $-m_{H_u}^2$ as discussed
above.

In the NMSSM, two different strategies can be pursued in order to
alleviate the fine tuning problem without assuming large values of
$m_T$: first, one can try to use the first positive term in the second
line of (\ref{3.2e}) to push the Higgs mass above the LEP bound. Then
$\l$ should be as large as possible, and $\tan\b$ must be small in order
to avoid a suppression by $\sin^2 2\b$. As discussed in Section~3,
within the NMSSM $\l$ is bounded from
above by $\l \lsim 0.7-0.8$ if one requires the absence of a Landau
singularity below the GUT scale; nevertheless, a reduction of the
required amount of fine tuning by the LEP bound has been observed in
\cite{BasteroGil:2000bw}. Otherwise one can enlarge the particle content
or reduce the range of validity of the NMSSM to scales far below
$M_\text{GUT}$ in order to justify larger values of $\l$
\cite{Harnik:2003rs,Chang:2004db, Birkedal:2004zx,Delgado:2005fq,
Chacko:2005ra, Nomura:2005rj,Barbieri:2006bg,Cavicchia:2007dp,
Cao:2008un}.

Instead of trying to push the Higgs mass above the LEP bound with the
help of the NMSSM specific terms, it is conceivable that the LEP bound
of 114~GeV on the mass of a SM-like Higgs scalar $H$ does not apply
within the NMSSM (see Section~\ref{sec:5}): if $H$ decays dominantly
into a pair of light pseudoscalars $H \to A\,A$, the branching fraction
for $H \to b\bar{b}$ can be much smaller than~1, and the Higgs mass
$M_H$ can be well below 114~GeV even for SM-like couplings of $H$ to the
$Z$~boson. If $A$ decays into $b\bar{b}$, combined LEP constraints still
require $M_H > 110$~GeV  \cite{Schael:2006cr}, hence $A$ should be
lighter than 10.5~GeV. Then, $M_H$ can be as low as 86~GeV
\cite{Abbiendi:2002in} which is easily compatible with (\ref{3.2e})
without large stop masses. A
reduction of the required fine tuning in the region of the parameter
space of the NMSSM corresponding to light pseudoscalars has been
discussed first in \cite{Dermisek:2005ar,Dermisek:2005gg} and was
elaborated further in \cite{Dermisek:2006wr,Dermisek:2006py,
Dermisek:2007yt,Dermisek:2007ah}. (However, the latest constraints from
ALEPH~\cite{Schael:2010aw} should be taken care of.)

At first sight, light pseudoscalars appear naturally in the NMSSM in the
case of an approximate global Peccei-Quinn- or R-symmetry (in the
Higgs sector, see Section~\ref{sec:2.2}). It is not always
straightforward, however, to realise a consistent scenario in these
limits. In the Peccei-Quinn limit ($\k \to 0$ in the scale invariant
NMSSM), two possible minima exist generically for the vev of the singlet
$s$: one is the ``large $s$ solution'' where $s \sim 1/\k$ (see
(\ref{2.21e})) would be very large. Then, in order to avoid an
excessively large
$\mu_\mathrm{eff} \equiv \l s$, $\l$ must be of the order of $\k$ which
corresponds to the decoupling limit (\ref{2.40e}). Since the light
pseudoscalar would be nearly a pure singlet, the coupling $H\,A\,A$ and
hence the desired branching ratio would disappear. In the other possible
minimum for $s$ where $s \simeq {\l A_\l \sin 2\b}/{\l^2+m_S^2/v^2}$
(from the third of eqs.~(\ref{2.15e})), it is difficult to have
$|\mu_\mathrm{eff}| \equiv |\l s|$ large enough (above 100~GeV) and to
ensure that it is the global minimum of the potential
\cite{Schuster:2005py}.

In the R-symmetry limit both trilinear soft terms $A_\l$ and $A_\k$
are assumed to be small; however, since the R-symmetry is broken by
the gaugino masses $M_i$, $A_\l$ and $A_\k$ receive radiative
corrections proportional to $M_i$ (cf. the RGEs in (\ref{b.3e})). Hence
it is unnatural to assume that $A_\l$ and $A_\k$ are smaller than these
radiative corrections, which requires some tuning in the $(5-10)\%$
range \cite{Dermisek:2006wr}. Hence, at least within the minimal NMSSM
without enlarged particle content up to the GUT scale, the present
experimental constraints also require some adjustment of parameters.

\medskip

Besides these qualitative arguments for the required amount of fine
tuning, probability distributions or likelihood analyses
in parameter space can be studied numerically using scans over input
parameters and suitable codes, which compute spectra and couplings and
allow to check experimental constraints. As stated above, the input
parameters should preferably correspond to the parameters of the
fundamental Lagrangian (at the GUT scale) up to an overall mass scale of
the soft terms, which can be fixed by $M_Z$. Then, quantities like
$\tan\b$ are obtained as output from the minimisation of the Higgs
potential. This approach had been pursued in the fully constrained NMSSM
in \cite{Ellwanger:1995ru,Ellwanger:1996gw,Ellwanger:1999ji} with the
result that $\sim 90\%$ of the points in the parameter space still
allowed by LEP1 are now eliminated by fruitless searches for the Higgs
boson, charginos and sleptons at LEP2, implying a necessary fine tuning
of at least $\sim 10\%$. (Also, large values $\tan\b \gsim 10$ require
fine tuned parameters at the GUT scale in the cNMSSM.)

Apart from the investigation of the required fine tuning, such analyses
serve to study the reach of future experiments in the parameter space
within specific scenarios. In the semi-constrained NMSSM (where the soft
singlet mass squared $m_S^2$ is not required to unify with $m_0^2$, see
Section~\ref{sec:ss.1.3}) such studies have recently been performed in
\cite{LopezFogliani:2009np,Balazs:2009su}, where also an LSP dark matter
relic density compatible with WMAP constraints was imposed. In these
approaches, the code NMSPEC \cite{Ellwanger:2006rn} was used, where
$\tan\b$ is an input parameter and $m_S^2$ and $\k$ are determined by
the minimisation equations of the Higgs potential, which can hide the
required amount of fine tuning, notably for large soft SUSY breaking
terms. Large values of $\tan\b$ and/or quite heavy sparticle masses
appear with relatively high probabilities in
\cite{LopezFogliani:2009np,Balazs:2009su}, which would imply that the
LHC cannot test large parts of the corresponding parameter space
(whereas the prospects for direct dark matter detection in the future
look better). It can be expected that, once fine tuning criteria are
taken into account, the prospects of sparticle detection at the LHC
would look much brighter.

\section{Variants of the NMSSM}

\subsection{CP violation}
\label{sec:var.1}

In the SM, CP is explicitly violated by charged current interactions:
complex Yukawa couplings lead to one physical phase in the
Cabibbo-Kobayashi-Maskawa (CKM) quark mixing matrix
($\delta_\text{CKM}$), which can successfully 
explain the observed direct and indirect CP violation (CPV) in the
neutral kaon sector
($\varepsilon^\prime/\varepsilon$ and $\varepsilon_K$, respectively) and
in the $B$-meson sector~\cite{Amsler:2008zzb}. On the other hand, the
non-observation of electric dipole moments (EDMs) of the electron,
neutron and atoms (like Hg) puts severe constraints  on flavour
\emph{conserving} phases, such as the strong CP phase, forcing
them to be extremely small.  

Supersymmetric extensions of the SM, like the MSSM and the NMSSM, 
introduce several potential new sources of explicit CPV: all couplings
and fermion masses appearing in the superpotential or in the soft SUSY
breaking Lagrangian can be complex, leading to non-trivial CPV phases.
SUSY CPV can manifest itself in several low energy
observables: in addition to the new mixings in the Higgs sector and the
implications for the Higgs spectrum and decays, one can expect
contributions to the EDMs, lepton polarisation asymmetries in
semi-leptonic decays, CPV in $B$-meson mixings and decays, among many
others. The measured values of (or bounds on) these observables lead to
severe constraints on the new phases, which is sometimes denoted as the
``SUSY CP problem''.

In particular, flavour conserving  SUSY phases can induce very large
contributions to the EDMs, several orders above the current bounds, even
if the SUSY spectrum is relatively heavy ($\sim 1$ TeV). In the
absence of some cancellation mechanism (as it would be natural in
minimal flavour violation, such as in the case of scenarios based on
mSUGRA or GMSB), complying with observation forces the new flavour
conserving phases to be very small, typically in the range $10^{-3} -
10^{-1}$. This holds both in the MSSM and the NMSSM.

In principle, generalizations of the Higgs sector of the SM allow for
spontaneous CPV at the electroweak scale, an attractive scenario at
first sight. However, in all supersymmetric models (including the NMSSM)
where $H_u$ couples to up-type and $H_d$ to down-type quarks only,
possible phases of $H_{u,d}$ can be rotated away by redefinitions of
quark fields, and the CKM matrix is real. (This remains true if
explicit CPV occurs in the Higgs sector only.) As a consequence, SM
contributions to the CPV observables are absent, and all the observed
values must be generated from SUSY contributions, which is difficult
(but not excluded \emph{\`a priori}). Furthermore, the structure of the
Higgs potential often forbids stable minima with spontaneous CPV as in
the MSSM and the $\mathbb{Z}_3$-invariant NMSSM (see below).

Independently from CPV in charged or neutral current interactions,
spontaneous or explicit CPV in the Higgs sector can have important
phenomenological implications: in the CP conserving case, scalar and
pseudoscalar Higgs bosons do not mix, leading to physical states with
well defined CP parity. CPV in the Higgs sector may lead to mixings
between scalar and pseudoscalar states (as the physical
states no longer are CP eigenstates): the full neutral Higgs mass matrix
will no longer be block diagonal, but a 5$\times$5 matrix which can be 
parametrized as
\beq\label{var.1e} 
{\cal M}_{H^0}^2 = \left(
\begin{array}{cc} {\cal M}_S^2 & 
{\cal M}_{SP}^2\\ 
({\cal M}_{SP}^2)^T & 
{\cal M}_P^2
\end{array} \right)\;, 
\eeq 
where ${\cal M}_{SP}^2\neq 0$ implies CPV. Actually, at tree level no CP
violation exclusively within the Higgs doublet sector occurs, since it
can be ``rotated away'' by field redefinitions; at tree level only
NMSSM specific doublet-singlet couplings can violate CP.

In general, the CPV couplings of the Higgs scalars to SM gauge bosons
and fermions, to superpartners, and to Higgs bosons themselves can
significantly differ from the CP conserving case, which requires to
re-interpret present bounds on the Higgs spectrum, and has consequences
for Higgs and sparticle detection at colliders. Finally, given the
possibility of having an enhanced first order electroweak phase
transition in the context of the NMSSM (see Section~4.3), additional
sources of CPV, beyond those of the SM, could play an important r\^ole
in obtaining successful electroweak baryogenesis, which is difficult
within the MSSM.

In what follows, we briefly review some relevant aspects of NMSSM-like
models
with spontaneous or explicit CPV, focusing on the Higgs sector and the
relevant literature. (For a review of CPV in connection with
non-standard Higgs sectors see \cite{Accomando:2006ga}.)

\subsubsection{Spontaneous CP violation}

Assuming a CP conserving Lagrangian before electroweak symmetry breaking implies
that all bilinear, trilinear and quartic terms can be chosen real (after
possible field redefinitions). The Higgs fields can
develop complex vevs, so that (\ref{2.10e}) generalizes to
\beq\label{var.2e} 
\langle H_u \rangle = v_u\, e^{i \varphi_u}\;, \quad
\langle H_d \rangle = v_d\, e^{i \varphi_d}\;, \quad 
\langle S \rangle = s\, e^{i \varphi_s}\;.   
\eeq 
The two physical phases (up to a $U(1)_Y$ gauge transformation), 
\beq\label{var.3e} 
\theta=\varphi_u+\varphi_d+\varphi_s\,, \quad \delta= 3 \varphi_s\,,
\eeq 
open the possibility of spontaneous CPV (SCPV). CP can be spontaneously
broken at the minimum of the tree level Higgs potential, or only once
radiative corrections to the effective potential are included --
the latter case is sometimes denoted as ``radiative SCPV''.

\bigskip
\noindent{\it Tree level SCPV in the $\mathbb{Z}_3$-invariant NMSSM}

\medskip
As stated in a no-go theorem in~\cite{Romao:1986jy}, spontaneous CP
violation is not possible in the $\mathbb{Z}_3$-symmetric NMSSM with a
superpotential as in (\ref{2.6e}) at tree level. Minimising the scalar
potential
\bea
V_\mathrm{Higgs} & = & \frac{g_1^2+g_2^2}{8} 
\left(v_u^2 - v_d^2\right)^2 + \left(m_{H_u}^2 + 
\lambda^2 s^2\right) v_u^2 +
\left(m_{H_d}^2 + \lambda^2 s^2\right) v_d^2 
+m_{S}^2\, s^2 \nn\\ 
&& +\kappa^2 \, s^4 + \lambda^2 \, v_u^2\,v_d^2 + 2\, 
\kappa\,\lambda \,v_u\,v_d\,s \cos(\theta-\delta) \nn
\\ && + 2\, \lambda A_\lambda\,v_u\,v_d\,s 
\cos \theta + \frac{2}{3}\kappa A_\kappa\, s^3 \cos \delta
\label{var.4e} 
\eea 
with respect to the absolute values of the vevs and their phases,  one
finds that CP violating extrema are always local maxima and never local
minima of the scalar potential, implying negative masses squared for at
least one Higgs boson.

\newpage

\bigskip
\noindent{\it Radiatively induced SCPV in the $\mathbb{Z}_3$-invariant
NMSSM}

\medskip
Radiatively induced SCPV is theoretically possible in the
NMSSM~\cite{Babu:1993qm,Haba:1995aw,Ham:1999zs}, but these scenarios are
severely constrained by the non-observation of Higgs pair production off
$Z$ bosons at LEP: analogous to the MSSM, a very light spin-0,
non-Goldstone boson appears in the spectrum. The presence of  such a
light boson is a consequence of the Georgi-Pais
theorem~\cite{Georgi:1974au}, and current experimental bounds seem to
lead to the exclusion of this class of models.

\bigskip
\noindent{\it SCPV in the general NMSSM}

\medskip
In the general NMSSM, where dimensionful terms are present in the
superpotential (\ref{2.1e}), the no-go theorem of~\cite{Romao:1986jy} no
longer holds and spontaneous CP violation is indeed viable already at
tree level~\cite{Branco:2000dq,Davies:2001uv}. Combining the
minimisation conditions for spontaneous CP violation with the
constraints from $\varepsilon_K$, it was noticed that the
theoretical upper bound on the lightest Higgs boson mass becames
stronger, leading to $M_{H_1} \lesssim 100$ GeV (still in agreement with
LEP bounds due to reduced couplings to SM gauge
bosons)~\cite{Branco:2000dq,Davies:2001uv}. However, these models
re-introduce a $\mu$-term in the superpotential so that a solution to
the $\mu$-problem of the MSSM is not obtained.

\bigskip
\noindent{\it SCPV in the NMSSM with a tadpole term}

\medskip

As discussed in Section~\ref{sec:ss.3}, the nMSSM contains singlet
tadpole terms $\sim\xi_S,\;\xi_F$ in the scalar potential and/or
superpotential~\cite{Panagiotakopoulos:1998yw,Panagiotakopoulos:1999ah,
Panagiotakopoulos:2000wp,Dedes:2000jp,Panagiotakopoulos:2001zy}, but $\k
= 0$. Then, as in the MSSM, neither spontaneous nor explicit CPV in the
Higgs sector is possible at tree level. Keeping $\k \neq 0$, but adding
just a soft SUSY breaking tadpole term $\sim\xi_S$ to the otherwise
$\mathbb{Z}_3$-invariant NMSSM, allows already to circumvent the no-go
theorem in~\cite{Romao:1986jy}, and one can find true minima of the
scalar potential associated with non-trivial spontaneous CP violating
phases which are phenomenologically  viable~\cite{Hugonie:2003yu}. Again
some Higgs states can be quite light, but not excluded by current bounds
due to their reduced coupling to gauge bosons. The new Higgs mass
matrices, together with the relevant radiative corrections, can be found
in~\cite{Hugonie:2003yu}. 

\bigskip
\noindent{\it Problems with SCPV}

\medskip

However, although elegant and apparently simple to realise, scenarios of
SCPV are seldom viable: recalling that the CKM matrix is real, there are
no SM contributions to either flavour conserving (EDMs) or flavour
violating CP violating observables. SUSY contributions, with $\delta$
and $\theta$ as the only sources of CPV, would be the only means to saturate
the observed values of $\varepsilon_K$,
$\varepsilon^\prime/\varepsilon$, the CP asymmetry of the $B_d$ meson
decay, etc.. In particular, the dominant contributions
to $\varepsilon_K$ should arise from the chargino-mediated box
diagrams~\cite{Pomarol:1992uu,Hugonie:2003yu}. In general, complying
with observation requires the phases $\delta$ and
$\theta$ to be quite large, as well as maximal
left-right squark mixing~\cite{Pomarol:1992uu,Lebedev:1999uc}. Large
values for flavour conserving CP violating phases nearly
inevitably lead to sizable contributions to the EDMs of the electron,
neutron and atoms. Complying with the EDM bounds, either via a 
cancellation mechanism, or a heavy SUSY spectrum (not always possible as
excessively heavy squarks would preclude saturating the values of
$\varepsilon_K$ etc.) represents the most serious challenge to the
survival of all models with SCPV in the Higgs sector. 

\bigskip

\noi{\it Spontaneous CP violation at finite temperature}
\medskip

As discussed in Section~\ref{sec:4.3}, additional sources of CP
violation in the Higgs sector would be desirable for baryogenesis. An
interesting hypothesis is to assume that CP is indeed conserved at
$T=0$, and to have spontaneous CP violation at finite temperature in the
$\mathbb{Z}_3$-invariant NMSSM. This would allow to evade constraints from EDMs,
and to avoid a light spectrum in the Higgs sector (arising in the case
of radiative CPV, as discussed before). Finite temperature effects can
indeed trigger spontaneous CP violation inside the walls of the
propagating bubbles present after the phase
transition~\cite{Comelli:1994ew,Comelli:1994rt}. An extremely tiny
explicit phase is nevertheless required in the effective potential, in
order to lift the degeneracy between vacua (each generating identical
baryon asymmetries, but of opposite signs). Such a phase, typically
$\mathcal{O}(10^{-6} - 10^{-5})$, would give rise to negligible
contributions to the EDMs.

\subsubsection{Explicit CP violation}

The NMSSM is the simplest SUSY extension of the SM where one can have 
explicit CP violation in the Higgs sector at tree
level~\cite{Ellis:1988er,Barr:1992if}. In the MSSM the explicit CP
violating phases in the Higgs sector (in $\mu$ and $B$) can be rotated
away by a redefinition of the $H_u$ and $H_d$ fields; this is not
possible in the NMSSM due to the presence of the additional singlet
couplings. Since now the standard CKM mechanism is assumed to be the
dominant source of CPV in the quark sector, we will not discuss the SUSY
contributions to flavour-dependent CP observables. MSSM-like phenomena
(e.g. CP violation in the $B$-mesons) will also not be discussed here,
as we will focus on the implications of CP violation for the NMSSM Higgs
sector. 

\bigskip
\noindent{\it Explicit CPV at tree level in the $\mathbb{Z}_3$-invariant
NMSSM}
\medskip

In the NMSSM with a scale invariant superpotential (\ref{2.6e}), the
parameters  $\lambda$, $\kappa$, $A_\lambda$ and $A_\kappa$ can be
complex. The associated phases only appear in a subset of terms of
$V_\mathrm{Higgs}$ (\ref{2.9e}), 
\beq\label{var.5e}
V_\mathrm{Higgs}^\mathrm{phase} = 
- \lambda \kappa^*\, H_u^0 H_d^0 S^{*2}  
- \lambda A_\lambda\, H_u^0 H_d^0 S 
+ \frac{1}{3} \kappa A_\kappa\, S^3\, + \text{h.c.}\;.
\eeq

By a redefinition of $H_u$, $H_d$ and $S$, only one physical phase
remains~\cite{Matsuda:1995ta,Haba:1996bg} which can be taken to be 
the phase of $\lambda \kappa^*$:
\beq\label{var.6e}
\phi \equiv \arg(\lambda \kappa^*) \;.
\eeq
The stationary conditions for the phases induce necessarily complex vevs
($\theta,\,\delta \neq 0$) provided $\phi \neq 0,\,
\pi$~\cite{Haba:1996bg}.

Regarding the tree level Higgs mass matrix (\ref{var.1e}),
the diagonal blocks ${\cal M}_S^2$ and ${\cal M}_P^2$ are those
already given in (\ref{2.22e}, \ref{2.24e}), once the appropriate
redefinitions have been performed:
\beq\label{var.7e}
\lambda A_\lambda \to \lambda A_\lambda \cos \theta, 
\quad
\kappa \lambda \to \kappa \lambda \cos(\phi +\theta-\delta),
\quad
\kappa A_\kappa \to \kappa A_\kappa \cos \delta\;.
\eeq
The entries of the new off-diagonal block ${\cal M}_{SP}^2$ in
(\ref{var.1e}) are all proportional to  $\sin(\phi +\theta-\delta)$, and
can be found  in~\cite{Matsuda:1995ta,Haba:1996bg,Funakubo:2004ka,
Ham:2001kf,Ham:2007mt}. However, it is important to stress that, at
tree level, explicit CP violation only induces scalar-pseudoscalar
mixings between $H_{u,d}$ and $S$; the scalar and
pseudoscalar components of the Higgs doublets $H_{u,d}$ do not mix.

Nevertheless the impact of explicit CP violation on low energy
phenomenology is not negligible. Although CPV effects are
small in limiting cases for the singlet vev ($s \ll v_u, v_d$ or $s \gg
v_u, v_d$) or very large $\tan \beta$, large scalar-pseudoscalar mixings
can occur in regions where $s \sim \mathcal{O}(v)$ and $\tan \beta \sim
\mathcal{O}(1)$~\cite{Haba:1996bg}. 
As a consequence, the lightest Higgs mass can decrease by 10--30~GeV
compared to the CP conserving case. However, avoiding an excessive
contribution to the neutron EDM would require comparatively heavy
gauginos and squarks with masses $\sim \mathcal{O}$(TeV).

\bigskip
\noindent{\it Explicit CPV in the $\mathbb{Z}_3$-invariant NMSSM with
radiative corrections}
\medskip

In order to obtain scenarios that are still in agreement with the bounds
on EDMs, while at the same time inducing a moderate amount of CP
violation (as desirable for electroweak baryogenesis), one can
consider NMSSM scenarios where explicit CP violation in the Higgs sector
is induced only through radiative corrections to the Higgs masses and
couplings~\cite{Garisto:1993ms}. Explicit CP violation in the Higgs
sector of the NMSSM at the one-loop level, including radiative
corrections from third generation squarks, has been studied
in~\cite{Ham:2001kf,Ham:2001wt,Ham:2003jf,Ham:2007mt}: due to the
possibly complex soft SUSY breaking terms in the squark sector, new CPV
terms will appear in the neutral and charged Higgs effective Lagrangian
via these radiative corrections. (The RGEs for the parameters of the
NMSSM with explicit CP violation are given
in~\cite{Asatrian:1995be,Asatrian:1996wn}.)

Considering only the effect of the (dominant) corrections from third
generation quark-squark loops, the new phases in addition to the three
tree level phases in (\ref{var.5e}) are $\phi_{A_t}$, $\phi_{A_b}$,
arising from  $A_t$, $A_b$, respectively~\cite{Ham:2001kf}. The
minimisation conditions (with respect to $H_{iI}$ and $S_I$) reduce 
the five phases to three. (Note that the complex phases of
the third generation quark Yukawa couplings can be reabsorbed by
redefinitions of the quark fields.) An important consequence of taking
these higher order effects into account is that scalar-pseudoscalar
mixings between the two Higgs \emph{doublets} (absent at tree-level) can
now occur.

CP violating effects originating from the chargino sector can also play
a r\^ole. Contributions from loops involving charged particles ($W^\pm$
bosons, charged Higgs and charginos) were taken into account 
in~\cite{Ham:2001wt}. It was found that  the relative phase between the
soft breaking $SU(2)$ gaugino mass $M_2$ and $\lambda$, $\phi_C$, yields
corrections to {\it every} element of the $5\times5$ neutral Higgs mass
matrix proportional to $\sin \phi_C$. Depending on the other low energy
NMSSM parameters, the chargino induced CP
violating corrections can lead to contributions to the lightest Higgs
boson mass in the range $-24$ GeV $ \lesssim
\delta_{\mathrm{CPV}}^{\chi^\pm}
M_{H_1} \lesssim 16$ GeV~\cite{Ham:2001wt}. Including all radiative
corrections, $M_{H_1, \mathrm{max}}$ increases with $\tan \beta$ (in
contrast to the case without CP violation, see Section~\ref{sec:3.2}),
and the upper bound of $M_{H_1, \mathrm{max}}\sim 150$ GeV is saturated
for $\tan \beta \sim 30$~\cite{Ham:2007kc}.

Explicit CPV can also have implications on the production and decay
rates of Higgs bosons at colliders~\cite{Ham:2001wt,Ham:2007mt}: at a
linear $e^+\,e^-$ collider ($\sqrt{s} =500$ GeV and 1 TeV), the
production cross sections for the neutral Higgs bosons in the NMSSM with
explicit CPV can be much smaller than in the CP conserving case.
Explicit CPV in the NMSSM can possibly be tested at the ILC with
$\sqrt{s} =500$ GeV, where at least one of the five neutral Higgs bosons
would be produced via Higgs-strahlung, $WW$- and $ZZ$-fusion with
corresponding cross sections of $\sim$ 12, 15 and 1.5~fb, respectively.
The decays of the neutral Higgs bosons can also differ from those of the
NMSSM without CP violation in the form of modified branching ratios of
decays into pairs of $s \bar s$ and $c\bar c$ quarks~\cite{Ham:2007kc}. 

\medskip

In the nMSSM (with tadpole terms and $\k=0$, see
Section~\ref{sec:ss.3}), explicit (as well as spontaneous) CPV is
impossible at tree level, but radiative corrections from sbottom and
stop squarks can also trigger CP violation in the Higgs sector. The
associated phenomenology (mass spectrum and collider prospects) is
analysed in~\cite{Ham:2008cg}.

\medskip
In all cases of explicit CP violation in the Higgs sector, EDMs are
expected to receive potentially large contributions. However, the
phase combination responsible for the EDMs can be distinct from the one
responsible for Higgs mixing, which potentially offers additional
sources of CP violation for electroweak baryogenesis. It has been
suggested~\cite{Funakubo:2004ka}  that the phases can be arranged in
such a way that the combination inducing the neutron EDM can be
suppressed, while the ones affecting the Higgs sector are allowed to be
sizable. However, this requires a substantial fine tuning between the
phases of the Higgs vevs and the phases of parameters in the Lagrangian.
In order to ensure that SUSY contributions to the EDMs are  below the
experimental bounds, the SUSY spectra (notably gaugino masses) should be
of  $\mathcal{O}$(1 TeV)~\cite{Matsuda:1995ta,Haba:1996bg,Boz:2005sf}.

Modifying the NMSSM by promoting the phases of gaugino masses and the
trilinear couplings $A_i$ to fields allows to address the strong and the
SUSY CP problems simultaneously~\cite{Demir:1999qj}. Then, however,
observable SUSY CPV requires non-minimal flavour structures beyond those
associated with the Yukawa couplings. 

\bigskip
\noi{\it CP violation in multi-singlet extensions of the NMSSM}
\medskip

In order to keep sizable CPV in the Higgs sector (contributing to
baryogenesis at the electroweak scale) without excessively large induced
EDMs, multi-singlet extensions of the NMSSM have been considered. In
\cite{Garisto:1993ms} the NMSSM is extended by two additional vev-less
singlets ($S^\prime$ and $S^{\prime\prime}$), and explicit CPV in the
form of complex couplings is allowed only between fields which have no
tree level couplings to quarks or leptons. Nevertheless, through
loop corrections involving the new singlets, small CP violating phases
can appear for the EDMs (consistent with current bounds). 

Spontaneous CP violation in the NMSSM with two singlets $S$ and $S'$,
taking into account radiative effects, has also been
proposed~\cite{Ham:2001ze}. Both Higgs doublets and singlets can develop
complex vevs, and one now has three independent phase combinations 
($\theta=\varphi_u+\varphi_d+\varphi_{s}+\varphi_{s^\prime}$, $  \delta=
3 \varphi_s$ and $\delta^\prime = 3 \varphi_{s^\prime}$). Contrary to
the NMSSM (with either spontaneous or explicit CPV),  the addition of
the second singlet implies that scalar-pseudoscalar mixings between the
Higgs doublets can occur already at tree level. Although the Higgs
spectrum still exhibits light states, it is possible to find regions in
parameter space where the lightest Higgs mass is as large as 80 GeV,
barely viable in view of LEP bounds. A general analysis of CP violation
at the one-loop level in the Higgs sector of the MSSM extended by an
arbitrary number singlet fields has been performed
in~\cite{Ham:2004pd}. 

\newpage

\subsection{R-parity violation and neutrino masses in the NMSSM}

The phenomenological implications of R-parity violation (RpV) in the
NMSSM are extensive (for a review of RpV in the MSSM,
see~\cite{Barbier:2004ez}). Here we will briefly summarise some of the
most relevant consequences, as neutrino mass generation.

R-parity is a discrete symmetry corresponding to the remnant of a
group of continuous $U(1)_\text{R}$ transformations,
whose charges $R_p$ are assigned as~\cite{Farrar:1978xj}
\beq\label{var.8e}
R_p =(-1)^R\,\quad \text{with} \,\,
R=\left\{
\begin{array}{l}
+1 \,\,\text{for SM particles including all Higgs bosons}\\
-1 \,\,\text{for the superpartners}
\end{array}
\right.
\eeq
Historically, R-parity was introduced in early supersymmetric extensions
of the SM in order to maintain the baryon ($B$) and lepton ($L$) number
conservation laws of the
SM~\cite{Fayet:1976cr,Fayet:1977yc,Farrar:1978xj,
Weinberg:1981wj,Sakai:1981pk}:
conserved R-parity naturally leads to conserved $B$ and $L$, while RpV
requires a violation of at least $B$ or $L$. If R-parity is conserved, 
R-odd particles are necessarily pair produced, the lightest R-odd
particle (the LSP) is stable and, if neutral and colourless, a good dark
matter candidate. 

In the MSSM, the most general renormalisable
superpotential including RpV terms is 
\beq\label{var.9e}
W_\mathrm{MSSM}\,=\,W_\mathrm{RpC} + \mu_i\,\widehat{H}_u \cdot 
\widehat{L}_i
+ \frac{1}{2}\,\lambda_{ijk}\,\widehat{L}_i \cdot \widehat{L}_j\; 
\widehat{E}_{Rk}^c 
+ \lambda_{ijk}^\prime\,\widehat{L}_i \cdot \widehat{Q}_j\; 
\widehat{D}_{Rj}^c 
+ \frac{1}{2}\,\lambda_{ijk}^{\prime\prime}\,\widehat{D}_i \cdot 
\widehat{D}_j\, \widehat{U}_{Rk}\;,
\eeq
where $W_\mathrm{RpC}$ denotes the R-parity conserving part, and 
$i,j,k=1,2,3$ are generation indices. The terms $\sim \mu_i$, $\sim
\lambda_{ijk}$ and $\sim \lambda_{ijk}^\prime$ violate $L$, and the term
$\sim \lambda_{ijk}^{\prime\prime}$ violates $B$. (RpV can also
occur spontaneously through the vev of a neutral R-odd scalar,
necessarily a sneutrino in the MSSM or NMSSM, which also violates $L$.)
Note that the four superfields $\widehat{H}_d$ and $\widehat{L}_i$ have
the same gauge quantum numbers, and in the presence of the bilinear
terms $\mu_i$ and $\mu$ it is no longer possible to distinguish them
unambiguously. 

If $B$ and $L$ are violated simultaneously, this can lead to fast proton
decay (and other nuclear conversions). Generally, RpV can induce single
production of neutralinos and charginos, LSP decays into SM particles,
flavour violating $Z$ couplings to leptons, violation of charged current
universality in the lepton and quark sectors, additional sources of
mixing (and CP violation, if the couplings are complex) in neutral
mesons, new rare decays as lepton flavour violation (LFV) in the charged
lepton sector. The baryon and/or lepton number violation induced by RpV
can be an ingredient useful for baryogenesis. 

One of the most desirable effects of the  $L$ number violation stemming
from both bilinear and trilinear R-parity violating couplings (if the
vevs of the sneutrinos are non-vanishing) is the mixing between Higgs bosons
$H_d$ and sleptons $L_i$ (in the scalar sector) and between leptons and
neutralinos/charginos. The latter implies enlarged chargino and
neutralino mass matrices (respectively $5 \times 5$ and $7\times7$, in
the MSSM). These new mixtures allow to obtain massive neutrinos without
the inclusion of right-handed neutrinos,  and open the possibility of
explaining neutrino data (mass squared differences and mixing angles)
with a minimal superfield content. Bilinear R-parity violating couplings
can induce, at the tree level, a single massive neutrino via a
``seesaw-like'' mechanism, where much heavier gauginos and higgsinos
play the r\^ole of right-handed neutrinos. In order to account for
realistic neutrino masses and mixings, higher order contributions
(1-loop), arising from trilinear $\lambda_{ijk}$, $\lambda_{ijk}^\prime$
and/or bilinear R-parity violating couplings, must be taken into
account.

In addition to the MSSM RpV terms, the NMSSM superpotential can contain
a trilinear coupling 
\beq\label{var.10e}
W^\mathrm{RpV}_\mathrm{NMSSM}\;=\;\lambda_i\,\widehat{S}\,\widehat{H}_u
\cdot \widehat{L}_i\;, 
\eeq
and a corresponding contribution $\sim A_{\lambda_i} S H_u \widetilde
L_i$ to the soft SUSY breaking Lagrangian. As in the MSSM with R-parity
violating $\mu_i$ terms, the couplings $\lambda_i$ in (\ref{var.10e})
induce numerous mixings (see e.g.~\cite{Chemtob:2006ur}): the charged
Higgs bosons mix with charged sleptons, and charginos with charged
sleptons, leading to enlarged mass matrices. The neutral CP-even Higgs
bosons mix with (left-handed) sneutrinos, now leading to $6\times 6$
mass matrices. In the NMSSM one finds 8 neutral fermions resulting from
the mixing of (left-handed) neutrinos and neutralinos. 

Note that in the presence of bilinear RpV soft breaking terms,
electroweak symmetry breaking generally leads to non-vanishing sneutrino
vevs, and these have to be taken into account in the conditions for
vacuum stability (UFB directions and CCB minima) in this class of
models~\cite{Chemtob:2007rg}. The RG equations of the R-parity violating
couplings in the NMSSM have been given (and analysed for fixed points)
in~\cite{Pandita:1999jd,Pandita:2001cv}, and all possible baryon and
lepton number violating operators are classified
in~\cite{Pandita:2001cv}.

\medskip
A common phenomenological feature of  all SUSY models violating R-parity
is the occurrence of an unstable LSP. The lightest neutralino (or the
lightest stau) can decay via its R-parity violating couplings into SM
particles. An unstable LSP can still remain a viable dark matter
candidate provided its lifetime is sufficiently long (of the order of
the age of the universe). This possibility has been investigated
recently in the NMSSM -- including constraints from neutrino data -- in
\cite{JeanLouis:2009du} with the result that a gravitino LSP can be
sufficiently long-lived.
 If the LSP is short-lived (hence irrelevant to
the solution of the dark matter problem), it must decay sufficiently
fast as not to affect the successful predictions of Big-Bang
nucleosynthesis. In both cases severe constraints on the R-parity
violating couplings can be derived~\cite{Barbier:2004ez}.

At colliders, one can expect displaced vertices from LSP
decays~\cite{Barbier:2004ez}. Its branching fractions (depending on its
composition) can shed light on the R-parity violating couplings: if
the LSP is a singlino dominated neutralino, it would be possible to
distinguish the R-parity violating NMSSM from the R-parity violating
MSSM at future colliders~\cite{Ghosh:2008yh}.

\bigskip
Let us review some phenomenological implications of RpV, including the
possibility of adding right-handed neutrino superfields to the NMSSM,
and mechanisms of neutrino mass generation with and without~RpV.

\subsubsection{Massive neutrinos in the NMSSM}

\medskip
\noindent{\it Neutrino masses from R-parity violation}
\medskip

Dirac neutrino-higgsino masses, which do not require the introduction of
right-handed neutrino fields, can be generated from
the R-parity violating MSSM-like first term $\sim \mu_i$ in
(\ref{var.9e}). However, the resulting neutrino mass matrix is only of
rank~1 at tree level, i.\,e. only one neutrino is massive, which must be
improved by loop effects. Furthermore the smallness of the neutrino
masses must be understood, which requires some fine tuning of
parameters.

Adding only the terms $\sim \mu_i$ in (\ref{var.9e}) to the NMSSM, the
required amount of fine tuning can be reduced due to possible
compensations between gaugino and singlino exchange diagrams (at tree
level)~\cite{Abada:2006qn}. Such a compensation is equally possible if
only the NMSSM specific terms (\ref{var.10e}) are
present~\cite{Abada:2006qn, Chemtob:2006ur}. Once loop corrections are
included, a hierarchical spectrum and mixing angles in agreement with
observation can be obtained (albeit still at the expense of some
fine tuning)~\cite{Chemtob:2006ur}. Natural values for the heaviest
neutrino mass lead to bounds on $\lambda_i \sim
\mathcal{O}(10^{-5}-10^{-2})$~\cite{Abada:2006qn, Chemtob:2006ur}.

If both terms $\sim \mu_i$ in (\ref{var.9e}) \emph{and} those of
(\ref{var.10e}) are present, two massive neutrino states can be
obtained  at tree-level (in contrast to the
MSSM)~\cite{Abada:2006qn,Abada:2006gh}. This offers the appealing
possibility of reproducing neutrino data without having to resort to
loop effects. However, again the observed mass hierarchy and mixing
angles can be obtained only at the price of substantial fine tuning of
the parameters: although neutrino data by itself would allow for ratios
$|\mu_i/\mu_\text{eff}|$ as large as $\sim 10^{-1}$, potential
contributions to LFV processes (such as  $\mu \to e \gamma$ or $\mu \to
e e e$) constrain the values of $|\mu_i/\mu_\text{eff}| \lesssim
10^{-2}$. 

In~\cite{Abada:2006qn} it has been pointed out that the NMSSM singlet
can generate thermal leptogenesis (which would later be converted into a
baryon asymmetry of the universe) via decays involving the R-parity
violating trilinear couplings. The neutrino refraction indices (relevant
for the neutrino propagation in matter, as in supernovae) have been
studied in the NMSSM, including radiative corrections,
in~\cite{Gava:2009gt}. 

\bigskip
\noindent{\it Right-handed neutrino sector}
\medskip

If one adds three generations of right-handed neutrino superfields
$\widehat{N}_{Ri}$ (with, by convention, negative R-parity of its scalar
components in contrast to the singlet superfield), the superpotential
can include the following additional terms:
\beq\label{var.11e}
W^{N^c_R}_\mathrm{NMSSM}\;= \;
{h_\nu}_{ij}\,\widehat{L}_i\cdot\widehat{H}_u\,\widehat{N}_{Rj}^c +
\frac{1}{2} M^\text{Maj}_{ij} \widehat{N}_{Ri}^c\,
\widehat{N}_{Rj}^c + \frac{\kappa_i}{3} \widehat{N}_{Ri}^{c3} 
+ \frac{1}{2}\,\lambda^\text{Maj}_{ij} \widehat{N}_{Ri}^c\,
\widehat{N}_{Rj}^c \,\widehat{S}
\eeq
and corresponding contributions to the soft SUSY breaking Lagrangian.
The first two R-parity conserving terms are identical to those present
in the MSSM with right-handed neutrinos, the third term $\sim \k_i$ is
R-parity violating, and only the last R-parity conserving term is
specific for the NMSSM. The fermion and scalar components of
$\widehat{N}_{Ri}$ will mix in general with all neutral fermions
and scalars.

Models with right-handed neutrino superfields $\widehat{N}_R$ and a
scale and R-invariant superpotential ($M^\text{Maj}=0$ and $\k_i=0$ in
(\ref{var.11e})) allow for the spontaneous generation of mass terms for
$\widehat{N}_R$ through the last term  $\sim \lambda^\text{Maj}$ in
(\ref{var.11e}): when the singlet gets a vev $s$, a Majorana mass 
$M^\text{Maj}_\text{eff}=  \lambda^\text{Maj}\, s$ (preferably of
$\cal{O}$(TeV)) is dynamically generated~\cite{Kitano:1999qb}. These
models have been shown to possess regions in parameter space where left-
and right-handed sneutrinos acquire vevs, so that R-parity is
spontaneously broken and effective $\mu_i^\text{eff}$-terms in
(\ref{var.9e}) are generated as well. In this case, the model contains
two distinct sources for neutrino mass generation (the ordinary seesaw
with right-handed neutrinos and contributions from the 
$\mu_i^{\text{eff}}$-terms) and both solar and atmospheric neutrino data
can be accommodated. 

\bigskip
\noindent{\it Solving the $\mu$-problem with right-handed neutrino
superfields and RpV} 
\medskip

From the point of view of particle content, an economic and interesting
possibility is to replace the singlet $\widehat{S}$ by the right-handed
neutrino superfields $\widehat{N}_{Ri}$ (which is not compatible with an
R-invariant superpotential). This class of models is also denoted as
the $\mu\nu$SSM~\cite{LopezFogliani:2005yw}. In addition to quark and
lepton Yukawa couplings -- including the first term in (\ref{var.11e})
-- the superpotential of the $\mu\nu$SSM contains the following terms:
\beq\label{var.12e}
W\;= \;
\tilde \lambda_i\,\widehat{N}_{Ri}^c\,\widehat{H}_u \cdot
\widehat{H}_d\; + \frac{\kappa_i}{3} \widehat{N}_{Ri}^{c3} \;.
\eeq
Then the vevs $\tilde \nu_{Ri}$ of the scalar components of
$\widehat{N}_{Ri}$ generate an effective $\mu$-term ($\mu_\text{eff} =
\sum_i \tilde \lambda_i \tilde \nu_{Ri}$), and the Higgs vevs as well as
the vevs $\tilde \nu_{Ri}$ generate both Dirac and Majorana masses for
the neutrinos. The observed neutrino masses originate from two
different sources: from the Yukawa couplings $h_{\nu ij}$ in
(\ref{var.11e}) roughly of ${\cal{O}}(10^{-6})$ (and a seesaw mechanism
with right-handed neutrino Majorana masses of the order of the
electroweak scale from the terms $\sim \k_i$), and an R-parity violating
mixing of neutrinos with neutralinos. 
(In an earlier study of the effects of a coupling $\sim
\widehat{N}_{Ri}^c \widehat{H}_u
\widehat{H}_d$~\cite{Allahverdi:1994pn}, an explicit Majorana mass term
for the right-handed neutrinos of the order of the GUT scale was added,
and cosmological constraints on the lifetime of the neutralino LSP --
unstable due to the R-parity violating couplings -- were derived.)

The phenomenology of this class of models has been extensively studied
in~\cite{LopezFogliani:2005yw,
Escudero:2008jg,Ghosh:2008yh,Bartl:2009an,Fidalgo:2009dm}; in large
regions in parameter space, viable solutions can be found (avoiding
Landau poles, and in agreement with collider constraints on Higgs and
sparticle masses, and current neutrino data). In spite of the additional
Higgs-sneutrino mixings, the upper bound on the lightest Higgs mass
turns out to be similar to the one obtained in the framework of the
general NMSSM ($M_{H_1} \lesssim 140$ GeV)~\cite{Escudero:2008jg}. 
Regarding the neutrino masses, both normal and inverted hierarchies can
be obtained, and two large and one small mixing angle can be
generated~\cite{Ghosh:2008yh} even with a simple  flavour diagonal
structure for the neutrino Yukawa couplings. In~\cite{Fidalgo:2009dm} it
was shown that bimaximal and tri-bimaximal neutrino mixing could be
accommodated with diagonal and degenerated Yukawas for $\mu-$ and 
$\tau$-neutrinos, with degenerated vevs for the left-handed sneutrinos.
Allowing for complex Higgs
and sneutrino vevs, spontaneous CP violation can occur in the
lepton sector, leading to non-vanishing phases in the lepton mixing
matrix. 

\bigskip
\noindent{\it R-parity conserving NMSSM with right-handed neutrinos}
\medskip

``Standard'' seesaw mechanisms (type I, II and III) can be also
incorporated in the R-parity conserving NMSSM, in which case additional
higher dimensional operators (of dimension 6 or 7) can arise from
integrating out heavy states~\cite{Gogoladze:2008wz}. Even with a
TeV-scale seesaw mechanism, these can have
sizable couplings to SM leptons, and may be detected at the LHC. Within
minimal left-right extensions of the NMSSM one can aim at a simultaneous
explanation of the smallness of neutrino masses via the seesaw
mechanism, and a solution of the tachyon slepton problem arising in
Anomaly Mediated SUSY breaking models~\cite{Mohapatra:2008gz}. 

Within the R-parity conserving NMSSM, the vev of the singlet can be the
only source for non-vanishing Majorana masses for the right-handed
neutrinos~\cite{Cerdeno:2008ep,Cerdeno:2009dv} via the last term in
(\ref{var.11e}). Remarkably, a right-handed sneutrino is a viable dark
matter candidate in such a model (see also Section \ref{sec:dm.ss}).

\subsection*{8.3\quad $\pmb{U(1)^\prime}$-extensions of the NMSSM}
\addtocounter{subsection}{1}
\addcontentsline{toc}{subsection}{\numberline{8.3} 
$U(1)^\prime$-extensions of the NMSSM}
\label{sec:var.3}

Extensions of the SM gauge group by one (or several) $U(1)^\prime$ gauge
symmetries can arise naturally from GUTs (see below), string-inspired
constructions~\cite{Suematsu:1994qm,Cvetic:1995rj,Babu:1996vt,
Cvetic:1997ky}, solutions of the $\mu$ problem in
GMSB~\cite{Langacker:1999hs} etc.; for recent reviews and a more
complete list of references
see~\cite{Barger:2006dh,Barger:2007ay, Langacker:2008yv,Kang:2009rd}.

The simplest class of $U(1)^\prime$ extended SUSY models involves one
additional SM singlet $S$, charged under the additional $U(1)^\prime$,
whose vev is responsible for the $U(1)^\prime$ breaking. The
$U(1)^\prime$ charges of the matter and $SU(2)$ doublet Higgs
supermultiplets are not unique, but typically bilinear Higgs couplings
(as the $\mu$-term) are forbidden whereas a $\lambda \widehat{S}
\widehat{H}_u \widehat{H}_d$ term is
allowed, in which case $U(1)^\prime$ models are similar to the NMSSM
(sometimes they are denoted as the UMSSM). 
The term $\sim \kappa \widehat{S}^3$ in the superpotential of the NMSSM
(which stabilises the potential for $s \to \infty$) is no longer
possible, but its r\^ole is now played by the $U(1)^\prime$ $D$-terms.
Note that the Peccei-Quinn symmetry, that appears in the NMSSM in the
absence of the cubic singlet term, is embedded in the gauged
$U(1)^\prime$ so that the would-be Peccei-Quinn axion is eaten by the
$Z^\prime$. Also the NMSSM domain wall problem is avoided, since the
discrete $\mathbb{Z}_3$-symmetry is embedded in the continuous
$U(1)^\prime$ as well. 

The singlet vev generates simultaneously an effective $\mu$-term
(which should not be too large in order to avoid fine tuning, see
Section~6.4) and the mass of the new $Z^\prime$ boson.
Limits on $M_{Z^\prime}$ (from direct searches at the Tevatron) are 
model-dependent, especially on the $Z^\prime$ couplings to quarks
and leptons (see~\cite{Langacker:2008yv,Amsler:2008zzb} and references
therein), and are typically $M_{Z^\prime} \gtrsim 600-900$ GeV
for a $Z^\prime$ that decays into SM fermions. 

Loops from matter, charged under both the SM and the
$U(1)^\prime$ gauge groups, generate a $Z-Z^\prime$ mixing mass term
parametrized by the angle $\alpha_{ZZ^\prime}$, which is constrained
by electroweak precision data to be smaller than
$\mathcal{O}(10^{-3})$, the exact value depending again on the 
chosen $U(1)^\prime$ charges. Both constraints from $M_{Z^\prime}$ and
$\alpha_{ZZ^\prime}$ require a quite large singlet vev $s \gsim {\cal
O}$(1~TeV), and hence a sufficiently negative (and large) value of
$m_S^2$ (see, e.\,g., \cite{Suematsu:2001mq}).

All possible $U(1)^\prime$ charge assignments allowing for a $\lambda
\widehat{S} \widehat{H}_u \widehat{H}_d$ term imply at least mixed
anomalies between the $U(1)^\prime$ and the SM symmetry groups. Their
cancellation requires the
introduction of new exotic fermions (hence superfields) which are
vector-like with respect to the SM, but chiral under $U(1)^\prime$
(see, e.\,g.,~\cite{Cvetic:1997ky,Erler:2000wu,Demir:2005ti,
Morrissey:2005uz,Lee:2007fw}). Then, unification of the gauge couplings
implies the introduction of additional exotics which are charged, but
non-chiral under both the SM and $U(1)^\prime$ gauge symmetries. These
new states do not contribute to the anomalies, but do affect the RGEs
for the gauge couplings. 

In what follows we briefly review some aspects of
NMSSM-like models with one SM singlet $S$. Then we refer to some
features of more general $U(1)^\prime$ extended models.

\medskip
The most relevant modifications of the Higgs potential with respect to
the $\mathbb{Z}_3$-invariant NMSSM are the absence of all terms $\sim \k$, but
the presence of a new $D$ term in the potential of the form
\beq
V_D^{U(1)^\prime}\,=\,\frac{1}{2}g^2_{Z^\prime} \left( Q_S |S|^2 +
Q_{H_d} |H_d|^2 + Q_{H_u} |H_u|^2 + \text{exotics} \right)^2\,, 
\label{var.13e}
\eeq
where $g_{Z^\prime}$ is the $U(1)^\prime$ gauge coupling, and 
$Q_i$ are the $U(1)^\prime$ charges which obey $Q_S+ Q_{H_d}+
Q_{H_u}=0$ in order to allow for the term $\l \widehat S \widehat H_u
\widehat H_d$ in the superpotential.

After symmetry breaking, the Higgs spectrum consists of a pair of
charged Higgs bosons, three neutral CP-even scalars and only one
pseudoscalar. A detailed overview of the Higgs mass matrices
for various models, including the dominant top-stop one-loop
corrections, can be found in~\cite{Barger:2006dh}.

Compared to the NMSSM, theoretically and phenomenologically allowed
Higgs mass ranges have to be reconsidered: due to the additional
quartic Higgs self-couplings from the $D$ terms in (\ref{var.13e}),
Higgs masses are generally larger; in particular, LEP bounds on Higgs
bosons with SM-like couplings to the $Z$ boson are easier to satisfy.
The theoretical upper bound $M_{H_1\text{max}}$ on the mass of the
lightest neutral Higgs boson increases to $M_{H_1\text{max}} \sim
170$~GeV~\cite{Demir:1998dk,Daikoku:2000ep}. Radiative corrections from
top-stop loops have been included
in~\cite{Demir:1998dk,Amini:2002jp,Ham:2009bu}, as well as constraints
from gauge coupling unification in~\cite{Maloney:2004rc}.

Recalling that constraints from $M_{Z^\prime}$ and $\alpha_{ZZ^\prime}$
favour large values of $s$, the Higgs spectrum is generally similar to
the MSSM with a large $\mu$-term~\cite{Cvetic:1997ky}: $H_1$ is SM-like,
the heavy pseudoscalar is approximately degenerate with $H_2$ and
$H^\pm$, completing a full $SU(2)$ doublet. The mass of the heaviest
singlet-like neutral Higgs is typically of the order of $M_{Z^\prime}$.
Then, the lower bound on the lightest CP even Higgs mass implied by LEP
constraints turns out to be in general similar to the one of the MSSM
($M_{H_1} \gtrsim 90$ GeV).

However, in particular cases one can obtain different configurations for
the Higgs spectrum as large doublet-singlet mixing, light singlet
dominated states, etc.~\cite{Erler:2002pr,Han:2004yd,Langacker:2008yv}. 
In~\cite{Demir:2005kg} a scenario with two light CP-even Higgs bosons
has been proposed in order to explain the (light) excesses of events in
Higgs searches at LEP (see
Section~\ref{sec:5}). Note that in the absence of a light pseudoscalar
one cannot relax the lower LEP bound on a doublet-like Higgs allowing
for its decays into a pair of light pseudoscalars as in the NMSSM, see
Section~\ref{sec:5.1}. 

The neutralino sector of the NMSSM is extended by the $Z^\prime$ gaugino
$\widetilde{Z}^\prime$ with a soft SUSY breaking mass
$M_{\widetilde{Z}^\prime}$, and possible 
$\widetilde{B}-\widetilde{Z}^\prime$ mixing
terms~\cite{Suematsu:1998wm}. A detailed discussion of the $6 \times 6$
neutralino mixing matrix can be found 
in~\cite{Suematsu:1997au,Hesselbach:2001ri,Barger:2005hb,Choi:2006fz}.
In the  limit $g_{Z^\prime}\, s \gg M_{\widetilde{Z}^\prime}$, the
singlino and the $Z^\prime$ gaugino mix to form an approximate Dirac
fermion with mass  $\sim M_{Z^\prime}$, which mixes very little with the
other four neutralinos. Conversely, in the large
$M_{\widetilde{Z}^\prime}$ regime one has a very heavy Majorana fermion
$\widetilde{Z}^\prime$, and a much lighter singlino with a ``seesaw''
mass $\sim M^2_{Z^\prime}/M_{\widetilde{Z}^\prime}$. In all cases the
new states will likely affect sparticle decay chains. 

\medskip
CPV has also been considered in $U(1)^\prime$ extended 
models~\cite{Suematsu:1997tv,Demir:2003ke,Ham:2007kc,Kang:2009rd}.
However, at the tree level the Higgs sector is CP conserving, and
spontaneous CPV (after including radiative corrections) cannot be
realised in the allowed parameter space~\cite{Ham:2007kc}. 
Explicit CPV in the Higgs sector can be possible via radiative
corrections, through the explicit CPV phases  present in the stop mass
matrix or the additional phases in the exotic quark 
sector of the model. These can significantly alter the mixing between
the pseudoscalar and the heaviest scalar~\cite{Demir:2003ke}. 
Due to the extended neutralino sector, 
phases in $\lambda$ and in the slepton trilinear
couplings can contribute to the electron EDM~\cite{Suematsu:1997tv}.
EDM bounds strongly constrain the effective phase in the chargino mass
matrix, but all other SUSY phases remain largely unconstrained, thus
alleviating the SUSY CP problem~\cite{Demir:2003ke}. 

\medskip
$U(1)^\prime$ extended  models have also important cosmological
implications, since the extra states can modify the nature of the LSP. 
Early studies~\cite{deCarlos:1997yv} 
suggested that $Z^\prime$ mediated neutralino annihilation
could provide important (or even dominant) contributions to the 
correct relic density. More recent analyses 
\cite{Barger:2004bz,Barger:2005hb,Suematsu:2005bc,Nakamura:2006ht,
Barger:2007nv,Kang:2009rd} (for a general discussion,
see~\cite{Kalinowski:2008iq} and references therein) have shown that 
compatibility with the WMAP bounds on the relic density can be 
achieved in the case of a singlino-like LSP, with a small higgsino
component,  via annihilation into a ${Z^\prime}$; enhanced couplings to
the $Z$
also allow annihilation via a $Z$ resonance~\cite{Barger:2004bz}. 
The prospects for observation of the LSP in dark matter direct detection
experiments have been considered in~\cite{Barger:2007nv}.
These models offer also non-neutralino dark matter 
candidates: scalar right-handed sneutrinos can be viable 
dark matter candidates 
due to the possibility of annihilation through the 
$Z^\prime$~\cite{Lee:2007mt}; other
possibilities include exotic LSPs~\cite{Hur:2007ur}, see also
Section~\ref{sec:dm.ss}.

\medskip
Dirac neutrino masses in the experimentally preferred range can also be
generated within $U(1)^\prime$-extended generalized SUSY
models~\cite{Kang:2004ix,Kang:2005ci,Demir:2007dt,Suematsu:2009ia}. If
the $U(1)^\prime$ symmetry forbids the Yukawa couplings to right-handed
neutrinos in the superpotential, effective Dirac masses can be generated
from (hard) SUSY breaking Yukawa couplings or, at the loop level, from
non-holomorphic trilinear couplings to the ``wrong'' Higgs boson, which
are naturally small~\cite{Demir:2007dt}. SUSY models of sterile
neutrinos can be naturally realised in $U(1)^\prime$ extended
models~\cite{Kang:2005ci}: sterile neutrino masses of $\mathcal{O}(1
\text{eV})$ and mixings among the active and sterile neutrinos can then
be obtained via the high-dimensional operators, generated by integrating
out the heavy fields. In $U(1)^\prime$ extended models where the vev
$s$ generates Dirac neutrino masses through a dimension-4 term in the
superpotential, the decays of the LSP and NLSP
(the two lightest sneutrinos) could explain the PAMELA
anomaly~\cite{Demir:2009kc}.

\medskip
We recall that, once R-parity is violated, proton decay can be
excessively
fast since both baryon number and lepton number are violated in general.
In $U(1)^\prime$ models with RpV, the proton can still be sufficiently
stable in a natural way, since the $U(1)^\prime$ symmetry can forbid the
renormalisable (dimension 3 and 4) and the most dangerous dimension~5
operators~\cite{Erler:2000wu}. In~\cite{Lee:2007fw} models are discussed
which can still contain either lepton number violating or baryon number
violating renormalisable interactions, but never simultaneously both.
In~\cite{Hundi:2009yf} models have been put forward where bilinear
lepton number violating terms in the superpotential allow to accommodate
the observed pattern of neutrino masses and mixings, but the most
dangerous dimension 3 and 4 baryon number violating operators (and,
possibly, all $\Delta B=1$ operators) are absent.

\medskip
Production (and decays) of the additional $Z^\prime$ gauge boson and
exotic states at colliders are reviewed 
in~\cite{Barger:2006sk,Langacker:2008yv}. The exotic quarks and squarks
could appear in decays of the $Z^\prime$, or may manifest themselves
indirectly, since the production of the heaviest Higgs boson (via gluon
fusion) can be dominantly  mediated by loops of these
exotics~\cite{Ham:2009bu}. The production of neutral Higgs bosons in
this class of models has been discussed in\cite{Ham:2008xf}. 

At $e^+\,e^-$ colliders of higher energy, the Higgs sectors of
$U(1)^\prime$ extended models and the NMSSM can be distinguished in the
limit of large trilinear Higgs couplings~\cite{Demir:1999mm}. At the
ILC, measurements of the $H^0_i\,Z\,Z$ couplings may help to
distinguish scenarios with a light, leptophobic $Z^\prime$ from the
NMSSM, since the couplings are expected to be substantially smaller in
the former case~\cite{Ham:2008ey}. In the case of TeV $Z^\prime$ bosons,
searches for $Z^\prime$ decays into higgsinos can also test if a gauge
symmetry is indeed responsible for the absence of a bare $\mu$-term in
the superpotential~\cite{Cohen:2008ni}.

\medskip
There are innumerous possibilities for the construction of more general
$U(1)^\prime$-extensions of the MSSM including, for instance, flavour
symmetries~\cite{Daikoku:2009pi}. A systematic analysis and
classification can be found in~\cite{Erler:2000wu}. In what follows we
consider some particular cases, some of which are motivated by GUTs or
string-inspired constructions. 

\bigskip
\noindent{\it $U(1)^\prime$ and a secluded singlet sector}
\medskip

A possible way to avoid the tension arising from a small enough
$\mu_\text{eff}$ and a sufficiently large $M_{Z^\prime}$ is to consider
the secluded $U(1)^\prime$-extended MSSM (sMSSM), where a separation
between $\mu_\text{eff}$ and $M_{Z^\prime}$ is achieved by the
introduction of four SM singlets $S$, $S_1$, $S_2$ and
$S_3$~\cite{Erler:2002pr}. The ordinary sector contains the NMSSM (or
USSM) singlet $S$, while the three additional singlets $S_i$ belong to a
secluded sector which couples to the ordinary sector only via
$U(1)^\prime$ gauge and possibly soft SUSY breaking terms. All singlets
are charged under $U(1)^\prime$ such that their vevs contribute to
$M_{Z^\prime}$, but only $S$ generates a $\mu_\text{eff}$ of the order
of the electroweak scale. The SM singlet-dependent part of the superpotential is
given by
\beq\label{var.u1.2e}
W^{U(1)}_H\,=\,\lambda\,\widehat S\, \widehat H_{d} \, \widehat H_{u}\,+
\lambda_{S}\,\widehat S_1\,\widehat S_2\,\widehat S_3\,.
\eeq
The soft SUSY breaking Lagrangian contains soft SUSY breaking masses 
and trilinear couplings for the
new singlets ($m^2_{S_i}$, $m^2_{S_i S_j}$, 
$A_{\lambda_{S}}$). Unwanted global symmetries have to be broken
explicitly  to avoid the potential occurrence of two massless
experimentally excluded Goldstone bosons in the spectrum.
For appropriate parameters, $S_{1,2,3}$ acquire large vevs along nearly
$F$-flat and $D$-flat directions of the scalar potential so that one
obtains $M_{Z^\prime} \gg \mu_\text{eff}$ in a natural way. $\l$
can be larger than in the NMSSM (without leading to a Landau
singularity), since the additional gauge coupling in its $\b$-function
slows down its increase towards high scales. As a consequence, the
lightest CP-even Higgs mass could be as large as $\sim 174$~GeV already
at tree level~\cite{Erler:2002pr}. In general the Higgs and neutralino
sectors of these models are quite complicated, and we refer
to~\cite{Han:2004yd} for a detailed discussion. A singlino-higgsino
LSP can account for an acceptable cold dark matter relic 
density~\cite{Barger:2004bz}. Regions in parameter
space exist, that can explain the experimental deviation of the muon
anomalous magnetic moment from the SM and yield an acceptable cold dark
matter relic density without conflict with collider experimental
constraints~\cite{Barger:2004mr}.

In the sMSSM, explicit CP violation in the Higgs sector can be induced
at the tree level by the non-zero phase of $m^2_{S_1
S_2}$~\cite{Chiang:2008ud}. Large values of the latter phase can allow
for a neutral Higgs boson lighter than $\sim 90$~GeV, compatible with
LEP constraints due to suppressed couplings to the $Z$ boson. Unlike the
MSSM and NMSSM, SCPV can occur at the tree level. The possibility of
electroweak baryogenesis in the sMSSM has been discussed
in~\cite{Kang:2004pp}.

\bigskip
\noindent{\it Additional $U(1)^\prime$ in string-inspired $E_6$}
\medskip

As mentioned before, gauge extensions of the SM gauge group by one (or
several) non-anomalous $U(1)^\prime$ gauge symmetries can naturally
arise from a (string-inspired) GUT gauge group as $E_6$
(see, e.\,g.,~\cite{Hewett:1988xc,Langacker:1998tc,Kang:2004bz}). 

In an $E_6$ GUT, the matter sector includes three 27-plets containing
the three ordinary families of quarks and leptons (including
right-handed neutrinos), three families of candidate Higgs doublets
($H_{di}$,
$H_{ui}$)  and singlets ($S_i$), and three families of extra colour
triplets ($D'_i$, $\bar D'_i$).  Anomalies are cancelled generation
by generation within each complete $27$ representation.  Gauge coupling
unification requires the addition of a further pair of Higgs-like
multiplets, $H^\prime$ and $\bar H^\prime$, arising from incomplete
$27^\prime + \overline{27}^\prime$ representations. The breaking of 
$E_6$ occurs via
\beq\label{var.e6.1e}
E_6 \, \to \, SO(10) \times U(1)_\psi \, \to
\, SU(5) \times U(1)_\chi \times U(1)_\psi \,.
\eeq
The low energy gauge group is assumed to be $SU(3)\times SU(2) \times
U(1)_Y \times U(1)^\prime$, where $U(1)^\prime = U(1)_\chi$ or a
specific linear combination of $U(1)_\chi$ and $U(1)_\psi$ motivated by
the breaking of $E_6$ by Wilson lines.

In~\cite{Hewett:1988xc,Langacker:1998tc},  patterns of $U(1)^\prime$ and
electroweak symmetry breaking were studied in this class of $E_6$ models. 
$E_6$-type relations between the Yukawa couplings are not imposed, and
only those conserving baryon and lepton number are kept. Then, the most
general superpotential allowed by gauge invariance is of the
form~\cite{Langacker:1998tc}
\beq\label{var.e6.2e}
W_H^{E_6}\,=\,h^u_{ijk} \widehat U^c_i\,\widehat Q_j \,\widehat H_{uk} +
h^d_{ijk} \widehat D^c_i\,\widehat Q_j \,\widehat H_{dk} +
\lambda_{ijk}\,\widehat S_i\, \widehat H_{dj}\, \widehat H_{uk}+ 
h^D_{ijk} \widehat S_i\,\widehat D_j' \,\widehat{D}'^c_{k}
\;,
\eeq
where $i,j,k$ are family indices. When some of the Higgs fields acquire
vevs, electroweak and $U(1)^\prime$ symmetries are broken. It is always possible
to work in a basis where only one family of Higgs doublets and singlets
develops a vev: one can define  $S=S_3$, $H_d=  H_{d3}$ and $H_u= 
H_{u3}$, while the remaining families are treated as additional exotic
doublets and singlets without vevs.  In order to obtain an acceptable
low energy phenomenology, universality at the GUT scale of the soft SUSY
breaking masses has to be relaxed. (Universal boundary conditions for
the soft breaking terms could be accommodated by identifying
$U(1)^\prime$ with a more general linear combination of $U(1)_\chi$ and
$U(1)_\psi$~\cite{Langacker:1998tc}.)

Another $E_6$ inspired version exhibiting an NMSSM-like low energy
spectrum and allowing for explicit CP violation in the Higgs sector
has been considered in~\cite{Ham:2008fx}. 
The electroweak gauge symmetry of the SM is enlarged to
$SU(2) \times U(1) \times U(1)_1 \times U(1)_2$, where both extra
$U(1)$ symmetries originate from $E_6$. The Higgs sector contains two
doublets and two singlets (only one being allowed to couple to the
doublets due to the underlying $E_6$ symmetry). 
Due to top-stop mediated radiative corrections the model allows for
explicit CP violation in the Higgs sector at the one-loop level, leading
to scalar-pseudoscalar mixings among the five neutral Higgs bosons. 

\bigskip
\noindent{\it The Exceptional SSM}
\medskip

Specific realisations of $E_6$-inspired $U(1)^\prime$ extensions offer
the possibility of incorporating a seesaw mechanism (to explain the
smallness of light neutrino masses) in a natural way. To this end the
extra $U(1)^\prime$ surviving at low energy (denoted by $U(1)_N$) must
correspond to a particular combination of $U(1)_\chi \times U(1)_\psi$
in (\ref{var.e6.1e}) under which the right-handed neutrinos (also
contained in the $27$ representation) transform trivially:
\beq\label{var.essm.3e}
U(1)_N\, \equiv \, U(1)^\prime \,=\,
 U(1)_\chi \cos \theta_N +  U(1)_\psi \sin \theta_N 
\quad \text{with} \quad  \theta_N = \arctan \sqrt{15}\,.
\eeq 
This is the so-called Exceptional SSM (E$_6$SSM,
see~\cite{King:2005jy,King:2005my,King:2007uj,Howl:2008xz,Athron:2009ue,
Athron:2009bs,Howl:2009ds,Hall:2009aj} and references therein).

As discussed in ~\cite{King:2005jy},
due to the vanishing charges under $U(1)_N$, right-handed neutrinos can
acquire very heavy Majorana masses suitable for the ``standard'' seesaw
mechanism. This also opens the possibility of baryogenesis  from
leptogenesis, and avoids constraints on the mass of the
$Z^\prime$ arising from nucleosynthesis. Although $B-L$ is automatically
conserved in $E_6$-inspired SUSY models, some of the new Yukawa
interactions could violate baryon number resulting in rapid proton
decay. Proton stability, successful leptogenesis, non-zero
neutrino masses and the absence of exotic coloured dark matter relics
can be guaranteed with the help of a
new conserved R-parity-like quantum number and corresponding
baryon and/or lepton number assignments for the exotic particles.
Furthermore, it is mandatory to impose a hierarchical structure for the
additional Yukawa couplings as well as a discrete $\mathbb{Z}_2$
symmetry as to avoid excessive flavour changing neutral currents.  

Due to the modified (two-loop) beta functions in the E$_6$SSM,
unification of the gauge couplings can be achieved for values of
$\alpha_s(M_Z)$ closer to the measured central
value than in the MSSM~\cite{King:2007uj}. 

The E$_6$SSM Higgs sector includes three CP-even, one CP-odd and two
charged states. In the MSSM limit $\lambda \ll 1$, the new states become
very heavy and decouple, rendering the model essentially
indistinguishable from the MSSM.
The analysis of RG flow of the gauge and Yukawa
couplings shows that the absence of a Landau singularity allows for
larger $\lambda$ and lower $\tan\b$ at the electroweak scale than in the
MSSM and NMSSM, which affects the Higgs, neutralino and chargino
spectra. When $\lambda \gtrsim g_1$, the lightest Higgs scalar can
be somewhat heavier than in the NMSSM; including two-loop corrections,
the maximally allowed value is around 150~GeV. Vacuum stability
constraints push the masses of the heaviest CP-even, CP-odd and charged
states above 1 TeV, so that only the lightest scalar would be within
reach of the LHC and an ILC~\cite{King:2005my}. 

The superpartners of the $Z^\prime$ and the singlet contribute to the
neutralino spectrum, while the chargino sector remains unchanged.
For $\lambda \gtrsim g_1$, the heaviest neutralino and
chargino states are the neutral and charged $H_{u,d}$
higgsinos; the lightest chargino is wino-like, while the LSP is a
bino-like neutralino~\cite{King:2005jy}. The dark matter prospects for
this class models have been addressed in~\cite{Hall:2009aj} (see
Section~\ref{sec:dm.ss}).

The many new additional exotics form vector-like multiplets which could
possibly be produced and detected at the LHC; the new $Z^\prime$ could
be visible if lighter than 4-5 TeV. Within the constrained E$_6$SSM
(assuming universal soft SUSY breaking parameters at high energy) a
detailed study of the phenomenology and impact for LHC discovery  has
been carried out in~\cite{Athron:2009ue,Athron:2009bs}.   Scenarios for
early LHC discovery were investigated in the low ($m_0, M_{1/2}$)
regime, and a set of benchmark points was proposed. 

E$_6$ models based on a $\Delta_{27}$ family symmetry broken
close to the GUT scale allow in addition to address the flavour
problem: quark and lepton masses and mixing angles can be accounted
for, with tri-bimaximal neutrino mixing~\cite{Howl:2008xz,Howl:2009ds}.

\section{Dark Matter in the NMSSM}
\label{sec:dm}

As in the MSSM, the lightest neutralino in the NMSSM is a candidate for
cold dark matter (DM) in the form of WIMPs (weakly interacting massive
particles), whose relic density should be in agreement with the
constraints from WMAP, $0.094 \lesssim \Omega h^2 \lesssim 0.136 $ at
the $2\,\sigma$  level~\cite{Spergel:2006hy,Tegmark:2006az}. (For a
review of DM in the MSSM see \cite{Jungman:1995df}.)
Due to the differences in the neutralino and Higgs sectors, the
properties of the LSP in the NMSSM can be significantly distinct from
those in the MSSM, both in its nature and/or in the processes relevant
for the LSP relic density and its detection: the LSP
can be dominantly singlino-like, and/or the additional Higgs bosons can
contribute to the LSP annihilation and detection processes.
A neutralino LSP in the form of a higgsino-singlino
mixed state in the NMSSM, giving rise to the required dark matter
density, was  considered first 
in~\cite{Greene:1986th,Olive:1990aj,Flores:1990bt,Gondolo:1990dk,
Abel:1992ts}, where often universality constraints on the soft SUSY
breaking terms were imposed.

The large number of processes relevant for the computation of the LSP
relic density (annihilation and coannihilation cross sections) and of
LSP detection in the NMSSM will not be discussed here in detail, but all
of them are included in the code MicrOMEGAS (see Appendix~D), and
described in~\cite{Belanger:2005kh,Belanger:2006is,Belanger:2008sj}.

Subsequently we discuss some aspects of the LSP relic density and
its detection in the NMSSM which differ from the MSSM: in
Section~\ref{dm:rd}, we discuss the processes relevant for the dark
matter relic density; in Section~\ref{dm:dd} we consider the
detection of an NMSSM dark matter candidate; finally, we will
address specific NMSSM scenarios for DM in Section~\ref{sec:dm.ss}.

\subsection{Neutralino relic density}\label{dm:rd}

The basic mechanisms that lead to a reduction of the
neutralino relic density are essentially the same as in the MSSM:
neutralino annihilation via $s$- and/or $t$-channel exchanges into gauge
boson, Higgs boson and fermion pairs, and coannihilation of the
neutralino with a heavier state (typically the NLSP). The cross section
for these mechanisms can be strongly enhanced by $s$-channel resonances,
occurring when $m_\text{LSP}+m_\text{(N)LSP}$ is close to the mass of the
particle exchanged in the $s$-channel.

Distinctive NMSSM scenarios can be manifest in several ways,
depending on the regions in the NMSSM parameter space: the singlino
component of the LSP can be sizable (e.g. mixed  higgsino-singlino LSP),
and one can even have (nearly) pure singlino LSPs. Secondly, the extra
scalar and pseudoscalar Higgs bosons can lead to rapid LSP annihilation
through $s$-channel Higgs resonances. Moreover, in the case where the
additional scalar/pseudoscalars are light, new final states can be
kinematically allowed: for example, annihilation into $Z\,H_1$,
$H_1\,H_1$,  $H_1\,A_1$ and $A_1\,A_1$ can  significantly contribute to
the reduction of the neutralino relic density, either via $s$-channel
$Z$, $H_1$, $A_1$, or $t$-channel (heavier) neutralino exchange. Here
the NMSSM specific couplings $\l$ and $\k$ can play an important~r\^ole.

A detailed discussion the viability of DM candidates in the
(unconstrained) NMSSM, using recent cosmological constraints and
numerical tools, is given in~\cite{Belanger:2005kh}. In what follows,  
we briefly discuss some illustrative examples of how the new features of
the NMSSM are manifest, separately considering annihilation and
coannihilation. 

\subsubsection{Annihilation}

Here we briefly consider neutralino annihilation processes that
reflect the new features of the NMSSM. For a detailed discussion, albeit
in the framework of the MSSM, we refer
to~\cite{Jungman:1995df}. Detailed formulae for (MSSM) annihilation
cross sections can be found, for example, in~\cite{Nihei:2002ij}.

\begin{itemize}
\item Annihilation via $s$-channel Higgs exchange (CP-even or CP-odd)

\noindent
The annihilation cross section is proportional to (neglecting
interference terms) 
\beq\label{dm.1e}
\sigma (\chi_1^0\,\chi_1^0 \overset{H_a}{\to} X\,X^\prime) \,
\propto\, \left| \sum_a 
\frac{(H_a \chi_1^0 \chi_1^0)\, (H_a X X^\prime)}{s - M_{H_a}^2 + i
\Gamma_{H_a} M_{H_a}}\right|^2 
f_s(s, m_{\chi_1^0}^2, m_X^2, m_{X^\prime}^2)\,,
\eeq
where $X,\,X^\prime$ denote SM fermions, gauge or Higgs bosons,
$H_a\chi_1^0 \chi_1^0$ the Higgs-neutra\-lino-neutralino couplings and $
H_a X X^\prime$ the Higgs couplings to the final states (see
Appendix~A). $\Gamma_{H_a}$ is the width of $H_a$, and we have
introduced a generic function  $f_s$ that depends on the $s$-channel 
momentum and on the masses of the initial and final
states~\cite{Nihei:2002ij}. A similar expression is obtained for the
exchange of a pseudoscalar (with the appropriate replacements in the
couplings, masses, and different $f_s$).

\item Annihilation via $s$-channel $Z$ exchange 

\noindent
In this case, one approximately has
\beq\label{dm.2e}
\sigma (\chi_1^0\,\chi_1^0 \overset{Z}{\to} X\,X^\prime) \,\propto \,
\left|\frac{(Z \chi_1^0 \chi_1^0)\, (Z X X^\prime)}{s - M_{Z}^2 + i
\Gamma_{Z} M_{Z}}\right|^2 
g_s(s, m_{\chi_1^0}^2, m_X^2, m_{X^\prime}^2, M_Z^2)\,,
\eeq
where the SM $Z X X^\prime$ couplings are proportional to electroweak
gauge couplings, and $(Z \chi_1^0 \chi_1^0) = g_2/(2 \cos \theta_W) \,
(N_{13}^2-N_{14}^2)$, with $N_{13}$ and $N_{14}$ (according to (A.7))
denoting the higgsino components of the lightest neutralino $\chi_1^0$.
For the function $g_s$ see~\cite{Nihei:2002ij}.

\item Annihilation via $t$-channel neutralino exchange 

\noindent
As in the MSSM, neutralino annihilation can also take place via
$t$-channel sfermion, chargino and neutralino exchange. In the latter
case, annihilation into a pair of singlet-like Higgs bosons ($H_1 H_1$,
$A_1 A_1$, or $H_1 A_1$) can play a significant r\^ole for singlino-like
LSPs~\cite{Belanger:2005kh}.  As an example, the cross section for 
$\chi_1^0\,\chi_1^0 \to H_1 A_1$ via $t$-channel neutralino exchange
reads
\beq\label{dm.3e}
\sigma (\chi_1^0\,\chi_1^0 \overset{\chi_i^0}{\to} H_1\, A_1) \,\propto
\, (H_1 \chi_1^0 \chi_i^0)^2 \,  (A_1 \chi_1^0 \chi_i^0)^2 
h_s(s,m^2_{\chi_1^0}, m^2_{H_1}, m^2_{A_1})\,;
\eeq
the scalar/pseudoscalar Higgs-neutralino-neutralino couplings are
given in (\ref{a.14e}) and $h_s$ denotes an auxiliary function encoding
the mass dependence~\cite{Nihei:2002ij}. In this case, $h_s$
decreases with the neutralino mass as $m^{-4}_{\chi_1^0}$.

\item Resonant $s$-channel annihilation

\noindent
LSP freeze-out occurs at temperatures $T \sim m_\text{LSP}/20$, where
the LSPs are non-relativistic and $s \sim 4\, m_{\text{LSP}}^2$. If
neutralino annihilation proceeds via the $s$-channel exchange of a
particle (Higgs or gauge boson) with mass $\approx 2 m_{\chi_1^0}$,  the
pole in the annihilation cross section can lead to a rapid decrease in
the neutralino
relic density. This resonance phenomenon can  allow for compatibility
with WMAP bounds in cases where the relic  density would be otherwise
too large. It is important to stress that for the case of resonant
annihilation, the relic density is very sensitive to the mass
differences of the LSPs and the exchanged particles.  Hence, regions in
the NMSSM (or MSSM) parameter space where WMAP compatibility is achieved
via resonant annihilation typically correspond to narrow bands (for the
NMSSM see, 
e.g.,~\cite{Belanger:2005kh,Hugonie:2007vd,Belanger:2008nt}).

\item Coannihilation LSP-NLSP 

\noindent
In the MSSM and in the NMSSM scenarios exist
where no efficient annihilation mechanism is available. However, the
correct relic density can still be obtained by coannihilation processes.
If the NLSP (typically the second lightest neutralino
$\chi_2^0$ or a sfermion) is not much heavier than the LSP, so that
it is still present in the thermal plasma at the time of LSP freeze-out,
coannihilation processes can efficiently reduce the LSP relic density.

In the NMSSM, 
coannihilations with heavier neutralinos or charginos
can occur via $s$-channel $Z$, $W^\pm$ or Higgs exchange, for example 
$\chi_1^0 + \chi_2^0 \to H \to X\, X^\prime$, or via 
$t$-channel $\chi^0_i, \chi_j^\pm$ exchange~\cite{Belanger:2005kh}. 
For a sfermion NLSP (usually a light, mostly right-handed
stau), $s$-channel and $t$-channel (the latter via neutralino or
chargino exchange) coannihilation lead to fermion final states.
Just like in the case of $s$-channel LSP-LSP annihilation, resonances
can also increase the LSP-NLSP annihilation cross section: in this
case the mass of the exchanged particle should be $\approx
m_\text{LSP}+m_\text{NLSP}$. 

\item NLSP-NLSP annihilation

\noindent
This process, usually also labeled coannihilation,  occurs when the
dominant annihilation process of R-odd particles is via NLSP
annihilation:  $\text{NLSP} + \text{NLSP} \to X$. Provided the NLSP-LSP
mass difference is small, the correct LSP relic density is achieved by 
processes that maintain the LSP and the NLSP in thermal equilibrium
(implying $n_\text{LSP} \sim n_\text{NLSP}$ for the corresponding
abundances): 
$\text{LSP} + X \rightleftarrows \text{NLSP} + X^\prime$, where  $X,
X^\prime$ are light SM particles (quarks and leptons).

The reaction rate for the NLSP annihilation process $\text{NLSP} +
\text{NLSP} \to X$ is given by $(n_\mathrm{NLSP})^2 \sigma$, where
$\sigma$ is the thermally averaged cross section. On the
other hand, the reaction rate for the processes that help to keep the
LSP and the NLSP in thermal equilibrium  ($\mathrm{LSP} + X \to
\mathrm{NLSP} + X'$ and its inverse) depends on $n_\mathrm{LSP}\,
n_{X}\, \sigma'$. Even if the latter cross section $\sigma'$ is possibly
suppressed (as, for instance, for a singlino-like LSP in the NMSSM), the
process  $\mathrm{LSP} + X \to \mathrm{NLSP} + X'$ is typically faster
than the annihilation process $\mathrm{NLSP} + \mathrm{NLSP} \to X$,
since near the  freeze-out temperature the abundance $n_X$ of quarks and
leptons is $\sim 10^9$ larger than the abundances of the LSP and NLSP 
(for $m_\mathrm{LSP} \sim m_\mathrm{NLSP}$)~\cite{Griest:1990kh}. This
allows to dilute the LSP density as fast as the NLSP density, and such
``assisted coannihilation'' can play a very important r\^ole in
reconciling the relic density of an NMSSM LSP with WMAP 
measurements~\cite{Belanger:2005kh,Hugonie:2007vd}.

\end{itemize}

In the MSSM, the LSP is usually a bino- or a higgsino-like neutralino
(or a bino-higgsino mixed state). These scenarios can also be found in
regions of the NMSSM parameter space where, apart from the enlarged
Higgs sector and heavy neutralinos, the spectrum is MSSM-like
(see~\cite{Belanger:2005kh,Hugonie:2007vd}). Before turning to
NMSSM specific annihilation processes, we briefly discuss these
MSSM-like scenarios.

The relic density of a nearly pure bino-like LSP is typically difficult
to reconcile with WMAP measurements: the only available annihilation
channel for a pure bino is via $t$-channel sfermion exchange (as all
couplings to gauge bosons require a higgsino component), or via
coannihilation with a sfermion (typically stau or stop) NLSP. The former
requires light sfermions (as the cross section is $\propto (m_{\tilde
f})^{-4}$), while the latter requires the sfermion to have a mass close
to the one of the bino-like LSP. 

If the LSP has a non-negligible higgsino component, it can
efficiently annihilate into a pair of gauge bosons ($WW$, $ZZ$) and
doublet-like Higgs bosons ($W^\mp H^\pm$, $Z H$, $H A$) via
$s$-channel $Z$ or Higgs exchange, as well as through $t$-channel
neutralino and chargino exchange. For sufficiently heavy LSPs
($m_{\chi_1^0} > m_t$), annihilation into a pair of top quarks via Higgs
exchange can significantly contribute. In the large $\tan \beta$ regime,
where the Higgs couplings to $b \bar b$ and $\tau^+ \tau^-$ are
enhanced, $b \bar b$ and $\tau^+ \tau^-$ final states are possible via
$s$-channel Higgs exchange (not necessarily resonant).

A tiny higgsino fraction already opens the possibility of annihilation
through resonant $s$-channel Higgs exchange. However, to overcome the
smallness of the Higgs-$\chi_1^0 \chi^0_1$ coupling in this case, a
significant fine tuning is required so that the corresponding Higgs
boson has the appropriate mass $\sim 2 m_{\chi_1^0}$, which is not
always possible in the MSSM. 

Higgsino-like LSPs can also coannihilate with neutralinos and charginos.
In the MSSM, it can happen that coannihilation is excessively efficient
in this case, so that the relic density is too small. In the NMSSM, a
non-vanishing singlino component can reduce the LSP couplings (and hence
the coannihilation cross sections), leading to viable values for the
relic density~\cite{Belanger:2005kh}. 

\subsubsection{NMSSM-specific annihilation processes}

In the NMSSM, the richer scalar/pseudoscalar Higgs sector  allows to
have the correct relic density in large regions of the parameter space,
from low to high $\tan \beta$~\cite{Belanger:2005kh,Hugonie:2007vd}.  
The possibility of light scalars and pseudoscalars consistent with LEP
constraints also implies that new final states are kinematically 
open~\cite{Belanger:2005kh,Hugonie:2007vd}. For example, nearly pure
binos can annihilate via $H_1$-resonances into $A_1 A_1$. 

A distinctive feature of the NMSSM is the possibility of a nearly pure
singlino-like LSP. However, due to the small couplings to SM particles,
singlino-like LSPs tend to have a too large relic density due to
comparatively small annihilation cross sections. Still, there are
several possible mechanisms that allow to render a singlino-like LSP
compatible with WMAP bounds~\cite{Belanger:2005kh,Hugonie:2007vd}:

The couplings of a singlino-like LSP (with $N_{15}^2 \sim 1$) to CP-even
Higgs bosons $H_a$ are approximately given by $\sim -\sqrt{2}\, S_{a3}\,
\kappa$ (see (\ref{a.14e})), with $S_{a3}$ replaced by $P_{a3}$ for
CP-odd Higgs bosons. Accordingly any CP-even (or CP-odd) Higgs state,
whose singlet component $S_{a3}$ (or $P_{a3}$) is sufficiently large,
can give an important contribution to LSP annihilation via $s$-channel
exchange, if $\k$ is not too small.

If the singlino-like LSP has a small higgsino component, resonant
$s$-channel annihilation via scalar/pseudoscalar Higgs or $Z$ bosons
allows for viable NMSSM scenarios if $m_{\chi_1^0} \lesssim 50$~GeV
(occurring in regimes of $\kappa \ll \lambda$, and not very large
$\mu_\text{eff}$), or even $m_{\chi_1^0} \sim$ a few GeV provided that
$2 m_{\chi_1^0} \approx M_{H_1}, M_{A_1}$. Annihilation can then proceed
into $\chi_1^0\,\chi_1^0 \to b \bar b$ or, for the case of $H_1$
exchange and a light pseudoscalar, into $\chi_1^0\,\chi_1^0 \to A_1
A_1$.

For a heavier mixed singlino/higgsino-like LSP ($m_{\chi_1^0} \gtrsim
100$ GeV) and large $\l$, the annihilation can occur via $t$-channel 
$\chi_1^0$ exchange into pairs of mostly singlet-like  $H_1$ and $A_1$
due to the enhanced $\chi_1^0 \chi_1^0 H_1 (A_1)$ couplings $\sim \l$.

An LSP with a large singlino component can also coannihilate efficiently
with heavier neutralinos (provided their mass difference is
$\lesssim 10$ GeV), especially with a higgsino-like $\chi_2^0$,
leading to $\chi_1^0 + \chi_2^0 \to t \bar t, \,b \bar b$ via heavy
doublet-like Higgs exchange. Such coannihilation cross sections can be
further enhanced by $s$-channel resonances: if the heavy states $H_3$
and $A_2$ belong to the heavy Higgs doublet and are nearly
degenerate, one may even obtain a  ``double resonance''
with two $s$-poles from both $H_3$ and $A_2$ exchange.

For a nearly pure singlino $(N_{15}^2 \gtrsim 99\%)$ in regimes of
very small $\lambda$ and $\kappa$ ($\kappa \lesssim \lambda \lesssim
0.01$), no efficient annihilation mechanisms are available, and even 
LSP-NLSP coannihilation (e.g. with heavier neutralinos/charginos, or
sfermions) fails to diminish the LSP relic density sufficiently.
In this case, the LSP density can be reduced by NLSP-NLSP annihilation
(assisted coannihilation) in cases of, for instance, a bino-like NLSP or
a sfermion NLSP. Assisted coannihilation is relevant for the fully
constrained NMSSM~\cite{Djouadi:2008yj,Djouadi:2008uj} as discussed in
Section~\ref{sec:ss.1}, in which case it requires nearly degenerate
$\chi_1^0$ and $\tilde \tau_1$ masses. 

However, to this end the reaction rate for the process $\mathrm{LSP} + X
\to \mathrm{NLSP} + X'$  ($\sim n_\mathrm{LSP}\, n_{X}\, \sigma'$) has
to be at least as large as the NLSP annihilation rate ($\sim
n_\mathrm{NLSP}^2\, \sigma$); otherwise the LSP will no longer be in
thermal equilibrium with the NLSP near the freeze-out temperature, but
can be considered as decoupled. This condition allows to derive a rough
lower bound on $\lambda$: the process $\mathrm{LSP} + X \to
\mathrm{NLSP} + X'$ is made possible only through the non-singlet
component of the LSP; for a dominantly singlet-like LSP, the analysis of
the neutralino mass matrix shows that this non-singlet component is
proportional to $\l$. Hence, the cross section $\sigma'$ is proportional
to $\l^2$, and $\sigma'/\sigma \sim \l^2/g_2^2$. As mentioned above, the
density $n_X$ of light SM particles is about $n_X \sim 10^9\,
n_\mathrm{NLSP}$ near the freeze-out temperature. Then, since
$n_\mathrm{NLSP} \sim n_\mathrm{LSP}$, the reaction rate for
$\mathrm{LSP} + X \to \mathrm{NLSP} + X'$ is larger than the NLSP
annihilation rate only if $\lambda^2/g_2^2 \gsim 10^{-9}$ or $\lambda
\gsim 10^{-5}$~\cite{Djouadi:2008yj,Djouadi:2008uj}.

\subsection{Dark matter detection in the NMSSM}\label{dm:dd}

In the NMSSM, the prospects for direct and indirect  detection of
neutralino DM can also be substantially different from what is expected
in the MSSM (for a review of WIMP detection see~\cite{Jungman:1995df});
subsequently we discuss direct and indirect detection separately.

\subsubsection{Direct dark matter detection}

Direct detection of neutralino dark matter (and generically WIMP dark
matter) proceeds via the measurement of the recoil energy deposited by
the scattering of WIMPs on nuclei in a detector. The energy transfer
arising from the elastic scatterings typically lies around
$\mathcal{O}$(10~keV), with predicted event rates below 0.1
events/kg/day. WIMP-nucleus interactions can be spin-independent (SI) or
spin-dependent (SD).  The SD contribution (via the axial-vector
interaction) is only non-zero if the target nucleus has a net
non-vanishing spin (unpaired nucleons), which is assumed to be carried
by ``odd-group'' nucleons: protons or neutrons, whichever is most
unpaired. The SI (scalar) contribution is proportional to the mass
squared of the nucleus, and almost always dominates for nuclei with $A
\gtrsim 30$, like those used in most modern detectors (e.\,g. Germanium
and Xenon). The total elastic cross section is the sum of both
contributions; below we will focus on the SI contribution. To obtain the
direct detection rate, one needs to take into account several
astroparticle physics ingredients such as the WIMP density and the WIMP
velocity distribution near the earth. 

With one remarkable exception, all dark matter direct detection
experiments have only set upper limits on the WIMP-nucleon scattering
cross section ($\sigma_N$, $N=n,p$). Xenon~\cite{Angle:2007uj} has
reported $\sigma_N^\text{SI} \lesssim 4.5\times 10^{-8}$~pb for a WIMP
mass of 30~GeV ($\sigma_N^\text{SI} \lesssim 8.8 \times 10^{-8}$~pb for
$m_\text{WIMP} \sim 100$~GeV) at 90\% CL, and CDMS~\cite{Ahmed:2008eu}
has set a limit $\sigma_N^\text{SI} \lesssim 4.6 \times 10^{-8}$~pb for
a WIMP mass of 60~GeV at 90\% CL. The best bounds on SD cross sections
are from  Xenon~\cite{Angle:2008we} ($\sigma_n^\text{SD} \lesssim 5
\times 10^{-3}$~pb) and KIMS~\cite{Lee.:2007qn} ($\sigma_p^\text{SD}
\lesssim 0.18 $~pb). 

On the other hand, a positive result has been reported by the DAMA
experiment (which looks for the so-called annual modulation signature):
$\sigma_N \approx  0.2 - 1.0 \times 10^{-5}$ pb for WIMP masses between
30-100~GeV~\cite{Bernabei:2000qi}. It has been noted
in~\cite{Gunion:2005rw} that, in the NMSSM, very light bino- or
singlino-like neutralinos (with $m_{\chi_1^0} \sim 6-9$ GeV) could
allow to reconcile the DAMA results with the negative searches by other
collaborations.

In the absence of CP violation, and in the zero momentum transfer
limit, the elastic scattering of a neutralino LSP with a quark in a
nucleon can be described by the effective low energy four-fermion
Lagrangian (in the Dirac fermion notation): 
\beq\label{dm.4e}
\mathcal{L}_\text{eff} \,=\, 
\alpha_q^\text{SI} \, \bar \chi^0_1 \chi_1^0 \,\bar q q +
\alpha_q^\text{SD} \, \bar \chi^0_1 \gamma_5 \gamma_\mu \chi_1^0 \, 
\bar q \gamma_5 \gamma^\mu q \,,
\eeq
where $\alpha_q^\text{SI (SD)}$  denotes the spin-independent
(spin-dependent) interaction with up- and down-type quarks (described by
spinors $q=u,d$). The contribution of the dominant SI term to the
$\chi^0_1$-nucleon cross section is given by (see,
e.\,g.,~\cite{Belanger:2008sj})
\beq\label{dm.5e}
\sigma_N^\text{SI} = \frac{4 m_N m_{\chi^0_1}}{\pi (m_N + m_{\chi^0_1})}
\, f_N ^2
\eeq
where $f_N/m_N \propto \sum_q \alpha_q^\text{SI}/m_q$, and each term in
the sum over quarks must be weighted by the appropriate hadronic
matrix elements.

The scalar neutralino-nucleon interaction $\alpha_q^\text{SI}$ arises
from $s$-channel squark and $t$-channel Higgs (or $Z$) exchange at the
tree level, while neutralino-gluon interactions contribute at the
one-loop level (where all (heavy) quark loops should be taken into
account). The squark exchange term is identical to the MSSM, the only
difference in the NMSSM arising if the LSP has an important singlino
component implying a significant reduction of
$\alpha_q^{\text{SI},\tilde q}$. 

Significant differences in the NMSSM
can arise from the $t$-channel Higgs contributions~\cite{Flores:1991rx,
Bednyakov:1998is, Bednyakov:1999yr, Cerdeno:2004xw,Cerdeno:2007sn,
Barger:2007nv, Belanger:2008nt}, which can dominate the SI interaction
provided the LSP higgsino component is large enough. One has (see,
e.g.,~\cite{Cerdeno:2004xw}) 
\beq\label{dm.6e}
\alpha_q^{\text{SI},H} \, \propto \,
- \frac{1}{M^2_{H_a}}\,(H_a\chi^0_1 \chi^0_1) \, (H_a q q) 
\eeq
where the quark Yukawa couplings ($H_a q q$) and the 
Higgs-neutralino-neutralino Yukawa couplings ($H_a\chi^0_1 \chi^0_1$) 
are given in Appendix A. The light Higgs doublet contributions are
often dominant, although in the large $\tan\beta$ regime the
contributions of the heavier doublet-like Higgs can also be important,
as their couplings to neutralinos and quarks are
enhanced~\cite{Belanger:2008nt}.

For a singlino-like LSP, the cross section for the scattering off,
e.\,g., a strange-quark is proportional to 
\beq
\sigma_N^\text{SI} \, \propto \,
(\alpha_s^{\text{SI},H})^2 \, \propto \,
 \frac{\kappa^2\, h_s^2}{M^4_{H_a}}\,S_{a3}^2\,S_{a1}^2\,. 
\eeq
Hence the exchange of a light $H_1$ can lead to large direct
detection cross sections within reach of the present generation of
detectors~\cite{Cerdeno:2004xw,Cerdeno:2007sn}. (For a singlino-like
LSP, the MSSM-like $s$-channel squark exchange diagram is in general too
small to lead to cross sections close to the experimental limits.
NMSSM specific contributions to the spin-dependent cross section can
occur from a higgsino-like LSP, scattering off nuclei via $Z$ boson
exchange~\cite{Belanger:2008nt}.) 

\begin{figure}[t!]
\begin{center}
\begin{tabular}{cc}
  \epsfig{file=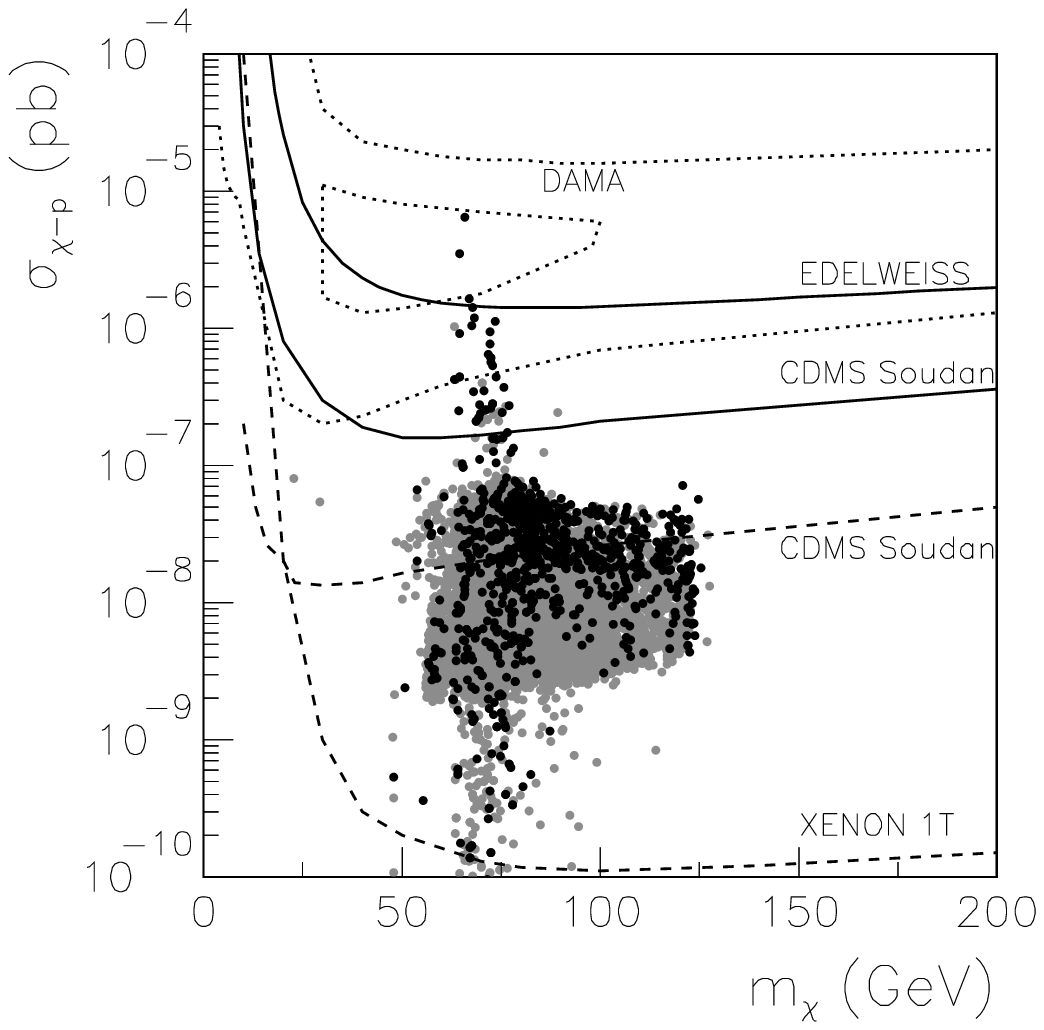, width=75mm}&
\epsfig{file=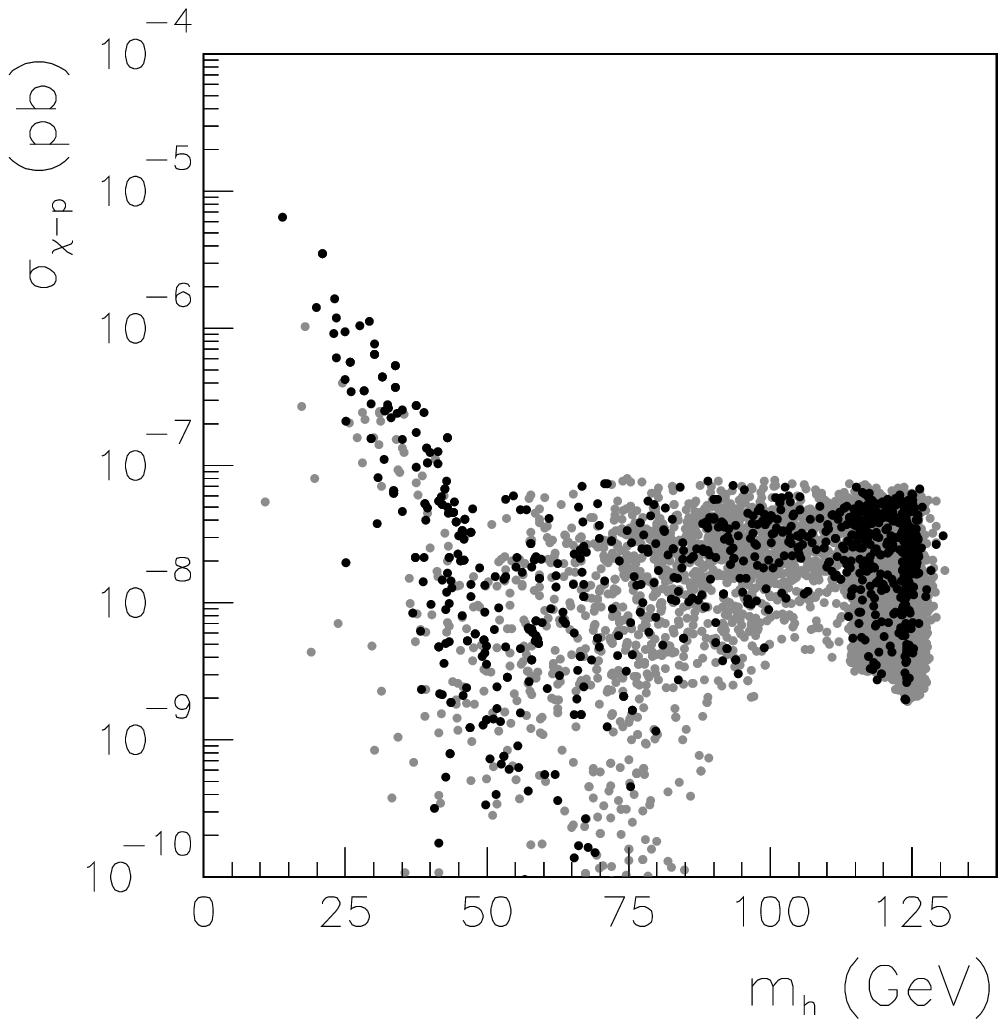, width=75mm}
\end{tabular}
  \caption{\small 
    Scatter plots of the spin-independent
    neutralino-nucleon cross section in the general NMSSM for
    $\tan \beta=5$, with the remaining parameters in the ranges 
    $0.01\le\lambda,\kappa\le0.7$, 
    $110\ \text{GeV} \lesssim M_2 \lesssim 430\ \text{GeV}$, 
    $110\ \text{GeV}<\mu_\text{eff}<300\ \text{GeV}$,  
    $-800\ \text{GeV} \lesssim A_\lambda \lesssim 800\ \text{GeV}$, and 
    $-300\ \text{GeV}\lesssim A_\kappa \lesssim 300\ \text{GeV}$.
    All the points  fulfil all the experimental constraints and have a
    relic density in agreement with $0.1 \lesssim \Omega h^2 \lesssim
    0.3$  (astrophysical bound, grey dots) or the WMAP constraint (black
    dots).
    On the left as a function of the neutralino mass (GeV), and on the
    right as a function of the lightest scalar Higgs mass (GeV).  
     From~\cite{Cerdeno:2007sn}.}\label{fig:dm:dd}
\end{center}
\end{figure}

As an example, we display in Fig.~\ref{fig:dm:dd}
(from~\cite{Cerdeno:2007sn}) the SI cross sections in a large region of
the unconstrained NMSSM parameter space, which is compatible with
collider, cosmological and $B$~physics constraints. (A heavy sfermion sector
$\sim 1$ TeV is assumed, except for slepton masses of $\sim 150$~GeV to
comply with $(g-2)_\mu$). Large detection cross sections can originate
from the presence of light singlet-like Higgs ($S_{13}^2 \gtrsim 0.9$)
with a mass $M_{H_1} \sim 50$~GeV, see the right-hand panel of
Fig.~\ref{fig:dm:dd}.  Neutralinos within the reach of dark matter
experiments are typically mixed singlino-higgsino states, with a mass
between  $50-130$~GeV. The upper bound on the neutralino mass is due to
the lightest stau becoming the LSP. If the slepton mass is increased,
heavier neutralinos can be the LSP, but the resulting SUSY contribution
to $(g-2)_\mu$ is generally too small.
 
\subsubsection{Indirect dark matter detection}

It is also possible to detect dark matter indirectly, looking for
distinctive signals of WIMP annihilation products: gamma-ray fluxes,
peculiar gamma-ray spectral features (like a sharp monochromatic peak at
$E_\gamma \sim m_{\chi_1^0}$ arising from the loop induced annihilation
$\chi_1^0 \chi_1^0 \to \gamma \gamma$), sizable amounts of antimatter
from pair annihilation in the galactic halo, and also energetic
neutrino fluxes from annihilations in the center of the sun or earth.
For a review in the MSSM we again refer to~\cite{Jungman:1995df}.

Indirect detection of NMSSM neutralino dark matter has been considered
in~\cite{Ferrer:2006hy}. Since the indirect detection rates generically
scale as inverse powers of $m_{\chi_1^0}$ (the pair annihilation rate
being fixed by the requirement of a correct relic abundance), light
neutralinos -- as possible in the NMSSM -- could give significantly
enhanced rates. Additionally, new contributions (annihilations through a
very light pseudoscalar) can enhance monochromatic $\chi_1^0\,\chi_1^0
\to \gamma\, \gamma$ gamma-ray lines. The neutrino production rates from
neutralino annihilation inside the earth and sun could also be distinct
from MSSM expectations (see, e.g.,~\cite{Ferrer:2006hy} and references
therein). 

NMSSM scenarios exist where the annihilation of sub-GeV neutralinos
could explain the 511 keV gamma-ray emission reported by
INTEGRAL/SPI~\cite{Jean:2003ci}. These gamma-rays can originate from
low energy positrons possibly arising from the annihilation of an
exceptionally light DM candidate. In the  NMSSM, a very light neutralino
($m_{\chi_1^0} \sim 100$~MeV) combined with  a light pseudoscalar, could
potentially produce the 511~keV gamma-rays~\cite{Gunion:2005rw}.
However, efficient LSP annihilation via resonant $s$-channel
pseudoscalar exchange would require $M_{A_1} \approx 2\, m_{\chi_1^0}\pm
10$~MeV. 

Recently, the PAMELA experiment~\cite{Adriani:2008zq,Adriani:2008zr} 
reported a cosmic ray positron excess, with a positron over electron
fraction that appears to rise at energies from 10 to 100~GeV.  However,
the same detector does not report any obvious antiproton excess in the
same energy range.  In order to interpret such a positron excess (and
the absence of a $\bar p$ excess) as dark matter annihilation products,
it would be required that the DM particles dominantly annihilate into
leptons (either due to dominant couplings to SM leptons, or annihilation
into intermediate particles, sufficiently light so that they cannot
decay into hadrons).   
Neutralino annihilation in the NMSSM (for $m_{\chi_1^0} \sim 160$ GeV,
and a very light pseudoscalar)~\cite{Bai:2009ka,Hooper:2009gm,
Wang:2009rj} has been put forward to explain the PAMELA excess:
relatively light neutralinos (typically bino-higgsino
mixtures) can annihilate as $\chi_1^0\chi_1^0 \to H_1 A_1$ via $A_2$
exchange, with $A_1$ so light that only decays into a pair of leptons
(muons, or even electrons, for $M_{A_1} \lesssim 2 m_\mu$)  are
kinematically allowed. In this case, the lightest scalar $H_1$ also
dominantly decays into $A_1 A_1$, so that the final state of the LSP
annihilation indeed consists of very energetic leptons. 
Extensions of the MSSM by a singlet and an additional heavy 
lepton~\cite{Huh:2008vj} (exhibiting two DM components) have also been
proposed in order to
accommodate the PAMELA data.

\subsection{Specific scenarios}\label{sec:dm.ss}

\bigskip
\noindent{\it The (semi)-constrained NMSSM}

\medskip
The relic density in the cNMSSM was first discussed in
\cite{Abel:1992ts,Stephan:1997rv,Stephan:1997ds}. In the meantime
constraints on Higgs and sparticle masses as well on the dark matter
relic density \cite{Spergel:2006hy,Tegmark:2006az} have become tighter
(and RGEs/radiative corrections are known to a higher accuracy), with
the result that some of the regions in the parameter space of the cNMSSM
considered in these early studies are no longer phenomenologically
viable. On the other hand, a singlino-like LSP was believed to be
excluded~\cite{Stephan:1997rv,Stephan:1997ds}, which turned out to be
incorrect after the inclusion of additional (co-)annihilation channels.

In more recent studies, the parameter space of the semi-constrained
NMSSM (relaxing the universality requirements for $m_S^2$ and $A_\kappa$
at the GUT scale, see Section~\ref{sec:ss.1.3}) has been thoroughly
investigated. Regions allowed by constraints from collider and flavour
physics and a WMAP compatible relic density have been identified
in~\cite{Hugonie:2007vd,Belanger:2008nt}. In the semi-constrained NMSSM,
less options for LSP annihilation than in the general NMSSM are
available:

\noindent
(i) MSSM-like mechanisms: annihilation of a bino- or bino-higgsino-like
LSP through sfermion or Higgs boson exchange; coannihilation of a
bino-like LSP with sfermions;

\noindent
(ii) annihilation near a pseudoscalar singlet-like Higgs resonance in
the $s$-channel ($\lambda \sim 0.1$, any $\tan \beta$); 

\noindent
(iii) (assisted) coannihilation of a singlino-like LSP with a
higgsino-like NLSP ($\lambda \ll 1$, large $m_0$ and $\tan \beta$);

\noindent
(iv) (assisted) coannihilation of singlino-like LSPs with bino-like
NLSPs, with the bino rapidly annihilating through a Higgs resonance
($\lambda \ll 1$, large $\tan \beta$); 

\noindent
(v) (assisted) coannihilation of a singlino-like LSP with a stau or stop
NLSP ($\lambda \ll 1$, small $m_0$).

\begin{figure}[t!]
\begin{tabular}{cc}
\psfig{file=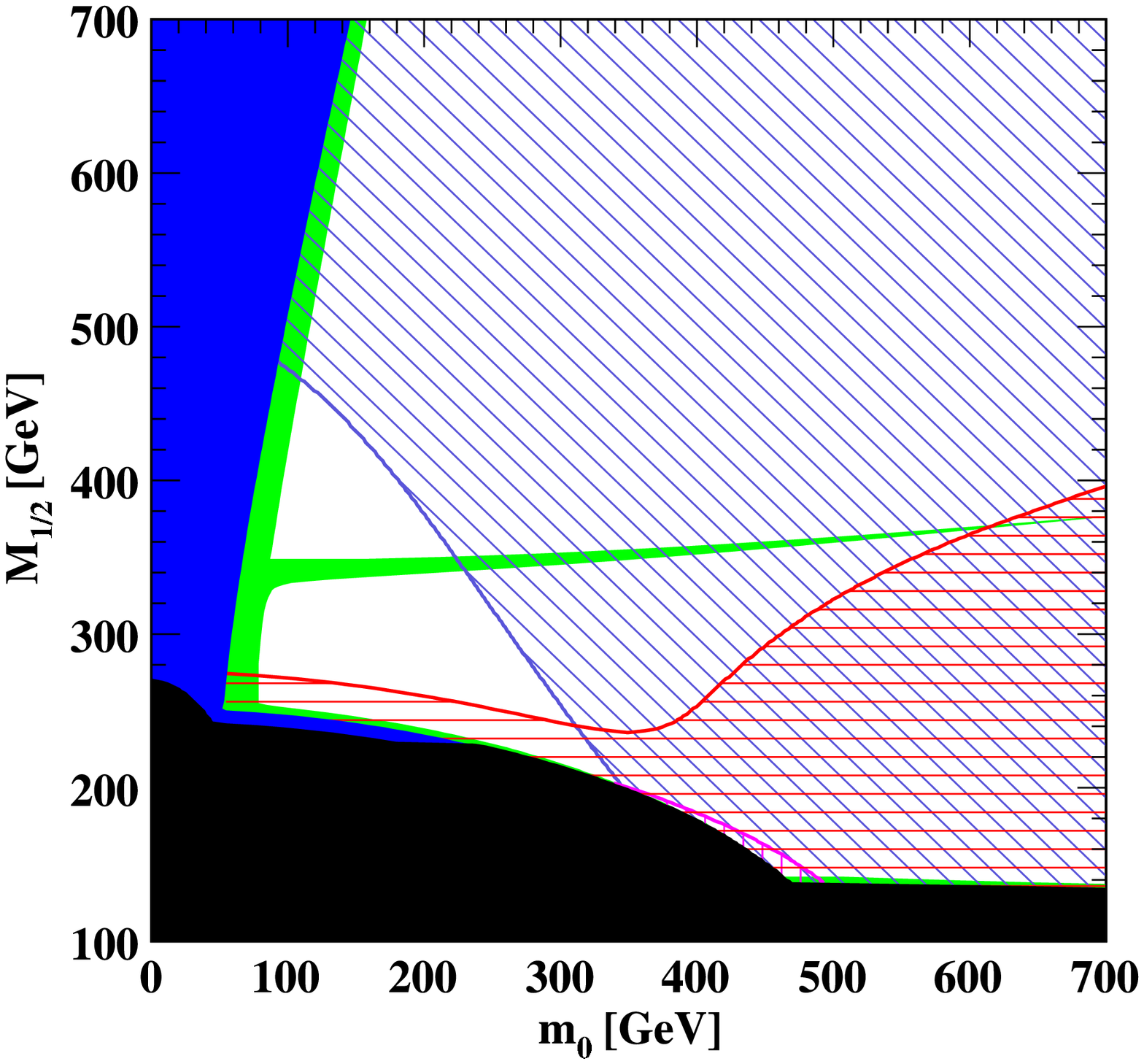,clip=,width=65mm,angle=90} 
\hspace*{2mm}&\hspace*{2mm}
\psfig{file=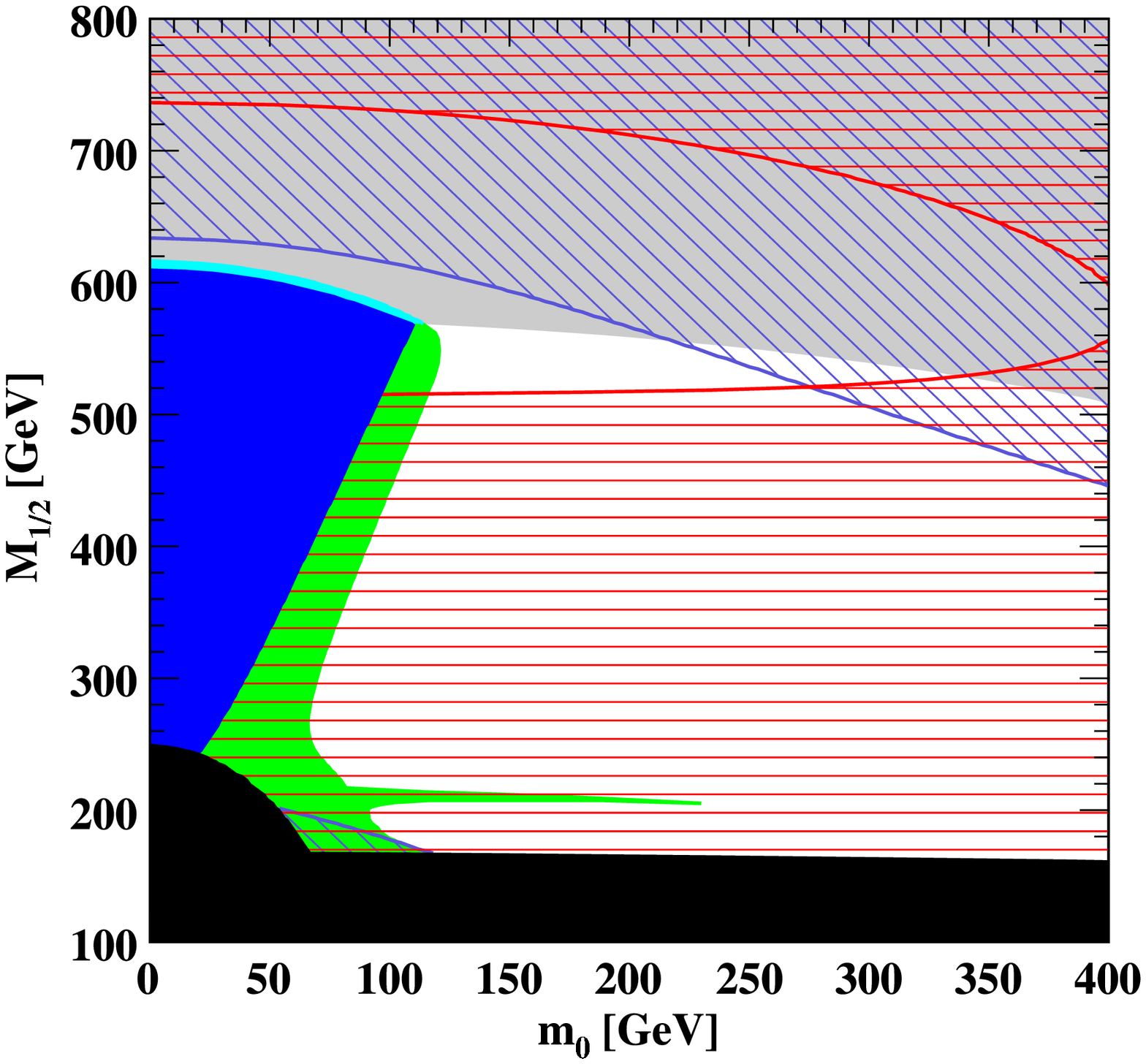,clip=,width=65mm,angle=90} 
\end{tabular}
\caption{Constraints in the $(m_0,M_{1/2})$ plane in the
semi-constrained NMSSM:  regions excluded by theoretical constraints or
by LEP/Tevatron searches  on sparticles are indicated in black, by
charged sfermion LSP in blue, from violation of LEP limits on Higgs
searches in red, by constraints from $B$-physics in pink and disfavoured
by $(g-2)_\mu$ in violet.  The regions allowed by the upper bound on the
DM relic density  are displayed in green. On the left: $\lambda=0.1$,
$A_0=-900$~GeV, $A_\kappa=-60$~GeV and $\tan \beta=5$. On the right:
$\lambda=0.01$, $A_0=300$~GeV, $A_\kappa=50$~GeV and $\tan \beta=10$.
Here, in addition, the regions with a  singlino LSP are shown in
grey/cyan (with a relic density above/below the WMAP limit).
From~\cite{Belanger:2008nt}.}
\label{fig:dm1}
\end{figure}

In Fig.~\ref{fig:dm1} (from~\cite{Belanger:2008nt}), we display some
regions in the $(m_0,M_{1/2})$ parameter space, where a DM relic density
within (or below) the WMAP bound is obtained through different
annihilation mechanisms. (Also indicated are other relevant theoretical
and phenomenological constraints.) The dominant annihilation mechanisms
depend on $(m_0,M_{1/2})$, and on the values of the remaining parameters
$\l$, $A_0$, $A_\k$ and $\tan\b$ of the semi-constrained NMSSM. On the
left, these are chosen as $\lambda=0.1$, $A_0=-900$~GeV,
$A_\kappa=-60$~GeV and $\tan \beta=5$, and the LSP is bino-like. Within
the green horizontal band around $M_{1/2}\sim 350$ GeV, the correct
relic density is achieved through a pseudoscalar resonance (ii).  In the
green region next to the blue region excluded by a stau LSP (at small
$m_0$), the dominant mechanism is bino-stau coannihilation (i), whereas
for small $m_0$ and $M_{1/2}\sim 250$ GeV the dominant mechanism is
bino-stop coannihilation. On the right we show an example with small
$\lambda$ and a singlino-like LSP. In the small $m_0$ regime efficient
singlino annihilation is achieved via coannihilation with a nearly
degenerate lightest stau (v).

The prospects for direct DM detection have also been considered
in~\cite{Belanger:2008nt}. Present bounds already constrain such regions
at large $\tan\b$ where the heavy Higgs doublet is relatively light
and/or its couplings to an LSP with a non-vanishing higgsino component
and quarks are enhanced. Large cross sections near or above present
limits can also be obtained for $\lambda \sim 0.1$, where the LSP
annihilates via a pseudoscalar singlet resonance (ii), and for bino-stau
coannihilation regions (i). Hence, the semi-constrained NMSSM parameter
space with a dominantly bino- or higgsino-like LSP has good prospects of
being probed by the future direct detection experiments. On the other
hand, for small $\tan \beta$ or TeV-range heavy Higgs and squarks, large
scale detectors will be required. The detection of a singlino LSP is
beyond the reach of (even far future) large scale detectors. 

\medskip

The cNMSSM where $A_\kappa=A_0$ is imposed at the GUT scale, but $m_S$
is allowed to differ from $m_0$, has been investigated
in~\cite{Balazs:2008ph} (see Section~\ref{sec:ss.1.3}). It is still
possible to have an LSP with a large bino or higgsino component which
satisfies WMAP constraints via neutralino-stop (and stau) coannihilation
or via Higgs resonances. For most points in this parameter space, direct
detection of DM would appear possible for upgraded large scale
detectors. For larger $\l$, a singlino-like LSP can annihilate via a
CP-odd Higgs resonance in the $s$-channel.  

\medskip

In the fully constrained NMSSM~\cite{Djouadi:2008yj,Djouadi:2008uj},
discussed in  Section~\ref{sec:ss.1}, the LSP is always singlino-like,
and $\l \lsim 10^{-2}$. The only available mechanism to reduce the relic
density is  ``assisted coannihilation'' (v) with the stau NLSP (provided
$\lambda \gtrsim 10^{-5}$), which requires a near degeneracy  of the
$\chi_1^0$ and $\tilde \tau_1$ masses.  Since direct (or indirect)
detection of a singlino LSP  relies on its non-singlet components of
${\cal O} (\lambda$), the prospects to detect the LSP in the cNMSSM are
quite dim: WIMP-nucleon cross sections are extremely small, and indirect
detection of the products of the annihilation  process $\chi_1^0
\chi_1^0 \to A_1$ (with a pseudoscalar sufficiently light to be produced
on-shell) also appears impossible, as  the $A_1 \chi_1^0 \chi_1^0$
coupling is proportional to $\kappa$ which is also tiny in the cNMSSM
($\kappa \lsim 10^{-3}$). Hence, the fully constrained NMSSM can be
excluded by the direct or indirect detection of a WIMP-like dark matter
candidate.

\bigskip
\noindent{\it The nMSSM}

\medskip
The neutralino relic density has also been studied in the framework of
the nMSSM (see Section~\ref{sec:ss.3}).
In~\cite{Menon:2004wv,Balazs:2007pf} it has been shown that, for large
$\l$, one can simultaneously achieve the correct $\Omega h^2$ and a
strong first order phase transition leading to successful EW 
baryogenesis. This occurs for a very light LSP with a mass 
$m_{\chi_1^0} \sim 30-40$~GeV, mostly singlino-like but with a sizable
higgsino component. The dominant annihilation modes are via (possibly
resonant) $s$-channel $Z$, $H_1$ and $A_1$
exchange~\cite{Menon:2004wv,Barger:2005hb, Balazs:2007pf,Cao:2009ad},
the latter being favoured by the constraint from
$(g-2)_\mu$~\cite{Cao:2009ad}. 

Regarding DM detection~\cite{Balazs:2007pf},  the dominant contributions
to SI and SD interactions arise typically from Higgs and $Z$ $t$-channel
exchange, respectively. Due to the smallness of the $\chi_1^0 \chi_1^0
Z$ coupling, the SD cross section lies beyond the reach of next
generation of DM  experiments. On the other hand, the SI cross section
could be within reach of the current and next generation of DM
experiments for sizable values of $\l$ (leading to an enhancement of the
$\chi_1^0 \chi_1^0 H_a$ coupling) and a for comparatively light Higgs
spectrum.

\bigskip
\noindent{\it $U(1)^\prime$ extensions}

\medskip
In $U(1)^\prime$ extensions (see Section~\ref{sec:var.3}),  the 
inclusion of an additional $U(1)^\prime$ gaugino (bino$'$) and the
absence of the singlet self-coupling term in the superpotential  can
significantly modify the nature and properties of neutralino dark matter
relative to the MSSM and
NMSSM~\cite{deCarlos:1997yv,Barger:2004bz,Barger:2005hb,
Suematsu:2005bc,Nakamura:2006ht, Barger:2007nv,
Kalinowski:2008iq}. Just like for the nMSSM, the absence of the $S^3$
term in the superpotential restricts the annihilation modes of the
singlino. Compared to
the NMSSM, new annihilation channels include $t$-channel $\chi^0_i$
mediated annihilation into a pair of $Z^\prime$s, and $Z^\prime$
(resonant) $s$-channel exchange. The analysis
of~\cite{Kalinowski:2008iq} shows that, depending on the nature of the
LSP (higgsino, mixed singlino-higgsino-bino$^\prime$, singlino) and on
the remaining spectrum, the correct DM relic density can be obtained via
$s$-channel resonances ($H_2/A/Z^\prime/H_3$). Pure singlino neutralinos
would not directly couple to the singlet Higgs boson, but they couple
strongly to the $Z^\prime$ so that even very heavy LSPs can efficiently
annihilate. The prospects for direct detection of neutralino DM in
$U(1)^\prime$ extensions have been addressed
in~\cite{deCarlos:1997yv,Kalinowski:2008iq}.

\medskip
Dark matter in supersymmetric $U(1)^\prime$ models with a secluded
$U(1)^\prime$ breaking sector (sMSSM) has also been
studied~\cite{Barger:2004bz}. For a light singlino-higgsino LSP with a
mass $\lsim 100$~GeV and an enhanced coupling to the $Z$,
annihilation through the $Z$-resonance leads to a viable relic density
over a large part of the parameter space without violating LEP
constraints.

\begin{figure}[t!]
\begin{tabular}{cc}\vspace*{-3mm}
\psfig{file=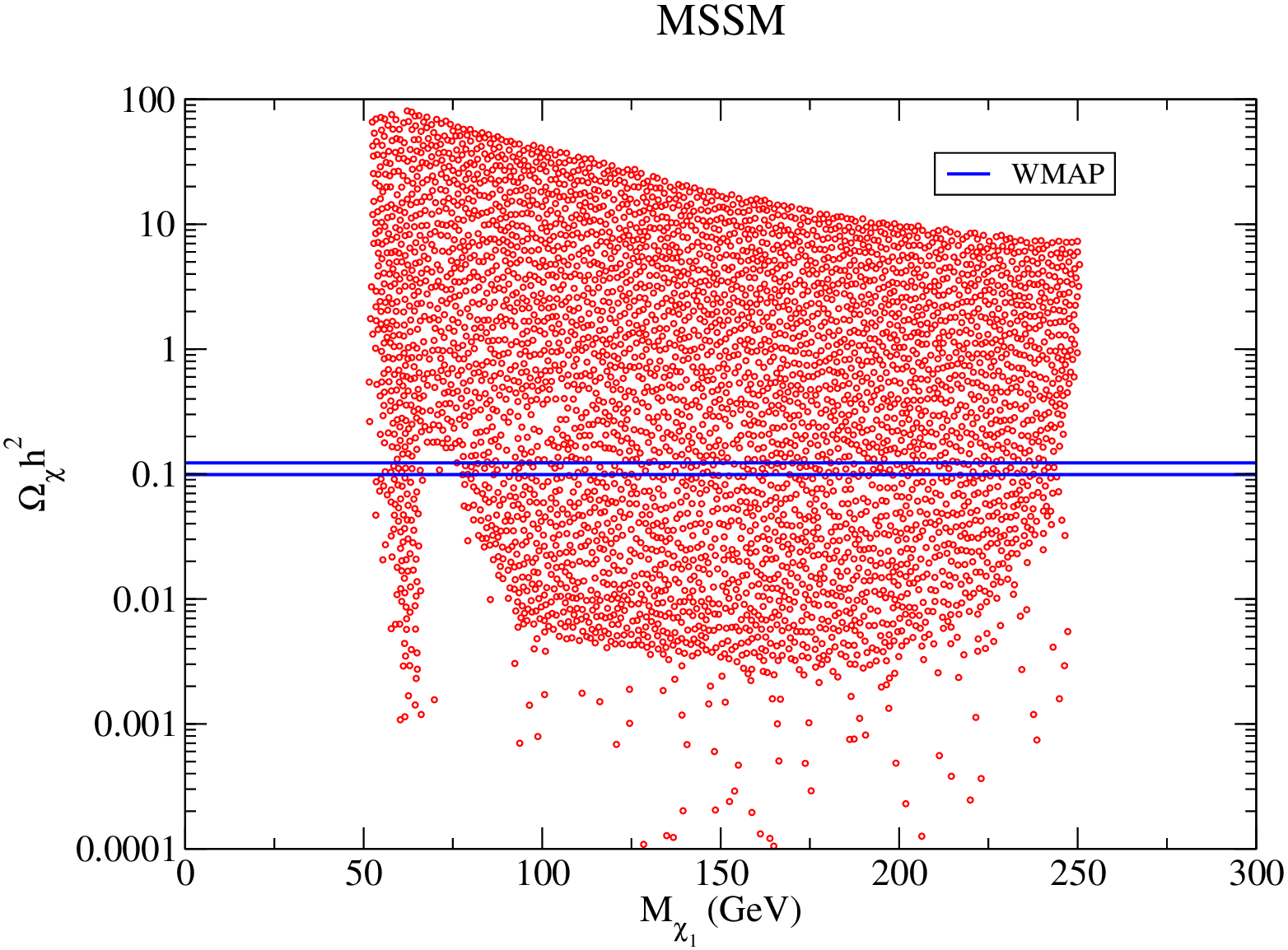,clip=,width=55mm,angle=-90} 
&\vspace*{-3mm}
\psfig{file=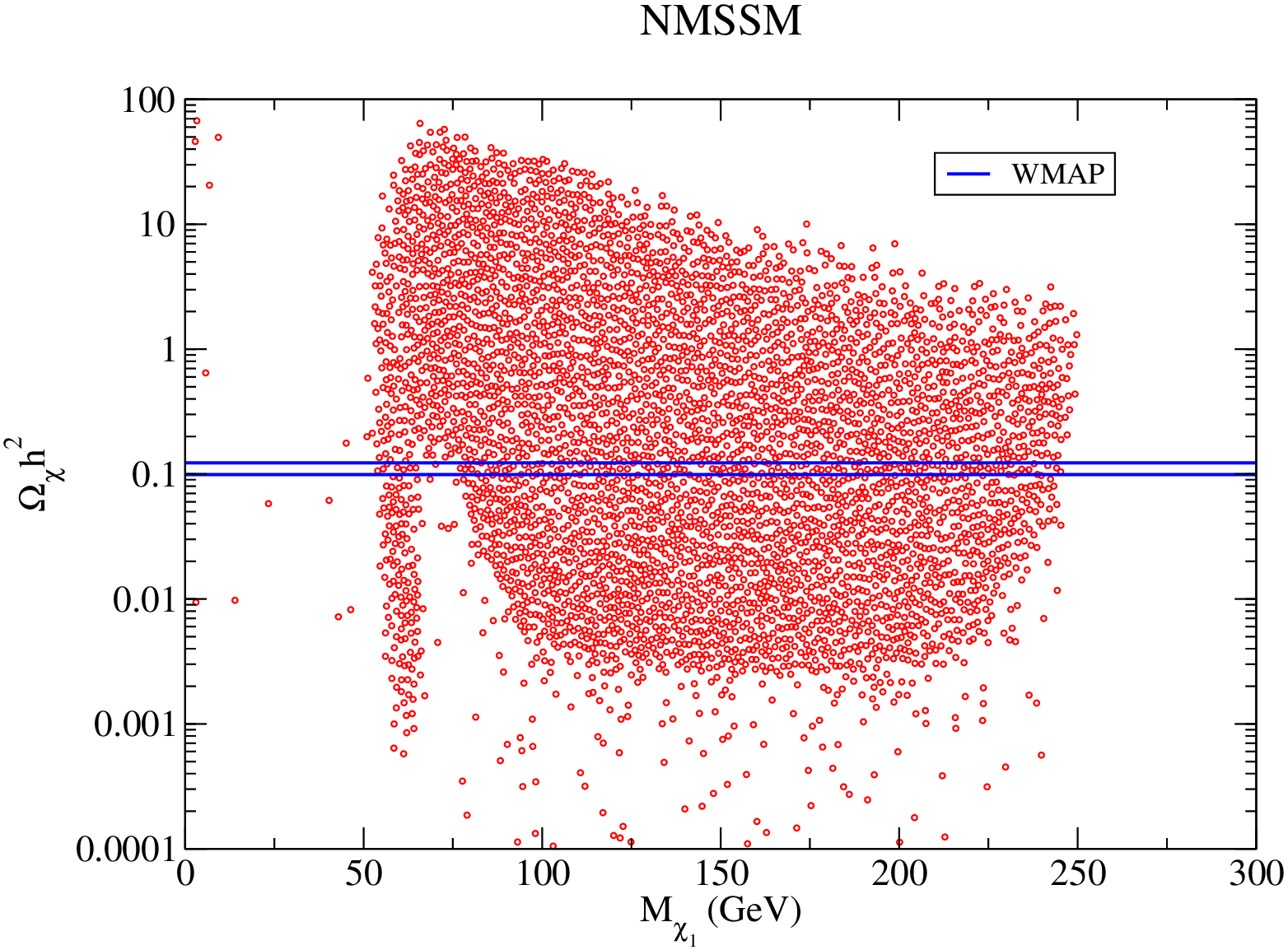,clip=,width=55mm,angle=-90} \\
\psfig{file=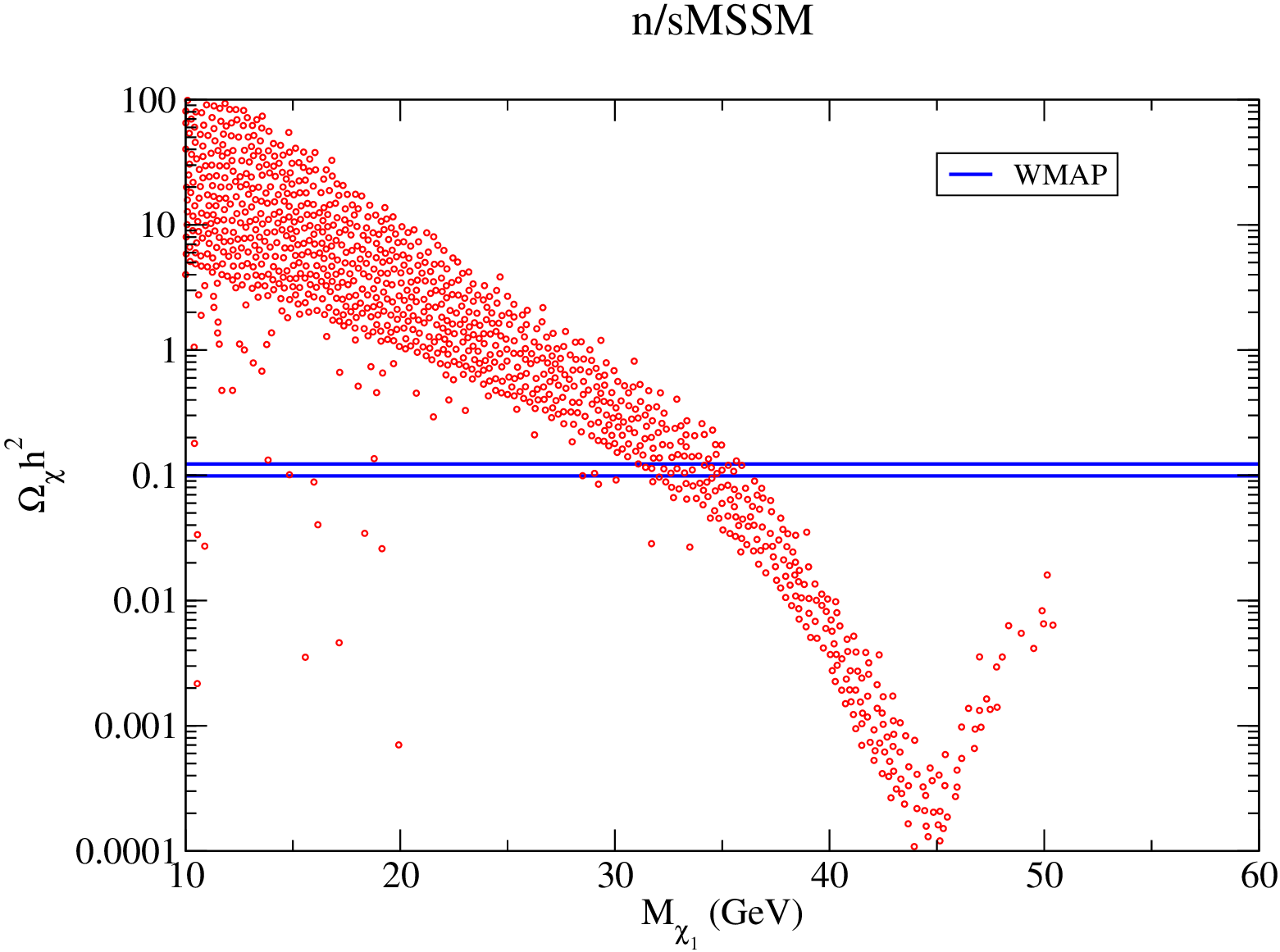,clip=,width=55mm,angle=-90} 
&
\psfig{file=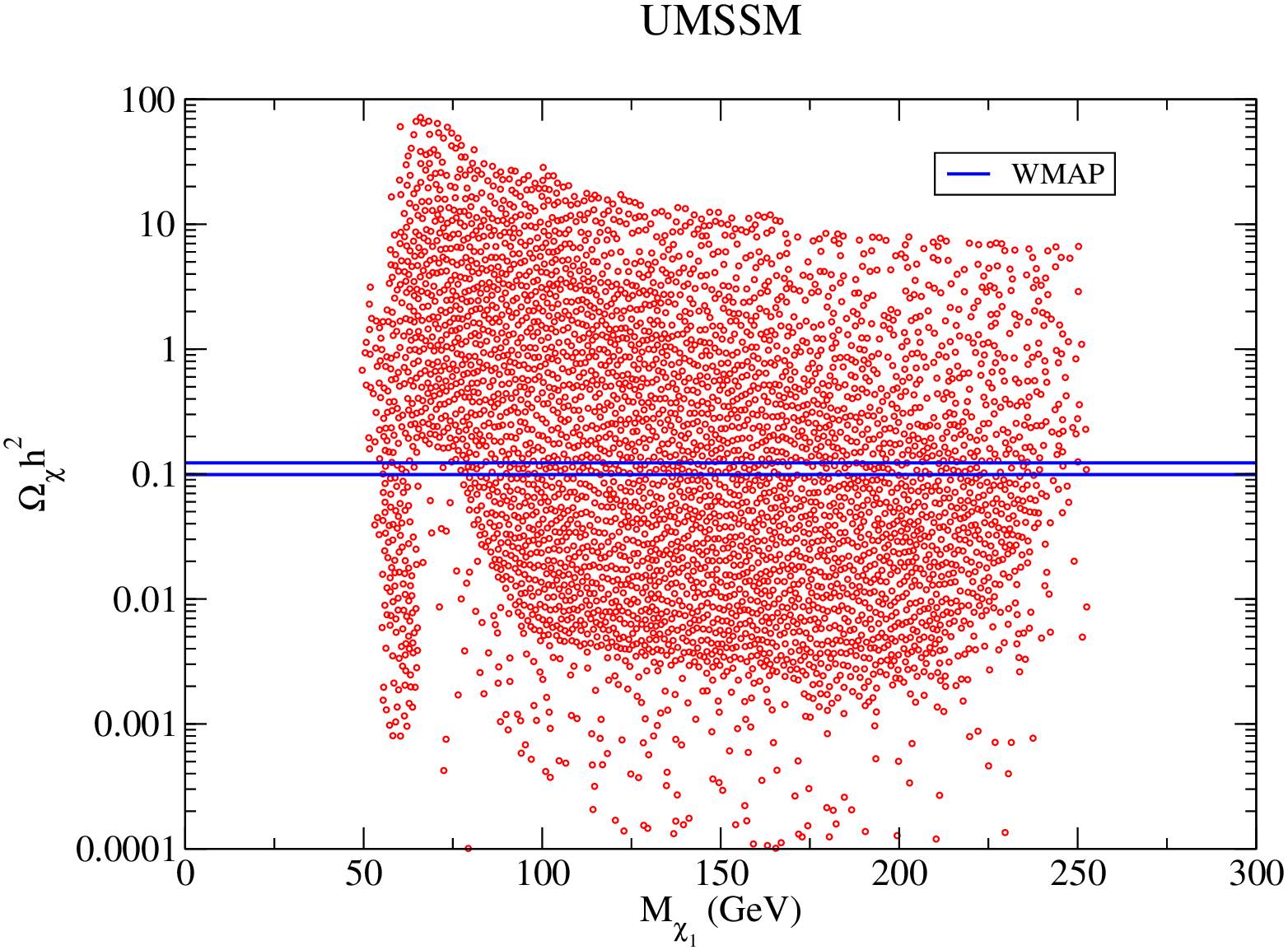,clip=,width=55mm,angle=-90}
\end{tabular}
\caption{Neutralino relic density versus the lightest neutralino mass
for the MSSM and NMSSM-like extensions, with $0.099 \lesssim\Omega
h^2\lesssim 0.123$ within the blue lines. Too large annihilation cross
sections (leading to a too small $\Omega h^2$) can originate from the
Higgs boson pole in the $s$-channel in the MSSM, NMSSM and UMSSM for
$m_{\chi_1^0} \sim M_{H_1}/2 \sim 60$~GeV, and from the $Z$ boson pole
at $m_{\chi_1^0}\sim M_Z / 2$ in the n/sMSSM.
(From~\cite{Barger:2007nv}.)}
\label{fig:dm2}
\end{figure}

\medskip
A comparative study of unconstrained versions of the MSSM, NMSSM, nMSSM,
$U(1)'$ extended models (UMSSM) and secluded models (sMSSM) has been
carried out
in~\cite{Barger:2005hb,Barger:2007nv}, identifying allowed ranges for
the mass of the LSP in each case as displayed in Fig.~\ref{fig:dm2}. 
Prospects for direct detection of neutralino DM were also compared
in~\cite{Barger:2007nv}: the cross sections for the $U(1)^\prime$
extended models are similar to those of the NMSSM, and can be within
reach of SuperCDMS and WARP. In view of the (comparatively) lighter LSPs
in secluded models as well as in the nMSSM (see Fig.~\ref{fig:dm2}),
spin-independent proton scattering cross sections could be detectable at
CDMS. 
 
\medskip
In the exceptional SSM (E$_6$SSM), which in addition to the $Z^\prime$
comprises three families of Higgs doublets and singlets, it has been
shown in~\cite{Hall:2009aj} that the two families of ``inert'' doublet
higgsinos and singlinos (i.e. whose scalar partners are vev-less) lead
to a decoupled neutralino sector with a naturally light LSP, that
successfully accounts for $\Omega h^2$ provided that $\tan\beta \lsim
2$.

\medskip
As referred to in Section~\ref{sec:var.3}, non-neutralino DM
candidates have also been put forward in the framework of  $U(1)^\prime$
extensions: exotic LSPs~\cite{Hur:2007ur}, and right-handed
sneutrinos~\cite{Lee:2007mt,Demir:2009kc}.  The correct relic density of
right-handed sneutrinos can be obtained via $t$-channel $Z^\prime$
exchange in large regions of the parameter space, and provided the
$Z^\prime$ mass is not excessively large, $s$-channel resonant
$Z^\prime$ exchange also provides an efficient annihilation mechanism.
The prospects for the detection of these DM candidates have been
investigated~\cite{Lee:2007mt}. $U(1)^\prime$~extended models with
sneutrinos as the LSP as well as the NLSP~\cite{Demir:2009kc} have been
proposed to explain the PAMELA observations, while accounting for a
relic density in agreement with the WMAP bounds.

\bigskip
\noindent{\it Other variants}

\medskip

In large-$\lambda$ models~\cite{Barbieri:2006bg} (see
Section~\ref{sec:3.2}), the LSP is higgsino-like in most of the
phenomenologically viable regions
of parameter space. Then, the relic density is in general too small;
only for $\tan \beta \lesssim 1.7$ regions with a substantial DM
abundance can be found: annihilation via $s$-channel $Z$ exchange can be
reduced by increasing the singlino component of the LSP, or via a
partial cancellation of the higgsino-$Z$ coupling. 

\medskip

In the NMSSM including right-handed neutrino superfields  with couplings
to the singlet Higgs (thus providing a dynamically generated
non-vanishing Majorana neutrino mass term), the properties of sneutrino
dark matter have been discussed in~\cite{Cerdeno:2009dv}.  For natural
values of the input parameters, right-handed sneutrinos with a mass in
the range of $5-200$~GeV can reproduce the observed dark matter relic
density without being excluded by direct dark matter searches. In this
case, the predicted direct detection cross sections are generally within
the reach of future experiments.

\section{Conclusions and outlook}
\label{sec:conc}

In this paper we reviewed the present status of knowledge of the
theoretical and phenomenological aspects of the NMSSM, including physics
at colliders, low energy precision observables and the properties of the
dark matter candidate, focusing on the possible differences with respect
to the MSSM.

We hope that the direct production of sparticles and/or Higgs bosons at
the Tevatron/LHC will provide us information about possible supersymmetric
extensions of the SM in the near future. Subsequently it will be
essential to study the precise nature of the supersymmetric extension of
the SM as thoroughly as possible. Depending on the experimental results,
it may not be easy to distinguish the MSSM from non-minimal extensions
as the NMSSM or variants thereof, but possibly the properties of the
Higgs sector and/or the neutralino sector (in particular the nature of
the LSP) will signal the presence of an additional singlet superfield.
Clearly, these signals must be well understood in order to be able to
interpret the data correctly. In view of the various scenarios that are
possible within the NMSSM (as unconventional Higgs decays, and/or a
neutralino with a large singlino component), additional efforts in the
form of simulations of events corresponding to such scenarios are still
required.

On the other hand, the properties of the Higgs and/or the neutralino
sectors could correspond to those expected within the MSSM -- which
would \emph{not} prove, however, that the MSSM is the correct
supersymmetric extension of the SM at the TeV~scale: at least in the
decoupling limit (see Section~\ref{sec:2.2.1}) -- and without a
singlino-like LSP -- the physics of the NMSSM becomes indistinguishable
from the one of the MSSM. The question, for which values of parameters
(like $\l$) the two models can be distinguished experimentally, is
complicated due to the large number of unknown parameters, and the
answer depends on the nature and the precision of future experimental
data. Also here, additional studies are desirable (at least once data is
available).

In any case it will be essential to combine the various pieces of
information
from low energy precision observables, physics at colliders and the
properties of dark matter; ideally, these will allow us
to pin down the fundamental theory at the TeV scale without too many
ambiguities.

\subsection*{Acknowledgements}

Some of the results presented here had been developed in cooperation
with colleagues, whose collaboration we acknowledge gratefully:
G.~Belanger, D.~G.~Cerde\~no, A.~Djouadi, F.~Domingo, J.~F.~Gunion,
C.-C.~Jean-Louis, D.~Lop\'ez-Fogliani, S.~Moretti, C.~Mu\~noz, A.~Pukhov
and M.~A.~San\-chis-Lozano. We are grateful to P. Richardson, F.
Roberto, C.~Spitzer, F. Staub and L. Suter for corrections of some
formulas.

\renewcommand{\theequation}{A.\arabic{equation}}
\setcounter{equation}{0}
\setcounter{section}{0}

\section*{Appendix A. Mixing matrices and tree level Higgs couplings}
\addcontentsline{toc}{section}
{Appendix A. Mixing matrices and tree level Higgs couplings}

The Feynman rules of the NMSSM have been described first in
\cite{Franke:1995tc} (including the quartic couplings not given here),
and subsequently in \cite{Ellwanger:2004xm}.

\subsection*{A.1 Mixing matrices}
\addcontentsline{toc}{subsection}
{A.1 Mixing matrices}

First we define the mixing matrices which diagonalize the Higgs,
neutralino and chargino mass matrices (after the addition of radiative
corrections).

The CP-even $3 \times 3$ mass matrix ${\cal M}_S^2$ (\ref{2.22e}) in the
basis $H_i^{\text{weak}} = (H_{dR}, H_{uR}, S_R)$ (in the SLHA2
conventions \cite{Allanach:2008qq}) is
diagonalized by an orthogonal $3 \times 3$ matrix $S_{ij}$ rotating the
basis $H_i^{\text{weak}}$,
\beq\label{a.1e}
H_i^{\text{mass}} = S_{ij} H_j^{\text{weak}}\; ,
\eeq
such that the mass eigenstates $H_i^{\text{mass}}$ are ordered in
increasing mass.

The CP-odd $2 \times 2$ mass matrix ${\cal M}_P^2$ (\ref{2.24e}) in the
basis (${A} = \cos\b\, H_{uI}+ \sin\b\, H_{dI}, S_I$) was obtained after
the rotation by the angle $\b$ (\ref{2.25e}), which allowed to omit the
Nambu-Goldstone boson. It can be diagonalized by an orthogonal $2 \times
2$ matrix $P_{ij}'$ such that the physical CP-odd states
$A_i^{\text{mass}}$ (ordered in mass) are 
\bea\label{a.2e}
A_1^{\text{mass}} &=& P_{11}' {A} + P_{12}' S_I\; , \nn \\
A_2^{\text{mass}} &=& P_{21}' {A} + P_{22}' S_I\; . \nn \\
\eea
In terms of the weak eigenstates $A_i^{\text{weak}} = (H_{dI}, H_{uI},
S_I)$, the mass eigenstates $(A_1^{\text{mass}},A_2^{\text{mass}})$ are
given by
\beq\label{a.3e}
A_i^{\text{mass}} = P_{ij} A_j^{\text{weak}}\; ,
\eeq
where
\beq\label{a.4e}
P_{i1}=\sin\b P_{i1}'\;,\qquad P_{i2}=\cos\b P_{i1}'\;,\qquad
P_{i3}=P_{i2}'\; .
\eeq

The inverse relation reads explicitly (omitting the Nambu-Goldstone
boson)
\bea\label{a.5e}
H_{dI} &=& P_{11} A^{\text{mass}}_1 + P_{21} A^{\text{mass}}_2 \; ,\nn \\
H_{uI} &=& P_{12} A^{\text{mass}}_1 + P_{22} A^{\text{mass}}_2 \; ,\nn \\
S_I &=& P_{13} A^{\text{mass}}_1 + P_{23} A^{\text{mass}}_2\; .
\eea

Omitting again the Goldstone boson, the charged weak eigenstates
$H^\pm_{u,d}$ contain a component of the physical charged Higgs boson
$H^\pm$ given by
\beq\label{a.6e}
H_u^\pm = \cos\b\, H^\pm\;,\qquad H_d^\pm = \sin\b\, H^\pm\; .
\eeq

The (symmetric) $5 \times 5$ neutralino mass matrix ${\cal M}_0$
(\ref{2.32e}) in the basis\nl $\psi^0 = (-i\l_1 , -i\l_2^3, \psi_d^0,
\psi_u^0, \psi_S)$ is diagonalized by an orthogonal real matrix
$N_{ij}$, such that the physical masses $m_{\chi^0_i}$ ordered in
$|m_{\chi^0_i}|$ are real, but not necessarily positive.
Denoting the 5 eigenstates by $\chi^0_i$, we have
\beq\label{a.7e}
\chi^0_i = N_{ij} \psi^0_j\; .
\eeq

The diagonalization of the (not symmetric) $2 \times 2$ chargino mass
matrix $X$ (\ref{2.35e}) in the basis $\psi^-$, $\psi^+$ (see
(\ref{2.33e}), (\ref{2.34e})) requires different rotations of $\psi^-$
and $\psi^+$ into the mass eigenstates $\chi^-$, $\chi^+$ as
\beq\label{a.8e}
\chi^- = U \psi^- , \qquad \chi^+ = V \psi^+
\eeq
with
\beq\label{a.9e}
U = \left(\ba{cc}
\cos\t_U & \sin\t_U \\ -\sin\t_U & \cos\t_U
\ea \right) , \qquad
V = \left(\ba{cc}
\cos\t_V & \sin\t_V \\ -\sin\t_V & \cos\t_V
\ea \right) .
\eeq

Finally the top squark mass matrix (\ref{2.36e}) is diagonalized in
terms of the mass eigenstates $\tilde{t}_1$, $\tilde{t}_2$ (with
$m_{\tilde{t}_1} < m_{\tilde{t}_2}$), the bottom squark mass matrix
(\ref{2.37e}) in terms of the mass eigenstates $\tilde{b}_1$,
$\tilde{b}_2$, and the tau slepton mass matrix (\ref{2.38e}) in terms of
the mass eigenstates $\widetilde{\tau}_1$, $\widetilde{\tau}_2$ by the
rotations
\beq\label{a.10e}
\ba{cccccc}
\tilde{t}_1 &=&\cos\t_T\, \tilde{t}_L + \sin\t_T\, \tilde{t}_R\,,
&\tilde{t}_2 &=& \cos\t_T\, \tilde{t}_R - \sin\t_T\, \tilde{t}_L\,, \\
\tilde{b}_1 &=&\cos\t_B\, \tilde{b}_L + \sin\t_B\, \tilde{b}_R\,,
&\tilde{b}_2 &=& \cos\t_B\, \tilde{b}_R - \sin\t_B\, \tilde{b}_L\,, \\
\widetilde{\tau}_1&=&\cos\t_\tau\, \widetilde{\tau}_L 
+ \sin\t_\tau\, \widetilde{\tau}_R\,,
&\widetilde{\tau}_2 &=& \cos\t_\tau\, \widetilde{\tau}_R - \sin\t_\tau\, 
\widetilde{\tau}_L\, .
\ea
\eeq

Now we proceed to give the Higgs couplings (Feynman rules)
in terms of physical mass eigenstates.

\subsection*{A.2 Higgs couplings}
\addcontentsline{toc}{subsection}
{A.2 Higgs couplings}

\subsubsection*{Higgs-quarks/leptons}

We consider Higgs couplings to quarks and leptons of the third
generation only, from which the remaining couplings can be deduced
easily. First we recall the relations between the quark and lepton
masses $m_t$, $m_b$, $m_\tau$ and the Yukawa couplings $h_t$, $h_b$ and
$h_\tau$, where $v^2 = v_u^2 + v_d^2$ (related to $M_Z$ by
(\ref{2.13e})):
\beq\label{a.11e}
h_t = \frac{m_t}{v\sin\b}\;,\qquad h_b = \frac{m_b}{v\cos\b}\;,
\qquad h_\tau = \frac{m_\tau}{v\cos\b}\;.
\eeq
(Corrections to $h_b$ are discussed in Section~\ref{sec:3.3}.)
Below, the left and right-handed top (bottom) quark and lepton
bi-spinors are denoted by $t_L,\; t_R$ ($b_L,\; b_R$) and
$\nu_{\tau_L},\; \tau_L,\; \tau_R$, and the quark couplings are diagonal
in the colour indices. (For simplicity, the index~$^{\text{mass}}$ of
the
CP-even mass eigenstates $H_i^{\text{mass}}$ and the CP-odd mass
eigenstates $A_i^{\text{mass}}$ will be omitted subsequently.) The
denominators $\sqrt{2}$ in the couplings originate from the rescaling of
the kinetic terms of real Higgs bosons with respect to complex Higgs
bosons.\newpage
\bea
H_i t_L t_R^c & : & -\frac{h_t}{\sqrt{2}} S_{i2} \nn \\
H_i b_L b_R^c & : & \frac{h_b}{\sqrt{2}} S_{i1} \nn \\
H_i \tau_L \tau_R^c & : & \frac{h_\tau}{\sqrt{2}} S_{i1} \nn \\
A_i t_L t_R^c & : & -i\frac{h_t}{\sqrt{2}} P_{i2} \nn \\
A_i b_L b_R^c & : & i\frac{h_b}{\sqrt{2}} P_{i1} \nn \\
A_i \tau_L \tau_R^c & : & i\frac{h_\tau}{\sqrt{2}} P_{i1} \nn \\
H^+ b_L t_R^c & : & h_t \cos\b \nn \\
H^- t_L b_R^c & : & -h_b \sin\b \nn \\
H^-\nu_{\tau_L} \tau_R^c & : & -h_\tau \sin\b
\label{a.12e}
\eea

\subsubsection*{Higgs-gauge bosons}

\bea
H_i Z_\mu Z_\nu & : & g_{\mu\nu} \frac{g_1^2 + g_2^2}{\sqrt{2}}
(v_d S_{i1} + v_u S_{i2})\nn \\
H_i W^+_\mu W^-_\nu & : & g_{\mu \nu} \frac{g_2^2}{\sqrt{2}} 
(v_d S_{i1} + v_u S_{i2}) \nn \\
H_i(p) H^+(p') W_\mu^- & : & \frac{g_2}{2} (\cos\b S_{i2} - 
\sin\b S_{i1}) (p - p')_\mu \nn \\
A_i(p) H^+(p') W_\mu^- & : & i\frac{g_2}{2} (\cos\b P_{i2} + 
\sin\b P_{i1}) (p - p')_\mu \nn \\
H_i(p) A_j(p') Z_\mu & : & -i\sqrt{\frac{g_1^2+g_2^2}{2}} (S_{i1}
P_{j1} -  S_{i2} P_{j2}) (p - p')_\mu \nn \\
H^+(p) H^-(p') Z_\mu & : & \frac{g_1^2-g_2^2}{\sqrt{g_1^2+g_2^2}}
(p - p')_\mu
\label{a.13e}
\eea

\subsubsection*{Higgs-neutralinos/charginos}

Here, the couplings to two neutralinos involve the combinations
$\Pi_{ij}^{ab} = N_{ia}N_{jb}+N_{ib}N_{ja}$ of the neutralino mixing
matrices:

\bea
H_a \chi^+_i \chi^-_j & : & \frac{\l}{\sqrt{2}} S_{a3} U_{i2} V_{j2} +
\frac{g_2}{\sqrt{2}} (S_{a2} U_{i1} V_{j2} + S_{a1} U_{i2} V_{j1}) \nn
\\
A_a \chi^+_i \chi^-_j & : & i\left(\frac{\l}{\sqrt{2}} P_{a3} U_{i2}
V_{j2} - \frac{g_2}{\sqrt{2}} (P_{a2} U_{i1} V_{j2} + P_{a1} U_{i2}
V_{j1})\right) \nn \\
H^+ \chi^-_i \chi^0_j & : & \l\cos\b U_{i2} N_{j5} -
\frac{\sin\b}{\sqrt{2}} U_{i2} (g_1 N_{j1} + g_2 N_{j2}) + g_2 \sin\b
U_{i1} N_{j3} \nn \\
H^- \chi^+_i \chi^0_j & : & \l\sin\b V_{i2} N_{j5} +
\frac{\cos\b}{\sqrt{2}} V_{i2} (g_1 N_{j1} + g_2 N_{j2}) + g_2 \cos\b
V_{i1} N_{j4} \nn \\
H_a \chi^0_i \chi^0_j & : & \frac{\l}{\sqrt{2}} (S_{a1} \Pi_{ij}^{45} +
S_{a2} \Pi_{ij}^{35} + S_{a3} \Pi_{ij}^{34}) - \sqrt{2} \k S_{a3}
N_{i5} N_{j5} \nn \\ 
& & + \frac{g_1}{2} (S_{a1} \Pi_{ij}^{13} - S_{a2}
\Pi_{ij}^{14}) - \frac{g_2}{2} (S_{a1} \Pi_{ij}^{23} - S_{a2}
\Pi_{ij}^{24}) \nn \\
A_a \chi^0_i \chi^0_j & : & i\left(\frac{\l}{\sqrt{2}} (P_{a1}
\Pi_{ij}^{45} + P_{a2} \Pi_{ij}^{35} + P_{a3} \Pi_{ij}^{34}) - \sqrt{2}
\k P_{a3} N_{i5} N_{j5}\right. \nn \\ 
& & \left. - \frac{g_1}{2} (P_{a1} \Pi_{ij}^{13} - P_{a2}
\Pi_{ij}^{14}) + \frac{g_2}{2} (P_{a1} \Pi_{ij}^{23} - P_{a2}
\Pi_{ij}^{24})\right)
\label{a.14e}
\eea

\subsubsection*{Triple Higgs couplings}

\bea
H_a H_b H_c & : &
\frac{\l^2}{\sqrt{2}} \left( v_d (\Pi_{abc}^{122}+\Pi_{abc}^{133}) +
v_u (\Pi_{abc}^{211}+\Pi_{abc}^{233}) + s
(\Pi_{abc}^{311}+\Pi_{abc}^{322}) \right) \nn \\
& & - \frac{\l\k}{\sqrt{2}} ( v_d \Pi_{abc}^{323} + v_u \Pi_{abc}^{313}
+ 2 s \Pi_{abc}^{123}) + \sqrt{2} \k^2 s \Pi_{abc}^{333} \nn \\
& & - \frac{\l A_\l}{\sqrt{2}} \Pi_{abc}^{123} + \frac{\k
A_\k}{3\sqrt{2}} \Pi_{abc}^{333} \nn \\
& & + \frac{g_1^2+g_2^2}{4\sqrt{2}} \left( v_d (\Pi_{abc}^{111} -
\Pi_{abc}^{122}) - v_u (\Pi_{abc}^{211} - \Pi_{abc}^{222}) \right)\nn \\
& & +\frac{\mu'}{\sqrt{2}}\left(\k \Pi_{abc}^{333} -\l\Pi_{abc}^{123}\right) 
\label{a.15e}
\eea

\noi where
\bea
\Pi_{abc}^{ijk} & = &
S_{ai} S_{bj} S_{ck} + S_{ai} S_{cj} S_{bk} + S_{bi} S_{aj} S_{ck} \nn
\\ & & + S_{bi} S_{cj} S_{ak} + S_{ci} S_{aj} S_{bk} + S_{ci} S_{bj}
S_{ak}\ .
\label{a.16e}
\eea

\bea
H_a A_b A_c & : &
\frac{\l^2}{\sqrt{2}} \left( v_d (\Pi_{abc}^{122}+\Pi_{abc}^{133}) +
v_u (\Pi_{abc}^{211}+\Pi_{abc}^{233}) + s
(\Pi_{abc}^{311}+\Pi_{abc}^{322}) \right) \nn \\
& & + \frac{\l\k}{\sqrt{2}} \left( v_d
(\Pi_{abc}^{233}-2\Pi_{abc}^{323}) + v_u
(\Pi_{abc}^{133}-2\Pi_{abc}^{313}) \right. \nn \\
& & \left. + 2 s (\Pi_{abc}^{312}-\Pi_{abc}^{123}-\Pi_{abc}^{213})
\right) + \sqrt{2} \k^2 s \Pi_{abc}^{333} \nn \\
& & + \frac{\l A_\l}{\sqrt{2}}
(\Pi_{abc}^{123}+\Pi_{abc}^{213}+\Pi_{abc}^{312}) - \frac{\k
A_\k}{\sqrt{2}} \Pi_{abc}^{333} \nn \\
& & + \frac{g_1^2+g_2^2}{4\sqrt{2}} \left( v_d (\Pi_{abc}^{111} -
\Pi_{abc}^{122}) - v_u (\Pi_{abc}^{211} - \Pi_{abc}^{222}) \right)\nn \\
& & +\frac{\mu'}{\sqrt{2}}\left(\l \left(\Pi_{abc}^{312}-\Pi_{abc}^{123}
-\Pi_{abc}^{213}\right) + \k \Pi_{abc}^{333}\right)
\label{a.17e}
\eea

\noi where
\beq
\Pi_{abc}^{ijk} = S_{ai} (P_{bj} P_{ck} + P_{cj} P_{bk})\ .
\label{a.18e}
\eeq

\bea
H_a H^+ H^- & : &  \frac{\l^2}{\sqrt{2}} ( s (\Pi_a^{311}+\Pi_a^{322})
- v_d \Pi_a^{212} - v_u \Pi_a^{112}) \nn \\
& & + \sqrt{2} \l \k s \Pi_a^{312} + \frac{\l A_\l}{\sqrt{2}}
\Pi_a^{312} \nn \\
& & + \frac{g_1^2}{4\sqrt{2}} \left( v_u (\Pi_a^{211}-\Pi_a^{222}) + 
v_d (\Pi_a^{122}-\Pi_a^{111}) \right)\nn \\
& & + \frac{g_2^2}{4\sqrt{2}} \left( v_d
(\Pi_a^{111}+\Pi_a^{122}+2\Pi_a^{212}) + v_u
(\Pi_a^{211}+\Pi_a^{222}+2\Pi_a^{112}) \right) \nn \\
& & + \frac{\l\mu'}{\sqrt{2}}\Pi_a^{312}
\label{a.19e}
\eea

\noi where
\beq\label{a.20e}
\Pi_{a}^{ijk} = 2 S_{ai} C_j C_k\ \mathrm{with}\ C_1 = \cos\b\;,
C_2 = \sin\b\; .
\eeq

In contrast to the previous couplings, the triple Higgs couplings in the
general NMSSM differ slightly (by the terms $\sim \mu'$) from the ones
in the scale invariant NMSSM. The quartic Higgs (and Higgs-gauge)
couplings, which are less relevant for the Higgs phenomenology (i.e.
Higgs-to-Higgs decays) can be found in \cite{Franke:1995tc}.

\subsubsection*{Higgs-squarks/sleptons}

Here we assume diagonal Yukawa couplings in family space which we
denote as $h_{ua}$, $h_{da}$, $h_{ea}$, $a = 1,2,3$, with $h_{u3} \equiv
h_t$, $h_{d3} \equiv h_b$, $h_{e3} \equiv h_\tau$. The soft trilinear
couplings are $A_{u a}$ with $A_{u 3}\equiv A_t$, $A_{d a}$ with
$A_{d 3}\equiv A_b$, and $A_{e a}$ with $A_{e 3}\equiv A_\tau$. 

The squarks are denoted as $U_{L a}$, $D_{L a}$, $U_{R a}$,
$D_{R a}$, the charged sleptons as 
$E_{L a}$, $E_{R a}$, and the sneutrinos as $\widetilde{\nu}_{L a}$.

In the case of the first two families
($a=1,2$), these states can be considered as mass eigenstates. For the
third family (apart from the tau sneutrino), the couplings of the mass
eigenstates defined in (\ref{a.10e}) depend on the corresponding squark
and slepton mixing angles. We denote the couplings of the squark/slepton
weak eigenstates (= mass eigenstates for $a=1,2$) by $g(\dots)$ for all
three families; the couplings of the mass eigenstates of the third
family are subsequently expressed in terms of $g(\dots)$ with $a=3$.

\vskip 3 mm

\noi Higgs-squarks:
\bea
g(H_i,U_{L a},U_{L a}) &=& \sqrt{2}\left(h_{ua}^2 v_u S_{i2}
+\left(\frac{g_1^2}{12}-\frac{g_2^2}{4}\right)
\left(v_uS_{i2}-v_dS_{i1}\right)\right)\;,\nn \\
g(H_i,D_{L a},D_{L a}) &=& \sqrt{2}\left(h_{da}^2 v_d S_{i1}
+\left(\frac{g_1^2}{12}+\frac{g_2^2}{4}\right)
\left(v_uS_{i2}-v_dS_{i1}\right)\right)\;,\nn \\
g(H_i,U_{R a},U_{R a}) &=& \sqrt{2}\left(h_{ua}^2 v_u S_{i2}
-\frac{g_1^2}{3}
\left(v_uS_{i2}-v_dS_{i1}\right)\right)\;,\nn \\
g(H_i,D_{R a},D_{R a}) &=& \sqrt{2}\left(h_{da}^2 v_d S_{i1}
+\frac{g_1^2}{6}
\left(v_uS_{i2}-v_dS_{i1}\right)\right)\;,\nn \\
g(H_i,U_{L a},U_{R a}) &=& \frac{-h_{ua}}{\sqrt{2}}
\left(A_{u  a}S_{i2}-\mu_\mathrm{eff}S_{i1}-\l v_dS_{i3}\right)\;,\nn \\
g(H_i,D_{L a},D_{R a}) &=& \frac{h_{da}}{\sqrt{2}}
\left(A_{d  a}S_{i1}-\mu_\mathrm{eff}S_{i2}-\l v_uS_{i3}\right)\;,\nn \\
g(A_i,U_{L a},U_{R a}) &=& \frac{-h_{ua}}{\sqrt{2}}
\left(A_{u  a}P_{i2}+\mu_\mathrm{eff}P_{i1}+\l v_dP_{i3}\right)\;,\nn \\
g(A_i,D_{L a},D_{R a}) &=& \frac{h_{da}}{\sqrt{2}}
\left(A_{d  a}P_{i1}+\mu_\mathrm{eff}P_{i2}+\l v_uP_{i3}\right)\;,\nn \\
g(H^+,U_{L a},D_{L a}) &=& 
\frac{v_u v_d}{v}(g_2^2-h_{ua}^2-h_{da}^2)\;,\nn \\
g(H^+,U_{L a},D_{R a}) &=& -h_{da}(\cos\b\ \mu + \sin\b\ A_{da})\;,\nn \\
g(H^+,U_{R a},D_{L a}) &=& -h_{ua}(\sin\b\ \mu + \cos\b\ A_{ua})\;,\nn \\
g(H^+,U_{R a},D_{R a}) &=&  -\frac{h_{ua} h_{da}}{v}\;.
\label{a.21e}
\eea
\noi Third family squarks, CP-even scalars $S_i$:
\bea
H_i \tilde{t}_1 \tilde{t}_1&: & \cos\t_T^2\, g(H_i,U_{L 3},U_{L 3})
+ \sin\t_T^2\, g(H_i,U_{R 3},U_{R 3})\nn \\ &&
+ 2 \cos\t_T \sin\t_T\, g(H_i,U_{L 3},U_{R 3})\;\nn \\
H_i \tilde{t}_2 \tilde{t}_2&: & \sin\t_T^2\, g(H_i,U_{L 3},U_{L 3})
+ \cos\t_T^2\, g(H_i,U_{R 3},U_{R 3}) \nn \\ &&
- 2 \cos\t_T \sin\t_T\, g(H_i,U_{L 3},U_{R 3})\;\nn \\
H_i \tilde{t}_1 \tilde{t}_2&: & \cos\t_T \sin\t_T\, (g(H_i,U_{R 3},U_{R
3}) - g(H_i,U_{L 3},U_{L 3}))\nn \\
&&+(\cos\t_T^2-\sin\t_T^2)\, g(H_i,U_{L 3},U_{R 3})\;\nn \\
H_i \tilde{b}_1 \tilde{b}_1&: & \cos\t_B^2\, g(H_i,D_{L 3},D_{L 3})
+ \sin\t_B^2\, g(H_i,D_{R 3},D_{R 3}) \nn \\ &&
+ 2 \cos\t_B \sin\t_B\, g(H_i,D_{L 3},D_{R 3})\;\nn \\
H_i \tilde{b}_2 \tilde{b}_2&: & \sin\t_B^2\, g(H_i,D_{L 3},D_{L 3})
+ \cos\t_B^2\, g(H_i,D_{R 3},D_{R 3}) \nn \\ &&
- 2 \cos\t_B \sin\t_B\, g(H_i,D_{L 3},D_{R 3})\;\nn \\
H_i \tilde{b}_1 \tilde{b}_2&: & \cos\t_B \sin\t_B\, (g(H_i,D_{R 3},D_{R
3}) - g(H_i,D_{L 3},D_{L 3}))\nn \\
&&+(\cos\t_B^2-\sin\t_B^2)\, g(H_i,D_{L 3},D_{R 3})\;
\label{a.22e}
\eea
\noi Third family squarks, CP-odd scalars $A_i$:
\bea
A_i \tilde{t}_1 \tilde{t}_1&:& 0\;\nn \\
A_i \tilde{t}_2 \tilde{t}_2&:& 0\;\nn \\
A_i \tilde{t}_1 \tilde{t}_2&:& g(A_i,U_{L 3},U_{R 3})\;\nn \\
A_i \tilde{b}_1 \tilde{b}_1&:& 0\;\nn \\
A_i \tilde{b}_2 \tilde{b}_2&:& 0\;\nn \\
A_i \tilde{b}_1 \tilde{b}_2&:& g(A_i,D_{L 3},D_{R 3})\;
\label{a.23e}
\eea
\noi Third family squarks, charged scalar $H^+$:
\bea
H^+ \tilde{t}_1 \tilde{b}_1&:& 
\cos\t_T\cos\t_B\, g(H^+,U_{L 3},D_{L 3})+
\sin\t_T\sin\t_B\, g(H^+,U_{R 3},D_{R 3})\nn \\
&& +\cos\t_T\sin\t_B\, g(H^+,U_{L 3},D_{R 3})+
\sin\t_T\cos\t_B\, g(H^+,U_{R 3},D_{L 3})\;,\nn \\
H^+ \tilde{t}_1 \tilde{b}_2&:& 
-\cos\t_T\sin\t_B\, g(H^+,U_{L 3},D_{L 3})+
\sin\t_T\cos\t_B\, g(H^+,U_{R 3},D_{R 3})\nn \\
&& +\cos\t_T\cos\t_B\, g(H^+,U_{L 3},D_{R 3})
-\sin\t_T\sin\t_B\, g(H^+,U_{R 3},D_{L 3})\;,\nn \\
H^+ \tilde{t}_2 \tilde{b}_1&:& 
-\sin\t_T\cos\t_B\, g(H^+,U_{L 3},D_{L 3})+
\cos\t_T\sin\t_B\, g(H^+,U_{R 3},D_{R 3})\nn \\
&& -\sin\t_T\sin\t_B\, g(H^+,U_{L 3},D_{R 3})
+\cos\t_T\cos\t_B\, g(H^+,U_{R 3},D_{L 3})\;,\nn \\
H^+ \tilde{t}_2 \tilde{b}_2&:& 
\sin\t_T\sin\t_B\, g(H^+,U_{L 3},D_{L 3})+
\cos\t_T\cos\t_B\, g(H^+,U_{R 3},D_{R 3})\nn \\
&& -\sin\t_T\cos\t_B\, g(H^+,U_{L 3},D_{R 3})
-\cos\t_T\sin\t_B\, g(H^+,U_{R 3},D_{L 3})\;.
\label{a.24e}
\eea
\noi Higgs-sleptons:
\bea
g(H_i, E_{La}, E_{L a}) &=& \sqrt{2}\left(h_{ea}^2v_uS_{i1}
+\left(-\frac{g_1^2}{4}+\frac{g_2^2}{4}\right)
\left(v_uS_{i2}-v_dS_{i1}\right)\right)\;,\nn \\
g(H_i, E_{Ra}, E_{R a}) &=& \sqrt{2}\left(h_{ea}^2v_dS_{i1}
+\frac{g_1^2}{2}
\left(v_uS_{i2}-v_dS_{i1}\right)\right)\;,\nn \\
g(H_i,E_{L a},E_{R a}) &=& \frac{h_{ea}}{\sqrt{2}}
\left(A_{e  a}S_{i1}-\mu_\mathrm{eff}S_{i2}-\l v_uS_{i3}\right)\;,\nn \\
g(H_i, \widetilde{\nu}_{La}, \widetilde{\nu}_{L a}) &=& 
\sqrt{2}\left(-\frac{g_1^2}{4}-\frac{g_2^2}{4}\right)
\left(v_uS_{i2}-v_dS_{i1}\right)\;,\nn \\
g(A_i,E_{L a},E_{R a}) &=& \frac{h_{ea}}{\sqrt{2}}
\left(A_{e  a}P_{i1}+\mu_\mathrm{eff}P_{i2}+\l v_uP_{i3}\right)\;,\nn\\
g(H^+,E_{L a},\widetilde{\nu}_{La}) &=& g_2^2 \frac{v_u v_d}{v}\;.
\label{a.25e}
\eea

\noi Third family sleptons (staus), CP-even scalars $H_i$
(the coupling $H_i \widetilde{\nu}_{L3} \widetilde{\nu}_{L 3}$ does
not differ from (\ref{a.25e})):
\bea
H_i \widetilde{\tau}_1 \widetilde{\tau}_1&: & 
\cos\t_L^2\, g(H_i,E_{L 3},E_{L 3})
+ \sin\t_L^2\, g(H_i,E_{R 3},E_{R 3})\nn \\ &&
+ 2 \cos\t_L \sin\t_L\, g(H_i,E_{L 3},E_{R 3})\;\nn \\
H_i \widetilde{\tau}_2 \widetilde{\tau}_2&: & 
\sin\t_L^2\, g(H_i,E_{L 3},E_{L 3})
+ \cos\t_L^2\, g(H_i,E_{R 3},E_{R 3}) \nn \\ &&
- 2 \cos\t_L \sin\t_L\, g(H_i,E_{L 3},E_{R 3})\;\nn \\
H_i \widetilde{\tau}_1 \widetilde{\tau}_2&: & \cos\t_L \sin\t_L\, 
(g(H_i,E_{R 3},E_{R 3})- g(H_i,E_{L 3},E_{L 3}))\nn \\
&&+(\cos\t_L^2-\sin\t_L^2)\, g(H_i,E_{L 3},E_{R 3})\; 
\label{a.26e}
\eea

\noi Third family sleptons (staus), CP-odd scalars $A_i$:
\bea
A_i \widetilde{\tau}_1 \widetilde{\tau}_1&: & 0\;\nn \\
A_i \widetilde{\tau}_2 \widetilde{\tau}_2&: & 0\;\nn \\
A_i \widetilde{\tau}_1 \widetilde{\tau}_2&: & g(A_i,E_{L 3},E_{R 3})\;
\label{a.27e}
\eea

\noi Third family sleptons (staus), charged scalar $H^+$:
\bea
H^+ \widetilde{\tau}_1 \widetilde{\nu}_{L3}&: & 
\cos\t_L \frac{v_u v_d}{v}(g_2^2-h_{e3}^2)
-\sin\t_L h_{e3}(\cos\b\ \mu + \sin\b\ A_{e3})\;\nn \\
H^+ \widetilde{\tau}_2 \widetilde{\nu}_{L3}&: & 
-\sin\t_L \frac{v_u v_d}{v}(g_2^2-h_{e3}^2)
-\cos\t_L h_{e3}(\cos\b\ \mu + \sin\b\ A_{e3})
\label{a.28e}
\eea

\subsubsection*{Yukawa induced radiative corrections to triple
Higgs couplings}

As in the MSSM \cite{Kunszt:1991qe,Barger:1991ed,
Brignole:1992zv,Heinemeyer:1996tg, Hollik:2001px,Boudjema:2001ii}, the
dominant radiative corrections to triple Higgs couplings originate from
top/bottom-quark loops, which are regularized in the ultraviolet by the
corresponding contributions from squark loops. Here we give these
radiative corrections in the NMSSM, which just require to generalize the
Higgs mixing matrices of the MSSM to the NMSSM, see Appendix~A.1. 

However, we confine ourselves to the contributions involving a
potentially large logarithm: we assume approximately degenerate
squark masses given by $M_\mathrm{SUSY}$. The infrared cutoff inside the
dominant logarithm is given by the mass $m_q$ of the quark inside the loop or
the mass of the decaying Higgs boson through this vertex (whichever is
larger); we will denote it by $M_H$ here. Then the potentially large
logarithm is
\beq\label{a.29e}
t \equiv \ln\left(M_\mathrm{SUSY}^2/\operatorname{max}(m_q^2,\,M_H^2)
\right)\; .
\eeq

The dominant corrections to the triple CP-even Higgs interactions (A.15)
are then given by
\beq\label{a.30e}
\Delta H_a H_b H_c\ : \frac{3\, t}{8\sqrt{2}\,\pi^2}
\left(h_t^4 v_u \Pi^{222}_{abc} + h_b^4 v_d \Pi^{111}_{abc}\right)
\eeq
with $\Pi^{ijk}_{abc}$ as in (A.16). The dominant corrections to the
CP-even/CP-odd Higgs interactions (A.17) read the same,
\beq\label{3.9e}
\Delta H_a A_b A_c\ : \frac{3\, t}{8\sqrt{2}\,\pi^2}
\left(h_t^4 v_u \Pi^{222}_{abc} + h_b^4 v_d \Pi^{111}_{abc}\right)
\eeq
where now $\Pi^{ijk}_{abc}$ is given in (A.18). The dominant
corrections to the CP-even/charged Higgs couplings (A.19) read
\bea
\Delta H_a H^+ H^- &:&  \frac{3\, t}{8\sqrt{2}\,\pi^2} 
\Big[ h_t^2 v_u \left(h_t^2 \Pi^{211}_a + h_b^2
(\Pi^{222}_a+\Pi^{112}_a)\right)\nn \\
&&+ h_b^2 v_d \left(h_b^2 \Pi^{122}_a + h_t^2
(\Pi^{212}_a+\Pi^{111}_a)\right)\Big]
\label{3.10e}
\eea
with $\Pi^{ijk}_a$ as in (A.20).

\renewcommand{\theequation}{B.\arabic{equation}}
\setcounter{equation}{0}
\setcounter{section}{0}

\section*{Appendix B. Renormalisation group equations}
\addcontentsline{toc}{section}
{Appendix B. Renormalisation group equations}

In this Appendix we give the renormalisation group equations in the
$\overline{\text{DR}}$ scheme, valid above the SUSY breaking scale, for
the parameters of the general NMSSM. The conventions are $t = \ln Q^2$,
and the $U(1)_Y$ gauge coupling $g_1^2$ is defined in the Standard Model
normalisation (related to the GUT or $SU(5)$ normalisation by $g_1^2 =
\frac{3}{5} \left(g_1^{\mathrm{GUT}}\right)^2$).

\subsection*{B.1 Gauge and Yukawa couplings}
\addcontentsline{toc}{subsection}{B.1 Gauge and Yukawa couplings}
The RGEs are known to two-loop order
\cite{Jones:1974pg,Jones:1983vk,West:1984dg} (for the NMSSM see 
\cite{King:1995vk,Masip:1998jc}); it suffices, however, to include the
Yukawa couplings of the third family only. (The CKM mixing matrix is not
considered here.)

\bea
16\pi^2 \frac{dg_1^2}{dt} &=& 11g_1^4 + \frac{g_1^4}{16\pi^2}
\bigg( \frac{199}{9}g_1^2 + 9g_2^2 +\frac{88}{3}g_3^2 -
\frac{26}{3}h_t^2 - \frac{14}{3}h_b^2 - 6h_\tau^2 - 2\l^2 \bigg)\; , \nn
\\
16\pi^2 \frac{dg_2^2}{dt} &=& g_2^4 + \frac{g_2^4}{16\pi^2}
\bigg( 3g_1^2 + 25g_2^2 + 24g_3^2 - 6h_t^2
- 6h_b^2 - 2h_\tau^2 - 2\l^2 \bigg)\; , \nn \\
16\pi^2 \frac{dg_3^2}{dt} &=& - 3 g_3^4 + \frac{g_3^4}{16\pi^2} \bigg(
\frac{11}{3}g_1^2 + 9g_2^2 + 14g_3^2 - 4h_t^2 - 4 h_b^2 \bigg)\; , \nn
\\
16\pi^2 \frac{dh_t^2}{dt} &=& h_t^2\bigg( 6h_t^2 + h_b^2 + \l^2
- \frac{13}{9}g_1^2 - 3g_2^2 - \frac{16}{3}g_3^2 \bigg) \nn \\
&+& \frac{h_t^2}{16\pi^2} \bigg( - 22h_t^4 - 5h_b^4 - 3\l^4 -
5h_t^2h_b^2 - 3h_t^2\l^2 - h_b^2h_\tau^2 - 4h_b^2\l^2 \nn \\
&-& h_\tau^2\l^2 - 2\l^2\k^2 +  2g_1^2h_t^2
+ \frac{2}{3}g_1^2h_b^2 + 6g_2^2h_t^2 + 16g_3^2h_t^2 \nn \\
&+& \frac{2743}{162}g_1^4 + \frac{15}{2}g_2^4 - \frac{16}{9}g_3^4
+ \frac{5}{3}g_1^2g_2^2 + \frac{136}{27}g_1^2g_3^2 + 8g_2^2g_3^2
\bigg)\; , \nn \\ 
16\pi^2 \frac{dh_b^2}{dt} &=& h_b^2\bigg( 6h_b^2 + h_t^2 + h_\tau^2 +
\l^2 - \frac{7}{9}g_1^2 - 3g_2^2 - \frac{16}{3}g_3^2 \bigg) \nn \\
&+& \frac{h_b^2}{16\pi^2} \bigg( - 22h_b^4 - 5h_t^4
- 3h_\tau^4 - 3\l^4 - 5h_b^2h_t^2 - 3h_b^2h_\tau^2 - 3h_b^2\l^2 \nn \\
&-& 4h_t^2\l^2 - 2\l^2\k^2 + \frac{2}{3}g_1^2h_b^2 +
\frac{4}{3}g_1^2h_t^2 + 2g_1^2h_\tau^2 + 6g_2^2h_b^2 + 16g_3^2h_b^2 \nn
\\
&+& \frac{1435}{162}g_1^4 + \frac{15}{2}g_2^4 - \frac{16}{9}g_3^4
+ \frac{5}{3}g_1^2g_2^2 + \frac{40}{27}g_1^2g_3^2 + 8g_2^2g_3^2\bigg)\;
, \nn \\ 
16\pi^2 \frac{dh_\tau^2}{dt} &=& h_\tau^2\Big( 4h_\tau^2 + 3h_b^2 + \l^2
- 3g_1^2 - 3g_2^2 \Big) \nn \\
&+& \frac{h_\tau^2}{16\pi^2} \bigg( - 10h_\tau^4 - 9h_b^4 - 3\l^4
- 9h_\tau^2h_b^2 - 3h_\tau^2\l^2 - 3h_t^2h_b^2 - 3h_t^2\l^2 \nn \\
&-& 2\l^2\k^2  + 2g_1^2h_\tau^2 - \frac{2}{3}g_1^2h_b^2
+ 6g_2^2h_\tau^2 + 16g_3^2h_b^2 + \frac{75}{2}g_1^4
+ \frac{15}{2}g_2^4 + 3g_1^2g_2^2 \bigg)\; , \nn
\eea
\bea
16\pi^2 \frac{d\l^2}{dt} &=& \l^2\Big(3h_t^2 + 3h_b^2 + h_\tau^2 +4\l^2
+ 2\k^2 - g_1^2 - 3g_2^2 \Big) \nn \\
&+& \frac{\l^2}{16\pi^2} \bigg( - 10\l^4  - 9h_t^4 - 9h_b^4
- 3h_\tau^4 - 8\k^4 - 9\l^2h_t^2 - 9\l^2h_b^2 \nn \\
&-& 3\l^2h_\tau^2 -12\l^2\k^2 - 6h_t^2h_b^2 + 2g_1^2\l^2 +
\frac{4}{3}g_1^2h_t^2 - \frac{2}{3}g_1^2h_b^2 + 2g_1^2h_\tau^2 \nn \\
&+& 6g_2^2\l^2 + 16g_3^2h_t^2 + 16g_3^2h_b^2
+ \frac{23}{2}g_1^4 + \frac{15}{2}g_2^4 + 3g_1^2g_2^2 \bigg)\; , \nn \\
16\pi^2 \frac{d\k^2}{dt} &=& \k^2\Big(6\l^2 +6\k^2\Big)
+ \frac{\k^2}{16\pi^2} \bigg( - 24\k^4 - 12\l^4 - 24\k^2\l^2 \nn \\
&-& 18h_t^2\l^2 - 18h_b^2\l^2 - 6h_\tau^2\l^2 + 6g_1^2\l^2 + 18g_2^2\l^2
\bigg)\; .
\label{b.1e}
\eea

\subsection*{B.2 Gaugino masses}
\addcontentsline{toc}{subsection}{B.2 Gaugino masses}

The RGEs to two-loop order can be found in
\cite{Martin:1993yx,Yamada:1993ga}.

\bea
16\pi^2 \frac{dM_1}{dt} &=& 11g_1^2M_1
+ \frac{g_1^2}{16\pi^2} \bigg( \frac{398}{9}g_1^2M_1
+ 9g_2^2\big(M_1+M_2\big) + \frac{88}{3}g_3^2\big(M_1+M_3 \big) \nn \\
&+& \frac{26}{3}h_t^2\big(A_t-M_1\big) +
\frac{14}{3}h_b^2\big(A_b-M_1\big)
+ 6h_\tau^2\big(A_\tau-M_1\big) + 2\l^2\big(A_\l-M_1\big)\bigg)\; , \nn
\\
16\pi^2 \frac{dM_2}{dt} &=& g_2^2M_2
+ \frac{g_2^2}{16\pi^2} \bigg( 3g_1^2\big(M_1+M_2\big)
+ 50g_2^2M_2 + 24g_3^2\big(M_2+M_3\big) \nn \\
&+& 6h_t^2\big(A_t-M_2\big) + 6h_b^2\big(A_b-M_2\big)
+ 2h_\tau^2\big(A_\tau-M_2\big) + 2\l^2\big(A_\l-M_2\big) \bigg)\; , \nn
\\
16\pi^2 \frac{dM_3}{dt} &=& -3g_3^2M_3 
+ \frac{g_3^2}{16\pi^2} \bigg( \frac{11}{3}g_1^2\big(M_1+M_3\big)
+9g_2^2\big(M_2+M_3\big) + 28g_3^2M_3 \nn \\
&+& 4h_t^2\big(A_t-M_3\big) + 4h_b^2\big(A_b-M_3\big) \bigg)\; .
\label{b.2e}
\eea

\subsection*{B.3 Trilinear couplings}
\addcontentsline{toc}{subsection}{B.3 Trilinear couplings}

The two-loop $\beta$-functions are known for the parameters of 
a general softly broken SUSY theory \cite{Martin:1993zk}. We concentrate
on the trilinear couplings involving the squarks and sleptons of the
third family; those of the first two families do not play an important
phenomenological r\^ole, except for the muon trilinear coupling $A_\mu$
($\equiv A_{e2}$)
which enters the muon anomalous magnetic momentum $(g-2)_\mu$.

\bea
16\pi^2 \frac{dA_t}{dt} &=& 6h_t^2A_t + h_b^2A_b  + \l^2A_\l
+ \frac{13}{9}g_1^2M_1 + 3g_2^2M_2 + \frac{16}{3}g_3^2M_3\nn \\
&+& \frac{1}{16\pi^2} \bigg( - 44h_t^4A_t - 10 h_b^4A_b - 6\l^4A_\l
- 5h_t^2h_b^2\big(A_t+A_b\big) \nn \\
&-& 3h_t^2\l^2\big(A_t+A_\l\big) - h_b^2h_\tau^2\big(A_b+A_\tau\big)
- 4h_b^2\l^2\big(A_b+A_\l\big) \nn \\
&-& h_\tau^2\l^2\big(A_\tau+A_\l\big) - 2\l^2\k^2\big(A_\l+A_\k\big)
+ 2g_1^2h_t^2\big(A_t-M_1\big) \nn \\
&+& \frac{2}{3}g_1^2h_b^2\big(A_b-M_1\big) + 6g_2^2h_t^2\big(A_t-M_2\big)
+ 16g_3^2h_t^2\big(A_t-M_3\big) \nn \\
&-& \frac{2743}{81}g_1^4M_1 - 15g_2^4M_2 + \frac{32}{9}g_3^4M_3
- \frac{5}{3}g_1^2g_2^2\big(M_1+M_2\big) \nn \\
&-& \frac{136}{27}g_1^2g_3^2\big(M_1+M_3\big)
- 8g_2^2g_3^2\big(M_2+M_3\big) \bigg)\; , \nn \\
16\pi^2 \frac{dA_b}{dt} &=& 6h_b^2A_b + h_t^2A_t + h_\tau^2A_\tau
+ \l^2A_\l +\frac{7}{9}g_1^2M_1 + 3g_2^2M_2 + \frac{16}{3}g_3^2M_3\nn \\
&+& \frac{1}{16\pi^2} \bigg( - 44h_b^4A_b - 10 h_t^4A_t
- 6h_\tau^4A_\tau - 6\l^4A_\l - 5h_b^2h_t^2\big(A_b+A_t\big) \nn \\
&-& 3h_b^2h_\tau^2\big(A_b+A_\tau\big) -  3h_b^2\l^2\big(A_b+A_\l\big)
- 4h_t^2\l^2\big(A_t+A_\l\big) \nn \\
&-& 2\l^2\k^2\big(A_\l+A_\k\big) + \frac{2}{3}g_1^2h_b^2\big(A_b-M_1\big)
+ \frac{4}{3}g_1^2h_t^2\big(A_t-M_1\big) \nn \\
&+& 2g_1^2h_\tau^2\big(A_\tau-M_1\big) + 6g_2^2h_b^2\big(A_b-M_2\big)
+ 16g_3^2h_b^2\big(A_b-M_3\big) \nn \\
&-& \frac{1435}{81}g_1^4M_1 - 15g_2^4M_2 + \frac{32}{9}g_3^4M_3
- \frac{5}{3}g_1^2g_2^2\big(M_1+M_2\big) \nn \\
&-& \frac{40}{27}g_1^2g_3^2\big(M_1+M_3\big)
- 8g_2^2g_3^2\big(M_2+M_3\big) \bigg)\; , \nn \\
16\pi^2 \frac{dA_\tau}{dt} &=& 4h_\tau^2A_\tau + 3h_b^2A_b
+ \l^2A_\l + 3g_1^2M_1 + 3g_2^2M_2 + \frac{1}{16\pi^2} \bigg( \nn \\
&-& 20h_\tau^4A_\tau - 18 h_b^4A_b - 6\l^4A_\l
- 9h_\tau^2h_b^2\big(A_\tau+A_b\big) - 3h_\tau^2\l^2\big(A_\tau+A_\l\big) \nn \\
&-& 3h_t^2h_b^2\big(A_t+A_b\big) - 3h_t^2\l^2\big(A_t+A_\l\big)
- 2\l^2\k^2\big(A_\l+A_\k\big)  \nn \\
&+& 2g_1^2h_\tau^2\big(A_\tau-M_1\big) - \frac{2}{3}g_1^2h_b^2\big(A_b-M_1\big)
+ 6g_2^2h_\tau^2\big(A_\tau-M_2\big) \nn \\
&+&  16g_3^2h_b^2\big(A_b-M_3\big) - 75g_1^4M_1 - 15g_2^4M_2
- 3g_1^2g_2^2\big(M_1+M_2\big) \bigg)\; , \nn \\
16\pi^2 \frac{dA_\mu}{dt} &=& 3h_b^2A_b + h_\tau^2A_\tau
+ \l^2A_\l + 3g_1^2M_1 + 3g_2^2M_2 + \frac{1}{16\pi^2} \bigg( \nn \\
&-& 18 h_b^4A_b - 6h_\tau^4A_\tau  - 6\l^4A_\l
- 3h_t^2h_b^2\big(A_t+A_b\big) - 3h_t^2\l^2\big(A_t+A_\l\big)  \nn \\
&-& 2\l^2\k^2\big(A_\l+A_\k\big) - \frac{2}{3}g_1^2h_b^2\big(A_b-M_1\big)
+ 2g_1^2h_\tau^2\big(A_\tau-M_1\big)  \nn \\
&+& 16g_3^2h_b^2\big(A_b-M_3\big) - 75g_1^4M_1 - 15g_2^4M_2
- 3g_1^2g_2^2\big(M_1+M_2\big) \bigg)\; , \nn
\eea
\bea
16\pi^2 \frac{dA_\l}{dt} &=& 4\l^2A_\l + 3h_t^2A_t + 3h_b^2A_b
+ h_\tau^2A_\tau + 2\k^2A_\k + g_1^2M_1 + 3g_2^2M_2 \nn \\
&+& \frac{1}{16\pi^2} \bigg( - 20 \l^4A_\l - 18h_t^4A_t  - 18h_b^4A_b
- 6h_\tau^4A_\tau - 16\k^4A_\k \nn \\
&-& 9\l^2h_t^2\big(A_\l+A_t\big) - 9\l^2h_b^2\big(A_\l+A_b\big)
- 3 \l^2h_\tau^2\big(A_\l+A_\tau\big) \nn \\
&-& 12\l^2\k^2\big(A_\l+A_\k\big) - 6h_t^2h_b^2\big(A_t+A_b\big)
+ 2g_1^2\l^2(A_\l-M_1) \nn \\
&+& \frac{4}{3}g_1^2h_t^2\big(A_t-M_1\big)
- \frac{2}{3}g_1^2h_b^2\big(A_b-M_1\big)
+ 2g_1^2h_\tau^2\big(A_\tau-M_1\big) \nn \\
&+& 6g_2^2\l^2\big(A_\l-M_2\big) + 16g_3^2h_t^2\big(A_t-M_3\big)
+ 16g_3^2h_b^2\big(A_b-M_3\big) \nn \\
&-& 23g_1^4M_1 - 15g_2^4M_2
- 3g_1^2g_2^2\big(M_1+M_2\big) \bigg)\; , \nn \\
16\pi^2 \frac{dA_\k}{dt} &=& 6\k^2A_\k  + 6\l^2A_\l
+ \frac{1}{16\pi^2} \bigg( - 48\k^4A_\k - 24 \l^4A_\l \nn \\
&-& 24\k^2\l^2\big(A_\k+A_\l\big)
- 18h_t^2\l^2\big(A_t+A_\l\big) - 18h_b^2\l^2\big(A_b+A_\l\big) \nn \\
&-& 6h_\tau^2\l^2\big(A_\tau+A_\l\big)
+ 6g_1^2\l^2\big(A_\l-M_1\big) + 18g_2^2\l^2\big(A_\l-M_2\big) \bigg)\;
.
\label{b.3e}
\eea

\subsection*{B.4 Squark and slepton masses}
\addcontentsline{toc}{subsection}{B.4 Squark and slepton masses}

Let us define the following quantities:
\bea
M_t^2 &= & m_{Q_3}^2+m_{U_3}^2+m_{H_u}^2+A_t^2\; , \nn \\
M_b^2 &= & m_{Q_3}^2+m_{D_3}^2+m_{H_d}^2+A_b^2\; , \nn \\
M_\tau^2 &= & m_{L_3}^2+m_{E_3}^2+m_{H_d}^2+A_\tau^2\; , \nn \\
M_\l^2 &= & m_{H_u}^2+m_{H_d}^2+m_S^2+A_\l^2\; , \nn \\
M_\k^2 &= & 3m_S^2+A_\k^2\; , \nn \\
\xi &=& {\rm Tr}\big[{\bf m}_Q^2 - 2{\bf m}_U^2 + {\bf m}_D^2
- {\bf m}_L^2 + {\bf m}_E^2\big]
+ m_{H_u}^2 - m_{H_d}^2\; , \nn \\
\xi' &=& h_t^2\big(-m_{Q_3}^2+4m_{U_3}^2-3m_{H_u}^2\big)
+ h_b^2\big(-m_{Q_3}^2-2m_{D_3}^2+3m_{H_d}^2\big) \nn \\
&+& h_\tau^2\big(m_{L_3}^2-2m_{E_3}^2+m_{H_d}^2\big)
+ \l^2\big(m_{H_d}^2-m_{H_u}^2\big) \nn \\
&+& g_1^2 \Big( {\rm Tr}\Big[\frac{1}{18}{\bf m}_Q^2
- \frac{16}{9}{\bf m}_U^2 + \frac{2}{9}{\bf m}_D^2 - \frac{1}{2}{\bf
m}_L^2 + 2{\bf m}_E^2\Big] + \frac{1}{2}\big(m_{H_u}^2 - m_{H_d}^2\big)
\Big) \nn \\
&+& \frac{3}{2}g_2^2 \Big( {\rm Tr}\big[{\bf m}_Q^2 - {\bf m}_L^2 \big]
+ m_{H_u}^2 - m_{H_d}^2 \Big) + \frac{8}{3}g_3^2{\rm Tr}\big[ {\bf
m}_Q^2 - 2{\bf m}_U^2 + {\bf m}_D^2 \big]\; , \nn \\
\sigma_1 &=& g_1^2 \Big( {\rm Tr}\Big[\frac{1}{3}{\bf m}_Q^2
+ \frac{8}{3}{\bf m}_U^2 + \frac{2}{3}{\bf m}_D^2 + {\bf m}_L^2 + 2{\bf
m}_E^2\Big]
+ m_{H_u}^2 + m_{H_d}^2 \Big)\; , \nn \\
\sigma_2 &=& g_2^2 \Big( {\rm Tr}\big[3{\bf m}_Q^2 + {\bf m}_L^2 \big]
+ m_{H_u}^2 + m_{H_d}^2 \Big)\; , \nn \\
\sigma_3 &=& g_3^2 {\rm Tr}\big[2{\bf m}_Q^2 + {\bf m}_U^2 + {\bf
m}_D^2\big]\; .
\eea

\noindent where $\bf m$ denote matrices in family space.
The two-loop RGEs then read

\bea
16\pi^2 \frac{dm_{Q_a}^2}{dt} &=&
\delta_{a3}h_t^2M_t^2 + \delta_{a3}h_b^2M_b^2 - \frac{1}{9}g_1^2M_1^2
- 3g_2^2M_2^2 - \frac{16}{3}g_3^2M_3^2 + \frac{1}{6}g_1^2\xi \nn \\
&+& \frac{1}{16\pi^2} \bigg( - 10\delta_{a3}h_t^4\big(M_t^2+A_t^2\big)
- 10\delta_{a3}h_b^4\big(M_b^2+A_b^2\big) \nn \\
&-& \delta_{a3}h_t^2\l^2\big(M_t^2+M_\l^2+2A_tA_\l\big)
- \delta_{a3}h_b^2h_\tau^2\big(M_b^2+M_\tau^2+2A_bA_\tau\big) \nn \\
&-& \delta_{a3}h_b^2\l^2\big(M_b^2+M_\l^2+2A_bA_\l\big)
+ \frac{4}{3}\delta_{a3}g_1^2h_t^2\big(M_t^2-2M_1(A_t-M_1)\big) \nn \\
&+& \frac{2}{3}\delta_{a3}g_1^2h_b^2\big(M_b^2-2M_1(A_b-M_1)\big)
+ \frac{199}{54}g_1^4M_1^2 + \frac{33}{2}g_2^4M_2^2
- \frac{64}{3}g_3^4M_3^2 \nn \\
&+& \frac{1}{3}g_1^2g_2^2(M_1^2+M_2^2+M_1M_2)
+ \frac{16}{27}g_1^2g_3^2(M_1^2+M_3^2+M_1M_3) \nn \\
&+& 16g_2^2g_3^2(M_2^2+M_3^2+M_2M_3)
+ \frac{1}{3}g_1^2\xi' + \frac{1}{18}g_1^2\sigma_1
+ \frac{3}{2}g_2^2\sigma_2 + \frac{8}{3}g_3^2\sigma_3 
\bigg)\; , \nn \\
16\pi^2 \frac{dm_{U_a}^2}{dt} &=&
2\delta_{a3}h_t^2M_t^2 - \frac{16}{9}g_1^2M_1^2
- \frac{16}{3}g_3^2M_3^2 - \frac{2}{3}g_1^2\xi \nn \\
&+& \frac{1}{16\pi^2} \bigg( - 16\delta_{a3}h_t^4\big(M_t^2+A_t^2\big)
- 2\delta_{a3}h_t^2h_b^2\big(M_t^2+M_b^2+2A_tA_b\big) \nn \\
&-& 2\delta_{a3}h_t^2\l^2\big(M_t^2+M_\l^2+2A_tA_\l\big)
- \frac{2}{3}\delta_{a3}g_1^2h_t^2\big(M_t^2-2M_1(A_t-M_1)\big) \nn \\
&+& 6 \delta_{a3}g_2^2h_t^2\big(M_t^2-2M_2(A_t-M_2)\big)
+ \frac{1712}{27}g_1^4M_1^2 - \frac{64}{3}g_3^4M_3^2 \nn \\
&+& \frac{256}{27}g_1^2g_3^2(M_1^2+M_3^2+M_1M_3)
- \frac{4}{3}g_1^2\xi' + \frac{8}{9}g_1^2\sigma_1
+ \frac{8}{3}g_3^2\sigma_3
\bigg)\; , \nn \\
16\pi^2 \frac{dm_{D_a}^2}{dt} &=&
2\delta_{a3}h_b^2M_b^2 - \frac{4}{9}g_1^2M_1^2
- \frac{16}{3}g_3^2M_3^2 + \frac{1}{3}g_1^2\xi \nn \\
&+& \frac{1}{16\pi^2} \bigg( - 16\delta_{a3}h_b^4\big(M_b^2+A_b^2\big)
- 2\delta_{a3}h_b^2h_t^2\big(M_b^2+M_t^2+2A_bA_t\big) \nn \\
&-& 2\delta_{a3}h_b^2h_\tau^2\big(M_b^2+M_\tau^2+2A_bA_\tau\big)
- 2\delta_{a3}h_b^2\l^2\big(M_b^2+M_\l^2+2A_bA_\l\big) \nn \\
&+& \frac{2}{3}\delta_{a3}g_1^2h_b^2\big(M_b^2-2M_1(A_b-M_1)\big)
+ 6 \delta_{a3}g_2^2h_b^2\big(M_b^2-2M_2(A_b-M_2)\big) \nn \\
&+& \frac{404}{27}g_1^4M_1^2 - \frac{64}{3}g_3^4M_3^2
+ \frac{64}{27}g_1^2g_3^2(M_1^2+M_3^2+M_1M_3) \nn \\
&+& \frac{2}{3}g_1^2\xi' + \frac{2}{9}g_1^2\sigma_1
+ \frac{8}{3}g_3^2\sigma_3
\bigg)\; , \nn \\
16\pi^2 \frac{dm_{L_a}^2}{dt} &=&
\delta_{a3}h_\tau^2M_\tau^2 - g_1^2M_1^2 - 3g_2^2M_2^2 - \frac{1}{2}g_1^2\xi
+ \frac{1}{16\pi^2} \bigg( - 6\delta_{a3}h_\tau^4\big(M_\tau^2+A_\tau^2\big) \nn \\
&-& 3\delta_{a3}h_\tau^2h_b^2\big(M_\tau^2+M_b^2+2A_\tau A_b\big)
- \delta_{a3}h_\tau^2\l^2\big(M_\tau^2+M_\l^2+2A_\tau A_\l\big) \nn \\
&+& 2\delta_{a3}g_1^2h_\tau^2\big(M_\tau^2-2M_1(A_\tau-M_1)\big)
+ \frac{69}{2}g_1^4M_1^2 + \frac{33}{2}g_2^4M_2^2 \nn \\
&+& 3g_1^2g_2^2(M_1^2+M_2^2+M_1M_2)
- g_1^2\xi' + \frac{1}{2}g_1^2\sigma_1
+ \frac{3}{2}g_2^2\sigma_2
\bigg)\; , \nn
\eea
\bea
16\pi^2 \frac{dm_{E_a}^2}{dt} &=&
2\delta_{a3}h_\tau^2M_\tau^2 - 4g_1^2M_1^2 + g_1^2\xi
+ \frac{1}{16\pi^2} \bigg( - 8\delta_{a3}h_\tau^4\big(M_\tau^2+A_\tau^2\big) \nn \\
&-& 6\delta_{a3}h_\tau^2h_b^2\big(M_\tau^2+M_b^2+2A_\tau A_b\big)
- 2\delta_{a3}h_\tau^2\l^2\big(M_\tau^2+M_\l^2+2A_\tau A_\l\big) \nn \\
&-& 2\delta_{a3}g_1^2h_\tau^2\big(M_\tau^2-2M_1(A_\tau-M_1)\big)
+ 6\delta_{a3}g_2^2h_\tau^2\big(M_\tau^2-2M_2(A_\tau-M_2)\big) \nn \\
&+& 156g_1^4M_1^2 + 2g_1^2\xi' + 2g_1^2\sigma_1
\bigg)\; .
\eea

\subsection*{B.5 Higgs masses}
\addcontentsline{toc}{subsection}{B.5 Higgs masses}

Similarly, the two-loop RGEs for the Higgs soft masses are

\bea
16\pi^2 \frac{dm_{H_u}^2}{dt} &=&
3h_t^2M_t^2 + \l^2M_\l^2 - g_1^2M_1^2 - 3g_2^2M_2^2 + \frac{1}{2}g_1^2\xi \nn \\
&+& \frac{1}{16\pi^2} \bigg( - 18h_t^4\big(M_t^2+A_t^2\big)
- 6\l^4\big(M_\l^2+A_\l^2\big) \nn \\
&-& 3h_t^2h_b^2\big(M_t^2+M_b^2+2A_tA_b\big)
- 3h_b^2\l^2\big(M_b^2+M_\l^2+2A_bA_\l\big) \nn \\
&-& h_\tau^2\l^2\big(M_\tau^2+M_\l^2+2A_\tau A_\l\big)
-  2\l^2\k^2\big(M_\l^2+M_\k^2+2A_\l A_\k\big) \nn \\
&+& \frac{4}{3}g_1^2h_t^2\big(M_t^2-2M_1(A_t-M_1)\big)
+ 16g_3^2h_t^2\big(M_t^2-2M_3(A_t-M_3)\big) \nn \\
&+& \frac{69}{2}g_1^4M_1^2 + \frac{33}{2}g_2^4M_2^2
+ 3g_1^2g_2^2(M_1^2+M_2^2+M_1M_2) \nn \\
&+& g_1^2\xi' + \frac{1}{2}g_1^2\sigma_1
+ \frac{3}{2}g_2^2\sigma_2
\bigg)\; , \nn \\
16\pi^2 \frac{dm_{H_d}^2}{dt} &=&
3h_b^2M_b^2 + h_\tau^2M_\tau^2 + \l^2M_\l^2 - g_1^2M_1^2
- 3g_2^2M_2^2 - \frac{1}{2}g_1^2\xi \nn \\
&+& \frac{1}{16\pi^2} \bigg( - 18h_b^4\big(M_b^2+A_b^2\big)
- 6h_\tau^4\big(M_\tau^2+A_\tau^2\big) \nn \\
&-& 6\l^4\big(M_\l^2+A_\l^2\big)
- 3h_b^2h_t^2\big(M_b^2+M_t^2+2A_bA_t\big) \nn \\
&-& 3h_t^2\l^2\big(M_t^2+M_\l^2+2A_tA_\l\big)
-  2\l^2\k^2\big(M_\l^2+M_\k^2+2A_\l A_\k\big) \nn \\
&-& \frac{2}{3}g_1^2h_b^2\big(M_b^2-2M_1(A_b-M_1)\big)
+ 2g_1^2h_\tau^2\big(M_\tau^2-2M_1(A_\tau-M_1)\big) \nn \\
&+& 16g_3^2h_b^2\big(M_b^2-2M_3(A_b-M_3)\big)
+ \frac{69}{2}g_1^4M_1^2 + \frac{33}{2}g_2^4M_2^2 \nn \\
&+& 3g_1^2g_2^2(M_1^2+M_2^2+M_1M_2)
- g_1^2\xi' + \frac{1}{2}g_1^2\sigma_1
+ \frac{3}{2}g_2^2\sigma_2
\bigg)\; , \nn \\
16\pi^2 \frac{dm_S^2}{dt} &=&
2\l^2M_\l^2 + 2\k^2M_\k^2
+ \frac{1}{16\pi^2} \bigg( - 8\l^4\big(M_\l^2+A_\l^2\big)
- 16\k^4\big(M_\k^2+A_\k^2\big) \nn \\
&-& 6\l^2h_t^2\big(M_\l^2+M_t^2+2A_\l A_t\big)
- 6\l^2h_b^2\big(M_\l^2+M_b^2+2A_\l A_b\big) \nn \\
&-& 2\l^2h_\tau^2\big(M_\l^2+M_\tau^2+2A_\l A_\tau\big)
- 8\l^2\k^2\big(M_\l^2+M_\k^2+2A_\l A_\k\big) \nn \\
&+& 2g_1^2\l^2\big(M_\l^2-2M_1(A_\l-M_1)\big)
+ 6g_2^2\l^2\big(M_\l^2-2M_2(A_\l-M_2)\big)
\bigg)\; .\
\label{b.6e}
\eea

\subsection*{B.6 Additional parameters of the general NMSSM}
\addcontentsline{toc}{subsection}
{B.6 Additional parameters of the general NMSSM}

The two-loop RGEs for the SUSY conserving $\mu$ and $\mu'$ terms are
\bea
32\pi^2 \frac{d\mu}{dt} &=&
\mu\Big( 3h_t^2 + 3h_b^2
+ h_\tau^2 + 2\l^2 - g_1^2 - 3g_2^2 \Big)
+ \frac{\mu}{16\pi^2} \Big( - 9h_t^4 - 9h_b^4 \nn \\
&-& 3h_\tau^4 - 6\l^4 - 6h_t^2h_b^2 - 3h_t^2\l^2 - 3h_b^2\l^2
- h_\tau^2\l^2 - 4\l^2\k^2 + \frac{4}{3}g_1^2h_t^2 \nn \\
&-& \frac{2}{3}g_1^2h_b^2 + 2g_1^2h_\tau^2 + 16g_3^2h_t^2
+ 16g_3^2h_b^2 + \frac{23}{2}g_1^4 + \frac{15}{2}g_2^4 + 3g_1^2g_2^2
\Big)\; , \nn \\
16\pi^2 \frac{d\mu'}{dt} &=&
\mu'\Big( 2\l^2 + 2\k^2 \Big)
+ \frac{\mu'}{16\pi^2} \Big( -4\l^4 - 8\k^4 - 6h_t^2\l^2 \nn \\
&-& 6h_b^2\l^2 - 2h_\tau^2\l^2 - 8\l^2\k^2 + 2g_1^2\l^2 + 6g_2^2\l^2
\Big)\; .
\eea
For the corresponding soft SUSY breaking terms $m_3^2$ and $m_S'^2$
(see (\ref{2.9e})) we have
\bea
32\pi^2 \frac{dm_3^2}{dt} &=&
3h_t^2\big(m_3^2 + 2\mu A_t\big) + 3h_b^2\big(m_3^2 + 2\mu A_b\big)
+ h_\tau^2\big(m_3^2 + 2\mu A_\tau\big) \nn \\
&+& 2\l^2\big(3m_3^2 + 2\mu A_\l\big) +  2\l\k m_S'^2
- g_1^2\big(m_3^2-2\mu M_1\big)
- 3g_2^2\big(m_3^2-2\mu M_2\big) \nn \\
&+& \frac{1}{16\pi^2} \Big( - 9h_t^4\big(m_3^2+4\mu A_t\big)
- 9h_b^4\big(m_3^2+4\mu A_b\big)
- 3h_\tau^4\big(m_3^2+4\mu A_\tau\big) \nn \\
&-& 2\l^4\big(7m_3^2+16\mu A_\l\big)
- 6h_t^2h_b^2\big(m_3^2+2\mu (A_t+A_b)\big) \nn \\
&-& 3h_t^2\l^2\big(5m_3^2+2\mu (3A_t+A_\l)\big)
- 3h_b^2\l^2\big(5m_3^2+2\mu (3A_b+A_\l)\big) \nn \\
&-& h_\tau^2\l^2\big(5m_3^2+2\mu (3A_\tau+A_\l)\big)
- 4\l^2\k^2\big(m_3^2+2\mu (A_\l+A_\k)\big) \nn \\
&-& 8\l^3\k\big(m_S'^2+\mu'A_\l\big)
- 8\l\k^3\big(m_S'^2+\mu'A_\k\big) \nn \\
&+& \frac{4}{3}g_1^2h_t^2\big(m_3^2+2\mu(A_t-M_1)\big)
- \frac{2}{3}g_1^2h_b^2\big(m_3^2+2\mu(A_b-M_1)\big) \nn \\
&+& 2g_1^2h_\tau^2\big(m_3^2+2\mu(A_\tau-M_1)\big)
+ 4g_1^2\l^2\big(m_3^2-\mu M_1\big) \nn \\
&+& 12g_2^2\l^2\big(m_3^2-\mu M_2\big)
+ 16g_3^2h_t^2\big(m_3^2+2\mu(A_t-M_3)\big) \nn \\
&+& 16g_3^2h_b^2\big(m_3^2+2\mu(A_b-M_3)\big)
+ \frac{23}{2}g_1^4\big(m_3^2-4\mu M_1\big) \nn \\
&+& \frac{15}{2}g_2^4\big(m_3^2-4\mu M_2\big)
+ 3g_1^2g_2^2\big(m_3^2-2\mu(M_1+M_2)\big)
\Big)\; , \nn \\
16\pi^2 \frac{dm_S'^2}{dt} &=&
2\l^2\big(m_S'^2 + 2\mu'A_\l\big) + 4\k^2\big(m_S'^2 + \mu'A_\k\big)
+ 4\l\k m_3^2 \nn \\
&+& \frac{1}{16\pi^2} \Big( - 4\l^4\big(m_S'^2+4\mu'A_\l\big)
- 8\k^4\big(2m_S'^2+5\mu'A_\k\big) \nn \\
&-& 8\l^2\k^2\big(2m_S'^2+\mu'(3A_\l+2A_\k)\big)
- 6\l^2h_t^2\big(m_S'^2+2\mu'(A_\l+A_t)\big) \nn \\
&-& 6\l^2h_b^2\big(m_S'^2+2\mu'(A_\l+A_b)\big)
- 2\l^2h_\tau^2\big(m_S'^2+2\mu'(A_\l+A_\tau)\big) \nn \\
&-& 8\l^3\k\big(m_3^2+\mu A_\l\big)
- 12\l\k h_t^2\big(m_3^2+\mu A_t\big)
- 12\l\k h_b^2\big(m_3^2+\mu A_b\big) \nn \\
&-& 4\l\k h_\tau^2\big(m_3^2+\mu A_\tau\big)
+ 4\l\k g_1^2\big(m_3^2-\mu M_1\big)
+ 12\l\k g_2^2\big(m_3^2-\mu M_2\big) \nn \\
&+& 2\l^2g_1^2\big(m_S'^2+2\mu'(A_\l-M_1)\big)
+ 6\l^2g_2^2\big(m_S'^2+2\mu'(A_\l-M_2)\big) \Big)\; .
\eea

\newpage

Finally for the singlet tadpole terms, the two-loop RGEs read

\bea
16\pi^2 \frac{d\xi_F}{dt} &=&
\xi_F\Big( \l^2 + \k^2 \Big)
+ \frac{\xi_F}{16\pi^2} \Big( -2\l^4 - 4\k^4 - 3h_t^2\l^2 \nn \\
&-& 3h_b^2\l^2 - h_\tau^2\l^2 - 4\l^2\k^2 + g_1^2\l^2 + 3g_2^2\l^2
\Big)\; ,\nn \\
16\pi^2 \frac{d\xi_S}{dt} &=&
\l^2\big(\xi_S+2A_\l\xi_F\big) + \k^2\big(\xi_S+2A_\k\xi_F\big) \nn \\
&+& 2\l\big(m_3^2(A_\l+\mu')+\mu(m_{H_u}^2+m_{H_d}^2)\big)
+ \k\big(m_S'^2(A_\k+\mu')+2\mu' m_S^2\big) \nn \\
&+& \frac{1}{16\pi^2} \bigg(
- 2\l^4\big(\xi_S+4A_\l\xi_F\big) - 4\k^4\big(\xi_S+4A_\k\xi_F\big) \nn \\
&-& 3\l^2h_t^2\big(\xi_S+2(A_\l+A_t)\xi_F\big)
- 3\l^2h_b^2\big(\xi_S+2(A_\l+A_b)\xi_F\big) \nn \\
&-& \l^2h_\tau^2\big(\xi_S+2(A_\l+A_\tau)\xi_F\big)
- 4\l^2\k^2\big(\xi_S+2(A_\l+A_\k)\xi_F\big) \nn \\
&-& 6\l h_t^2\Big(m_3^2\big(A_\l+A_t+\mu'\big)
+ \mu\big(M_t^2+A_t(A_\l+\mu')+m_{H_u}^2+m_{H_d}^2\big)\Big) \nn \\
&-& 6\l h_b^2\Big(m_3^2\big(A_\l+A_b+\mu'\big)
+ \mu\big(M_b^2+A_b(A_\l+\mu')+m_{H_u}^2+m_{H_d}^2\big)\Big) \nn \\
&-& 2\l h_\tau^2\Big(m_3^2\big(A_\l+A_\tau+\mu'\big)
+ \mu\big(M_\tau^2+A_\tau(A_\l+\mu')+m_{H_u}^2+m_{H_d}^2\big)\Big) \nn \\
&-& 4\l^3\Big(m_3^2 \big(2 A_\l+ \mu'\big)
+ \mu\big(M_\l^2+A_\l(A_\l+\mu')+m_{H_u}^2+m_{H_d}^2\big)\Big) \nn \\
&-& 4\l^2\k\Big(m_S'^2\big(A_\l+A_\k+\mu'\big)
+ \mu'\big(M_\l^2+A_\l(A_\k+\mu')+2m_S^2\big)\Big) \nn \\
&-& 4\k^3\Big(m_S'^2\big(2 A_\k+\mu'\big)
+ \mu'\big(M_\k^2+A_\k(A_\k+\mu')+2m_S^2\big)\Big)\nn \\
&+& \l g_1^2\Big(3 m_3^2 \big(A_\l+\mu'-M_1\big) 
  +2\mu\big(m_{H_u}^2+m_{H_d}^2-A_\l M_1 -\mu' M_1 +2M_1^2\big)\nn \\
&&  +\l\big(2 \xi_F (A_\l-M_1)+\xi_S\big)\Big)\nn \\
&+& 3 \l g_2^2\Big(3 m_3^2 \big(A_\l+\mu'-M_2\big) 
  +2\mu\big(m_{H_u}^2+m_{H_d}^2-A_\l M_2 -\mu' M_2 +2M_2^2\big)\nn \\
&&  +\l\big(2 \xi_F (A_\l-M_2)+\xi_S\big)\Big)
\bigg)\; .
\eea

\renewcommand{\theequation}{C.\arabic{equation}}
\setcounter{equation}{0}
\setcounter{section}{0}

\section*{Appendix C. Radiative corrections to the Higgs masses}
\addcontentsline{toc}{section}
{Appendix C. Radiative corrections to the Higgs masses}

As explained in Section~\ref{sec:3.2}, the following steps are required
for a systematic calculation of the radiative corrections to the Higgs
masses in the present approach:
i) the
gauge and Yukawa couplings at the scale $M_\mathrm{SUSY}$ must be
determined; ii) the effective potential and the Higgs wave function
normalisation constants must be computed with an ultraviolet cutoff
$M_\mathrm{SUSY}$, and iii) the pole masses have to be evaluated from
the effective action. Here we give the necessary formulae for the
calculation of the radiative corrections to the Higgs masses to the
order described in Section~\ref{sec:3.2}.

\subsection*{C.1 Yukawa and gauge couplings at the scale 
$M_\mathrm{SUSY}$} 
\addcontentsline{toc}{subsection}{C.1 Yukawa and
gauge couplings at the scale  $M_\mathrm{SUSY}$}

First the Yukawa couplings (we confine ourselves to the quarks of the
third family) have to be deduced from the pole masses. The running top
quark Yukawa coupling at the scale $m_{t}$ is given in terms of the top
quark pole mass  as\footnote{Here we use the $\overline{\text{MS}}$
relation between the pole mass and the running mass. In the
$\overline{\text{DR}}$ scheme the second factor $4/3\pi$ would read
$5/3\pi$, which would decrease $h_t(m_{t})$, and hence the lightest
Higgs mass, by $\sim 1\%$. Since we have not included subdominant
(single) logarithms in the two-loop corrections to the effective
potential we are not sensitive to the scheme in which the running top
quark mass is defined, which leads to a theoretical error of $\gsim 1\%$
on the mass of the lightest Higgs.}
\beq\label{c.1e}
h_t(m_{t}) = \frac{m_{t}^{\text{pole}}}{v_u}\left( 1+
\frac{4\alpha_s(m_{t})}{3\pi}
+\frac{11\alpha_s^2(m_{t})}{\pi^2}\right)^{-1}\; .
\eeq

For the computation of $h_t(M_\mathrm{SUSY})$, we consider only
contributions involving potentially large logarithms (the leading
logarithmic approximation or LLA) as $\ln(M_\mathrm{SUSY}^2/m_{t}^2)$
and more, which depend on the masses of the various particles
appearing in the loops. For instance, a complete $SU(2)$~multiplet of
Higgs bosons can have (nearly degenerate) masses given by $M_A$ (the
mass of the MSSM-like CP-odd scalar, cf. Section~\ref{sec:2}), if $M_A^2
\equiv {\cal M}_{P,11}^2 \sim {\cal M}_\pm^2 \gg M_Z^2$. These states
will only contribute to the radiative corrections to $h_t$ from scales
$Q^2 > M_A^2$. In addition, charginos and neutralinos can
have masses below $M_\mathrm{SUSY}$ and induce threshold effects; within
the LLA it is sufficient, however, to approximate their masses by the
diagonal elements of the corresponding mass matrices (\ref{2.35e}) and
(\ref{2.32e}). Then, pure gaugino and higgsino states appear
simultaneously in the Higgs wave function renormalisation diagrams, and
only the mass of the heavier state appears in the potentially large
logarithm.
(On the other hand, effects from squark/gluino loops can be neglected in
the LLA, since the squark masses define the ultraviolet cutoff
$M_\mathrm{SUSY}$.) One obtains (with $\nu' \equiv \k s +\mu'/2$,
and the scales of couplings without arguments are arbitrary in the LLA)
\bea
h_t(M_\mathrm{SUSY}) &=&
h_t(m_{t})\left(1+\frac{7}{4\pi}\alpha_s(m_{t})
\ln(M_\mathrm{SUSY}^2/m_{t}^2)\right)^{-4/7}
\nn \\
&&\times\Big(1+\frac{1}{64\pi^2}\Big[\left(
9h_t^2+h_b^2-\frac{17}{6}g_1^2-\frac{9}{2}g_2^2\right) 
\ln(M_\mathrm{SUSY}^2/m_{t}^2)\nn \\
&& +\left(-9\cos^2\b h_t^2 +(3\cos^2\b -1)h_b^2 +2\cos^2\b
h_\tau^2\right) \ln(M_A^2/m_{t}^2)\nn \\
&& -2\l^2 \ln(\operatorname{max}(\mu_\mathrm{eff}^2,4\nu'^2)
/M_\mathrm{SUSY}^2) 
-g_1^2 \ln(\operatorname{max}(\mu_\mathrm{eff}^2,M_1^2)/
M_\mathrm{SUSY}^2)\nn \\
&& -3g_2^2
\ln(\operatorname{max}(\mu_\mathrm{eff}^2,M_2^2)
/M_\mathrm{SUSY}^2)\Big]\Big)\; .
\label{c.2e}
\eea

In the case of $h_b$, we assume a given value of the running $b$ quark
mass $m_b(M_Z)$ (typically $\sim 2.9$~GeV), from which $h_b(M_Z)$ can be
obtained as in (\ref{a.11e}). Integrating the one-loop (QCD)
RGEs, $h_b(m_{t})$ is given by
\beq\label{c.3e}
h_b(m_{t}) = h_b(M_Z)\left( 1-\frac{23}{12\pi}\alpha_s(m_{t})
\ln{\frac{m_{t}^2}{M_Z^2}}\right) ^{12/23}\;,
\eeq
and subsequently $h_b(M_\mathrm{SUSY})$ by
\bea
h_b(M_\mathrm{SUSY}) &=&
h_b(m_{t})\left(1+\frac{7}{4\pi}\alpha_s(m_{t})
\ln(M_\mathrm{SUSY}^2/m_{t}^2)\right)^{-4/7}
\nn \\
&&\times\Big(1+\frac{1}{64\pi^2}\Big[\left(
9h_b^2+h_t^2 +2h_\tau^2-\frac{5}{6}g_1^2-\frac{9}{2}g_2^2\right) 
\ln(M_\mathrm{SUSY}^2/m_{t}^2)\nn \\
&& +\left(-9\sin^2\b h_b^2 +(3\sin^2\b -1)h_t^2 -2\sin^2\b
h_\tau^2\right) \ln(M_A^2/m_{t}^2)\nn \\
&& -2\l^2 \ln(\operatorname{max}(\mu_\mathrm{eff}^2,4\nu'^2)
/M_\mathrm{SUSY}^2) 
-g_1^2 \ln(\operatorname{max}(\mu_\mathrm{eff}^2,M_1^2)
/M_\mathrm{SUSY}^2)\nn \\
&& -3g_2^2
\ln(\operatorname{max}(\mu_\mathrm{eff}^2,M_2^2)
/M_\mathrm{SUSY})\Big]\Big)\; .
\label{c.4e}
\eea

The electroweak gauge couplings are assumed as given at the scale $M_Z$
e.\,g. in the on-shell scheme in terms of the Fermi coupling $G_F$,
$M_Z$ and $M_W$ as $g_2^2=4\sqrt{2}
G_F M_W^2$, $g_1^2=4\sqrt{2} G_F(M_Z^2-M_W^2)$ (the renormalisation
scheme is irrelevant in the LLA). For
their computation at the scale $M_\mathrm{SUSY}$, we include threshold
effects of the potentially heavy $SU(2)$ Higgs multiplet with a mass
$M_A$, higgsinos with mass $\mu_\mathrm{eff}$, $SU(2)$ gauginos with
mass $M_2$, and sleptons of a common mass $M_{\tilde{l}}$ (in the LLA).
Squarks with masses $\sim M_\mathrm{SUSY}$ do not induce threshold
effects at scales below $M_\mathrm{SUSY}$, and the top quark threshold
is not considered since $m_t \sim M_Z$ inside logarithms in the LLA.
\bea
g_1^2(M_\mathrm{SUSY}) &=& g_1^2(M_Z)\Big(1-\frac{g_1^2}{16\pi^2}
\Big[\frac{41}{6}\ln(M_\mathrm{SUSY}^2/M_Z^2) 
+\frac{1}{6}\ln(M_\mathrm{SUSY}^2/M_A^2)\nn \\
&+&\frac{2}{3}\ln(M_\mathrm{SUSY}^2/\mu_\mathrm{eff}^2)
+\frac{3}{2}\ln(M_\mathrm{SUSY}^2/M_{\tilde{l}}^2)
\Big]\Big)^{-1}\; ,
\label{c.5e}
\eea

\bea
g_2^2(M_\mathrm{SUSY}) &=& g_2^2(M_Z)\Big(1+\frac{g_2^2}{16\pi^2}
\Big[\frac{19}{6}\ln(M_\mathrm{SUSY}^2/M_Z^2) 
-\frac{1}{6}\ln(M_\mathrm{SUSY}^2/M_A^2)\nn \\
&-&\frac{2}{3}\ln(M_\mathrm{SUSY}^2/\mu_\mathrm{eff}^2)
-\ln(M_\mathrm{SUSY}^2/M_\mathrm{slept}^2)
-\ln(M_\mathrm{SUSY}^2/M_2^2)
\Big]\Big)^{-1}\; .
\label{c.6e}
\eea

The strong gauge coupling at $M_\mathrm{SUSY}$ is computed somewhat more
precisely; the one-loop RGE is solved exactly below and above the top
quark threshold:
\beq\label{c.7e}
g_3^2(m_{t}) = \frac{g_3^2(M_Z)}
{1+\frac{23g_3^2(M_Z)}{48\pi^2} \ln(m_{t}^2/M_Z^2)}\; ,
\eeq

\beq\label{c.8e}
g_3^2(M_\mathrm{SUSY}) = g_3^2(m_{t}) \Big(1
+\frac{g_3^2(m_{t})}{16\pi^2}\Big[ 7 \ln(M_\mathrm{SUSY}^2/m_{t}^2) 
-2 \ln(M_\mathrm{SUSY}^2/M_3^2)\Big] \Big)^{-1}\; .
\eeq

\subsection*{C.2 Higgs wave function renormalisation constants}
\addcontentsline{toc}{subsection}
{C.2 Higgs wave function renormalisation constants}

Within the present approximation it suffices to consider wave function
renormalisation constants for the weak eigenstates $H_u$, $H_d$ and $S$.
In the case of contributions from quarks and leptons of the third family
we use an infrared cutoff $m_{t}^2$, which simplifies the remaining
top- and bottom-quark induced corrections to the pole masses in
section~C.4 below. In the case of contributions from gauge
bosons, charginos and neutralinos we confine ourselves to potentially
large logarithms, i.e. we use $M_W \approx M_Z$ and the diagonal
elements of the chargino/neutralino mass matrices inside the logarithms.

If $M_A^2 \gg m_{t}^2$, we include the corresponding threshold
effects: the wave function renormalisation constants of heavy Higgs
states on their mass shell $\sim M_A^2$ do not receive contributions
from momenta $< M_A^2$ in the LLA, and subsequently this effect is
translated into the wave function renormalisation constants for the weak
eigenstates $H_u$, $H_d$ via a rotation by the angle $\b$. Then we
obtain for $Z_{H_u}$, $Z_{H_d}$ and $Z_S$ in the Landau gauge (we
recall $\nu' \equiv \k s +\mu'/2$):
\bea
Z_{H_u} &=& 1+\frac{1}{16\pi^2}\Big(
3h_t^2 \ln(M_\mathrm{SUSY}^2/m_{t}^2)
-\frac{3}{4}(g_1^2+3g_2^2) \ln(M_\mathrm{SUSY}^2/M_Z^2)
\nn \\
&+&\cos^2\b (3h_b^2 +h_\tau^2 -3h_t^2) \ln(M_A^2/m_{t}^2)
+\frac{g_1^2}{2} \ln(M_\mathrm{SUSY}^2/\operatorname{max}(\mu_\mathrm{eff}^2,M_1^2))\nn \\
&+&\frac{3g_2^2}{2}
\ln(M_\mathrm{SUSY}^2/\operatorname{max}(\mu_\mathrm{eff}^2,M_2^2))
+\l^2
\ln(M_\mathrm{SUSY}^2/\operatorname{max}(\mu_\mathrm{eff}^2,4\nu'^2))
\Big)\;,
\label{c.9e}\\
Z_{H_d} &=& 1+\frac{1}{16\pi^2}\Big(
(3h_b^2+h_\tau^2) \ln(M_\mathrm{SUSY}^2/m_{t}^2)
-\frac{3}{4}(g_1^2+3g_2^2)
\ln(M_\mathrm{SUSY}^2/M_Z^2)
\nn \\
&+&\sin^2\b (3h_t^2 -h_\tau^2 -3h_b^2) \ln(M_A^2/m_{t}^2)
+\frac{g_1^2}{2}
\ln(M_\mathrm{SUSY}^2/\operatorname{max}(\mu_\mathrm{eff}^2,M_1^2))\nn
\\
&+&\frac{3g_2^2}{2}
\ln(M_\mathrm{SUSY}^2/\operatorname{max}(\mu_\mathrm{eff}^2,M_2^2))
+\l^2
\ln(M_\mathrm{SUSY}^2/\operatorname{max}(\mu_\mathrm{eff}^2,4\nu'^2))
\Big)\; ,
\label{c.10e}\\
Z_S &=& 1+\frac{1}{8\pi^2}\Big(\l^2
\ln(M_\mathrm{SUSY}^2/\mu_\mathrm{eff}^2) +\k^2
\ln(M_\mathrm{SUSY}^2/4\nu'^2)\Big)\; .
\label{c.11e}
\eea

These wave function renormalisation constants will multiply the
corresponding kinetic terms in the effective action, which results from
the addition of quantum effects with
\break 
$Q^2 < M_\mathrm{SUSY}^2$ to the
``tree level'' Lagrangian, wherein all couplings are defined at
$Q^2 = M_\mathrm{SUSY}^2$. In the next subsection, the contributions of
quantum effects with $Q^2 < M_\mathrm{SUSY}^2$ to the Higgs mass
matrices are considered, in which -- to start with -- the Higgs fields
are \emph{not} (yet) properly normalised. We will denote the Higgs vevs
before the rescaling by square roots of the wave function
renormalisation constants by $v_u(Q)$, $v_d(Q)$ and $s(Q)$, which are
related to the vevs $v_u$, $v_d$ and $s$ of the properly normalised
Higgs fields by
\beq
v_u(Q) = v_u/\sqrt{Z_{H_u}}\; ,\qquad v_d(Q) = v_d/\sqrt{Z_{H_d}}\; ,
\qquad s(Q) = s/\sqrt{Z_{S}}\; .
\label{c.12e}
\eeq
In the LLA, $v_u$ and $v_d$ are related to $M_Z$ and the to quark and
lepton masses as in (\ref{2.13e}) and~(\ref{a.11e}).

\subsection*{C.3 Corrections to the Higgs mass matrices}
\addcontentsline{toc}{subsection}
{C.3 Corrections to the Higgs mass matrices}

According to the philosophy employed here, we denote subsequently by
``tree level'' mass matrices those of Section~\ref{sec:2} where i) all
couplings are defined at the scale $Q^2=M_\mathrm{SUSY}^2$, and ii) all
Higgs vevs are given by $v_u(Q)$, $v_d(Q)$ and $s(Q)$ as
defined in (\ref{c.12e}).

The one-loop corrections to the effective potential are given by the
Coleman-Weinberg formula (with an ultraviolet cutoff
$M_\mathrm{SUSY}^2$, and in the $\overline{\text{DR}}$ scheme in
agreement with the two-loop RGEs for the soft SUSY breaking terms in
Appendix~B):
\beq\label{c.13e}
\Delta V_\mathrm{eff} = \frac{1}{64\pi^2} {\text{STr}}\, M^4 
\left[ \ln\left(M^2/M_\mathrm{SUSY}^2\right) -\frac{3}{2} \right]
\eeq
where the couplings and Higgs vevs in the mass matrices $M$ of the
particles in the loops are still those at the scale $M_\mathrm{SUSY}^2$.
As a result of this procedure, the two-loop contributions
to $\Delta V_\mathrm{eff}$ and hence to the Higgs mass matrices -- at
least in the approximation considered here -- become quite simple.

The radiative corrections to the effective potential affect also the
three minimisation equations with respect to $v_u(Q)$, $v_d(Q)$ and
$s(Q)$, which serve to replace the three para\-meters
$m_{H_u}^2$, $m_{H_d}^2$ and $m_{S}^2$
($\mu_\mathrm{eff}$, $B_\mathrm{eff}$ and $\widehat{m}_3^2$ were defined
in (\ref{2.7e}) and (\ref{2.14e})):
\bea
&v_u(Q)&\Big(m_{H_u}^2+\mu_\mathrm{eff}^2+\l^2\,v_d^2(Q)
+\frac{g_1^2+g_2^2}{4}(v_u^2(Q)-v_d^2(Q))\Big) \nn \\
&&-v_d(Q)\left(\l s(Q) B_\mathrm{eff} +\widehat{m}_3^2\right)
+\frac{1}{2} \Delta V_{\mathrm{eff},v_u(Q)}= 0\; ,\nn \\
&v_d(Q)&\Big(m_{H_d}^2+\mu_\mathrm{eff}^2+\l^2\,v_u^2(Q)
+\frac{g_1^2+g_2^2}{4}(v_d^2(Q)-v_u^2(Q))\Big) \nn \\
&&-v_u(Q)\left(\l s(Q) B_\mathrm{eff} +\widehat{m}_3^2\right)
+\frac{1}{2}\Delta V_{\mathrm{eff},v_d(Q)} = 0\; ,\nn \\
&s(Q)&\Big(m_{S}^2 +m_{S}'^2 +\mu'^2 +2\k\xi_F 
+\k A_\k s(Q) +2\k^2 s^2(Q)  \nn \\
&&+\l^2(v_u^2(Q)+v_d^2(Q))-2\l\k v_u(Q)v_d(Q) \Big)\nn \\
&&+\xi_S+\xi_F \mu' -\l v_u(Q)v_d(Q)(A_\l+\mu') 
+\frac{1}{2}\Delta V_{\mathrm{eff},s(Q)} = 0\; .
\label{c.14e}
\eea
The contributions of $\Delta V_\mathrm{eff}$ to the Higgs mass matrices
are obtained by appropriate second derivatives of $\Delta
V_\mathrm{eff}$ (with respect to $v_u(Q)$, $v_d(Q)$ and $s(Q)$ in the
CP-even case), taking care of the modified replaced mass terms:
\bea
\Delta{\cal M}_{ij}^2 &=& \frac{1}{2} \Delta V_{\mathrm{eff},v_i\,v_j}
\ \mathrm{for} \ i\neq j\; , \nn \\
\Delta{\cal M}_{ii}^2 &=& \frac{1}{2} \Delta V_{\mathrm{eff},v_i\,v_i}
-\frac{1}{2v_i} \Delta V_{\mathrm{eff},v_i}\; , \nn \\
v_i &\equiv& v_u(Q),\; v_d(Q),\; s(Q)\; .
\label{c.15e}
\eea

Below we give the corrections to the Higgs mass matrices
for the CP-even, CP-odd and charged Higgs masses, separately for each
particle species in the loops.

\subsubsection*{CP-even scalars}

\subsubsection*{Top and bottom (s)quarks}

The top and bottom squark mass matrices have been given in (\ref{2.36e})
and (\ref{2.37e}) in Section~\ref{sec:2}. They are diagonalized as in
(\ref{a.10e}) with the help of angles $\t_T$ and $\t_B$ such that the
eigenstates have masses $M_{\tilde{t}_1}$ and $M_{\tilde{t}_2}$ with
$M_{\tilde{t}_1} < M_{\tilde{t}_2}$ (and corresponding eigenstates
$M_{\tilde{b}_1}$ and $M_{\tilde{b}_2}$). It is useful to define the
following quantities in terms of the mass eigenvalues and mixing angles:
\bea
f_t &=& \frac{1}{M_{\tilde{t}_2}^2-M_{\tilde{t}_1}^2}\left(
M_{\tilde{t}_2}^2
\ln{\left(\frac{M_{\tilde{t}_2}^2}{M_\mathrm{SUSY}^2}\right)} - 
M_{\tilde{t}_1}^2\ln{\left(\frac{M_{\tilde{t}_1}^2}{M_\mathrm{SUSY}^2}
\right)} \right)-1\; ,\nn \\
g_t &=& \sin^2 2\t_T
\left(\frac{M_{\tilde{t}_2}^2 + M_{\tilde{t}_1}^2}{M_{\tilde{t}_2}^2 -
M_{\tilde{t}_1}^2}
\ln{\left(\frac{M_{\tilde{t}_2}^2}{M_{\tilde{t}_1}^2}\right)}
-2\right)\; ,\nn \\
e_t &=& -m_{t} \sin 2\t_T
\ln{\left(\frac{M_{\tilde{t}_2}^2}{M_{\tilde{t}_1}^2}\right)}\; ,\nn  \\
f_b &=& \frac{1}{M_{\tilde{b}_2}^2-M_{\tilde{b}_1}^2} \left(
M_{\tilde{b}_2}^2
\ln{\left(\frac{M_{\tilde{b}_2}^2}{M_\mathrm{SUSY}^2}\right)} - 
M_{\tilde{b}_1}^2\ln{\left(\frac{M_{\tilde{b}_1}^2}{M_\mathrm{SUSY}^2}
\right)} \right)-1\; ,\nn \\
g_b &=& \sin^2 2\t_B
\left(\frac{M_{\tilde{b}_2}^2 + M_{\tilde{b}_1}^2}{M_{\tilde{b}_2}^2 -
M_{\tilde{b}_1}^2}
\ln{\left(\frac{M_{\tilde{b}_2}^2}{M_{\tilde{b}_1}^2}\right)}
-2\right)\; ,\nn \\
e_b &=& -m_{b} \sin 2\t_B
\ln{\left(\frac{M_{\tilde{b}_2}^2}{M_{\tilde{b}_1}^2}\right)}\;.
\label{c.16e}
\eea

Some of the radiative corrections due to top/bottom squark loops to
CP-even, CP-odd and charged Higgs masses can be described by
a shift of the trilinear soft SUSY breaking coupling $A_\l$ (at the
scale $M_\mathrm{SUSY}^2$) everywhere in the tree level mass
matrices:
\beq\label{c.17e}
A_\l \to A'_\l = A_\l + \frac{3h_t^2}{16\pi^2}A_t\, f_t
+ \frac{3h_b^2}{16\pi^2}A_b\, f_b\; .
\eeq

The remaining corrections $\sim h_t^2 \equiv h_t^2(M_\mathrm{SUSY}^2)$
and $\sim h_b^2 \equiv h_b^2(M_\mathrm{SUSY}^2)$ to the CP-even mass
matrix ${\cal M}_S^2$ (\ref{2.20e}) in the weak basis $(H_{dR}, H_{uR},
S_R)$ are given by

\bea
{\Delta \cal M}_{S,11}^2 & = & 
\frac{3 h_b^2}{32\pi^2}\left(-A_b^2\, g_b + 4A_b\, e_b
+ 4 m_{b}^2\ln\left({\frac{M_{\tilde{b}_1}^2 M_{\tilde{b}_2}^2}
{m_{b}^4}}\right) \right)\nn\\
&&- \frac{3 h_t^2}{32\pi^2}(\lambda s(Q))^2\ g_t\; , \nn\\
{\Delta \cal M}_{S,22}^2 & = & 
\frac{3 h_t^2}{32\pi^2}\left(-A_t^2\, g_t + 4A_t\, e_t
+ 4 m_{t}^2\ln\left({\frac{M_{\tilde{t}_1}^2
M_{\tilde{t}_2}^2}{m_{t}^4}}\right) \right)\nn\\
&& - \frac{3 h_b^2}{32\pi^2}(\lambda s(Q))^2\ g_b\; , \nn\\
{\Delta \cal M}_{S,33}^2 & = & 
- \frac{3 h_t^2}{32\pi^2}\lambda^2v_d^2(Q)\ g_t
- \frac{3 h_b^2}{32\pi^2}\lambda^2v_u^2(Q)\ g_b
\; , \nn\\
{\Delta \cal M}_{S,12}^2 & = & 
\lambda s(Q) \left(\frac{3 h_t^2}{32\pi^2}(A_t\, g_t -2\, e_t)
+ \frac{3 h_b^2}{32\pi^2}(A_b\, g_b -2\, e_b)\right)
\; , \nn\\ 
{\Delta \cal M}_{S,13}^2 & = & 
\frac{3 h_b^2}{32\pi^2} \lambda v_u(Q) (A_b\, g_b-2\, e_b)
+\frac{3 h_t^2}{32\pi^2}\lambda^2 s(Q)\ v_d(Q)(4\, f_t-g_t) 
\; ,\nn\\
{\Delta \cal M}_{S,23}^2 & = & 
\frac{3 h_t^2}{32\pi^2}\lambda v_d(Q) (A_t\, g_t-2\, e_t)
+\frac{3 h_b^2}{32\pi^2}\lambda^2 s(Q)\ v_u(Q)(4\, f_b-g_b)
\; .\nn\\
\label{c.18e}
\eea
Additional contributions proportional to the electroweak gauge couplings
$g_1^2$ and $g_2^2$ do not generate large logarithms and would be the
same as in the MSSM (see \cite{Pierce:1996zz} and refs. therein).

\subsubsection*{Dominant two loop corrections}

As mentioned above, the two-loop corrections to $\Delta V_\mathrm{eff}$
and hence to the Higgs mass matrices are relatively simple within the
present approach and within the approximation where we neglect all
terms without two powers of large logarithms; the ones given below can
be obtained by integrating the relevant RGEs. The only complication
arises, as above, from diagrams where Higgs bosons
are connected to two top or bottom-quark lines which generate a
(logarithmic) dependence on $M_A$. As infrared cutoff (in the case of
diagrams with $b$ quark lines) we use $m_{t} \approx M_Z$ (in the LLA,
inside the logarithms), since the running Higgs masses at this scale are
closer to the Higgs pole masses. Then one obtains
\bea
{\Delta \cal M}_{S,11}^2 & = & \frac{3 h_b^4 v_d^2(Q)}{128\pi^4} \Big(
\ln^2\left(\frac{M_\mathrm{SUSY}^2}{m_{t}^2}\right)
(16g_3^2-\frac{2}{3}g_1^2+3\sin^2\b\, h_t^2 -3\cos^2\b\, h_b^2)\nn \\
&+&\Big[\ln^2\left(\frac{M_A^2}{m_{t}^2}\right)-
\ln^2\left(\frac{M_\mathrm{SUSY}^2}{m_{t}^2}\right)\Big]
(3\sin^2\b\, h_b^2 +(3\sin^2\b+1)\,h_t^2)\Big)\;,\nn \\
{\Delta \cal M}_{S,22}^2 & = & \frac{3 h_t^4 v_u^2(Q)}{128\pi^4} \Big(
\ln^2\left(\frac{M_\mathrm{SUSY}^2}{m_{t}^2}\right)
(16g_3^2+\frac{4}{3}g_1^2-3\sin^2\b\, h_t^2 +3\cos^2\b\, h_b^2)\nn \\
&+&\Big[\ln^2\left(\frac{M_A^2}{m_{t}^2}\right)-
\ln^2\left(\frac{M_\mathrm{SUSY}^2}{m_{t}^2}\right)\Big]
(3\cos^2\b\, h_t^2 +(3\cos^2\b+1)\,h_b^2)\Big)\;.\
\label{c.19e}
\eea
(Contributions $\sim g_2^2$ cancel.)

\subsubsection*{Dominant slepton contributions}

Contributions proportional to the electroweak gauge couplings $g_1^2$
and $g_2^2$ from squark loops do not give large logarithms, since we
assume the squarks to have masses close to the ultraviolet cutoff
$M_\mathrm{SUSY}$. Sleptons could be considerably lighter, and here we
assume a common slepton mass $M_{\tilde{l}}$. The corresponding
contributions are best expressed in terms of the weak mixing angle
$\t_W$ and the auxiliary quantity
\beq\label{c.20e}
\Delta_{\tilde{l}} = - \frac{g_1^2+g_2^2}{32\pi^2} M_Z^2 (9\sin^4\t_W
+3\cos^4\t_W) \ln\left(\frac{M_\mathrm{SUSY}^2}{M_{\tilde{l}}^2}\right)
\;.
\eeq
Then we have
\bea
{\Delta \cal M}_{S,11}^2 &=& \Delta_{\tilde{l}} \cos^2\b\;, \nn \\
{\Delta \cal M}_{S,22}^2 &=& \Delta_{\tilde{l}} \sin^2\b\;, \nn \\
{\Delta \cal M}_{S,12}^2 &=& -\Delta_{\tilde{l}} \sin\b\cos\b\;.
\label{c.21e}
\eea

\subsubsection*{Chargino/neutralino contributions}

\def\lm2mu{\ln(\operatorname{max}
(M_{1,2}^2,\mu_\mathrm{eff}^2)/M_\mathrm{SUSY}^2)}
\def\lmunu{\ln(\operatorname{max}
(4\nu'^2,\mu_\mathrm{eff}^2)/M_\mathrm{SUSY}^2)}
\def\lnu{\ln(4\nu'^2/M_\mathrm{SUSY}^2)}
\def\lmu{\ln(\mu_\mathrm{eff}^2/M_\mathrm{SUSY}^2)}

Here we assume that the gaugino mass terms $M_1$ and $M_2$ are of
similar order of magnitude, $M_1 \sim M_2 \equiv M_{1,2}$ (inside
logarithms). We define the following potentially large logarithms:
\bea
L_{\mu} &=& \lmu\; , \nn \\
L_{\nu} &=& \lnu\; , \nn \\
L_{M_2\mu} &=& \lm2mu\; , \nn \\
L_{\mu\nu} &=& \lmunu\; .\label{c.22e} 
\eea

Again, some of the radiative corrections due to
chargino/neutralino loops to all CP-even, CP-odd and charged Higgs
masses can be described by an additional shift of the trilinear soft
SUSY breaking coupling $A_\l$ on top of the corrections
in~(\ref{c.17e}):
\beq\label{c.23e}
A'_\l \to A''_\l = A'_\l + \frac{1}{16\pi^2} (g_1^2 M_1 +3g_2^2
M_2) L_{M_2\mu}\; .
\eeq

The remaining contributions are given by
\bea
{\Delta \cal M}_{S,11}^2 &=& \frac{1}{16\pi^2} \Big[M_Z^2
(g_1^2+g_2^2)\cos^2\b (-10+16\sin^2\t_W-8\sin^4\t_W) L_{M_2\mu}\nn \\
&&-4\left(\l^2\mu_\mathrm{eff} \nu'\tan\b
+\frac{2M_Z^2\l^4\cos^2\b}{g_1^2+g_2^2}\right) L_{\mu\nu} \Big]\; ,\nn \\
{\Delta \cal M}_{S,22}^2 &=& \frac{1}{16\pi^2} \Big[M_Z^2
(g_1^2+g_2^2)\sin^2\b (-10+16\sin^2\t_W-8\sin^4\t_W) L_{M_2\mu}\nn \\
&&-4\left(\frac{\l^2\mu_\mathrm{eff} \nu'}{\tan\b}
+\frac{2M_Z^2\l^4\sin^2\b}{g_1^2+g_2^2}\right) L_{\mu\nu} \Big]\; ,\nn \\
{\Delta \cal M}_{S,33}^2 &=& \frac{1}{16\pi^2} \Big[ 16\k^2 
(-2\nu'^2 +\frac{1}{2}\mu'\nu' +\frac{1}{4}\mu'^2 
+\frac{\mu'^3}{8 \k s(Q)}) L_{\nu} \nn \\
&&-8\l^2\mu_\mathrm{eff}^2 L_{\mu} +\frac{\l^3\mu'
M_Z^2}{\mu_\mathrm{eff} (g_1^2+g_2^2)}(8\k-4\l\sin\b \cos\b) L_{\mu\nu}
\Big]\; ,\nn 
\eea\bea
{\Delta \cal M}_{S,12}^2 &=& \frac{1}{16\pi^2} \Big[
4\left(\l^2\mu_\mathrm{eff}\nu' -\frac{2\l^4M_Z^2\sin\b
\cos\b}{g_1^2+g_2^2}\right) L_{\mu\nu} \nn \\
&&-2M_Z^2(g_1^2+g_2^2)\sin\b \cos\b L_{M_2\mu} \Big]\; , \nn \\
{\Delta \cal M}_{S,13}^2 &=& \frac{1}{16\pi^2}
\frac{\sqrt{2}M_Z}{\sqrt{g_1^2+g_2^2}} \Big[ \l (g_1^2+g_2^2) \mu_\mathrm{eff}
\cos\b(-6+4\sin^2\t_W) L_{M_2\mu} \nn \\
&&+4\l^2(\l\sin\b(2\nu'-\frac{\mu'}{2}) -\cos\b(\l\mu_\mathrm{eff}+4\k \nu'))
L_{\mu\nu} \Big] \; , \nn \\
{\Delta \cal M}_{S,23}^2 &=& \frac{1}{16\pi^2}
\frac{\sqrt{2}M_Z}{\sqrt{g_1^2+g_2^2}} \Big[ \l (g_1^2+g_2^2) \mu_\mathrm{eff}
\sin\b(-6+4\sin^2\t_W) L_{M_2\mu} \nn \\
&&+4\l^2(\l\cos\b(2\nu'-\frac{\mu'}{2}) -\sin\b(\l\mu_\mathrm{eff}+4\k \nu'))
L_{\mu\nu} \Big]\; .
\label{c.24e}
\eea

\subsubsection*{Higgs contributions}

The contributions from Higgs loops are quite involved even in the LLA:
the eigenvalues of the tree level mass matrices (\ref{2.22e}),
(\ref{2.26e}) and (\ref{2.28e}) can be anywhere between $M_Z^2$ and
$M_\mathrm{SUSY}^2$, and consequently Higgs loops can generate many
different possibly large logarithms. Here we confine ourselves to the
loop corrections to the diagonal element of the CP-even mass matrix
\emph{after} the rotation by the angle $\b$, which corresponds at tree
level to (\ref{2.23e}), the upper bound on the lightest eigenvalue of
${\cal M}_S^2$. These Higgs induced loop corrections will be denoted as
$\Delta_\text{Higgs}$. Subsequently the rotation by the angle $\b$ can
be reversed, and the contributions can be written in terms of
corrections to the upper left $2\times 2$ CP-even submatrix. (This is
the procedure employed equally in \cite{Ellwanger:2005fh}; the present
expression for $\Delta_\text{Higgs}$ differs however from the one in
\cite{Ellwanger:2005fh}, since the conventions for the tree level mass
matrices are different.) Hence, first we define
\begin{align}
&\Delta_\text{Higgs} = \frac{M_Z^2}{8\pi^2(g_1^2+g_2^2)} \Biggl[ \Big\{
\frac{(g_1^2+g_2^2)^2}{4}
(-4+\sin^22\b+2\sin^2\t_W(1+\sin^22\b)-2\sin^4\t_W)\nn \\
&+\frac{g_1^2+g_2^2}{2}\l^2(2+\sin^22\b(1-2\sin^2\t_W))
-4\l^4-2\l^2\sin^22\b(\l^2+\k^2)\Big\} 
\ln\left(\frac{M_\mathrm{SUSY}^2}{M_Z^2}\right)
\displaybreak[0]
\nn \\ &
+\Big\{\frac{(g_1^2+g_2^2)^2}{2}(\sin^4\t_W-\sin^2\t_W(1+\sin^22\b)
-\frac{11}{4}\sin^42\b+5\sin^22\b+\frac{3}{4})\nn \\
&+\frac{g_1^2+g_2^2}{4}\l^2
(4\sin^2\t_W\sin^22\b+11\sin^42\b-15\sin^22\b-2)\nn \\
&+\l^4(-\frac{11}{4}\sin^42\b+\frac{5}{2}\sin^22\b+1) \Big\}
\ln\left({\cal M}_{P,11}^2/M_Z^2\right)\nn \\
\displaybreak[0]
&+\Big\{\l^2(\l-\k\sin 2\b)^2 
-\frac{\l^4}{{\cal M}_{S,33}^4}
\left(2\mu_\mathrm{eff}-\sin 2\b(A_\l+2\nu')\right)^4\nn  \\
&+\frac{3\l^2}
{{\cal M}_{S,33}^2}
\left(\frac{g_1^2+g_2^2}{2}+(\l^2-\frac{g_1^2+g_2^2}{2})\sin^22\b\right)
\left(2\mu_\mathrm{eff}-\sin 2\b(A_\l+2\nu')\right)^2 \Big\}
\ln\left({\cal M}_{S,33}^2/M_Z^2\right)\nn \\
\displaybreak[0]
&+\Big\{\frac{(g_1^2+g_2^2)^2}{16}(1-\sin^42\b) 
+\frac{(g_1^2+g_2^2)^2}{4}\l^2(\sin^42\b+\sin^22\b-2) \nn \\
&+\l^4(-\frac{1}{4}\sin^42\b-\frac{1}{2}\sin^22\b+1)
+\l^2(\l+\k\sin 2\b)^2 \Big\} 
\ln\left(({\cal M}_{P,11}^2+{\cal M}_{P,22}^2)/M_Z^2\right)\nn \\
\displaybreak[0]
&-\Big\{ \frac{{\cal M}_{P,22}^2}{2({\cal M}_{P,11}^2+{\cal
M}_{P,22}^2)}
\left(\frac{g_1^2+g_2^2}{2}-\l^2\right)^2 \sin^22\b(1-\sin^22\b)\nn \\
&-\l\frac{{\cal M}_{P,11}^2 {\cal M}_{P,22}^2} 
{({\cal M}_{P,11}^2 + {\cal M}_{P,22}^2)^2} (\l+\k\sin 2\b) 
\left(\frac{g_1^2+g_2^2}{2}(1-\sin^22\b)-\l^2(2-\sin^22\b)\right) 
\nn \\ &
+ \frac{1}{({\cal M}_{P,11}^2+{\cal M}_{P,22}^2)^2}
\Big(
\frac{{\cal M}_{P,11}^2+{\cal M}_{P,22}^2}{2}
\left(\frac{g_1^2+g_2^2}{2}(1-\sin^22\b)-\l^2(2-\sin^22\b)\right)\nn \\ &
-\l {\cal M}_{P,11}^2(\l+\k\sin 2\b) -\l^2(A_\l-2\nu')^2 
\Big)^2
\Big\}
\ln\left(\frac{({\cal M}_{P,11}^2+{\cal M}_{P,22}^2)^2}
{{\cal M}_{P,11}^2 {\cal M}_{P,22}^2- {\cal M}_{P,12}^4} \right)
\Biggl]\,. 
\label{c.25e}
\end{align}
Then we have
\bea
{\Delta \cal M}_{S,11}^2 &=& \Delta_\text{Higgs} \cos^2\b\;, \nn \\
{\Delta \cal M}_{S,22}^2 &=& \Delta_\text{Higgs} \sin^2\b\;, \nn \\
{\Delta \cal M}_{S,12}^2 &=& \Delta_\text{Higgs} \sin\b\cos\b\;.
\label{c.26e}
\eea

\subsubsection*{Gauge boson contributions}

These are relatively simple in the Landau gauge; first we define
\beq\label{c.27e}
\Delta_\text{Gauge}=\frac{M_Z^2 (g_1^2+g_2^2)}{32\pi^2}
(-9+12\sin^2\t_W-6\sin^4\t_W) \ln(M_\mathrm{SUSY}^2/M_Z^2)\; ;
\eeq
then we obtain
\bea
{\Delta \cal M}_{S,11}^2 &=& \Delta_\text{Gauge} \cos^2\b\;, \nn \\
{\Delta \cal M}_{S,22}^2 &=& \Delta_\text{Gauge} \sin^2\b\;, \nn \\
{\Delta \cal M}_{S,12}^2 &=& \Delta_\text{Gauge} \sin\b\cos\b\;.
\label{c.28e}
\eea

\subsubsection*{Rescaling}

After adding up all previous corrections to ${\Delta \cal M}_{S}^2$ (and
performing the shifts (\ref{c.17e}) and (\ref{c.23e}) of $A_\l$ in the
tree level expressions for ${\cal M}_{S}^2$) we obtain the Higgs mass
matrix in terms of Higgs fields which are not yet properly normalised,
since the kinetic terms in the effective action are multiplied by the
wave function renormalisation constants. Hence, the Higgs mass matrix
elements have to be rescaled by appropriate powers of wave function
renormalisation constants $Z_i$, $i = H_u,\;H_d,\;S$:
\beq\label{c.29e}
{\cal M'}_{S,ij}^2 = {\cal M}_{S,ij}^2/\sqrt{Z_i\,Z_j}\; .
\eeq
Herewith also dominant double logarithms (two-loop terms
corresponding to an RG improved one-loop calculation) are generated on
top of the ones already present in ${\cal M}_{S}^2$.
 
In the basis $H_i^\text{weak} = (H_{dR}, H_{uR}, S_R)$, ${\cal
M'}_{S,ij}^2$ is diagonalized by an orthogonal $3 \times 3$ matrix
$S_{ij}$ in analogy to the tree level procedure (\ref{a.1e}). The
eigenvalues of ${\cal M'}_{S,ij}^2$ (ordered in mass) will be denoted as
${\cal M'}_{S,i}^2$. This concludes the evaluation of the radiative
corrections to the CP-even Higgs mass matrix; the pole masses will be
computed in the subsection~C.4.

\subsubsection*{CP-odd scalars}

Here we confine ourselves to radiative corrections of the order
$h_{t,b}^4$ and $g^2\, h_{t,b}^2$ to the mass matrix, except for the
chargino/neutralino contributions $\sim g^4$, which are included as
well. These radiative corrections are quite simple: all effects of the
order $h_{t,b}^4$ are included by the shift (\ref{c.17e}) of
$A_\l$, all effects induced by chargino/neutralino loops by the
additional shift (\ref{c.23e}) of $A_\l$, and all contributions of the
order $g^2\, h_{t,b}^2$ -- as before -- by the rescaling by the
wave function renormalisation constants where it is sufficient to
consider the terms $\sim h_{t,b}^2$ in $Z_{H_u,H_d}$ (which we will not
indicate explicitly).

The only complication originates from the fact that the rescaling by the
wave function renormalisation constants is performed for the $3\times 3$
mass matrix before the identification of the Goldstone mode, and has to
be translated into the $2\times 2$ mass matrix after dropping the
Goldstone mode. The result for the \emph{rescaled} CP-odd mass matrix
elements ${\cal M'}_{P,ij}^2$ in the basis specified before
(\ref{2.26e}) is
\bea
{\cal M'}_{P,11}^2 & = & (\mu_\mathrm{eff}\, B_\mathrm{eff} +
\widehat{m}_3^2)\left(\frac{v_u(Q)}{Z_{H_d}v_d(Q)} +
\frac{v_d(Q)}{Z_{H_u}v_u(Q)}\right)\; , \nn\\
{\cal M'}_{P,22}^2 & = & \l (B_\mathrm{eff}+3\k s(Q))
\frac{v_u(Q) v_d(Q)}{s(Q)} -3\k A_\k s(Q)  -2 m_{S}'^2 
-\k \mu' s(Q)\nn\\
&& -\xi_F\left(4\k + \frac{\mu'}{s(Q)}\right)
-\frac{\xi_S}{s(Q)}\; , \nn\\
{\cal M}_{P',12}^2 & = &\l (A''_\l - 2\k s(Q) - \mu')
\sqrt{v_u^2(Q)/Z_{H_d} + v_d^2(Q)/Z_{H_u}}\; ,
\label{c.30e}
\eea
where $\mu_\mathrm{eff}$, $B_\mathrm{eff}$ and $\widehat{m}_3^2$ are
defined as in (\ref{2.7e}, \ref{2.14e}), but in terms of parameters (and
$s$) at the scale $Q=M_\mathrm{SUSY}$ and with $A''_\l$ instead of
$A_\l$.

The conventions for the diagonalization of the radiatively corrected
CP-odd mass matrix are the same as those at tree level corresponding to
(\ref{a.2e} - \ref{a.5e}) above, leading to two eigenvalues ${\cal
M'}_{P,i}^2$ (ordered in mass).

\subsubsection*{Charged scalar}

The precision of the radiative corrections is the same as in the CP-odd
case. However, both the corrections of the order $h_{t,b}^4$ and those
induced by chargino/neutralino, gauge boson and slepton loops give rise to some
additional terms on top of the shifts of $A_\l$. Altogether one obtains
\bea
{\cal M'}_\pm^2 &=& \left(\mu_\mathrm{eff}\, B_\mathrm{eff} +
\widehat{m}_3^2 + v_u(Q) v_d(Q) \left(\frac{g_2^2}{2} - \l^2\right)\right)
\left(\frac{v_u(Q)}{Z_{H_d}v_d(Q)} +
\frac{v_d(Q)}{Z_{H_u}v_u(Q)}\right)\nn \\
&&+\frac{v_u^2(Q)+ v_d^2(Q)}{16\pi^2} \Bigg(
6 h_t^2 h_b^2 \ln(M_\mathrm{SUSY}^2/m_{t}^2)
-\frac{3}{4} g_2^4 \ln (M_\mathrm{SUSY}^2/M_{\tilde{l}}^2)\nn \\
&& \left. + \frac{7g_1^2g_2^2 - g_2^4}{4}
\ln(M_\mathrm{SUSY}^2/M_Z^2) + 2(g_1^2g_2^2 - g_2^4)L_{M_2\mu}
\right)\; ,
\label{c.31e}
\eea
where $L_{M_2\mu}$ was defined in (\ref{c.22e}).

\subsection*{C.4 Pole masses}
\addcontentsline{toc}{subsection}{C.4 Pole masses}

In order to obtain the Higgs pole masses with the precision
defined in Section~\ref{sec:3.2} (exact to the order $h_t^4$, $h_b^4$,
$g^2 h_t^2,$ etc., but only in the LLA in the case of four powers
of electroweak gauge couplings and Yukawa couplings $\l$, $\k$) it
suffices to consider top- and bottom-quark corrections to the pole
masses, which can be deduced from \cite{Pierce:1996zz}. There, the
corrections $\Delta \Pi(p^2,Q^2)$ to the inverse tree level propagators
$\Pi(p^2) = p^2-M^2_\text{tree}$ were computed. In order to make contact
with our approach, we first have to write
\beq
\Delta \Pi(p^2,Q^2) = \Delta \widehat{\Pi}(p^2,Q^2)+\Delta
\Pi(p^2=0,Q^2)\;.
\label{c.32e}
\eeq

The quantity $\Delta \Pi(p^2=0,Q^2)$ corresponds to (the negative of)
our radiative corrections to the masses squared before the rescaling by
the wave function renormalisation constants. The quantity $\Delta
\widehat{\Pi}(p^2,Q^2)$ proportional to $p^2$ still contains potentially
large logarithms. We have already included these large logarithms in our
wave function renormalisation constants in (\ref{c.9e}) --
(\ref{c.11e}), but there remains a missing part $\delta \Pi(p^2)$ which
is defined as
\bea
\Pi(p^2)+\Delta \Pi(p^2,Q^2) &\equiv&  p^2+\Delta
\widehat{\Pi}(p^2,Q^2)- M^2_\text{tree}  +\Delta \Pi(p^2=0,Q^2)\nn \\
&=& Z(Q^2)\, p^2- M^2 +\delta \Pi(p^2)
\label{c.33e}
\eea
where $M^2 \equiv M^2_\text{tree} - \Delta \Pi(p^2=0,Q^2)$ are the radiatively corrected masses squared before the
rescaling by the wave function renormalisation constants. Hence we
obtain
\beq\label{c.34e}
\delta \Pi(p^2) = p^2+\Delta \widehat{\Pi}(p^2,Q^2) - Z(Q^2)\; p^2
\equiv \Delta \widehat{\Pi}(p^2,m_{t}^2) 
\eeq
where we have used the fact that the relevant terms $\sim h_{t,b}^2$ in
our wave function renormalisation constants $Z(Q^2)$ in (\ref{c.9e}) --
(\ref{c.11e}) were computed with an infrared cutoff $m_{t}^2$. Hence,
no large logarithms appear in the remaining pole corrections $\delta
\Pi(p^2)$, which justifies the present method. Finally, one obtains for
the pole mass from the last expression in (\ref{c.33e}) within the
present approximation
\beq\label{c.35e}
M^2_\text{pole} = M^2/Z(Q^2) - \delta \Pi(M^2/Z(Q^2))
\eeq
where the first term corresponds to the results in Section~C.3 (where
$M^2/Z(Q^2)\equiv M'^2$), and
$\delta \Pi$ can be deduced from \cite{Pierce:1996zz} according to
the formulae above. In the case of mass matrices, $M^2_\text{pole}$ is
diagonalized by the same rotation which diagonalizes $M'^2$
within the present approximation.

In the CP-even case, one obtains in terms of  the eigenvalues ${\cal
M'}_{S,i}^2$ and of the elements of the diagonalization matrix~$S_{ij}$
\bea
{\cal M}^{\text{pole}\ 2}_{S,i} &=& {\cal M'}_{S,i}^2 
-\frac{3h_t^2}{16\pi^2}
S_{i2}^2\left({\cal M'}_{S,i}^2-4m_{t}^2\right) B({\cal
M'}_{S,i}^2,m_{t}^2)\nn \\
&-&\frac{3h_b^2}{16\pi^2}
S_{i1}^2\left({\cal M'}_{S,i}^2\ln(m_{t}^2/m_{b}^2) +
\left({\cal M'}_{S,i}^2-4m_{b}^2\right) 
B({\cal M'}_{S,i}^2,m_{b}^2)\right)
\label{c.36e}
\eea
where the logarithm $\ln(m_{t}^2/m_{b}^2)$ in the second line
originates from the infrared cutoff $m_{t}^2$ in $Z_{H_d}$ in
(\ref{c.10e}), and leads to a finite pole mass correction for $m_{b}
\to 0$. $B(M^2,m^2)$ is defined by
\bea
B(M^2,m^2) &=& 2-\sqrt{1-\frac{4m^2}{M^2}}
\ln\left(\frac{1+\sqrt{1-\frac{4m^2}{M^2}}}
{1-\sqrt{1-\frac{4m^2}{M^2}}}\right)\ \mathrm{if}\ M^2 > 4m^2\; ;
\nn \\
B(M^2,m^2) &=&
2-2\sqrt{\frac{4m^2}{M^2}-1}\arctan
\left(\sqrt{\frac{M^2}{4m^2-M^2}}\right)\ \mathrm{if}\ M^2 < 4m^2\; .
\label{c.37e}
\eea

A similar expression holds for the CP-odd Higgs pole masses in terms of
the eigenvalues ${\cal M'}_{P,i}^2$ and the elements of the
diagonalization matrix~$P_{ij}$ defined in (\ref{a.3e}):
\bea
{\cal M}^{\text{pole}\ 2}_{P,i} &=& {\cal M'}_{P,i}^2 
-\frac{3h_t^2}{16\pi^2}
P_{i2}^2{\cal M'}_{P,i}^2 B({\cal M'}_{P,i}^2,m_{t}^2)\nn \\
&-&\frac{3h_b^2}{16\pi^2}
P_{i1}^2\left({\cal M'}_{P,i}^2\ln(m_{t}^2/m_{b}^2) +
{\cal M'}_{P,i}^2 B({\cal M'}_{P,i}^2,m_{b}^2)\right)\; .
\label{c.38e}
\eea

For the charged Higgs pole mass one obtains (neglecting $m_{b}^2$
inside logarithms)
\bea\label{c.39e}
{\cal M}_\pm^{\text{pole}\ 2} &=& {\cal M'}_\pm^2 
+\frac{3}{16\pi^2}\Bigg\{
(h_t^2\cos^2\b+h_b^2\sin^2\b)\Bigg( \\
&&{\cal M'}_\pm^2
\left[\left(1-\frac{m_{t}^2}{{\cal M'}_\pm^2}\right)
\ln\left|\frac{{\cal M'}_\pm^2-m_{t}^2}{m_{t}^2}\right|-2\right]\nn
\\ &&
+(m_{t}^2+m_{b}^2)
\left[\left(1-\frac{m_{t}^2}{{\cal M'}_\pm^2}\right)
\ln\left|\frac{m_{t}^2}{{\cal M'}_\pm^2-m_{t}^2}\right|+1\right]
\Bigg) \nn \\
&& +4h_t m_{t} h_b m_{b} \sin\b \cos\b
\left[\left(1-\frac{m_{t}^2}{{\cal M'}_\pm^2}\right)
\ln\left|\frac{m_{t}^2}{{\cal M'}_\pm^2-m_{t}^2}\right|+1\right]
\Bigg\}\; .
\nn
\eea

\section*{Appendix D. Public computer tools}
\addcontentsline{toc}{section}
{Appendix D. Public computer tools}

\noi {\bf NMSSMTools}
\cite{Ellwanger:2004xm,Ellwanger:2005dv,Ellwanger:2006rn} allows to
compute the Higgs and sparticle spectrum, mixing angles and Higgs
branching fractions. NMSSMTools includes
NMHDECAY (where the parameters can be specified at the weak/SUSY scale),
NMSPEC (where the parameters can be specified at the GUT scale) and a
version where the soft terms satisfy GMSB-like boundary conditions. A
connection to MicrOMEGAS is provided.

\noi Link: {\sf http://www.th.u-psud.fr/NMHDECAY/nmssmtools.html}

\bigskip

\noi{\bf CalcHEP} allows to compute cross sections in the NMSSM (used,
e.\,g., by MicrOMEGAS).

\noi Link: {\sf http://theory.sinp.msu.ru/$\sim$pukhov/calchep.html}

\bigskip

\noi {\bf MicrOMEGAS} \cite{Belanger:2006is,Belanger:2008sj} allows to
compute all relevant cross sections for LSP annihilation and
coannihilation, integrates the density evolution equations, determines
the LSP relic density and detection cross sections. It can be called
from NMSSMTools.

\noi Link: {\sf http://wwwlapp.in2p3.fr/lapth/micromegas}

\bigskip

\noi {\bf Spheno} \cite{Porod:2003um} allows to
compute the sparticle two-body decay amplitudes in the NMSSM;
a computation of the Higgs and sparticle spectrum is in progress.

\noi Link: {\sf http://ific.uv.es/$\sim$porod/SPheno.html}

\bigskip

\noi {\bf WHIZARD} is a multi-purpose Monte Carlo
event generator, into which the NMSSM has been implemented
\cite{Reuter:2009ex}.

\noi Link: {\sf http://projects.hepforge.org/whizard}

\bigskip

\noi {\bf SuperIso}~\cite{Mahmoudi:2009zz} allows to compute flavour
physics observables in the NMSSM such as $B\to X_s \gamma$, the isospin
asymmetry of $B \to K^*\gamma$, and the branching ratios $B_s \to
\mu^+\mu^-$, $B \to \tau\, \nu_\tau$, $B \to D\, \tau\, \nu_\tau$, $K
\to \mu\, \nu_\mu$, $D_s \to \tau\, \nu_\tau$ and $D_s \to \mu\,
\nu_\mu$.

\noi Link: {\sf http://superiso.in2p3.fr}

\end{document}